\documentclass[12pt]{article}
\pdfoutput=1

\usepackage[a4paper,text={16.8cm,22.4cm}]{geometry}
\PassOptionsToPackage{dvipsnames}{xcolor}
\usepackage{hf-tikz,amsmath,amsfonts,braket,slashed,amssymb,bm,psfrag,graphicx,color,colortbl,dsfont,euscript,ctable,bbm,hyperref} 
\usepackage[small,labelfont=bf]{caption}
\usetikzlibrary{arrows.meta}
\RequirePackage[sort&compress,square,comma,numbers]{natbib}
\usepackage{stmaryrd} 
\usepackage{comment}
\usepackage{soul}
\allowdisplaybreaks
\renewcommand{\arraystretch}{1.3}

\numberwithin{equation}{section}

\newcommand{\vsl}{\rlap{\hspace{0.25mm}/}{v}}
\newcommand{\nsl}{\rlap{\hspace{0.25mm}/}{n}}

\newcommand{\psl}{\rlap{\hspace{0.25mm}/}{p}}
\newcommand{\qsl}{\rlap{/}{q}}
\newcommand{\nbsl}{\rlap{\hspace{0.25mm}/}{\nb}}
\newcommand{\Asl}{\rlap{\hspace{0.8mm}/}{\A}}
\newcommand{\Gsl}{\rlap{\hspace{0.3mm}/}{\G}}
\newcommand{\Dsl}{\rlap{\hspace{0.75mm}/}{D}}
\newcommand{\delsl}{\rlap{\hspace{0.2mm}/}{\partial}}

\newcommand{\hc}{{hc}}

\newcommand{\cb}{{\overline{c}}}
\newcommand{\nb}{{\bar n}}
\newcommand{\myR}{R^{(\ell,B)}}

\newcommand{\bmunu}{$B^-\to\mu^-\spac\bar\nu_\mu$}

\newcommand{\blnu}{$B^-\to\ell^-\bar{\nu}_{\ell}$}

\newcommand{\cA}{{\mathcal{A}}}

\newcommand{\lambdaqcd}{\Lambda_\mathrm{QCD}}
\newcommand{\OurLambda}{\Lambda_c}
\newcommand{\abs}[1]{\left| #1 \right|}

\newcommand{\gdirac}[2]{#1 P_L\otimes #2 P_L}
\newcommand{\barn}{{\bar n}}
\newcommand{\del}{\partial}
\newcommand{\order}[1]{\mathcal{O}\!\left(#1\right)}

\def\A{{\EuScript A}}
\def\F{{\EuScript F}}
\def\G{{\EuScript G}}
\def\H{{\EuScript H}}
\def\P{{\EuScript P}}
\def\Q{{\EuScript Q}}
\def\V{{\EuScript V}}
\def\X{{\EuScript X}}

\newcommand{\spac}{{\hspace{0.3mm}}}

\definecolor{LightGray}{rgb}{0.85,0.85,0.85}

\definecolor{AccentColor}{rgb}{0.0,0.47,.90}
\newcommand{\shadecolor}[1]{AccentColor!#1!white}

\begin{document}

\begin{titlepage}

\begin{flushright}
\normalsize
CERN-TH-2026-006\\
MITP-25-075\\
June 17, 2026
\end{flushright}

\vspace{5mm}
\begin{center}
\LARGE\bf
The Simplest {\boldmath $B$} Decay, Precisely
\end{center}

\vspace{0.4cm}
\begin{center}
\textsc{Claudia Cornella$^a$, Max Ferr\'e\spac$^b$, Matthias K\"onig\spac$^b$ and Matthias Neubert\spac$^{b,c}$}

\vspace{6mm}
\textsl{${}^a$CERN, Theory Department, 1211 Geneva 23, Switzerland\\[0.3cm]
${}^b$PRISMA$^{++}$ Cluster of Excellence \& Mainz Institute for Theoretical Physics\\
Johannes Gutenberg University, Staudingerweg 9, D-55128 Mainz, Germany\\[0.3cm]
${}^c$Department of Physics \& LEPP, Cornell University, Ithaca, NY 14853, U.S.A.}
\end{center}

\vspace{0.8cm}
\begin{abstract}
 We derive the QCD$\times$QED factorization theorem governing the leptonic decay $B^-\to\mu^-\bar\nu_\mu(\gamma)$ at all orders in $\alpha_s$ and $\alpha$. Electromagnetic corrections to this decay probe multiple scales, which we disentangle through a sequence of effective field theories (EFTs). The resulting state-of-the-art prediction for the photon-vetoed rate includes the complete structure-dependent component and is accurate at the percent level, establishing the theoretical framework required for future high-precision measurements of this channel, which will allow for a clean determination of $|V_{ub}|$ and powerful tests of new physics. Our work presents the first complete study of QED effects to an exclusive $B$-meson decay at next-to-leading power (NLP) in the heavy-quark expansion. Important milestones are (i) the construction of the complete NLP operator basis in soft-collinear effective theory (SCET); (ii) the proposal of a ``SCET-friendly'' reduction scheme for the Dirac structures of four-fermion operators in dimensional regularization, which avoids power-enhanced evanescent operators; (iii) the consistent refactorization of endpoint-divergent convolution integrals and the first complete resummation of ``rapidity logarithms'' arising at the boundary between the contributions involving soft and hard-collinear quarks; (iv) the systematic discussion of the EFT below the scale of QCD confinement and the non-perturbative matching of SCET onto this low-energy theory; (v) the decoupling of pseudoscalar mesons in the context of heavy-hadron chiral perturbation theory, so that they can be integrated out for processes in which they do not appear as external particles. We perform a phenomenological analysis of direct and indirect contributions to the decay rate and radiation-energy spectrum, highlighting the importance of the chiral anomaly.
\end{abstract}

\end{titlepage}
\spac
\tableofcontents
\hspace{1cm}

\section{Introduction and strategy}
\label{sec:intro}

Weak decays of $B$ mesons are fundamental probes of the flavor structure of the Standard Model (SM) and its extensions. As the Belle\,II and LHCb experiments accumulate data, the precision in the measurements of several such decays is expected to reach the percent level. Achieving a comparable precision in theoretical predictions is essential for meaningful comparisons, in particular to accurately determine Cabibbo--Kobayashi--Maskawa (CKM) matrix elements and extract possible new-physics signals. To reach this level of accuracy, electroweak (EW) and QED effects must be incorporated into theoretical predictions. Electroweak corrections, being short-distance effects, are straightforward to calculate. In contrast, the inclusion of QED corrections introduces additional challenges.  

In Monte-Carlo generators like \texttt{PHOTOS}~\cite{Davidson:2010ew}, the standard way of treating electromagnetic corrections is to dress the charged final states with eikonal emissions. A proper assessment of electromagnetic effects, however, requires including also radiation from the initial state as well as virtual corrections. This is commonly achieved by modeling the decaying meson as a point-like particle, as was done for $B\to\ell\nu$ in~\cite{Dai:2021lei}. Explicit studies of other $B$-meson decays in this approximation include, e.g., the decays $\bar B\to\bar K\spac\ell^+\ell^-$ \cite{Bordone:2016gaq,Isidori:2020acz,Isidori:2022bzw} and $B\to M_1 M_2$ \cite{Baracchini:2005wp}. This treatment allows one to describe and exponentiate all large logarithms of the type $\ln(m_B/E_{\mathrm{cut}})$, where $E_{\mathrm{cut}}$ denotes the cut on the total energy of final-state real radiation. These structure-independent QED corrections are typically dominant, because they appear as double logarithms of the form $\frac{\alpha}{\pi}\ln(m_B/E_{\mathrm{cut}})\ln(m_B/m_f)$, where $m_f$ denotes the mass of a charged final-state particle. 

Once experimental uncertainties reach the percent level, structure-dependent QED effects associated with the finite size of the $B$ meson can no longer be neglected. For the inclusive decay $\bar B\to X_c\spac\ell\spac\nu_\ell$, a complete calculation of $\mathcal{O}(\alpha)$ corrections to the total decay width and the moments of the electron energy spectrum has been performed in \cite{Bigi:2023cbv} in the context of the heavy-quark expansion. Exclusive processes are, however, notoriously more difficult to describe theoretically. For the \blnu\ process, as we will show, the structure-dependent contributions contain single logarithms of the mass ratio $m_B/m_\ell$, as well as single and double logarithms of the ratio $\lambdaqcd/m_B$. These corrections can naturally be at the few percent level, since $\frac{\alpha}{\pi} \ln(m_B/m_\ell)\approx 1.9\%$ (for $\ell=\mu$) and $\frac{\alpha}{\pi} \ln^2(m_B/\lambdaqcd)\sim(1.5\!-\!5.4)\%$ for $\Lambda_{\rm QCD}\sim(0.5\!-\!1.5)$\,GeV. Unlike for lighter mesons, where dedicated lattice studies of structure-dependent QED effects exist \cite{Carrasco:2015xwa,Giusti:2017dwk,Sachrajda:2019uhh,Desiderio:2020oej, Frezzotti:2020bfa,DiCarlo:2021apt,Boyle:2022lsi,Gagliardi:2022szw,Frezzotti:2023ygt,DiPalma:2025iud}, lattice calculations of QED corrections for $B$-meson decays are not yet available. On the continuum side, a systematic way to separate structure-dependent from point-like QED effects is to use effective field theory (EFT) techniques to factorize the contribution of photons associated to different energy scales. The investigation of such effects along these lines and the derivation of the corresponding factorization theorems is a fairly recent endeavor, so far limited to hadronic two-body decays of $B$ mesons \cite{Beneke:2020vnb,Beneke:2021jhp,Beneke:2021pkl,Beneke:2022msp} and the leptonic decays $B_q\to\mu^+\mu^-$ \cite{Beneke:2017vpq,Beneke:2019slt} and $B^-\to\mu^-\spac\bar\nu_\mu$ \cite{Cornella:2022ubo}, with the latter forming the focus of the present work. As an alternative to this  EFT approach, a model based on gauge-invariant interpolating operators for charged hadrons has been developed in \cite{Nabeebaccus:2022jhu}, and applied in a sum-rule calculation of QED corrections to $B^-\to\ell^-\spac\bar\nu_\ell$ decays in \cite{Rowe:2024jml}. This method is less rigorous than our framework and does not capture the fine details of the structure-dependent virtual corrections. While it may still lead to reasonable numerical results, it does not allow for a systematic estimate of theoretical uncertainties. For completeness, we also note earlier studies focusing on structure-dependent effects arising from real photons only \cite{Becirevic:2009aq,Aditya:2012im}. 

The leptonic decay $B^-\to\ell^-\spac\bar\nu_\ell$ (with $\ell=e,\mu,\tau$) is arguably the simplest $B$-meson decay. Phenomenologically, it is relevant for several reasons. First, once measured precisely, it can provide an independent determination of $|V_{ub}|$ with minimal hadronic uncertainties compared to semileptonic channels such as $B\to X_u\spac\ell^-\spac\bar\nu_\ell$ or $B\to\pi\spac\ell^-\spac\bar\nu_\ell$. Second, it is a sensitive probe of new scalar and pseudoscalar interactions, whose contributions would not feature the chiral suppression of the decay amplitude in the SM. Third, the comparison of different lepton channels could provide a valuable probe of lepton-flavor universality. To date, only the branching ratio for the $\tau$ channel has been measured, with a rather large uncertainty, by the Babar and Belle collaborations \cite{BaBar:2009wmt,BaBar:2012nus,Belle:2012egh,Belle:2015odw} and, more recently, by Belle\,II \cite{Belle-II:2025ruy}. For the $e$ and $\mu$ channels only upper bounds are available. The Particle Data Group (PDG) quotes the averaged values \cite{ParticleDataGroup:2024cfk}\footnote{The averages quoted here do not include the Belle\,II result for $\text{Br}(B^-\to\tau^-\spac\bar\nu_\tau)$ \cite{Belle-II:2025ruy}. Including it shifts the central value upward by about $10\%$, with negligible impact on the quoted uncertainty.}
\begin{equation}
\begin{aligned}
   \text{Br}(B^-\to\tau^-\spac\bar\nu_\tau)
   &= (1.09\pm 0.24)\cdot 10^{-4} \,, \\
   \text{Br}(B^-\to\mu^-\spac\bar\nu_\mu) 
   &< 8.6\cdot 10^{-7} \quad (90\% \spac \text{CL}) \,, \\
   \text{Br}(B^-\to e^-\spac\bar\nu_e) 
   &< 9.8\cdot 10^{-7} \quad (90\% \spac \text{CL}) \,.
\end{aligned}
\end{equation}
This picture will change in the next decades, as Belle\,II aims at measuring the $B^-\to\mu^-\spac\bar\nu_\mu$ and $B^-\to\tau^-\spac\bar\nu_\tau$ branching fractions with $\mathcal{O}(5-6\%)$ accuracy, assuming $50\,\mathrm{ab}^{-1}$ of integrated luminosity \cite{Belle-II:2018jsg,Belle-II:2022cgf}. On a longer timescale, a future high-energy lepton collider such as FCC-ee and/or CEPC would further improve these measurements (for dedicated feasibility studies on $B^-_{(c)}\to\tau^-\spac\bar\nu_\tau$ at FCC-ee, see \cite{Amhis:2021cfy,Zuo:2023dzn}). These prospects motivate a systematic assessment of QED corrections to these channels, including structure-dependent effects. 

At first glance, the decay $B^-\to\ell^-\spac\bar\nu_\ell$ appears to be a very simple process. Neglecting QED effects, the non-perturbative contributions to the rate are completely accounted for by the decay constant of the $B$ meson, defined by the (QCD-only) matrix element
\begin{equation}\label{eq:fB}
   \langle 0|\,\bar u\spac\gamma^\mu\gamma_5\spac b\,|B^-(p)\rangle
   = i  f_{B}\spac p^\mu \,,
\end{equation}
which is currently known to sub-percent precision from lattice simulations of QCD. Including QED effects significantly complicates the picture. First, as anticipated above, there is a loss of universality in the non-perturbative quantities entering a given process. The reason is simple: whenever external states carry electric charges, photons -- unlike gluons -- can mediate interactions among them, and in particular spoil the naive factorization of the amplitude in a hadronic and a leptonic contribution. As a result, quantities which are universal in QCD, such as meson decay constants, light-cone distribution amplitudes (LCDAs) and form factors, must be generalized to quantities that depend on the directions and charges of all initial- and final-state particles. As discussed in \cite{Beneke:2019slt,Beneke:2020vnb,Beneke:2021jhp,Beneke:2022msp} for the case of the $B$-meson decay constant and LCDAs, this generalization is non-trivial. For example, it requires a rearrangement of infrared divergences between soft and collinear functions and leads to LCDAs with support for both positive and negative light-cone momenta. A further complication arises from the fact that the chirally-suppressed decay $B^-\to\ell^-\spac\bar\nu_\ell$ is genuinely a ``next-to-leading power (NLP) observable''; its amplitude is suppressed by a factor $m_\ell/m_B$ relative to that for the decay $B^{*-}\to\ell^-\spac\bar\nu_\ell$. This fact notoriously complicates the derivation of a factorization theorem. The analysis of QCD corrections to exclusive non-leptonic and radiative weak decays of $B$ mesons \cite{Beneke:2000ry,Beneke:2001ev,Beneke:2001at,Beneke:2008pi} has shown that, while decay amplitudes can be factorized at leading order in $\Lambda_\mathrm{QCD}/m_b$, some power-suppressed corrections contain divergent convolution integrals of hard-scattering kernels with meson LCDAs, which spoil a naive separation of scales. Endpoint-divergent convolutions were also encountered in the analysis of non-local power corrections to the inclusive radiative decay $B\to X_s\spac\gamma$ \cite{Benzke:2010js,Benzke:2010tq}. These endpoint divergences are now understood to be a generic feature of subleading-power factorization theorems, see e.g.\ \cite{Ebert:2018gsn,Moult:2019mog,Beneke:2019kgv,Moult:2019uhz, Beneke:2019oqx,Moult:2019vou,Liu:2019oav,Beneke:2020ibj,Liu:2020tzd,Liu:2020wbn} for some early references. In order to establish scale factorization, they must be regularized and removed in a systematic way. A consistent framework to do so is provided by the refactorization-based subtraction (RBS) scheme, in which operators with energetic fields are related, in their low-energy limit, to corresponding operators involving soft fields \cite{Liu:2019oav,Liu:2020wbn,Beneke:2020ibj,Beneke:2022obx}. In recent years, this scheme has been applied to derive factorization theorems for numerous NLP observables \cite{Bell:2022ott,Feldmann:2022ixt,Liu:2022ajh,Hurth:2023paz}. In the present study, we apply the RBS method in a non-perturbative context, which requires a redefinition of hadronic parameters related to the $B$-meson decay constant \cite{Cornella:2022ubo}. 

Electromagnetic corrections to $B^-\to\ell^-\spac\bar\nu_\ell$ arise from several energy scales. The largest one is the electroweak scale, $\mu_\mathrm{EW}\sim m_Z$, which sets the characteristic scale of the weak interactions underlying the process in the SM. The internal dynamics of the $B$ meson is governed by the scale $m_b$ set by the mass of the heavy $b$ quark, and the confinement scale $\lambdaqcd\sim 500$\,MeV, which determines the momentum fluctuations of light partons inside hadronic bound states. Below this scale, the appropriate degrees of freedom of QCD are hadrons rather than quarks and gluons. The kinematics of the decay is controlled by the mass of the decaying $B$ meson and the mass of the charged lepton, as well as by experimental cuts. In what follows we assume that events are selected by imposing an upper cut on the total energy of additional real-photon radiation in the $B$-meson rest frame, $E_{\rm rad}^{\rm tot}<E_\mathrm{cut}$, with the cut chosen such that $E_\mathrm{cut}\ll\Lambda_{\rm QCD}$.  

\begin{figure}[t]
\centering
\centering
 \def\y0{12}        
 \def\ya{10}        
 \def\yb{7}         
 \def\yc{2.7}       
 \def\yt{6.5}       
 \def\yh{3.2}       
 \def\yd{1.3}       
 \def\ye{0}         
 \def\yl{0.2}       
 \def\ylo{-1.2}     
 \def\ylabel{-1.6}  
 \def\vs{2}         
\begin{tikzpicture}
 \fill[\shadecolor{10}](\ya,0)rectangle(\y0,1);  
 \fill[\shadecolor{15}](\yb,0)rectangle(\ya,1);  
 \fill[\shadecolor{20}](\yh,0)rectangle(\yb,1);  
 \fill[\shadecolor{25}](\yh,0)rectangle(\ylo,1); 
 \draw[thick,-Latex]  (\ylo,0) -- (\y0,0);
 \node at (\y0,-.4){$\mu$};

 \draw[thick] (\ya,-.1)--(\ya,.1); 
 \draw[thick, dotted] (\ya, 0)--(\ya,1); 
 \node at (\ya, -.4){$m_B$};
 
 \draw[thick] (\yb,-.1)--(\yb,.1); 
 \draw[thick, dotted] (\yb, 0)--(\yb,1);
 \node at (\yb, -.4){$\sqrt{m_B\spac \Lambda_{\rm QCD}}$};

 \draw[thick] (\yh,-.1)--(\yh,.1); 
 \draw[thick, dotted] (\yh, 0)--(\yh,1);
 \node at (\yh+0.3, -.4){$m_\mu\sim\Lambda_\mathrm{QCD}$};

 \draw[thick] (\yd,-.1)--(\yd,.1); 
 \node at (\yd, -.35){$E_{\rm cut}$};
 
 \draw[thick] (\ye,-.1)--(\ye,.1); 
 \node at (\ye, -.4){$\frac{E_{\rm cut} m_\mu}{m_B}$};
 
 \node at (\ya*0.5+\y0*0.5,.5){\footnotesize{LEFT}};
 \node at (\ya*0.5+\yb*0.5,.5){\footnotesize{SCET-1}};
 \node at (\yh*0.5+\yb*0.5,.5){\footnotesize{SCET-2}}; 
 \node at (\yh*0.5+\ylo*0.5,.5){\footnotesize{HH$\chi$PT $\otimes$ bHLET}};

 \node[text width=10cm, align=left, anchor=west] at (\ylabel,1.5){\textbf{Muon channel}};
\end{tikzpicture} 
\begin{tikzpicture}
 \fill[\shadecolor{10}](\ya,0)rectangle(\y0,1);  
 \fill[\shadecolor{15}](\yb,0)rectangle(\ya,1);  
 \fill[\shadecolor{20}](\yh,0)rectangle(\yb,1);  
 \fill[\shadecolor{25}](\yh,0)rectangle(\ylo,1); 
 \draw[thick,-Latex]  (\ylo,0) -- (\y0,0);
 \node at (\y0,-.4){$\mu$};

 \draw[thick] (\ya,-.1)--(\ya,.1); 
 \draw[thick, dotted] (\ya, 0)--(\ya,1); 
 \node at (\ya, -.4){$m_B$};
 
 \draw[thick] (\yb,-.1)--(\yb,.1); 
 \draw[thick, dotted] (\yb, 0)--(\yb,1);
 \node at (\yb+0.2, -.4){$m_\tau\sim\sqrt{m_B\spac \Lambda_{\rm QCD}}$};

 \draw[thick] (\yh,-.1)--(\yh,.1); 
 \draw[thick, dotted] (\yh, 0)--(\yh,1);
 \node at (\yh+0.3, -.4){$\Lambda_\mathrm{QCD}$};
 
 \draw[thick] (\yd,-.1)--(\yd,.1); 
 \node at (\yd, -.35){$E_{\rm cut}$};
 
 \draw[thick] (\ye+0.4,-.1)--(\ye+0.4,.1); 
 \node at (\ye+0.4, -.4){$\frac{E_{\rm cut} m_\tau}{m_B}$};
 
 \node at (\ya*0.5+\y0*0.5,.5){\footnotesize{LEFT}};
 \node at (\ya*0.5+\yb*0.5,.5){\footnotesize{SCET-1}};
 \node at (\yh*0.5+\yb*0.5,.5){\footnotesize{HQET $\otimes$ bHLET}}; 
 \node at (\yh*0.5+\ylo*0.5,.5){\footnotesize{HH$\chi$PT $\otimes$ bHLET}};  
 
 \node[text width=10cm, align=left, anchor=west] at (\ylabel,1.5){\textbf{Tau channel}};
 \fill[\shadecolor{10}](\ya,0 - \vs)rectangle(\y0,1 - \vs);  
 \fill[\shadecolor{20}](\yh,0 - \vs)rectangle(\ya,1 - \vs);  
 \fill[\shadecolor{25}](\yh,0 - \vs)rectangle(\ylo,1 - \vs); 
 \draw[thick,-Latex]  (\ylo,0 - \vs) -- (\y0,0 - \vs);
 \node at (\y0,-.4 - \vs){$\mu$};

 \draw[thick] (\ya,-.1 - \vs)--(\ya,.1 - \vs); 
 \draw[thick, dotted] (\ya, 0 - \vs)--(\ya,1 - \vs); 
 \node at (\ya, -.4 - \vs){$m_B$};
 
 \draw[thick] (\yh,-.1 - \vs)--(\yh,.1 - \vs); 
 \draw[thick, dotted] (\yh, 0 - \vs)--(\yh,1 - \vs);
 \node at (\yh+0.3, -.4 - \vs){$\Lambda_\mathrm{QCD}$};

 \draw[thick] (\yb,-.1 - \vs)--(\yb,.1 - \vs); 
 \node at (\yb -0.0 , -.4 - \vs){$m_\tau$};
 
 \draw[thick] (\yd,-.1 - \vs)--(\yd,.1 - \vs); 
 \node at (\yd, -.35 - \vs){$E_{\rm cut}$};
 
 
 \node at (\ya*0.5+\y0*0.5,.5 - \vs){\footnotesize{LEFT}};
 \node at (\yh*0.5+\ya*0.5,.5 - \vs){\footnotesize{HQET $\otimes$ HLET}}; 
 \node at (\yh*0.5+\ylo*0.5,.5 - \vs){\footnotesize{HH$\chi$PT $\otimes$ HLET}};
\end{tikzpicture} 
\begin{tikzpicture}
 \fill[\shadecolor{10}](\ya,0)rectangle(\y0,1);  
 \fill[\shadecolor{15}](\yb,0)rectangle(\ya,1);  
 \fill[\shadecolor{20}](\yh,0)rectangle(\yb,1);  
 \fill[\shadecolor{25}](\yh,0)rectangle(\ylo,1); 
 \draw[thick,-Latex]  (\ylo+0.5,0) -- (\y0,0);
 \draw[thick,dotted]  (\ylo,0) -- (\ylo+0.5,0);
 \node at (\y0,-.4){$\mu$};

 \draw[thick] (\ya,-.1)--(\ya,.1); 
 \draw[thick, dotted] (\ya, 0)--(\ya,1); 
 \node at (\ya, -.4){$m_B$};
 
 \draw[thick] (\yb,-.1)--(\yb,.1); 
 \draw[thick, dotted] (\yb, 0)--(\yb,1);
 \node at (\yb, -.4){$\sqrt{m_B\spac \Lambda_{\rm QCD}}$};

 \draw[thick] (\yh,-.1)--(\yh,.1); 
 \draw[thick, dotted] (\yh, 0)--(\yh,1);
 \node at (\yh+0.3, -.4){$\Lambda_\mathrm{QCD}$};

 \draw[thick] (\yd,-.1)--(\yd,.1); 
 \node at (\yd, -.35){${E_\mathrm{cut}}$};
 
 \node at (\ya*0.5+\y0*0.5,.5){\footnotesize{LEFT}};
 \node at (\ya*0.5+\yb*0.5,.5){\footnotesize{SCET-1}};
 \node at (\yh*0.5+\yb*0.5,.5){\footnotesize{SCET-2}}; 
 \node at (\yh*0.5+\ylo*0.5,.5){\footnotesize{HH$\chi$PT}};
 
 \node[text width=10cm, align=left, anchor=west] at (\ylabel,1.5){\textbf{Electron channel}};
\end{tikzpicture}
\vspace{-6mm}
\caption{Illustration of the appropriate effective theories below the electroweak scale to treat the $B^- \to \ell^- \bar{\nu}$ process for the cases $\ell=\mu,\tau,e$. In the case of the $\tau$ lepton, two options are shown, in which the lepton is either treated as a hard-collinear or a hard particle. The dots in the electron case indicate the scales $m_e$ and $E_{\rm cut}\spac(m_e/m_B)$, which are both much smaller than $E_{\rm cut}$.}
\label{fig:scales_all_leptons}
\end{figure}
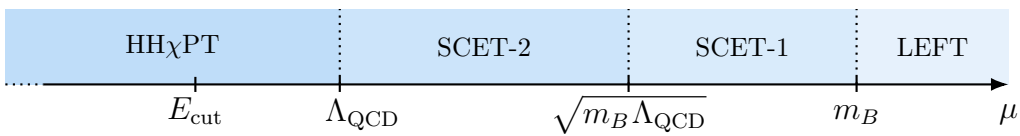

The effects associated with these scales can be disentangled by an appropriate sequence of effective field theories, schematically illustrated in Figure~\ref{fig:scales_all_leptons}, whose construction is based on expansions in several small scale ratios. In the effective theory below the weak scale, the relevant expansion parameters are
\begin{equation}\label{eq:lambdadef} 
   \lambda\sim\frac{\Lambda_\mathrm{QCD}}{m_B} \,, \qquad 
   \lambda_\ell \sim\frac{m_\ell}{m_B} \,, \qquad 
   \zeta\sim\frac{E_\mathrm{cut}}{\Lambda_\mathrm{QCD}} \,, 
\end{equation}
where we use the fact that $m_b$ and $m_B$ are of the same order. The first two ratios provide the basis for systematic expansions at scales above $\lambdaqcd$, while the third controls the structure of the low-energy theory below $\lambdaqcd$. 

Below the electroweak scale, for all lepton flavors the appropriate description of $B^-\to\ell^-\spac\bar\nu_\ell$ is provided by the effective weak Hamiltonian \cite{Buchalla:1995vs}, referred to as Low-Energy Effective Theory (LEFT) in more recent literature \cite{Jenkins:2017jig,Jenkins:2017dyc}. The starting point of our investigation is the LEFT Lagrangian at the scale $\mu_h\sim m_B$, which we refer to as the ``hard" scale. In the SM only a single $(V-A)\times(V-A)$ four-fermion operator contributes, and  we denote its Wilson coefficient by $K_{\mathrm{EW}}$. At scales below $\mu_h$, the $B$ meson and the heavy $b$ quark therein are described in heavy-quark effective theory (HQET) \cite{Eichten:1989zv,Georgi:1990um,Grinstein:1990mj,Neubert:1993mb}. The rest of the construction depends on the mass of the charged lepton. For $\ell=\mu,e$, QED interactions of the light spectator quark with the final-state charged lepton probe an intermediate ``hard-collinear'' scale $\mu_{hc}\sim\sqrt{m_B\spac\Lambda_\mathrm{QCD}}$, which can resolve the inner structure of the meson \cite{Cornella:2022ubo}. The appropriate effective field theory describing these effects is the soft-collinear effective theory (SCET) \cite{Bauer:2000yr,Bauer:2001yt,Bauer:2002nz,Beneke:2002ph}, in the variant called SCET-1. The Wilson coefficients arising in the matching of the LEFT onto SCET-1 are referred to as {\em hard functions\/} and denoted by $H_i$. These functions depend on the hard scales $m_B$ and $m_b$ as well as on the large light-cone energies of hard-collinear particles. In the next step, the hard-collinear modes are integrated out by matching SCET-1 onto a second variant of SCET, commonly called SCET-2. The matching coefficients arising in this step are referred to as {\em jet functions\/} and denoted $J_i$. These functions depend on the large light-cone energies of the hard-collinear fluctuations integrated out, but also on soft light-cone momenta of the light constituents (light quarks and gluons) of the $B$ meson. The hierarchy $m_\ell\ll\mu_\hc$ implies that the charged-lepton mass is a power-suppressed parameter in SCET-1. In SCET-2, the soft hadronic dynamics and the collinear leptonic dynamics are factorized, because interactions between soft and collinear particles are forbidden by momentum conservation. The $B$-meson matrix elements of the (non-local) SCET-2 operators define the {\em soft functions\/} $S_i$. These can be parameterized in terms of two HQET decay constants $F_{\mp}$ as well as the two-particle and three-particle LCDAs of the $B$ meson defined in HQET. The matrix elements of the leptonic current operators in SCET-2, on the other hand, can be calculated in perturbation theory and expressed in terms of functions $K_i$ of the lepton mass $m_\ell$.

The situation is different for the $\tau$ channel, for which the parameter $\lambda_\ell\approx 0.34$ is not particularly small. Here two approaches seem reasonable. Either one treats $m_\tau$ as a hard-collinear scale, $m_\tau\sim\sqrt{m_B\spac\Lambda_\mathrm{QCD}}$, in which case it enters as an additional parameter in the jet functions $J_i$. Below the scale $\mu_\hc$, the tau lepton is then treated in heavy-lepton effective theory (HLET). Or one treats $m_\tau$ as a hard scale, $m_\tau\sim m_B$, in analogy with the standard treatment of charm mesons in $B\to D^{(*)}$ transitions in HQET \cite{Isgur:1990yhj,Falk:1990yz,Neubert:1991td}, in which case $m_\tau$ enters in the hard functions $H_i$ instead. In this case, the effective theory below $\mu_h$ consists of HQET$\,\times\,$HLET, and it does not involve any hard-collinear or collinear modes. 

Below the hadronic scale $\lambdaqcd$, the strong and electromagnetic dynamics of $B$ mesons and their excitations can be described in a systematic way using heavy-hadron chiral perturbation theory (HH$\chi$PT) \cite{Wise:1992hn,Yan:1992gz,Burdman:1992gh}. Hadrons containing a $b$ quark can be classified as multiplets under the heavy-quark spin symmetry, with the ground state consisting of the doublet $(B,B^\ast)$ of the lowest-lying pseudoscalar and vector mesons, whose masses are degenerate in the heavy-quark limit \cite{Shuryak:1981fza,Isgur:1990yhj}. Photons with energies $E_\gamma\ll\lambdaqcd$ cannot resolve the internal structure of the mesons, and their interactions can therefore be described by treating the mesons as point-like objects. The corresponding eikonal emissions are then encoded in Wilson lines associated with the mesons. This forbids transitions of the form $B\to X_b\spac\gamma$, since $m_{X_b}-m_B\gtrsim\mathcal{O}(\Lambda_{\rm QCD})$ for excited states containing a $b$ quark.\footnote{Virtual hard-collinear photons in SCET-1 can excite such transitions. This effect is calculated in our framework using quark and gluon degrees of freedom. Specifically, the hard-collinear $\bar u$-quark propagator in diagrams such as the second graph in Figure~\ref{fig:O1Aren_mixing_diags}, combined with the heavy-quark, is dual to the sum over the entire tower of excited states $X_b$.} 
Importantly, the only exception is the case $X_b=B^*$, which is special since the mass splitting $m_{B^*}-m_B\sim\Lambda_{\rm QCD}^2/m_B$ vanishes in the heavy-quark limit. The transition $B\to B^*\spac\gamma$ is unsuppressed for  $E_\gamma\sim\lambdaqcd^2/m_B$. A second possibility is the transition $B\to B^*\spac\pi^0$ with an on-shell pion, which then decays to two photons. In both cases, the $B^\ast$ meson can subsequently decay leptonically, $B^{*-}\to\ell^-\spac\bar\nu_\ell$, without suffering the chiral suppression of the pseudo-scalar channel. The relevance of these contributions relative to the direct $B^-\to\ell^-\spac\bar\nu_\ell$ decay scales with $(m_B/m_\ell)^2\times(E_{\rm cut}/\Lambda_{\rm QCD})^2$ times a phase-space function, which also depends on $E_{\rm cut}$. As a consequence, they turn out to be relevant for the cases where $\ell=\mu,e$. Real photons emitted in the low-energy effective theory come with two different momentum scalings, which we refer to as ``ultrasoft'' and ``ultrasoft-collinear''. Ultrasoft photons have homogeneously small momentum components of order $E_{\rm cut}$ in the rest frame of the $B$ meson, while ultrasoft-collinear photons have homogeneously small momentum components of order $E_{\rm cut}\spac(m_\ell/m_B)$ in the rest frame of the charged lepton. When boosted to the $B$-meson rest frame, they become collinear, with their largest light-cone momentum component of order $E_{\rm cut}$. The charged leptons in the low-energy effective theory are described by a boosted version of HLET, since they appear heavy on the energy scale $E_{\rm cut}\spac(m_\ell/m_B)$ of the ultrasoft-collinear radiation. In the factorization theorem, the real-photon emissions give rise to the ultrasoft and ultrasoft-collinear functions $W_{us}$ and $W_{usc}$, living at the scales $E_\mathrm{cut}$ and $E_\mathrm{cut}\spac(m_\ell/m_B)$, respectively. The matching of SCET-2 onto HH$\chi$PT is non-perturbative. The soft functions $S_i$ mentioned above appear as the matching coefficients in this transition.

\begin{figure}
\centering
\centering 

 \def\ywidth{14}    
 \def\ylo{0}     
 \def\yhi{\ylo+\ywidth}        
 
 \def\wtick{-0.4}   
 \def\ticks{0.1}  
 \def\boxv{0.35}   
 \def\boxh{0.65}   
 \def\arrowspace{0.02} 
 
 \def\xEW{0.94}  
 \def\xhard{0.79} 
 \def\xhardcollinear{0.6} 
 \def\xzero{0.5}  
 \def\xcollinear{0.28} 
 \def\xsoft{0.375} 
 \def\xultrasoft{0.175} 
 \def\xusc{0.05}  
 
 \def\muzero{\xzero*\ywidth+\ylo}
 \def\vzero{\vEW+0.75+1.5*\boxv}
 
 \def\yEW{\xEW*\ywidth+\ylo}
 \def\vEW{2.25}

 \def\yhard{\xhard*\ywidth+\ylo}
 \def\vhard{1.5}
 
 \def\yjet{\xhardcollinear*\ywidth+\ylo}
 \def\vjet{0.75}

 \def\ysoft{\xsoft*\ywidth+\ylo}
 \def\vsoft{\vjet}

 \def\ycol{\xcollinear*\ywidth+\ylo}
 \def\vcol{\vhard}

 \def\yusoft{\xultrasoft*\ywidth+\ylo}
 \def\vusoft{\vEW}

 \def\yusc{\xusc*\ywidth + \ylo}
 \def\vusc{\vEW+0.75}

\begin{tikzpicture}
  \draw[thick,-Latex]  (\ylo,0) -- (\yhi,0);    
  \node at (\yhi,\wtick){$\mu$};                
  
  \node at     (\muzero, \wtick){$\mu_0$};           
  \draw[thick, dotted] (\muzero, 0)--(\muzero,\vzero);  
  \draw[thick] (\muzero,-\ticks)--(\muzero,\ticks);   

  \node at     (\yEW, \wtick){$m_Z$};           
  \draw[thick] (\yEW,-\ticks)--(\yEW,\ticks);   
  \draw[thick, dotted] (\yEW, 0)--(\yEW,\vEW-\boxv);  
  \draw[thick,-Latex] (\yEW - \boxh, \vEW)--(\muzero + \arrowspace, \vEW);  
  \filldraw[draw=black,fill=\shadecolor{20},thick,rounded corners=1ex] (\yEW - \boxh, \vEW - \boxv) rectangle (\yEW + \boxh,\vEW + \boxv); 
  \node at     (\yEW, \vEW){$K_\mathrm{EW}$};   
  
  \node at     (\yhard, \wtick){$m_B$, $m_b$};           
  \draw[thick] (\yhard,-\ticks)--(\yhard,\ticks);   
  \draw[thick, dotted] (\yhard, 0)--(\yhard,\vhard-\boxv);  
  \draw[thick,-Latex] (\yhard - \boxh, \vhard)--(\muzero + \arrowspace, \vhard);  
  \filldraw[draw=black,fill=\shadecolor{20},thick,rounded corners=1ex] (\yhard - \boxh, \vhard - \boxv) rectangle (\yhard + \boxh,\vhard + \boxv); 
  \node at     (\yhard, \vhard){$H_i$};   

  \node at     (\yjet, \wtick){$\sqrt{m_B\Lambda_\mathrm{QCD}}$};           
  \draw[thick] (\yjet,-\ticks)--(\yjet,\ticks);   
  \draw[thick, dotted] (\yjet, 0)--(\yjet,\vjet-\boxv);  
  \draw[thick,dashed,-Latex] (\yjet - \boxh, \vjet)--(\muzero + \arrowspace, \vjet);  
  \filldraw[draw=black,fill=\shadecolor{20},thick,rounded corners=1ex] (\yjet - \boxh, \vjet - \boxv) rectangle (\yjet + \boxh,\vjet + \boxv); 
  \node at     (\yjet, \vjet){$J_i$};   

  
  \node at     (\ysoft, \wtick){$\Lambda_\mathrm{QCD}$};           
  \draw[thick] (\ysoft,-\ticks)--(\ysoft,\ticks);   
  \draw[thick, dotted] (\ysoft, 0)--(\ysoft,\vsoft-\boxv);  
  \draw[thick,dashed,-Latex] (\ysoft + \boxh, \vsoft)--(\muzero - \arrowspace, \vsoft);  
  \filldraw[draw=black,fill=\shadecolor{20},thick,rounded corners=1ex] (\ysoft - \boxh, \vsoft - \boxv) rectangle (\ysoft + \boxh,\vsoft + \boxv); 
  \node at     (\ysoft, \vsoft){$S_i$};   

  \node at     (\ycol, \wtick){$m_\mu$};           
  \draw[thick] (\ycol,-\ticks)--(\ycol,\ticks);   
  \draw[thick, dotted] (\ycol, 0)--(\ycol,\vcol-\boxv);  
  \draw[thick,-Latex] (\ycol + \boxh, \vcol)--(\muzero - \arrowspace, \vcol);  
  \filldraw[draw=black,fill=\shadecolor{20},thick,rounded corners=1ex] (\ycol - \boxh, \vcol - \boxv) rectangle (\ycol + \boxh,\vcol + \boxv); 
  \node at     (\ycol, \vcol){$K_i$};   

  \node at     (\yusoft, \wtick){$E_\mathrm{cut}$};           
  \draw[thick] (\yusoft,-\ticks)--(\yusoft,\ticks);   
  \draw[thick, dotted] (\yusoft, 0)--(\yusoft,\vusoft-\boxv);  
  \draw[thick,-Latex] (\yusoft + \boxh, \vusoft)--(\muzero - \arrowspace, \vusoft);  
  \filldraw[draw=black,fill=\shadecolor{20},thick,rounded corners=1ex] (\yusoft - \boxh, \vusoft - \boxv) rectangle (\yusoft + \boxh,\vusoft + \boxv); 
  \node at     (\yusoft, \vusoft){$W_{us}$};   

  \node at (\yusc, \wtick){$\frac{E_\mathrm{cut} m_\mu}{m_B}$};           
  \draw[thick] (\yusc,-\ticks)--(\yusc,\ticks);   
  \draw[thick, dotted] (\yusc, 0)--(\yusc,\vusc-\boxv);  
  \draw[thick,-Latex] (\yusc + \boxh, \vusc)--(\muzero - \arrowspace, \vusc);  
  \filldraw[draw=black,fill=\shadecolor{20},thick,rounded corners=1ex] (\yusc - \boxh, \vusc - \boxv) rectangle (\yusc + \boxh,\vusc + \boxv); 
  \node at     (\yusc, \vusc){$W_{usc}$};   

  \end{tikzpicture}
\caption{Ingredients of the \bmunu\ factorization formula and their respective natural scales. Solid arrows denote RG evolution to a common renormalization scale $\mu_0$. Dashed arrows indicate that the corresponding component functions are evaluated at $\mu=\mu_0$ without resummation.}
\label{fig:scales_rge}
\end{figure}
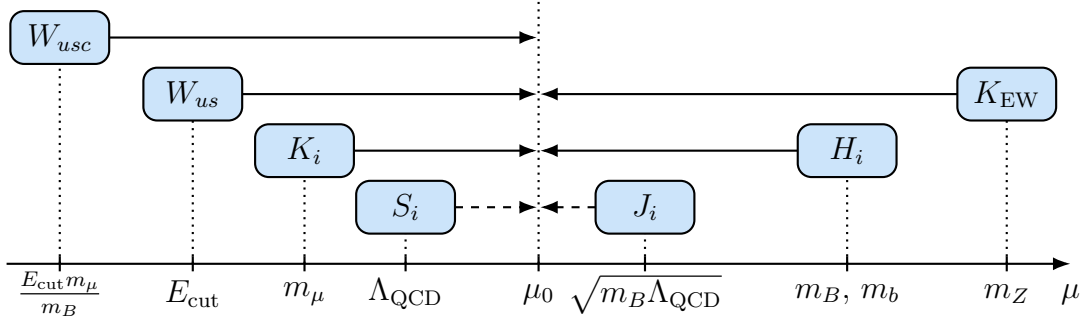

In the present work we focus on the decay \bmunu, postponing a detailed treatment of the $\tau$ channel to future work. Our analysis of the muon channel is general enough to carry over to the electron case. However, due to the very strong chiral suppression in this case, the ``direct'' contribution to the decay rate is completely negligible, and only ``indirect'' decay modes such as $B^-\to B^{*-}\spac\gamma\to e^-\spac\bar\nu_e\spac\gamma$ would be observable. We derive the general multi-scale QCD$\,\times\,$QED factorization theorem for these decays, valid to all orders in QCD and QED interactions. The functions entering this factorization formula, together with their natural scales, are summarized in Figure~\ref{fig:scales_rge}. We provide explicit expressions for the component functions needed for the evaluation of the decay amplitude at one-loop order. We also derive the renormalization-group (RG) evolution equations of the hard functions, which allows us to resum the leading logarithmic corrections of order $(\alpha_s L^2)^n$, $(\alpha_s L)^n$, and $(\alpha L^2)^n$ (with $L$ a logarithm of a large scale ratio) to all orders of perturbation theory. We also include the most relevant mixed corrections of $\mathcal{O}(\alpha\spac\alpha_s)$. All component functions, including the non-perturbative hadronic decay constants and LCDAs,  are evaluated at a common perturbative scale $\mu_0=1.5$\,GeV, chosen sufficiently large to be in the perturbative domain, but sufficiently low to ensure that all large logarithms are resummed. With the hard-collinear scale being of order 1.6\,GeV (for $\lambdaqcd\sim 0.5$\,GeV), the RG evolution of the jet functions from the hard-collinear scale to the scale $\mu_0$ can be safely neglected. For all other functions we perform RG evolution from their natural scale to $\mu_0$.

This paper is organized as follows. We begin in Section~\ref{sec:LEFT} by describing the relevant operators in the LEFT, which contribute to leptonic $B$ decays in the effective theory below the electroweak scale. In Section~\ref{sec:SCET-1} we then construct the effective theory SCET-1. Since we need to work beyond leading order in SCET power counting, the general operator basis required for our analysis is rather large, containing 26 physical, plus six evanescent operators needed at one-loop order. We calculate the bare Wilson coefficients of these operators -- the hard functions -- to the perturbative order needed for our analysis and show how the $1/\epsilon$ pole terms in the bare coefficients are removed by renormalization. The latter step is non-trivial, since different hard functions mix under renormalization, and some of the mixing terms contain endpoint-divergent convolution integrals. In Section~\ref{subsec:3.7}, we solve the RG evolution equations for the renormalized hard functions and present our final results for these objects. In Section~\ref{sec:SCET-2}, we present the analogous construction of the effective theory SCET-2. After explaining the systematics of the SCET-1$\,\to\,$SCET-2 matching procedure, which exploits the symmetries of the effective theories, we construct the SCET-2 operator basis relevant to our problem, which consists of 12 operators. We then present the bare expressions for the Wilson coefficients of these operators -- the jet functions -- to one-loop order. The convolution integrals of the hard and jet functions exhibit endpoint divergences in the regions where some of the hard-collinear particles carry soft momenta. We explain in Section~\ref{sec:endpoint} how these divergences are treated in the RBS scheme, and how this leads to a redefinition of four of the basis operators and requires a novel, two-scale RG evolution of the related hard functions. We then discuss the renormalization of the jet functions and the derivation of their RG evolution equations. Finally, the matrix elements of the SCET-2 operators, which factorize into independent matrix elements of hadronic and leptonic currents, are either expressed in terms of well-defined hadronic quantities in HQET -- form factors and LCDAs -- (hadronic currents), or calculated at one-loop order in perturbation theory (leptonic currents). In Section~\ref{sec:vir_ampl}, we then present our final, RG-improved result for the virtual \blnu\ decay amplitude. In Section~\ref{sec:HHChiPT}, we discuss in detail the low-energy effective theory below the hadronic scale, in which the $B$ meson is described as a point-like object interacting with low-energetic photons and light pseudoscalar mesons. We perform the non-perturbative matching of SCET-2 onto this theory. We also show that the pseudoscalar mesons can be integrated out (for processes in which they do not appear as final-state particles), but this generates non-local effective $B^{(*)}B^{(*)}\gamma\gamma$ vertices from the chiral anomaly, which has an important impact on phenomenology. We present a detailed phenomenological study of our findings in Section~\ref{sec:pheno}. Finally, we summarize our main findings and give our conclusions in Section~\ref{sec:conclusions}. Additional technical details of the calculations are relegated to four appendices. 

Our discussion in Sections~\ref{sec:SCET-1} and \ref{sec:SCET-2} is necessarily   technical, since we are dealing with a NLP SCET factorization theorem affected by endpoint divergences. The reader not interested in the details can skip these sections and consult directly the final result for the virtual \blnu\ decay amplitude given in \eqref{eq:4.98}, \eqref{eq:total_T} and \eqref{eq:Rvirt_final}, and then proceed with Section~\ref{sec:HHChiPT}.

\newpage
\section{Effective weak interactions below the weak scale}
\label{sec:LEFT}

Since $B$-meson decays occur at energies well below the electroweak scale, they can be described in terms of operators involving only light SM fields. Using the LEFT operator basis \cite{Jenkins:2017jig}, the most general effective Lagrangian describing \blnu\ decays can be written as 
\begin{equation}\label{eq:Lleft}
\begin{aligned}
   \mathcal{L}_{\mathrm{LEFT}} 
   &\ni L^{V,LL}_\ell \left( \bar\ell\spac\gamma^\mu P_L\spac\nu_\ell \right)\!
    \big( \bar u\spac\gamma_\mu P_L b \big) 
    + L^{V,LR}_\ell \left( \bar\ell\spac\gamma^\mu P_L\spac\nu_\ell \right)\!
    \big( \bar u\spac\gamma_\mu P_R\spac b \big) \\
   &\quad + L^{S,RL}_\ell \left( \bar\ell\spac P_L\spac\nu_\ell \right)\!
    \big( \bar u\spac P_R\spac b \big)
    + L^{S,RR}_\ell \left( \bar\ell\spac P_L\spac\nu_\ell \right)\!
    \big( \bar u P_L b \big) 
    + L_\ell^{T,RR} \left( \bar\ell\spac\sigma^{\mu\nu} P_L\spac\nu_\ell \right)\!
    \big( \bar u\spac\sigma_{\mu\nu} P_L\spac b \big) \\
   &\quad + \big[ L_{\nu e}^{V,LL} \big]_{2112} \spac  
    \big( \bar\nu_{\mu}\spac\gamma_\mu P_L\spac\nu_{e} \big)\!
    \left( \bar e\spac\gamma^\mu P_L\spac\mu \right) ,
\end{aligned}
\end{equation}
where we have defined $L_\ell^X\equiv\left[L_{\spac\nu e d u}^X\right]_{\spac\ell\ell 31}^*$ for compactness, and for convenience we included also the operator mediating the decay of the muon, from which the Fermi constant is extracted. In scenarios with new exotic states heavier than the EW scale, all coefficients in \eqref{eq:Lleft} can be non-zero. At tree level, the \blnu\ decay rate obtained from this Lagrangian reads 
\begin{equation}
   \Gamma_{\mathrm{LEFT}}
   = \frac{m_\ell^2\spac m_B\spac f_B^2}{64\pi} 
    \left( 1 - \frac{m_\ell^2}{m_B^2} \right)^2 
    \left|\spac L_\ell^{V, LL} - L_\ell^{V, LR} + \frac{m_B^2}{m_\ell\spac(m_b+m_u)} 
    \left( L_\ell^{S, RL} - L_\ell^{S, RR} \right) \right|^2 ,
\end{equation}
where the Wilson coefficients and running quark masses are evaluated at a scale $\mu_b\sim m_B$. Note that the scalar operators give chirally unsuppressed contributions to the rate. 

In this work, we are primarily interested in QED corrections to the prediction in the SM, where only the two vector-current operators in \eqref{eq:Lleft} have non-vanishing Wilson coefficients. These read \cite{Dekens:2019ept}
\begin{equation}\label{eq:electroweak}
\begin{aligned}
   \big[ L_{\nu e}^{V,LL} \big]_{1212} 
   &= - \frac{4\spac G_F}{\sqrt2}\,K_\ell
    \equiv - \frac{4\spac G_F^{(\mu)}}{\sqrt2} \,, \\
   L_\ell^{V,LL}  
   &= - \frac{4\spac G_F}{\sqrt2}\,V_{ub}\,K_q 
    \equiv -\frac{4\spac G_F^{(\mu)}}{\sqrt2}\,V_{ub}\,K_\mathrm{EW}(\mu) \,,
\end{aligned}
\end{equation}
where $G_F$ denotes the Fermi constant obtained from integrating out the $W$ boson at tree level, and the parameters $K_\ell$ and $K_q$ contain the one-loop matching corrections. In practice, one extracts the Fermi constant from the lifetime of the muon and identifies it with the Wilson coefficient of the purely leptonic operator in the last line of \eqref{eq:Lleft}, yielding $G_F^{(\mu)}=G_F\spac K_\ell$. This implies the relation
\begin{equation}\label{eq:KEW}
   K_\mathrm{EW}(\mu) 
   = \frac{K_q(\mu)}{K_\ell} 
   = 1 + \frac{Q_\ell\spac\alpha}{4\pi}
    \left[ 3\spac Q_u \left( \ln\frac{\mu^2}{m_Z^2} + \frac{11}{6} \right) 
    + (Q_b+Q_u) \left( 1 + \frac{\kappa}{4} \right) \right] 
    + \mathcal{O}(\alpha^2) \,,
\end{equation}
which reproduces the well-known result of \cite{Marciano:1993sh} up to scheme-dependent constant terms. 

Throughout this paper we work in the naive dimensional regularization scheme with anti-commuting $\gamma_5$, in which the vector and axial-vector currents obey analogous matching relations and hence the chirality of the states is preserved. The parameter $\kappa$ is introduced to encode the scheme choice when extending four-dimensional Dirac reductions to $d=4-2\epsilon$ dimensions, using the identity
\begin{equation}\label{eq:reduction_kappa}
   \gdirac{\gamma_\mu \gamma_\nu \gamma_\rho}{\gamma^\mu \gamma^\nu \gamma^\rho} 
   = (16+\kappa\spac\epsilon)\,\gdirac{\gamma_\mu}{\gamma^\mu} \,.
\end{equation} 
As we will show in Section~\ref{subsec:evanescent}, this parameter cancels out in the matching of the LEFT onto SCET-1. 

The scale dependence of $K_\mathrm{EW}(\mu)$ is governed by the RG equation 
\begin{equation}\label{eq:gammaEW}
   \frac{dK_\mathrm{EW}(\mu)}{d\ln\mu} 
   = \gamma_\mathrm{EW}\,K_\mathrm{EW}(\mu)
    \quad \text{with} \quad
   \gamma_\mathrm{EW}
   = 3\spac Q_\ell\spac Q_u\,\frac{\alpha}{2\pi} + \mathcal{O}(\alpha^2) \,,
\end{equation}
whose solution can be written as
\begin{equation}
   K_\mathrm{EW}(\mu) = U_\mathrm{EW}(\mu,m_Z)\spac K_\mathrm{EW}(m_Z) \,.
\end{equation}
For $m_b<\mu<m_Z$, the evolution function takes the simple form 
\begin{equation}
   U_\mathrm{EW}(\mu,m_Z) 
   = \left( \frac{\alpha(m_Z)}{\alpha(\mu)} \right)^{\gamma_0^\mathrm{EW}/2\beta^\mathrm{QED}_0} 
   = \left( \frac{\alpha(m_Z)}{\alpha(\mu)} \right)^{\frac{9}{40}} \,,
\end{equation}
where $\gamma_0^\mathrm{EW}=6\spac Q_\ell\spac Q_u$ and $\beta^\mathrm{QED}_0=-\frac43\sum_f N_c^f\spac Q_f^2$ are the one-loop coefficients of the anomalous dimension of $K_\mathrm{EW}$ and the QED $\beta$-function, respectively, where the sum includes all electrically charged fermions except the top quark. Here and in the following, we use the conventions 
\begin{equation}\label{eq:betagammaexp}
   \gamma(\alpha) 
   = \sum_{n=0}^\infty\,\gamma_n \left( \frac{\alpha}{4\pi} \right)^{n+1} , \qquad 
   \beta(\alpha) = - 2\spac\alpha \sum_{n=0}^\infty\spac \beta_n 
    \left( \frac{\alpha}{4\pi} \right)^{n+1} . 
\end{equation}
Given the values of the QED running coupling at the scales of relevance to our analysis, namely $\alpha^{-1}(m_Z)\simeq 127.9$, $\alpha^{-1}(m_B)\simeq 132.0$ and $\alpha^{-1}(\mu_0)\simeq 133.6$, it is sufficient for our purposes to include the running of $\alpha$ between the weak scale $m_Z$ and the hard scale $m_B$, while treating the coupling as fixed in the low-energy theory. For $\mu<m_B$ we therefore approximate the evolution factor as
\begin{equation}\label{eq:electroweak_ev}
   U_\mathrm{EW}(\mu,m_Z) 
   = \left( \frac{\alpha(m_Z)}{\alpha(m_B)} \right)^{\frac{9}{40}}
    \left[ 1 + 3\spac Q_\ell\spac Q_u\,\frac{\alpha(\mu_0)}{4\pi}\spac \ln\frac{\mu^2}{m_B^2} \right] .
\end{equation}

\newpage
\section{Matching to Soft-Collinear Effective Theory}
\label{sec:SCET-1}

At the scale $\mu_h \sim m_B$ we match the LEFT onto SCET, which is the appropriate effective theory describing $B$-meson decays into light particles \cite{Bauer:2000yr,Bauer:2001yt,Bauer:2002nz,Beneke:2002ph}. These light particles can either be soft, with energies of order the QCD scale, or carry large energies scaling with the $B$-meson mass. In the following discussion we focus on the SM, in which the decay \blnu\ is mediated solely by the operator proportional to $L_\ell^{V,LL}$ in \eqref{eq:Lleft}, 
\begin{equation}\label{eq:OVLLdef}
   \mathcal{O}^{V, LL}_{\ell} 
   = \big( \bar u\spac\gamma_\mu P_L\spac b \big)\,
    \big( \bar\ell\spac\gamma^\mu P_L\spac\nu_\ell \big) \,,
\end{equation}
and both the $u$-quark and the neutrino are described by left-handed fields. As mentioned in the introduction, the matching onto SCET is performed in two steps \cite{Bauer:2002aj,Beneke:2003pa,Becher:2005fg}: 
\begin{enumerate}
\item
At the hard scale $\mu_h\sim m_B$, modes with momenta of order $m_B$ are integrated out and the theory is matched onto SCET-1, an effective theory containing soft and hard-collinear particles. The latter have virtualities of order $p_\hc^2\sim m_B\spac\Lambda_\mathrm{QCD}$, which is parametrically higher than the QCD scale and the charged-lepton mass (for $\ell=\mu,e$). The Wilson coefficients arising in this first matching step are the hard functions $H_i$, and they can depend on all scales of order $m_B$. 
\item
After the effective theory SCET-1 has been evolved down to the hard-collinear scale $\mu_\hc\sim\sqrt{m_B\spac\Lambda_\mathrm{QCD}}$ (sometimes referred to as the ``jet'' scale), the hard-collinear modes are integrated out and one matches to the final effective theory SCET-2 containing soft and collinear particles. The Wilson coefficients arising in this second matching step are the jet functions $J_i$. The low-energy matrix elements remaining in SCET-2 capture all long-distance hadronic dynamics in the decay process as well as the dependence on the charged-lepton mass.
\end{enumerate}
In our discussion, we will closely follow the detailed treatments of the two-step matching procedure presented in \cite{Beneke:2003pa,Becher:2005fg}. The power counting in the two versions of SCET is controlled by the small parameters $\lambda$ and $\lambda_\ell$ defined in \eqref{eq:lambdadef}, where $\ell=\mu$ or $e$, so that $\lambda_\ell$ is at most of order $\lambda$. Without loss of generality we work in the $B$-meson rest frame, where the 4-velocity $v^\mu$ of the $B$ meson (with $v^2=1$) is chosen as $v^\mu=(1,\bm{0})$. In SCET it is convenient to define two light-like reference vectors $n^\mu$ and $\nb^\mu$, with $n^2=\nb^2=0$ and $n\cdot\nb=2$, which, up to power corrections, are aligned with the momenta of the charged lepton ($p_\ell$) and anti-neutrino ($p_{\bar\nu_\ell}$), respectively. Any 4-vector can be decomposed in the light-cone basis spanned by $n$ and $\nb$, such that
\begin{equation}\label{eq:light-cone}
   p^\mu = (n\cdot p)\,\frac{\nb^\mu}{2} + (\nb\cdot p)\,\frac{n^\mu}{2} + p_\perp^\mu 
   \equiv p_+^\mu + p_-^\mu + p_\perp^\mu \,.
\end{equation}
In our reference frame, it follows that $v_\perp=p_{\ell\perp}=p_{\bar\nu_\ell\perp}=0$ and $(n\cdot v)(\nb\cdot v)=1$. One often chooses the reference vectors such that $n\cdot v=\nb\cdot v=1$, but we will refrain from making this choice. All operators and Wilson coefficients in SCET must be invariant under the reparameterization transformation \cite{Chay:2002vy,Manohar:2002fd}
\begin{equation}\label{eq:RPI}
   n^\mu\to\zeta\spac n^\mu \,, \qquad \nb^\mu\to\zeta^{-1}\spac\nb^\mu \,,
\end{equation}
with $\zeta=\mathcal{O}(1)$, which is a consequence of the fact that the effective theory must be independent of the choice of the reference vectors $n$ and $\nb$, as long as $n^2=\nb^2=0$ and $n\cdot\nb=2$. This invariance enforces important constraints on the form of the operators.

Below the weak scale the neutrino is effectively a sterile particle, which does not interact and does not propagate in any loop diagram. Therefore, it is possible to choose an operator basis where the neutrino momentum does not appear as a hard scale. The relevant hard scales are then given by
\begin{equation}\label{eq:hardscales}
   m_b \,, \qquad 
   \frac{\nb\cdot p_\ell}{\nb\cdot v} = m_B \,, \qquad 
   2v\cdot p_\ell = \frac{m_B^2+m_\ell^2}{m_B} 
   = \frac{\nb\cdot p_\ell}{\nb\cdot v} + \mathcal{O}(\lambda_\ell^2) \,,
\end{equation}
which are all of similar magnitude. 

For a detailed discussion of the effective Lagrangian of SCET-1, including the leading power corrections, we refer the reader to the founding papers \cite{Bauer:2000yr,Bauer:2001yt,Bauer:2002nz,Beneke:2002ph} (see \cite{Becher:2014oda} for a review). A brief summary can be found in Appendix~\ref{app:SCETbasics}.

\subsection{Ingredients of SCET-1}
\label{subsec:ingredients_SCET1}

The effective field theory SCET-1 describes physics at or below the hard scale $\mu_h$. It contains interacting hard-collinear ($hc$) and soft ($s$) particles, which differ in the scalings of their momenta. Using the decomposition \eqref{eq:light-cone}, and collecting the components in a tuple $(p_+^\mu, p_-^\mu, p_\perp^\mu)$ with well-defined scaling properties, hard-collinear and soft momenta scale as
\begin{equation} 
   p_\hc^\mu \sim m_B\,(\lambda,1,\lambda^{\frac12}) \,, \qquad
   p_s^\mu \sim m_B\,(\lambda,\lambda,\lambda) \,.
\end{equation}
From here on we will drop the hard scale $m_B$ in such relations. 

We will express the relevant SCET-1 operators in terms of so-called hard-collinear building blocks, which are invariant under hard-collinear gauge transformations and transform covariantly under soft gauge transformations \cite{Bauer:2002nz,Hill:2002vw}. The corresponding effective fermion and gauge-boson fields have scalings 
\begin{equation}\label{eq:fieldscale}
\begin{aligned}
  \X_\hc &\sim \lambda^{\frac12} \,, & \A_\hc^\mu &\sim (\lambda,1,\lambda^{\frac12}) \,, \\
  u_s, b_v &\sim \lambda^{\frac32} \,, \quad & A_s^\mu &\sim(\lambda,\lambda,\lambda) \,.
\end{aligned}
\end{equation}
Here $u_s$ is the ordinary QCD spinor field for a massless up-quark, and $b_v$ is the effective $b$-quark field in HQET \cite{Georgi:1990um,Neubert:1993mb}, defined as
\begin{equation}
   b_v(x) = e^{im_b\spac v\cdot x}\,\frac{1+\vsl}{2}\,b(x) \,, 
\end{equation}
where $b(x)$ denotes the $b$-quark field in full QCD. In our kinematics, the Fourier components of the soft fields $b_v$ and $u_s$ are restricted to carry soft momenta. The hard-collinear building block for the charged lepton is defined as
\begin{equation}
   \X_\hc^{(\ell)}(x) 
   = \frac{\nsl\nbsl}{4}\,W_{hc}^{(\ell)\dagger}(x)\spac\psi_\hc^{(\ell)}(x) \,,
\end{equation}
where the hard-collinear Wilson line 
\begin{equation}
   W_{hc}^{(\ell)}(x)
   = \exp\left[ i\spac Q_\ell\spac e\!\int_{-\infty}^0\!ds\,\nb\cdot A_\hc(x+s\nb) \right]
\end{equation}
contains the photon field with charge $Q_\ell=-1$, and $\psi_\hc^{(\ell)}(x)$ denotes the field for the charged lepton with its Fourier components restricted to the region of hard-collinear momenta. In the construction of the SCET-1 operator basis we also need a hard-collinear field for the light spectator quark in the $B$ meson (an up-quark in the present case), which is defined in an analogous way. In this case, the hard-collinear Wilson line contains both gluon and photon fields, i.e.\
\begin{equation}
   W_{hc}^{(u)}(x)
   = \exp\left[ i\spac Q_u\spac e\!\int_{-\infty}^0\!ds\,\nb\cdot A_\hc(x+s\nb) \right]
    \bm{P}\exp\left[ i\spac g_s\spac t^a\!\int_{-\infty}^0\!ds\,\nb\cdot G_\hc^a(x+s\nb) \right] .
\end{equation}
The hard-collinear gauge fields can also appear in isolated form. The corresponding building blocks for the hard-collinear gluon and photon fields are defined as \cite{Bauer:2002nz,Hill:2002vw}
\begin{equation}
\begin{aligned}
   \G_\hc^\mu(x) 
   &= W_\hc^\dagger(x)\spac\big[ iD^\mu\spac W_\hc(x) \big] 
    = g_s \int_{-\infty}^0\!ds\,\nb_\alpha\spac
     \big[ W_\hc^\dagger\spac G_\hc^{\alpha\mu}\spac W_\hc \big](x+s\spac\nb) 
    \equiv t^a\,\G_\hc^{\mu,a}(x) \,, \\
   \A_\hc^{(\ell)\mu}(x) 
   &= W_\hc^{(\ell)\dagger}(x)\spac\big[ iD^\mu\spac W_\hc^{(\ell)}(x) \big] 
    = Q_\ell\,e \int_{-\infty}^0\!ds\,\nb_\alpha\spac F_\hc^{\alpha\mu}(x+s\spac\nb) 
    \equiv Q_\ell\,\A_\hc^\mu(x) \,,
\end{aligned}
\end{equation}
where $W_\hc$ without a superscript denotes the hard-collinear Wilson line including the gluon field only. It follows from these definitions that the hard-collinear fields are subject to the constraints
\begin{equation}
   \nsl\,\X_\hc^{(\ell)} = \nsl\,\X_\hc^{(u)} = 0 \,, \qquad 
   \nb\cdot\G_\hc = \nb\cdot\A_\hc = 0 \,, 
\end{equation}
while the effective heavy-quark field satisfies $\vsl\spac b_v=b_v$. The scalings of the fields indicated in \eqref{eq:fieldscale} can be derived from the scaling of the corresponding propagators. The spinor field $\nu_\cb$ for the neutrino is non-interacting and satisfies $\nbsl\spac\nu_\cb=0$ with our choice of reference frame. It is not necessary to assign a power counting to this field.

The effective Lagrangian of SCET-1 has the generic form
\begin{equation}
   {\cal L}_\mathrm{eff}^{\text{SCET-1}} 
   = {\cal L}_\hc + {\cal L}_s + {\cal L}_{\rm HQET} + {\cal L}_{hc+s} \,.
\end{equation}
The first term on the right-hand side is known in exact form, while the soft Lagrangian is given by the ordinary QCD Lagrangian for the soft quark and the HQET Lagrangian for the heavy quark, where the latter contains an infinite series of operators in a $1/m_b$ expansion. The interaction terms connecting the hard-collinear sector to the soft sector can be systematically expanded in powers of $\lambda$. 

When a LEFT operator is matched onto SCET-1, the hard-collinear fields in the effective theory can be smeared out along the light-like direction $\nb$, because the corresponding momentum components $\nb\cdot p_\hc$ are $\mathcal{O}(1)$ in power counting. A generic term in the SCET-1 effective Lagrangian for our problem is thus of the form
\begin{equation}
\label{eq:3.14}
   \mathcal{L}_\mathrm{eff}^{\text{SCET-1}} 
   \ni \int\!dt_1\ldots\int\!dt_n\,\widetilde{H}(m_b,t_1,\ldots,t_n,\mu)\,
    \phi_\hc^{(1)}(x+t_1\spac\nb)\ldots\phi_\hc^{(n)}(x+t_n\spac\nb)\,\prod_j \Phi_j(x) \,,
\end{equation}
where $\Phi_j$ represent the soft fields, and $\widetilde H$ is the hard matching coefficient. Using translational invariance, we have
\begin{equation}
   \phi_\hc^{(i)}(x+t_i\spac\nb) = e^{t_i\spac\nb\cdot\partial_x}\,\phi_\hc^{(i)}(x)
   \equiv e^{i\spac t_i\spac\nb\cdot\P_\hc^{(i)}}\,\phi_\hc^{(i)}(x) \,,
\end{equation}
where the label operator $\nb\cdot\P_\hc^{(i)}$ projects out the large momentum component $\nb\cdot p_i$ of the $i^{\rm th}$ hard-collinear particle. We then obtain 
\begin{equation}
   \mathcal{L}_\mathrm{eff}^{\text{SCET-1}} 
   \ni H(m_b,\nb\cdot\P_\hc^{(1)},\ldots,\nb\cdot\P_\hc^{(n)},\mu)\,
    \phi_\hc^{(1)}(x)\ldots\phi_\hc^{(n)}(x)\,\prod_j \Phi_j(x) \,,
\end{equation}
where we have introduced the Fourier-transformed hard matching coefficient
\begin{equation}
   H(m_b,\nb\cdot\P_\hc^{(1)},\ldots,\nb\cdot\P_\hc^{(n)},\mu)
   = \int\!dt_1\,e^{i\spac t_1\spac \nb\cdot\P_\hc^{(1)}} \ldots 
    \int\!dt_n\,e^{i\spac t_n\spac \nb\cdot\P_\hc^{(n)}}\,\widetilde{H}(m_b,t_1,\ldots,t_n,\mu) \,.
\end{equation}
If a SCET-1 operator contains $n>1$ hard-collinear fields, these fields share the large component of the total hard-collinear momentum, given by $\nb\cdot p_\ell$ in our case. Since the large components of hard-collinear momenta are always positive, we can assign variables $y_i\in[0,1]$ with $i=1,\dots,n$ to the fields, which specify the fraction of the total large moment carried by the individual fields. Specifically, we define \cite{Alte:2018nbn} 
\begin{equation}
   \phi_{\hc\spac[y_i]}^{(i)}(x) 
   = \delta\bigg( y_i - \frac{\nb\cdot\P_\hc^{(i)}}{\nb\cdot\P_\hc} \bigg)\,\phi_\hc^{(i)}(x) \,,
\end{equation}
where the label operator $\nb\cdot\P_\hc$ projects out the large component of the total hard-collinear momentum carried by an operator. We write the Wilson coefficients as functions of $\nb\cdot\P_\hc$ and the variables $\{y_i\}$, and in the effective Lagrangian one must integrate over these variables. The condition $\sum_{i=1}^n\spac\nb\cdot\P_\hc^{(i)}=\nb\cdot\P_\hc$ implies that one of these integrations is trivial, since the last $\delta$-distribution can be rewritten as
\begin{equation}
   \delta\bigg( y_n - \frac{\nb\cdot\P_\hc^{(n)}}{\nb\cdot\P_\hc} \bigg)
   = \delta\bigg( \sum_{i=1}^n\,y_i - 1 \bigg) \,.
\end{equation}
This leaves us with the final form (for $n>1$)
\begin{equation}
\begin{aligned}
   \mathcal{L}_\mathrm{eff}^{\text{SCET-1}} 
   &\ni \int_0^1\!dy_1\ldots\int_0^1\!dy_{n-1}\,\theta\Big(1-\sum_{i=1}^{n-1} y_i\Big)\,
    H(m_b,\nb\cdot\P_\hc,y_1,\ldots,y_{n-1},\mu) \\
   &\quad\times \phi_{\hc\spac[y_1]}^{(1)}(x)\ldots\phi_{\hc\spac[y_{n-1}]}^{(n-1)}(x)\,
    \phi_\hc^{(n)}(x) \prod_j \Phi_j(x) \,.
\end{aligned}
\end{equation}
Reparameterization invariance requires that, besides the scaling variables $y_i$, the hard functions can depend on the hard scales $m_b$ and $\nb\cdot\P_\hc/\nb\cdot v$. 

The Dirac basis can be spanned by the 16 matrices 
\begin{equation}
   1 \,,~ \gamma_5 \,,~ \gamma_\perp^\mu \,,~ \gamma_\perp^\mu\gamma_5 \,,~
   \nsl \,,~ \nsl\gamma_5 \,,~ \nsl\gamma_\perp^\mu \,, ~
   \nbsl \,,~ \nbsl\gamma_5 \,,~ \nbsl\gamma_\perp^\mu \,, ~
   \nsl\spac\nbsl \,,~ [\gamma_\perp^\mu,\gamma_\perp^\nu] \,,
\end{equation}
where the transverse Lorentz indices can take two distinct values. We will express all SCET operators in terms of fermion fields with definite chirality, and hence there is no need to write out factors of $\gamma_5$. Lorentz invariance requires that all open transverse Lorentz indices in the SCET operators must be contracted using the symbols (we use the convention $\epsilon^{0123}=-1$)
\begin{equation}\label{eq:gperp}
   g_\perp^{\mu\nu} = g^{\mu\nu} - \frac{n^\mu\spac\nb^\nu+\nb^\mu\spac n^\nu}{2} \,, \qquad
   \epsilon_\perp^{\mu\nu} = \frac12\,\epsilon^{\mu\nu\alpha\beta}\,\nb_\alpha\spac n_\beta \,.
\end{equation}
For $n^\mu=(1,0,0,1)$ and $\nb^\mu=(1,0,0,-1)$ these definitions imply $g_\perp^{11}=g_\perp^{22}=-1$ and $\epsilon_\perp^{12}=-\epsilon_\perp^{21}=1$, and all other entries vanish. When $\epsilon_\perp^{\mu\nu}$ is contracted with a Dirac matrix next to a hard-collinear or anti-hard-collinear spinor, it can be traded for $g_\perp^{\mu\nu}$ using the relations \cite{Lange:2003pk}
\begin{equation}\label{eq:nicerelations2}
\begin{aligned}   
   \nsl\,i\epsilon_\perp^{\mu\nu} \gamma_{\perp\nu}
   &= - \nsl\,g_\perp^{\mu\nu} \gamma_{\perp\nu}\gamma_5 
    = - \nsl\spac\gamma_\perp^\mu\gamma_5 \,, \\
   i\epsilon_\perp^{\mu\nu}\gamma_{\perp\nu}\,\nbsl
   &= - g_\perp^{\mu\nu}\gamma_{\perp\nu}\gamma_5\,\nbsl
    = - \gamma_\perp^\mu\gamma_5\,\nbsl \,, 
\end{aligned}
\end{equation}
which hold in four spacetime dimensions. As an important corollary of these relations, we note the remarkable identities
\begin{equation}\label{eq:nicerelations3}
\begin{aligned}
   \nsl\spac\gamma_\perp^\mu P_{L,R} \otimes \nsl\spac\gamma_{\perp\mu} P_{L,R}
   &= 0 \,, \\
   \nbsl\spac\gamma_\perp^\mu P_{L,R} \otimes \nbsl\spac\gamma_{\perp\mu} P_{L,R} 
   &= 0 \,, \\
   \nsl\spac\gamma_\perp^\mu P_{L,R} \otimes \nbsl\spac\gamma_{\perp\mu} P_{R,L}
   &= 0 \,,
\end{aligned}
\end{equation}
which can be used to eliminate several operators from the basis. In our analysis, we will employ a projection scheme in which these relations are maintained even in $d=4-2\epsilon$ spacetime dimensions (see Section~\ref{subsec:evanescent}). Note also that there is no need to allow for the commutator $[\gamma_\perp^\mu,\gamma_\perp^\nu]$ next to a hard-collinear or anti-collinear spinor, because the identities \cite{Lange:2003pk}
\begin{equation}\label{eq:nicerelations1}
   \nsl\,[\gamma_\perp^\mu,\gamma_\perp^\nu] 
   = \nsl\,2i\epsilon_\perp^{\mu\nu}\spac\gamma_5 \,, \qquad
   [\gamma_\perp^\mu,\gamma_\perp^\nu]\,\nbsl 
   = 2i\epsilon_\perp^{\mu\nu}\spac\gamma_5\,\nbsl
\end{equation}
can be used to eliminate these structures. 

We now lay out the construction of the SCET-1 operator basis needed to describe the decay \blnu, working at leading non-trivial order in the expansion parameters $\lambda\sim\Lambda_\mathrm{QCD}/m_B$ and $\lambda_\ell\sim m_\ell/m_B$. (For most of the discussion we have in mind the process $B^-\to\mu^-\bar\nu_\mu$, for which it seems reasonable to adopt a power counting where $m_\ell=m_\mu\sim\Lambda_\mathrm{QCD}$, but our analysis does not rely on this counting.) We will show that the problem at hand is intrinsically a SCET factorization problem at next-to-leading power, and therefore the discussion will be rather technical. 

\subsection{Construction of the SCET-1 operator basis}
\label{subsec:SCET1basis}

The lepton current in SCET-1 must be of the form (modulo additional gauge fields or derivatives, see below)
\begin{equation}
   \bar\X_\hc^{(\ell)}\,\Gamma\spac\nu_\cb \,,
\end{equation}
where the Dirac basis is spanned by $\Gamma\in\{1,\gamma_\perp^\alpha\}$. For the case where $\Gamma=1$ the charged lepton is right-handed, and the chirality flip introduces a factor $m_\ell\sim \lambda_\ell$, which we include in the definition of the operators. This factor must be compensated by a hard scale in the denominator. In SCET-1, the spectator quark in the $B$ meson is described either by a soft field or a hard-collinear field. In the latter case, the large momentum carried by the field needs to be transferred to the charged lepton via the exchange of a hard-collinear photon (in the matching onto SCET-2) or a collinear photon (in the evaluation of the SCET-2 matrix elements), and a soft spectator quark remains. 

The simplest operator one can write down in SCET-1 is
\begin{equation}
   O_0^A = \big(\bar u_s\spac\gamma_\perp^\alpha P_L\spac b_v\big)\,
    \big(\bar\X_\hc^{(\ell)}\spac\gamma_\alpha^\perp\spac P_L\spac\nu_\cb\big) 
    \sim \lambda^{\frac72} \,,
\end{equation}
where for simplicity we count the sterile neutrino field as $\mathcal{O}(\lambda^0)$. The analogous operator in which the soft quark field is replaced by a hard-collinear field vanishes due to the first identity in \eqref{eq:nicerelations3}. Converting the hard-collinear lepton field into a collinear one in SCET-2 costs a factor $\lambda_\ell/\lambda^{\frac12}$, and hence $O_0^A$ matches onto a SCET-2 operator scaling as $\sim\lambda_\ell\,\lambda^3$. This operator can mediate the decay $B^{*-}\to\ell^-\spac\bar\nu_\ell$ with an amplitude of $\mathcal{O}(\lambda^{\frac32})$, corresponding to the scaling of the decay constant of the $B^*$ meson.\footnote{The external $B$-meson and lepton states in SCET-2 ``eat up'' factors of $\lambda^{-\frac32}$ and $\lambda_\ell^{-1}$, respectively.}
However, it has a vanishing projection onto the $B$-meson state and thus does not contribute to the \blnu\ decay amplitude, the leading contributions to which arise from SCET-2 operators with scaling $\sim\lambda_\ell^2\,\lambda^3$. This observation shows that our process of interest requires the construction of a SCET-1 operator basis beyond the leading power, which as we will see is a rather challenging task. 

Concretely, our task is to find all relevant SCET-1 operators which can match onto $\mathcal{O}(\lambda_\ell^2\,\lambda^3)$ operators in SCET-2. If these operators do not include the lepton mass explicitly, they must contain additional hard-collinear fields or derivatives, so that a factor of $m_\ell$ can be produced either in the matching onto SCET-2, when hard-collinear loop graphs are evaluated, or in the calculation of collinear matrix elements in SCET-2. In deriving the structure of the relevant operators, one needs to understand the systematics of the matching onto SCET-2, which will be discussed in detail in Section~\ref{sec:SCET-2}. These considerations lead to the following four simple rules: 
\begin{enumerate}
\item
For $\ell=\mu,e$ the lepton mass is a power-suppressed parameter in SCET-1, and a mass insertion costs a factor $\lambda_\ell/\lambda^{\frac12}$.
\item 
The conversion of a hard-collinear into a soft quark field costs a factor $\lambda^{\frac12}$.
\item 
The conversion of a hard-collinear into a collinear lepton field costs a factor $\lambda_\ell/\lambda^{\frac12}$. The same is true for the conversion of a hard-collinear gauge field into a collinear gauge field.
\item
The conversion of a soft into a collinear lepton field costs a factor $\lambda_\ell$. Obtaining a mass term from a soft lepton propagator costs another factor $\lambda_\ell/\lambda$.
\end{enumerate}
The first rule implies that operators not containing the lepton mass can have one additional ``transverse object'' (either a transversely polarized gauge field or a transverse derivative) compared with the corresponding operators containing a factor of $m_\ell$. Since a soft quark field has power counting $\lambda^{\frac32}$ and a hard-collinear quark field has power counting $\lambda^{\frac12}$, the second rule further implies that operators with a hard-collinear quark field may contain one additional transverse object compared to operators with a soft quark field. Operators with a hard-collinear quark field and no factor of $m_\ell$ can therefore contain up to two additional transverse objects. The second and the fourth rule imply that operators containing a soft lepton field along with a hard-collinear quark field must at most be of $\mathcal{O}(\lambda^{\frac72})$ in power counting and hence cannot contain any additional fields or derivatives. It follows from these observations that we can distinguish six different classes of operators:

\paragraph{Type-$\bm{A}$ operators containing a hc lepton field:}
\begin{equation}\label{typeA} 
\begin{aligned}
O_1^A 
   &= \frac{m_\ell}{\nb\cdot\P_\hc}\,
    \big(\bar u_s\spac\nbsl\spac P_L\spac b_v\big)\,
    \big(\bar\X_\hc^{(\ell)} \spac P_L\spac\nu_\cb\big) &&
    \sim \lambda_\ell\,\lambda^{\frac72} \\
   O_2^A 
   &= \frac{m_\ell\,(\nb\cdot v)^2}{\nb\cdot\P_\hc}\,
    \big(\bar u_s\spac\nsl P_L\spac b_v\big)\,
    \big(\bar\X_\hc^{(\ell)} \spac P_L\spac\nu_\cb\big) &&
    \sim \lambda_\ell\,\lambda^{\frac72} \\
   O_3^A 
   &= \frac{1}{\nb\cdot\P_\hc}\,
    \big(\bar u_s\spac\nbsl\spac P_L\spac b_v\big)\,
    \big[ \bar\X_\hc^{(\ell)}\spac(-i\!\overleftarrow{\delsl}_{\!\!\perp}) \spac P_L\spac\nu_\cb \big] &&
    \sim \lambda^4 \\
   O_4^A 
   &= \frac{(\nb\cdot v)^2}{\nb\cdot\P_\hc}\,
    \big(\bar u_s\spac\nsl P_L\spac b_v\big)\,
    \big[ \bar\X_\hc^{(\ell)}\spac(-i\!\overleftarrow{\delsl}_{\!\!\perp}) \spac P_L\spac\nu_\cb \big] &&
    \sim \lambda^4
\end{aligned}
\end{equation}

\paragraph{Type-$\bm{B}$ operators containing a hc lepton field and a hc gauge field:}
\begin{equation}\label{typeB}
\begin{aligned}
   O_1^B 
   &= \frac{1}{\nb\cdot\P_\hc}\,
    \big(\bar u_s\spac\nbsl\spac P_L\spac b_v\big)\,
    \big(\bar\X_\hc^{(\ell)}\,\Asl_{\hc\spac[y]}^\perp \spac P_L\spac\nu_\cb\big) &&
    \sim \lambda^4 \\
   O_2^B 
   &= \frac{(\nb\cdot v)^2}{\nb\cdot\P_\hc}\,
    \big(\bar u_s\spac\nsl P_L\spac b_v\big)\,
    \big(\bar\X_\hc^{(\ell)}\,\Asl_{\hc\spac[y]}^\perp \spac P_L\spac\nu_\cb\big) &&
    \sim \lambda^4 \\
   O_3^B 
   &= \frac{1}{\nb\cdot\P_\hc}\,
    \big(\bar u_s\spac\nbsl\spac P_L\spac\G_{\hc\spac[y]}^{\perp\alpha}\spac b_v\big)\,
    \big(\bar\X_\hc^{(\ell)}\spac\gamma_\alpha^\perp \spac P_L\spac\nu_\cb\big) &&
    \sim \lambda^4 \\
   O_4^B 
   &= \frac{(\nb\cdot v)^2}{\nb\cdot\P_\hc}\,
    \big(\bar u_s\spac\nsl P_L\spac\G_{\hc\spac[y]}^{\perp\alpha}\spac b_v\big)\,
    \big(\bar\X_\hc^{(\ell)}\spac\gamma_\alpha^\perp \spac P_L\spac\nu_\cb\big) &&
    \sim \lambda^4
\end{aligned}
\end{equation}

\paragraph{Type-$\bm{C}$ operators containing a hc lepton field and a hc quark field:}
\begin{equation}\label{typeC}
\begin{aligned}
   O_1^C
   &= \frac{m_\ell}{\nb\cdot\P_\hc}\,
    \big(\bar\X_{\hc\spac[y]}^{(u)}\spac\nbsl\spac P_L\spac b_v\big)\,
    \big(\bar\X_\hc^{(\ell)} \spac P_L\spac\nu_\cb\big) && 
    \sim \lambda_\ell\,\lambda^{\frac52} \\
   O_2^C
   &= \frac{m_\ell\,\nb\cdot v}{(\nb\cdot\P_\hc)^2}\,
    \big[ \bar\X_{\hc\spac[y]}^{(u)}\spac (-i\!\overleftarrow{\delsl}_{\!\!\perp}) \spac
    \nbsl\spac P_R\,b_v\big]\,
    \big(\bar\X_\hc^{(\ell)} \spac P_L\spac\nu_\cb\big) && 
    \sim \lambda_\ell\,\lambda^3 \\
   O_3^C
   &= \frac{1}{\nb\cdot\P_\hc}\,
    \big[\bar\X_{\hc\spac[y]}^{(u)}\spac(-i\overleftarrow{D}_{\!\!s\perp}^\alpha)
    \spac\nbsl\spac P_L\spac b_v\big]\,
    \big(\bar\X_\hc^{(\ell)}\spac\gamma_\alpha^\perp \spac P_L\spac\nu_\cb\big) && 
    \sim \lambda^3\spac,\,\lambda^{\frac72} \\
   O_4^C
   &= \frac{1}{\nb\cdot\P_\hc}\,
    \big(\bar\X_{\hc\spac[y]}^{(u)}\spac\nbsl\spac P_L\spac b_v\big)\,
    \big[ \bar\X_\hc^{(\ell)}\spac(-i\overleftarrow{\Dsl}_{\!\!s\perp}) \spac P_L\spac\nu_\cb\big] && 
    \sim \lambda^3\spac,\,\lambda^{\frac72} \\
   O_5^C
   &= \frac{\nb\cdot v}{(\nb\cdot\P_\hc)^2}
    \big[\bar\X_{\hc\spac[y]}^{(u)}\spac
    (-i\!\overleftarrow{\delsl}_{\!\!\perp})\spac\nbsl\spac P_R\,b_v\big]\,
    \big[ \bar\X_\hc^{(\ell)}\spac(-i\!\overleftarrow{\delsl}_{\!\!\perp}) \spac P_L\spac\nu_\cb \big] && 
    \sim \lambda^{\frac72}
\end{aligned}
\end{equation}
\paragraph{Type-$\bm{D}$ operators containing three hc fields:}
\begin{equation}\label{typeD}
\begin{aligned}
   O_1^D 
   &= \frac{m_\ell\,\nb\cdot v}{(\nb\cdot\P_\hc)^2}\,
    \big(\bar\X_{\hc\spac[y_1]}^{(u)}\,\Asl_{\hc\spac[y_2]}^\perp\spac
    \nbsl\spac P_R\,b_v\big)\,
    \big(\bar\X_\hc^{(\ell)} \spac P_L\spac\nu_\cb\big) && 
    \sim \lambda_\ell\,\lambda^3 \\
   O_2^D 
   &= \frac{m_\ell\,\nb\cdot v}{(\nb\cdot\P_\hc)^2}\,
    \big(\bar\X_{\hc\spac[y_1]}^{(u)}\spac\Gsl_{\hc\spac[y_2]}^\perp\spac
    \nbsl\spac P_R\,b_v\big)\,
    \big(\bar\X_\hc^{(\ell)} \spac P_L\spac\nu_\cb\big) && 
    \sim \lambda_\ell\,\lambda^3 \\
   O_3^D 
   &= \frac{1}{\nb\cdot\P_\hc}\,\big(\bar\X_{\hc\spac[y_1]}^{(u)}\spac\nbsl\spac P_L\spac b_v\big)\,
    \big(\bar\X_\hc^{(\ell)}\,\Asl_{\hc\spac[y_2]}^\perp \spac P_L\spac\nu_\cb\big) && 
    \sim \lambda^3 \\
   O_4^D 
   &= \frac{1}{\nb\cdot\P_\hc}\,\big(\bar\X_{\hc\spac[y_1]}^{(u)}\spac\G_{\hc\spac[y_2]}^{\perp\alpha}
    \spac\nbsl\spac P_L\spac b_v\big)\,
    \big(\bar\X_\hc^{(\ell)}\spac\gamma_\alpha^\perp \spac P_L\spac\nu_\cb\big) && 
    \sim \lambda^3 \\
   O_5^D 
   &= \frac{\nb\cdot v}{(\nb\cdot\P_\hc)^2}\,
    \big[\bar\X_{\hc\spac[y_1]}^{(u)}\spac
    (-i\!\overleftarrow{\delsl}_{\!\!\perp})\spac\nbsl\spac P_R\,b_v\big]\,
    \big(\bar\X_\hc^{(\ell)}\,\Asl_{\hc\spac[y_2]}^\perp \spac P_L\spac\nu_\cb\big) \hspace{-1.0cm} && 
    \sim \lambda^{\frac72} \\
   O_6^D 
   &= \frac{\nb\cdot v}{(\nb\cdot\P_\hc)^2}\,
    \big[\bar\X_{\hc\spac[y_1]}^{(u)}\spac
    (-i\!\overleftarrow{\delsl}_{\!\!\perp})\spac\G_{\hc[y_2]}^{\perp\alpha}\spac\nbsl\spac P_R\,b_v\big]\,
    \big(\bar\X_\hc^{(\ell)}\spac\gamma_\alpha^\perp \spac P_L\spac\nu_\cb\big) && 
    \sim \lambda^{\frac72} \\
   O_7^D 
   &= \frac{\nb\cdot v}{(\nb\cdot\P_\hc)^2}\,
    \big(\bar\X_{\hc\spac[y_1]}^{(u)}\,
    \Asl_{\hc\spac[y_2]}^\perp\spac\nbsl\spac P_R\,b_v\big)\,
    \big[ \bar\X_\hc^{(\ell)}\spac(-i\!\overleftarrow{\delsl}_{\!\!\perp}) \spac P_L\spac\nu_\cb \big] && 
    \sim \lambda^{\frac72} \\
   O_8^D 
   &= \frac{\nb\cdot v}{(\nb\cdot\P_\hc)^2}\,
    \big(\bar\X_{\hc\spac[y_1]}^{(u)}\spac
    \Gsl_{\hc\spac[y_2]}^\perp\spac\nbsl\spac P_R\,b_v\big)\,
    \big[ \bar\X_\hc^{(\ell)}\spac(-i\!\overleftarrow{\delsl}_{\!\!\perp}) \spac P_L\spac\nu_\cb \big] && 
    \sim \lambda^{\frac72}
\end{aligned}
\end{equation}

\paragraph{Type-$\bm{E}$ operators containing four hc fields:}
\begin{equation}\label{typeE}
\begin{aligned}
   O_1^E
   &= \frac{\nb\cdot v}{(\nb\cdot\P_\hc)^2}\,
    \big(\bar\X_{\hc\spac[y_1]}^{(u)}\,
    \Asl_{\hc\spac[y_2]}^\perp\spac\nbsl\spac P_R\,b_v\big)\,
    \big(\bar\X_\hc^{(\ell)}\,\Asl_{\hc\spac[y_3]}^\perp \spac P_L\spac\nu_\cb\big) &&
    \sim \lambda^{\frac72} \\
   O_2^E 
   &= \frac{\nb\cdot v}{(\nb\cdot\P_\hc)^2}\,
    \big(\bar\X_{\hc\spac[y_1]}^{(u)}\spac
    \Gsl_{\hc\spac[y_2]}^\perp\spac\nbsl\spac P_R\,b_v\big)\,
    \big(\bar\X_\hc^{(\ell)}\,\Asl_{\hc\spac[y_3]}^\perp \spac P_L\spac\nu_\cb\big) &&
    \sim \lambda^{\frac72} \\
   O_3^E 
   &= \frac{\nb\cdot v}{(\nb\cdot\P_\hc)^2}\,
    \big(\bar\X_{\hc\spac[y_1]}^{(u)}\spac\G_{\hc[y_2]}^{\perp\alpha}\,
    \Gsl_{\hc\spac[y_3]}^\perp\spac\nbsl\spac P_R\,b_v\big)\,
    \big(\bar\X_\hc^{(\ell)}\spac\gamma_\alpha^\perp \spac P_L\spac\nu_\cb\big) &&
    \sim \lambda^{\frac72} \\
   O_4^E
   &= \frac{\nb\cdot v}{(\nb\cdot\P_\hc)^2}\,
    \big(\bar\X_{\hc\spac[y_1]}^{(u)}\spac\G_{\hc[y_2]}^{\perp\alpha,a}\,
    \Gsl_{\hc\spac[y_3]}^{\perp,a}\spac\nbsl\spac P_R\,b_v\big)\,
    \big(\bar\X_\hc^{(\ell)}\spac\gamma_\alpha^\perp \spac P_L\spac\nu_\cb\big) &&
    \sim \lambda^{\frac72} 
\end{aligned}
\end{equation}

\paragraph{Type-$\bm{F}$ operator containing a soft lepton field:}
\begin{equation}\label{typeF}
\begin{aligned}
   O_1^F
   &= \big(\bar\X_\hc^{(u)}\spac\nbsl\spac P_L\spac b_v\big)\,
    \big(\bar\ell_s\spac\nsl\spac P_L\spac\nu_\cb\big) &&
    \sim \lambda^{\frac72} 
\end{aligned}
\end{equation}
After renormalization, the charged-lepton mass appearing in the definition of some of the operators is defined as the running mass $m_\ell(\mu)$ in the $\overline{\mathrm{MS}}$ scheme. It will later be converted into the physical pole mass (see Section~\ref{subsec:SCET2_matrix}).

The label operator $\nb\cdot\P_\hc$ projects out the large component of the total hard-collinear momentum carried by an operator, which in our case evaluates to $\nb\cdot p_\ell$. As mentioned earlier, when an operator contains two or more hard-collinear fields, one needs to indicate how these fields share the large component of the total hard-collinear momentum. For type-$B$ operators we assign the momentum fraction $y$ to the gauge field, which implies that the lepton field carries the momentum fraction $(1-y)$. Similarly, for type-$C$ operators we assign the momentum fraction $y$ to the hard-collinear quark field. For type-$D$ operators the large component of the total hard-collinear momentum is shared among three fields. We assign momentum the fractions $y_1$ and $y_2$ to the up-quark field and the gauge field, respectively, which implies that the charged lepton carries the momentum fraction $(1-y_1-y_2)$. For type-$E$ operators, finally, the large component of the total hard-collinear momentum is shared among four fields as indicated, and the charged lepton carries momentum fraction $(1-y_1-y_2-y_3)$.

After each operator, we give its power counting with the expansion parameters $\lambda_\ell$ and $\lambda$. Using the above rules, one finds that these operators match onto SCET-2 operators scaling (at least) as $\lambda_\ell^2\,\lambda^3$ and thus potentially can give leading contributions to the \blnu\ decay amplitude in the SM. There are five exceptions, namely
\begin{equation}
   O_1^C \,, \quad O_3^C \,, \quad O_4^C \,, \quad O_3^D \,, \quad O_4^D \,, 
\end{equation}
which naively appear to give ``super-leading'' contributions scaling as $\lambda_\ell\,\lambda^{\frac72}$ in SCET-2. However, for these operators an additional power-suppressed Lagrangian insertion of $\mathcal{O}(\lambda^{\frac12})$ is needed to perform the matching onto SCET-2.

Some comments are in order concerning the construction of the operators in the different classes:
\begin{itemize}
\item
Operators containing a transverse derivative acting on the (non-interacting) neutrino field can be omitted, since with our choice of reference frame $p_\nu^\perp=0$.
\item  
In general, Dirac structures in the quark and lepton bilinears involving $\gamma_\perp^\alpha$ matrices can be simplified using the identities \eqref{eq:nicerelations2} and \eqref{eq:nicerelations1}. For all other operators containing a pair of contracted transverse Lorentz indices, relation \eqref{eq:nicerelations2} implies that contracting the two indices with $\epsilon_\perp^{\mu\nu}$ rather than $g_\perp^{\mu\nu}$ yields nothing new. This statement applies to all other operators with a pair of contracted transverse indices.  
\item 
There is no need to include type-$A$ or type-$B$ operators in which the quark current has the form $\bar u_s\spac P_R\,b_v$. Using that $\vsl\spac b_v=b_v$, it is straightforward to show that (with $v^2=n\cdot v\,\nb\cdot v=1$)
\begin{equation}\label{trick}
   \bar u_s\spac P_R\,b_v 
   = \frac{n\cdot v}{2}\,\bar u_s\spac\nbsl\spac P_L\spac b_v
    + \frac{\nb\cdot v}{2}\,\bar u_s\spac\nsl\spac P_L\spac b_v \,,
\end{equation}
which implies that any such operator can be reduced to our basis operators. 
\item
Several of the basis operators of type-$C$, $D$ and $E$ contain a transverse Dirac matrix in both the quark and the lepton current. The first relation in \eqref{eq:nicerelations3} implies that
\begin{equation}\label{eq:magic}
   \big(\bar\X_\hc^{(u)}\spac\gamma_\perp^\mu P_L\spac\Gamma\,b_v\big)\,
    \big(\bar\X_\hc^{(\ell)}\spac\gamma_{\perp\mu} P_L\spac\nu_\cb\big) = 0 \,,
\end{equation}
where $\Gamma$ can be an arbitrary Dirac structure. We will refer to this equality as the ``magic identity''. Setting $\Gamma=\gamma_\perp^\alpha\spac\gamma_\perp^\beta$, and using that in $d=4$ spacetime dimensions
\begin{equation}
   \gamma_\perp^\mu\spac\gamma_\perp^\alpha\spac\gamma_\perp^\beta
   = \gamma_\perp^\mu\spac g_\perp^{\alpha\beta} + \gamma_\perp^\beta\spac g_\perp^{\mu\alpha}
    - \gamma_\perp^\alpha\spac g_\perp^{\mu\beta} \,,
\end{equation}
we find the identity
\begin{equation}\label{eq:Schouten}
   \big(\bar\X_\hc^{(u)}\spac\gamma_\perp^\alpha P_L\spac b_v\big)\,
    \big(\bar\X_\hc^{(\ell)}\spac\gamma_\perp^\beta P_L\spac\nu_\cb\big) 
   = \big(\bar\X_\hc^{(u)}\spac\gamma_\perp^\beta P_L\spac b_v\big)\,
    \big(\bar\X_\hc^{(\ell)}\spac\gamma_\perp^\alpha P_L\spac\nu_\cb\big) \,.
\end{equation}
\item 
Operators containing two hard-collinear transverse objects with Lorentz indices $\alpha$ and $\beta$, i.e.\ $O_5^C$, $O_{5,6,7,8}^D$ and $O_{1,2,3,4}^E$, need to involve the Dirac structure
\begin{equation}
   \gamma_\perp^\alpha P_L \otimes \gamma_\perp^\beta P_L \,.
\end{equation}
If we contract this with $g_{\alpha\beta}^\perp$ or $\epsilon_{\alpha\beta}^\perp$, the result vanishes by the magic identity \eqref{eq:magic}. Hence, each index must be contracted with one of the transverse objects. Relation \eqref{eq:Schouten} allows us to exchange the indices $\alpha$ and $\beta$ in order to reduce the set of operators. Using this freedom, the operators 
\begin{equation}
\begin{aligned}
   \frac{\nb\cdot v}{(\nb\cdot\P_\hc)^2}\,
    \big[\bar\X_{\hc\spac[y_1]}^{(u)}\spac
    (-i\!\overleftarrow{\partial}_{\!\!\perp}^\alpha)\,
    \Asl_{\hc\spac[y_2]}^\perp\spac\nbsl\spac P_R\,b_v\big]\,
    \big(\bar\X_\hc^{(\ell)}\,\gamma_\alpha^\perp \spac P_L\spac\nu_\cb\big)
   &= O_5^D \,, \\
   \frac{\nb\cdot v}{(\nb\cdot\P_\hc)^2}\,
    \big[\bar\X_{\hc\spac[y_1]}^{(u)}\spac
    (-i\!\overleftarrow{\partial}_{\!\!\perp}^\alpha)\spac
    \Gsl_{\hc\spac[y_2]}^\perp\spac\nbsl\spac P_R\,b_v\big]\,
    \big(\bar\X_\hc^{(\ell)}\,\gamma_\alpha^\perp \spac P_L\spac\nu_\cb\big) 
   &= O_6^D \,, \\
   \frac{\nb\cdot v}{(\nb\cdot\P_\hc)^2}\,
    \big[\bar\X_{\hc\spac[y_1]}^{(u)}\spac
    \big(i\delsl_\perp\spac\A_{\hc\spac[y_2]}^{\perp\alpha}\big)\spac\nbsl\spac P_R\,b_v\big]\,
    \big(\bar\X_\hc^{(\ell)}\spac\gamma_\alpha^\perp \spac P_L\spac\nu_\cb\big) 
   &= O_5^D + O_7^D + \mathcal{O}(\lambda^4) \,, \\
   \frac{\nb\cdot v}{(\nb\cdot\P_\hc)^2}\,
    \big[\bar\X_{\hc\spac[y_1]}^{(u)}\spac
    \big(i\delsl_\perp\spac\G_{\hc\spac[y_2]}^{\perp\alpha}\big)\spac\nbsl\spac P_R\,b_v\big]\,
    \big(\bar\X_\hc^{(\ell)}\spac\gamma_\alpha^\perp \spac P_L\spac\nu_\cb\big)
   &= O_6^D + O_8^D + \mathcal{O}(\lambda^4) \,,
\end{aligned}
\end{equation}
as well as 
\begin{equation}
\begin{aligned}
   \frac{\nb\cdot v}{(\nb\cdot\P_\hc)^2}\,
    \big(\bar\X_{\hc\spac[y_1]}^{(u)}\,
    \Asl_{\hc\spac[y_2]}^\perp\spac\G_{\hc[y_3]}^{\perp\alpha}\spac\nbsl\spac P_R\,b_v\big)\,
    \big(\bar\X_\hc^{(\ell)}\spac\gamma_\alpha^\perp \spac P_L\spac\nu_\cb\big)
   &= O_2^E(y_2\leftrightarrow y_3) \,, \\
   \frac{\nb\cdot v}{(\nb\cdot\P_\hc)^2}\,
    \big(\bar\X_{\hc\spac[y_1]}^{(u)}\spac
    \Gsl_{\hc\spac[y_2]}^\perp\,\G_{\hc[y_3]}^{\perp\alpha}\spac\nbsl\spac P_R\,b_v\big)\,
    \big(\bar\X_\hc^{(\ell)}\spac\gamma_\alpha^\perp \spac P_L\spac\nu_\cb\big)
   &= O_3^E 
\end{aligned}
\end{equation}
can be removed from the basis. Moreover, it follows that
\begin{equation}
   O_4^E = O_4^E(y_2\leftrightarrow y_3) \,,
\end{equation}
implying that the coefficient $H_4^E$ can be symmetrized in the variables $y_2$ and $y_3$.
\item 
The operators $O_3^C$ and $O_4^C$, which involve the derivative $iD_{s\perp}^\alpha=i\partial_\perp^\alpha+g_s\spac G_{s\perp}^{\alpha,a}\spac t^a+Q_u\spac e\spac A_{s\perp}^{\alpha,a}\spac t^a$ acting on the hard-collinear up-quark field, have inhomogeneous power counting, as indicated. As shown in \cite{Beneke:2002ph,Beneke:2002ni}, it would be possible to rewrite them as sums of manifestly gauge-invariant operators with different (but homogeneous) power counting using field redefinitions. Here we keep the compact form in \eqref{typeC}. An additional operator
\begin{equation}
   \frac{1}{\nb\cdot\P_\hc}\,\big(\bar\X_{\hc\spac[y]}^{(u)}\spac\nbsl\spac P_L\spac
    iD_{s\perp}^\alpha\spac b_v \big)\,
    \big(\bar\X_\hc^{(\ell)}\spac\gamma_\alpha^\perp \spac\nu_\cb\big) 
   = O_3^C + O_4^C \sim \lambda^{\frac72}
\end{equation}
can be reduced to $O_3^C$ and $O_4^C$ using an integration by parts (recall that we set the transverse momentum of the neutrino to be zero). The other possible single-derivative operator (recall that $iv\cdot D_s\spac b_v=0$ in HQET)
\begin{equation}
   \frac{1}{\nb\cdot\P_\hc}\,\big( \bar\X_{\hc\spac[y]}^{(u)}\spac\gamma_\perp^\alpha P_L\spac
    i\nb\cdot D_s\spac b_v \big)\,\big(\bar\X_\hc^{(\ell)}\spac\gamma_\alpha^\perp \spac\nu_\cb\big) = 0
\end{equation}
vanishes due to the magic identity \eqref{eq:magic}. The same is true for operators containing a derivative $in\cdot\partial$ acting on a hard-collinear field or the small component $n\cdot\A_\hc$ or $n\cdot\G_\hc$ of a hard-collinear gauge field. 
\end{itemize}

\subsection{Evanescent operators in SCET}
\label{subsec:evanescent}

In the matching from the LEFT to SCET-1 (and also from SCET-1 to SCET-2) several reducible Dirac structures appear. Their reduction in $d=4-2\epsilon$ spacetime dimensions typically generates $O(\epsilon)$ remnants. When inserted into divergent loop integrals, these ``evanescent'' terms can yield finite contributions. To work with a physical basis containing only non-evanescent operators, such contributions must be computed and absorbed into the matching coefficients of the physical basis. The general procedure for how to do this consistently is well known \cite{Buras:1989xd,Dugan:1990df,Herrlich:1994kh} (see also  \cite{Fuentes-Martin:2022vvu} and references therein) and can be summarized as follows:  Suppose an identity that is valid only in $d=4$ is used to reduce a redundant operator $R_i$ to $O_i$, i.e.\
\begin{equation}
   R_i\,\stackrel{d=4}{=}\,O_i \,.
\end{equation}
In $d\ne 4$ dimensions, the difference between the two operators must be retained,
\begin{equation}
   R_i\,\stackrel{d\ne 4}{=}\, O_i + (R_i-O_i)
   \equiv O_i + \tilde O_i \,.
\end{equation}
By construction, the evanescent operator $\tilde O_i$ has a vanishing tree-level matrix element. At loop order, however, it can yield finite contributions, which can be absorbed into shifts of the Wilson coefficients of the physical operators $O_j$. For example, at one-loop order we define
\begin{equation}
   E_i^{\text{tree}}\,\langle \tilde O_i \rangle^{\text{1-loop}}  
   \equiv \sum_j\,\delta H_j^{\text{1-loop}}\,\langle O_j \rangle^{\text{tree}} \,. 
\end{equation}
The explicit form of  $\langle \tilde O_i \rangle$ depends on how the original reduction identity, valid only in $d=4$ dimensions, is extended to $d=4-2\epsilon$. The specific choice of this extension does not matter as long as it reduces to the original identity for $d=4$, and it is applied consistently throughout the calculation. Some choices, however, are more convenient than others, as they may help to avoid the appearance of spurious intermediate contributions.  

To define a consistent reduction scheme, the $\mathcal{O}(\epsilon)$ terms can be chosen freely only for a set of linearly independent Dirac structures.  We identify such structures by enforcing a canonical ordering on the transverse objects: slashed vectors always comes first, followed by repeated transverse Lorentz indices ordered alphabetically from left to right (i.e.\ $\mu$, $\nu$, $\rho$, $\sigma$). In the lepton current we always use the projector $\frac14\spac\nbsl\spac\nsl$. With this ordering prescription, we define our reduction scheme by imposing the validity of the $d=4$ identities also in $d\ne 4$ dimensions for the canonically ordered Dirac structures, for instance
\begin{equation}
\begin{aligned}
   \gdirac{\frac{\nsl}{2}\spac\gamma_\mu^\perp\spac\gamma_\nu^\perp}%
          {\frac{\nbsl\spac\nsl}{4}\spac \gamma_\perp^\mu\spac\gamma_\perp^\nu}  
   &= 0 \,, \\ 
   \gdirac{\gamma_\mu^\perp\spac\gamma_\nu^\perp\spac\gamma_\rho^\perp}%
          {\frac{\nbsl\spac\nsl}{4}\spac\gamma_\perp^\mu\spac\gamma_\perp^\nu\spac\gamma_\perp^\rho}
   &= 4\spac\gdirac{\gamma_\mu^\perp}{\frac{\nbsl\spac\nsl}{4} \gamma_\perp^\mu} \,,
\end{aligned}
\end{equation}
where we omitted the explicit spinors for simplicity. The full list of reductions needed in our calculations is given in Appendix~\ref{app:all_reductions}. Structures that are not in canonical form can always be rewritten as combinations of those that are. Note that for the relation shown in \eqref{eq:reduction_kappa} our scheme implies $\kappa=0$, unlike the more common choice  $\kappa=-4(1+\epsilon)$ obtained using the ``greek trick'' \cite{Tracas:1982gp}. Therefore, care needs to be taken to convert any literature result used here to our scheme choice. 

Our ``SCET-friendly'' reduction scheme offers several advantages. First of all, it preserves the identities \eqref{eq:nicerelations3} also in $d=4-2\epsilon$ dimensions, which greatly simplifies the construction of the evanescent operators and eliminates power-enhanced contributions at intermediate steps, including the LEFT to SCET-1 matching, the SCET-1 to SCET-2 matching, and the SCET-2 matrix elements. Such spurious power-enhanced contributions cancel out in the sum of all terms (see Appendix~\ref{app:all_reductions} for an explicit example), but avoiding them altogether is of great advantage. Second, with our scheme no evanescent shifts $\delta J$ arise in the SCET-1 to SCET-2 matching. Third, at one-loop order in QED, it is not necessary to consider contributions involving a soft lepton loop. 

Already at tree level, the LEFT operator $O^{V,LL}_{\ell}$ matches onto several evanescent SCET-1 operators, whose one-loop matrix elements we need to consider. Among these, only operators of type-$B$, $C$ and $F$ can contribute to the \blnu\  process at $\mathcal{O}(\alpha)$. They are  
\begin{equation}\label{eq:Otilde1}
\begin{aligned}
   \tilde O^B_1 
   &= \frac{1}{\nb\cdot\P_\hc}\,\big(\bar u_s\spac\nbsl\spac\gamma^\alpha_\perp\spac
    \Asl^{\perp}_{hc[y]}\spac P_L\spac b_v\big)\,
    \big(\bar\X_\hc^{(\ell)}\spac\gamma_\alpha^\perp P_L\spac\nu_\cb\big) 
   &&\sim \lambda^4 \,,\\[-1mm]
   \tilde O^B_2 
   &= \frac{(\nb\cdot v)^2}{\nb\cdot\P_\hc}\,\Big(\bar u_s\spac\nsl\,
    \Asl^{\perp}_{hc[y]}\spac\gamma^\alpha_\perp P_L\spac b_v\Big)\,
    \big(\bar\X_\hc^{(\ell)}\spac\gamma_\alpha^\perp\spac P_L\spac\nu_\cb\big)
   &&\sim \lambda^4 \,, \\
   \tilde O^C_1 
   &= \big(\bar\X_{\hc[y]}^{(u)}\spac\gamma^\alpha_\perp\spac P_L\spac b_v\big)\,
    \big(\bar\X_\hc^{(\ell)}\spac\gamma_\alpha^\perp\spac P_L\spac\nu_\cb\big)
   &&\sim \lambda^{\frac52} \,,\\
   \tilde O^C_2 
   &= \frac{1}{\nb\cdot\P_\hc}\,\big[\bar\X_{\hc[y]}^{(u)}\spac\nbsl\spac
    \gamma^\alpha_\perp\spac(-i\!\overleftarrow{\delsl}_{\!\!\perp})\spac P_L\spac b_v\big]\,
    \big(\bar\X_\hc^{(\ell)}\spac\gamma_\alpha^\perp\ P_L\spac\nu_\cb\big) 
   &&\sim \lambda^3 \,,\\[-1mm]
   \tilde O^C_3 
   &= \frac{1}{\nb\cdot\P_\hc}\,\big[\bar\X_{\hc[y]}^{(u)}\spac\nbsl\spac
    \gamma^\alpha_\perp\,i{\Dsl}_{\!s}\spac P_L\spac b_v\big]\,
    \big(\bar\X_\hc^{(\ell)}\spac\gamma_\alpha^\perp P_L\spac\nu_\cb\big)
   &&\sim \lambda^{\frac72} \,,\\
   \tilde O^F_1 
   &= \big(\bar\X_{\hc[y]}^{(u)}\spac\gamma^\alpha_\perp\spac P_L\spac b_v\big)\,
    \big(\bar l_s\spac\gamma_\alpha^\perp P_L\spac\nu_\cb \big) \quad
   &&\sim \lambda^{\frac72} \,.
\end{aligned}
\end{equation}
Since by definition these operators have non-vanishing matrix elements only at loop order, it is sufficient for our purposes to compute their hard matching coefficients at tree level.

To account for the contributions from the evanescent operators, we compute the convolutions of their one-loop matrix elements with the corresponding tree-level matching coefficients, working with an off-shell lepton as an infrared regulator to ensure that only the ultraviolet (UV)  poles of the matrix elements contribute. We then match the results onto the physical SCET-1 operator basis constructed in Section~\ref{subsec:SCET1basis}. This yields a result of the form
\begin{equation}
   \sum_i \int_0^1\!dy\,E_i(y)\,\langle\tilde O_i(y)\rangle
   = \delta H_1^A\,\langle O_1^A\rangle + \delta H_2^A\,\langle O_2^A\rangle + \dots \,.
\end{equation}
At $\mathcal{O}(\alpha)$, we only need the evanescent shifts to the hard matching coefficients $H_{1,2}^A$, because only the operators $O_{1,2}^A$ give tree-level matrix elements for the \blnu\ process. Applying the reduction scheme defined above, we obtain
\begin{equation}\label{eq:deltaH1A}
   \delta H_1^A = \frac{Q_\ell\spac\alpha}{2\pi} \int_0^1\!dy\,
    \big[ 2y E^B_1(y) + 4 Q_u\spac y(1-y)\spac E^C_2(y) \big] 
    = - Q_\ell\spac Q_u\,\frac{\alpha}{\pi} \,,
\end{equation}
and $\delta H_2^A=0$, where we have used the tree-level expressions 
\begin{equation}\label{eq:evanescentHi}
   E_1^B(y) = - \frac{Q_u}{2y} + \mathcal{O}(\alpha_s) \,, \qquad 
   E_2^C(y) = - \frac{1}{2y} + \mathcal{O}(\alpha_s) \,.
\end{equation}
The contributions of evanescent operators other than $\tilde O_1^B$ and $\tilde O_2^C$ vanish in our projection scheme.

\subsection{Results for the bare hard matching coefficients}

Based on the operator basis constructed in Section~\ref{subsec:SCET1basis}, we write the SCET-1 representation for the operator $\mathcal{O}_\ell^{V,LL}$ in \eqref{eq:OVLLdef} in the form 
\begin{equation}\label{eq:LeffSCET1}
\begin{aligned}
\mathcal{O}_\ell^{V,LL} 
   & \to \sum_{i=1}^4 H_i^A\,O_i^A
    + \int_0^1\!dy\,\bigg[ \sum_{i=1}^4 H_i^B(y)\,O_i^B(y) 
    + \sum_{i=1}^5 H_i^C(y)\,O_i^C(y) \bigg] \\
   &\quad + \int_0^1\!dy_1\!\int_0^1\!dy_2\,\theta(1-y_1-y_2)\,
    \sum_{i=1}^8 H_i^D(y_1,y_2)\,O_i^D(y_1,y_2) \\
   &\quad + \int_0^1\!dy_1\!\int_0^1\!dy_2\!\int_0^1\!dy_3\,\theta(1-y_1-y_2-y_3)\,\sum_{i=1}^4
    H_i^E(y_1,y_2,y_3)\,O_i^E(y_1,y_2,y_3) \\
   &\quad + H_1^F\spac O_1^F \,,
\end{aligned}
\end{equation}
where for simplicity we suppress the dependence of the Wilson coefficients on the hard scales $m_b$ and $\nb\cdot\P_\hc$. Without loss of generality, we evaluate all operators at the spacetime point $x=0$. Otherwise, there would be a phase factor $e^{-im_b\spac v\cdot x}$ in front of each term, and soft fields would need to be multipole expanded in interactions with hard-collinear fields \cite{Beneke:2002ph,Beneke:2002ni}. It is straightforward to calculate the tree-level matching conditions for the hard functions in \eqref{eq:LeffSCET1} by performing the substitutions 
\begin{align}\label{eq:match_fields}
   b &\to \bigg[ 1 - \frac{1}{m_b\,v\cdot n}\,\frac{\nsl}{2} 
    \left( Q_b\,\Asl_\hc^\perp + \Gsl_\hc^\perp \right) 
    + \left( \frac{i\slashed D_s}{2m_b} + \dots \right)
    + \mathcal{O}(\lambda^{3/2}) \bigg]\,b_v \,, \notag\\
   u &\to \left[ 1 - \frac{1}{i\nb\cdot\partial}\,\frac{\nbsl}{2} 
    \left( Q_u\,\Asl_\hc^\perp + \Gsl_\hc^\perp \right) \right] u_s \notag\\
   &\quad + \left[ 1 - \frac{1}{i\nb\cdot\partial}\,\frac{\nbsl}{2} 
    \left( i\Dsl_s^\perp + Q_u\,\Asl_\hc^\perp + \Gsl_\hc^\perp \right) \right] \X_\hc^{(u)} \,, \\
   \ell 
   &\to \bigg[ 1 + \mathcal{O}(\lambda^{\frac12}) \bigg]\,\ell_s 
    + \left[ 1 - \frac{1}{i\nb\cdot\partial}\,\frac{\nbsl}{2} 
    \left( i\Dsl_s^\perp + Q_\ell\,\Asl_\hc^\perp - m_\ell \right) \right] \X_\hc^{(\ell)} \,, \notag\\ 
   \nu_\ell &\to \nu_\cb \notag
\end{align}
in the four-fermion operator $\mathcal{O}_\ell^{V,LL}$, where all fields are evaluated at $x=0$. These replacements account for the hard matching corrections resulting from integrating out the small components of the heavy-quark spinor in HQET and the hard-collinear spinors in SCET-1. The contributions of hard-collinear gauge fields to the matching relations for the heavy-quark and up-quark fields have been derived in \cite{Beneke:2002ph}. While the $\mathcal{O}(\lambda)$ corrections in the relation for the heavy-quark spinor give rise to SCET-1 operators that vanish by virtue of the magic identity \eqref{eq:magic}, the term involving $i\slashed D_s/2m_b$ is needed for the correct treatment of evanescent operators, see the operator $\tilde O_3^C$ in \eqref{eq:Otilde1}. The dots refer to further $\mathcal{O}(\lambda)$ terms involving collinear gauge fields, which vanish in our reduction scheme at least to one-loop order. The $\mathcal{O}(\lambda^{\frac12})$ corrections to the soft lepton field would give rise to power-suppressed SCET-1 operators.

\begin{figure}
\centering
\includegraphics[scale=0.45]{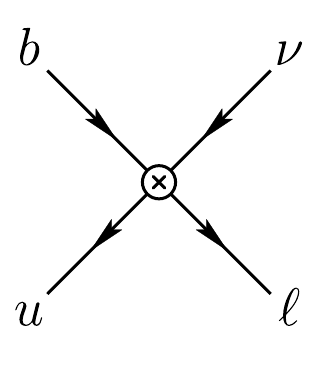} \quad
\includegraphics[scale=0.45]{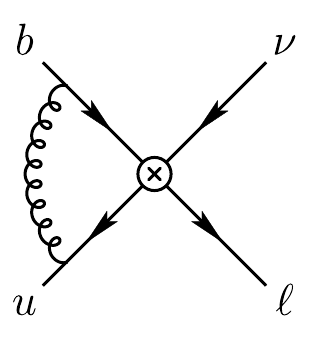} \quad
\includegraphics[scale=0.45]{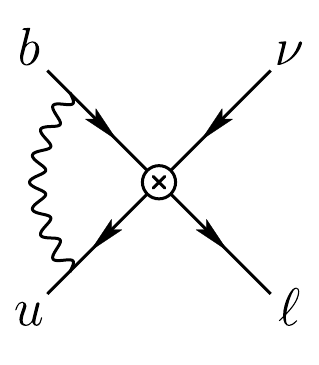} \quad
\includegraphics[scale=0.45]{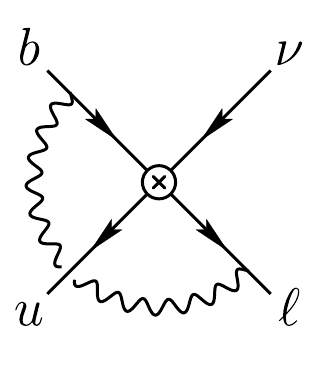} 
\caption{Tree-level and one-loop Feynman diagrams for the LEFT to SCET-1 matching for the type-$A$ four-fermion operators. There is also a contribution from the wave-function renormalization of the $b$ quark.}
\label{fig:graphs_left}
\end{figure}

For the type-$A$ operators, it is possible to include the first-order QCD corrections using known results from the literature. For the case of the operators $O_1^A$ and $O_2^A$, it is necessary to also include the one-loop QED corrections, which we compute here for the first time. The relevant diagrams are shown in Figure~\ref{fig:graphs_left}. For all other operators, a photon loop is required when performing the matching onto SCET-2, and hence we give their coefficients at zeroth order in $\alpha$. Working in dimensional regularization with $d=4-2\epsilon$ spacetime dimensions, we obtain the bare expressions 
\begin{align}\label{eq:HAibare}
   H_1^A
   &= \mathcal{C}_1^{\rm HQET} + \frac12\,\mathcal{C}_2^{\rm HQET}
    - \frac{Q_b\spac\alpha}{4\pi}\,\bigg\{ 
    Q_\ell\,\bigg[ \frac{1}{\epsilon^2} + \frac{L}{\epsilon} 
    - \frac{1}{\epsilon} + \frac{L^2}{2} - L - \frac{z\ln z}{z-1} 
    + 2\spac\text{Li}_2(1-z) + 1 + \frac{\pi^2}{12} \bigg] \notag\\
   &\hspace{5.45cm} + Q_b \left( \frac{3}{2\epsilon} + \frac32\,L_m + 2 \right) 
    + Q_u \bigg\} \notag\\
   &\quad - Q_\ell\spac(Q_b+Q_u)\,\frac{\alpha}{16\pi}\,\kappa 
    + \delta H_1^A \,, \notag\\
   H_2^A 
   &= \frac12\,\mathcal{C}_2^{\rm HQET} 
    + \frac{Q_b\spac\alpha}{4\pi} \left[ 
    z\spac Q_\ell \left( \frac{1}{\epsilon} + L + \ln z + \frac{z-1}{z} \right) 
    + Q_u \right] , \\
   H_3^A 
   &= - \mathcal{C}_1^{\rm HQET} - \frac12\,\mathcal{C}_2^{\rm HQET} 
    + \mathcal{O}(\alpha) \,, \notag\\
   H_4^A 
   &= - \frac12\,\mathcal{C}_2^{\rm HQET} + \mathcal{O}(\alpha) \,, \notag
\end{align}
where the scale $\mu$ is defined through dimensional transmutation in the $\overline{\mathrm{MS}}$ scheme, and we have introduced 
\begin{equation}\label{eq:Ldef}
   L = \ln\frac{\mu^2\,(\nb\cdot v)^2}{(\nb\cdot\P_\hc)^2} \,, \qquad
   L_m = \ln\frac{\mu^2}{m_b^2} \,, \qquad
   z = \frac{\nb\cdot\P_\hc}{m_b\,\nb\cdot v} \,.
\end{equation}
The coefficients 
\begin{equation}\label{eq:HQETcoefs}
\begin{aligned}
   \mathcal{C}_1^{\rm HQET} 
   &= 1 + \frac{C_F\spac\alpha_s}{4\pi} 
    \left( - \frac{3}{2\epsilon} - \frac32\,L_m - 4 \right) 
    + \mathcal{O}(\alpha_s^2) \,, \\
   \mathcal{C}_2^{\rm HQET} 
   &= \frac{C_F\spac\alpha_s}{2\pi} + \mathcal{O}(\alpha_s^2) \,.
\end{aligned}
\end{equation}
appear in the leading-order matching relations for the heavy-light vector current in HQET \cite{Ji:1991pr,Neubert:1993mb}. Note that the scheme-dependent terms proportional to $\kappa$ cancel the $\kappa$-dependent terms in the matching coefficient $K_\mathrm{EW}$ in \eqref{eq:KEW}. The quantity $\delta H_1^A$ is the remnant shift from eliminating evanescent operators and has been given in \eqref{eq:deltaH1A}. For our process of interest, the variable $z$ evaluates to $z=m_B/m_b$, where $m_b$ denotes the pole mass of the $b$ quark. This ratio equals~1 up to power corrections of $\mathcal{O}(\lambda)$, but we prefer to work with the general expressions given above.

The coefficients of the type-$B$ operators take the form 
\begin{equation}
\begin{aligned}\label{eq:HiBres}
   H_1^B(y) 
   &= - Q_\ell \left( \mathcal{C}_1^{\rm HQET} 
    + \frac12\,\mathcal{C}_2^{\rm HQET} \right)
    + \left[ \frac{Q_u}{y} + \mathcal{O}(\alpha_s) \right] 
    = - Q_\ell + \frac{Q_u}{y} + \mathcal{O}(\alpha_s) \,, \\   
   H_2^B(y) 
   &= - \frac{Q_\ell}{2}\,\mathcal{C}_2^{\rm HQET} 
    + \big[ z\spac Q_b + \mathcal{O}(\alpha_s) \big]
    = z\spac Q_b + \mathcal{O}(\alpha_s) \,, \\   
   H_3^B(y) 
   &= \frac{1}{y} + \mathcal{O}(\alpha_s) \,, \\   
   H_4^B(y) 
   &= z + \mathcal{O}(\alpha_s) \,,  
\end{aligned}
\end{equation}
where the QCD corrections proportional to $Q_\ell$ are again determined in terms of the coefficients in \eqref{eq:HQETcoefs}. The yet unknown QCD corrections are associated with the terms involving hard-collinear gauge fields in the $\mathcal{O}(\lambda)$ contribution to the matching relations for the $b$ quark in \eqref{eq:match_fields} (marked by the dots). For two of the type-$C$ operators one can use known expressions for the matching relations of heavy-light current operators in SCET-1 to derive the matching coefficients including QCD corrections. For the remaining coefficients only the tree-level expressions are known, because they can receive contributions from yet unknown second-order ($\sim\lambda$) power corrections to the SCET-1 heavy-light current operators. We find 
\begin{equation}\label{eq:HiCres}
\begin{aligned}
   H_1^C(y)
   &= \frac{1}{1-y} \left[ \mathcal{C}_1^A(y) + \frac12\,\mathcal{C}_2^A(y) 
    + \mathcal{C}_3^A(y) \right]
    = \frac{1}{1-y} + \mathcal{O}(\alpha_s) \,, \\   
   H_2^C(y) 
   &= - \frac{1}{y\spac(1-y)} \left[ \,\frac12\,\mathcal{C}_2^A(y) 
    + \mathcal{C}_3^A(y) \right] 
    = \mathcal{O}(\alpha_s) \,, \\ 
   H_3^C(y) 
   &= \frac{1}{y} + \mathcal{O}(\alpha_s) \,, \\   
   H_4^C(y) 
   &= - \frac{1}{1-y} + \mathcal{O}(\alpha_s) \,, \\   
   H_5^C(y) 
   &= \mathcal{O}(\alpha_s) \,,
\end{aligned}
\end{equation}
where \cite{Bauer:2003pi,Bosch:2004th} 
\begin{equation}
\begin{aligned}
   \mathcal{C}_1^A(y) 
   &= 1 + \frac{C_F\spac\alpha_s}{4\pi}\,\bigg[ 
    - \frac{1}{\epsilon^2} - \frac{1}{\epsilon}\,L_y - \frac{5}{2\epsilon} 
    - \frac12\,L_y^2 - L_y - \frac32\,L_m \\
   &\hspace{2.63cm} - \frac{yz \ln yz}{1-yz} - 2\spac\mbox{Li}_2(1-yz) - 6 
    - \frac{\pi^2}{12} \bigg] + \mathcal{O}(\alpha_s^2) \,, \\
   \mathcal{C}_2^A(y) 
   &= \frac{C_F\spac\alpha_s}{4\pi}\,\frac{2}{1-yz} 
    \left( \frac{yz\ln yz}{1-yz} + 1 \right) + \mathcal{O}(\alpha_s^2) \,, \\
   \mathcal{C}_3^A(y) 
   &= \frac{C_F\spac\alpha_s}{4\pi}\,\frac{yz}{1-yz} 
    \left( \frac{1-2yz}{1-yz}\,\ln yz - 1 \right) + \mathcal{O}(\alpha_s^2) \,,
\end{aligned}
\end{equation}
where 
\begin{equation}
   L_y = \ln  \frac{\mu^2\,(\nb\cdot v)^2}{\left(y\,\nb\cdot\P_\hc\right)^2} \,.
\end{equation}

\begin{figure}
\centering
\includegraphics[scale=0.5]{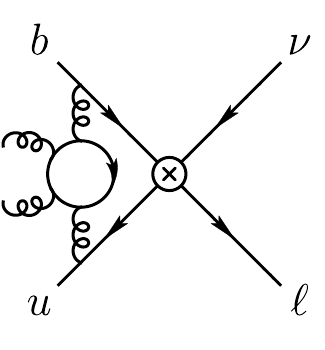} \quad
\includegraphics[scale=0.5]{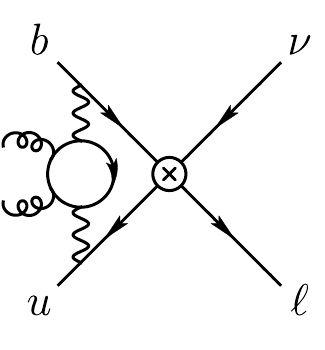} \quad
\includegraphics[scale=0.5]{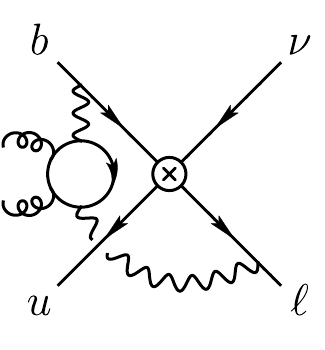} 
\caption{Hard two-loop diagrams which can generate the operator $O_4^E$ in SCET-1. The light quark, the charged lepton and the external gluons carry hard-collinear momenta.}
\label{fig:new}
\end{figure}

The coefficients of the remaining operators will only be needed at tree level. For the type-$D$ operators, we obtain 
\begin{align}
   H_1^D(y_1,y_2) 
   &= - \frac{z\spac Q_b}{1-y_1-y_2} + \mathcal{O}(\alpha_s) \,, \notag\\ 
   H_2^D(y_1,y_2)  
   &= - \frac{z}{1-y_1-y_2} + \mathcal{O}(\alpha_s) \,, \notag\\
   H_3^D(y_1,y_2)  
   &= - \frac{Q_\ell}{1-y_1} + \frac{Q_u}{y_1+y_2} + \mathcal{O}(\alpha_s) \,, \notag\\
   H_4^D(y_1,y_2)   
   &= \frac{1}{y_1+y_2} + \mathcal{O}(\alpha_s) \,, \notag\\ 
   H_5^D(y_1,y_2) 
   &= - \frac{z\spac Q_b}{y_1} + \mathcal{O}(\alpha_s) \,, \\ 
   H_6^D(y_1,y_2) 
   &= - \frac{z}{y_1} + \mathcal{O}(\alpha_s) \,, \notag\\ 
   H_7^D(y_1,y_2) 
   &= \frac{z\spac Q_b}{1-y_1-y_2} + \mathcal{O}(\alpha_s) \,, \notag\\ 
   H_8^D(y_1,y_2) 
   &= \frac{z}{1-y_1-y_2} + \mathcal{O}(\alpha_s) \,. \notag
\end{align}
At $\mathcal{O}(\alpha_s)$, these coefficients receive contributions from yet unknown higher-order power corrections to the heavy-light currents in SCET-1. For the coefficients of the type-$E$ operators, we find 
\begin{equation}
\begin{aligned}
   H_1^E(y_1,y_2,y_3)
   &= \frac{z\spac Q_\ell\spac Q_b}{1-y_1-y_2} - \frac{z\spac Q_b\spac Q_u}{y_1+y_3} 
    + \mathcal{O}(\alpha_s) \,, \\
   H_2^E(y_1,y_2,y_3) 
   &= \frac{z\spac Q_\ell}{1-y_1-y_2} - \frac{z\spac Q_b}{y_1+y_2} 
    - \frac{z\spac Q_u}{y_1+y_3} + \mathcal{O}(\alpha_s) \,, \\ 
   H_3^E(y_1,y_2,y_3) 
   &= - \frac{z}{y_1+y_2} + \mathcal{O}(\alpha_s) \,, \\ 
   H_4^E(y_1,y_2,y_3) 
   &= \mathcal{O}(\alpha_s^2, \alpha\spac\alpha_s) \,.
\end{aligned}
\end{equation}
The coefficient $H_4^E$ can first be generated at two-loop order from diagrams such as those shown in Figure~\ref{fig:new}. For the coefficient of the type-$F$ operator, we get
\begin{equation}
   H_1^F(\mu) = \frac12 + \mathcal{O}(\alpha_s) \,.
\end{equation}

Finally, we list the tree-level matching coefficients for the evanescent operators defined in \eqref{eq:Otilde1}. We obtain
\begin{align}
   E^B_1(y) &= - \frac{Q_u}{2y} + \mathcal{O}(\alpha_s) \,, \notag\\
   E^B_2(y) &= - \frac{z\spac Q_b}{2} + \mathcal{O}(\alpha_s) \,, \notag\\
   E^C_1(y) &= 1 + \mathcal{O}(\alpha_s) \,, \notag\\
   E^C_2(y) &= - \frac{1}{2y} + \mathcal{O}(\alpha_s) \,, \\
   E^C_3(y) &= \frac{z}{4} + \mathcal{O}(\alpha_s) \,, \notag\\
   E^F_1(y) &= 1 + \mathcal{O}(\alpha_s) \,. \notag
\end{align}
Two of these relations have already been given in \eqref{eq:evanescentHi}.

\subsection[\texorpdfstring{Simplifications arising at $\mathcal{O}(\alpha)$ in perturbation theory}{Simplifications arising at O(alpha) in perturbation theory}]{\boldmath Simplifications arising at $\mathcal{O}(\alpha)$ in perturbation theory}\label{sec:order_alpha_simp}

\begin{table}[tp]
\centering
\begin{tabular}{ccccccc}
\hline
\rowcolor{\shadecolor{30}}
Operator & Hard function\ & Jet function$^*$\ & \multicolumn{4}{c}{Contribution to decay amplitude} \\ 
\hline
\rowcolor{\shadecolor{15}}
\multicolumn{3}{l}{Perturbative approximation:} & $\mathcal{O}(\alpha^0)$ & $\mathcal{O}(\alpha)$ & $\mathcal{O}(\alpha\spac\alpha_s)$ & \!\!$\mathcal{O}(\alpha\spac\alpha_s^2,\alpha^2)$\!\! \\
\hline
$O_1^A$ & 1 & 1 & \checkmark & \checkmark & \checkmark & \checkmark \\
$O_2^A$ & $\alpha_s$ & 1 & \checkmark & \checkmark & \checkmark & \checkmark \\
$O_3^A$ & 1 & 0 & --- & --- & --- & --- \\
$O_4^A$ & $\alpha_s$ & 0 & --- & --- & --- & --- \\
\hline
$O_{1,2}^B$ & 1 & $\alpha$ & --- & \checkmark & \checkmark & \checkmark \\
$O_{3,4}^B$ & 1 & 0 & --- & --- & --- & --- \\
\hline
$O_{1,3,4}^C$ & 1 & $\alpha$ & --- & \checkmark & \checkmark & \checkmark \\
$O_2^C$ & $\alpha_s$ & $\alpha$ & --- & --- & \checkmark & \checkmark \\
$O_5^C$ & $\alpha_s$ & $\alpha^\dagger$ & --- & --- & --- & --- \\
\hline
$O_{1,3,5,7}^D$ & 1 & $\alpha^2$ & --- & --- & --- & \checkmark \\
$O_{2,4,6,8}^D$ & 1 & $\alpha\spac\alpha_s$ & --- & --- & \checkmark & \checkmark \\
\hline
$O_1^E$ & 1 & $\alpha^3$ & --- & --- & --- & --- \\
$O_2^E$ & 1 & $\alpha^2\spac\alpha_s$ & --- & --- & --- & --- \\
$O_3^E$ & 1 & $\alpha\spac\alpha_s^2$ & --- & --- & --- & \checkmark \\
$O_4^E$ & $\alpha_s^2$ & $\alpha\spac\alpha_s^2$ & --- & --- & --- & --- \\
\hline
$O_1^F$ & 1 & $\alpha^\dagger$ & --- & --- & --- & \checkmark \\
\hline
\end{tabular}
\caption{\label{tab:scalings}
Perturbative scaling of the hard functions of the SCET-1 basis operators defined in Section~\ref{subsec:SCET1basis}, and of the jet functions arising in their matching onto SCET-2. The star indicates that for the type-$B$, $D$ and $E$ operators the indicated order refers to the product of the jet function and the low-energy matrix element, as in some cases the low-energy matrix element may include a collinear photon loop. The product of the entries in the first two columns gives the (largest possible) scaling behavior of the corresponding contributions to the $B^-\to\mu^-\spac\bar\nu_\mu$ decay amplitude. Entries in the third column marked with a dagger are naively of the given order, but vanish at that order upon explicit calculation.}
\end{table}

While the detailed matching onto SCET-2 will be discussed in Section~\ref{sec:SCET-2}, a simple inspection of the relevant Feynman graphs can be used to deduce at which order in the couplings $\alpha_s$ and $\alpha$ a given basis operator might contribute to the $B^-\to\ell^-\spac\bar\nu_\ell$ decay amplitude (in some cases these contributions may still turn out to be zero upon explicit calculation). The results of such an analysis are shown in Table~\ref{tab:scalings}. When QED corrections are neglected, only the operators $O_1^A$ and $O_2^A$ contribute to the decay amplitude, and to all orders in QCD perturbation theory the sum of their contributions can be expressed in terms of the $B$-meson decay constant $f_{B}$ defined in QCD. In the approximation where one works consistently at first order in $\alpha$ or $\alpha_s$, but neglects two-loop mixed QCD--QED corrections, the operators $O_{1,2}^A$, $O_{1,2}^B$, and $O_{1,3,4}^C$ need to be retained. When the goal is to work consistently up to $\mathcal{O}(\alpha\spac\alpha_s)$, in addition the type-$C$ operator $O_2^C$ and the type-$D$ operators $O_{2,4,6,8}^D$ need to be included. The remaining type-$D$ operators and the type-$F$ operator could first contribute at $\mathcal{O}(\alpha^2)$, while the leading type-$E$ operator $O_3^E$ could first enter at $\mathcal{O}(\alpha\spac\alpha_s^2)$. For all practical purposes, these latter three contributions will play no role for the foreseeable future.

In Section~\ref{sec:SCET-2}, we will construct the most general SCET-2 operator basis, so that the analysis of the $B^-\to\mu^-\spac\bar\nu_\mu$ decay amplitude can in principle be performed at any order in perturbation theory. For practical applications, we will then limit ourselves to $\mathcal{O}(\alpha)$ accuracy, including however the leading logarithmic QCD and QED corrections to all orders in perturbation theory.

\subsection{Renormalization of the hard functions}

We now discuss the renormalization of the hard functions $H_i^X$ at one-loop order in both QCD and QED. From the discussion in the previous section we know that in this approximation the operator basis can be restricted to the operators $O_{1,2}^A$, $O_{1,2}^B$, and $O_i^C$, where the type-$B$ and $C$ operators depend on the momentum variable $y$. Only the type-$A$ hard functions contain divergences from QED, as exhibited in \eqref{eq:HAibare} (type-$B$ and type-$C$ hard functions are only needed at tree level). 

\begin{figure}
\centering
\includegraphics[scale=0.65]{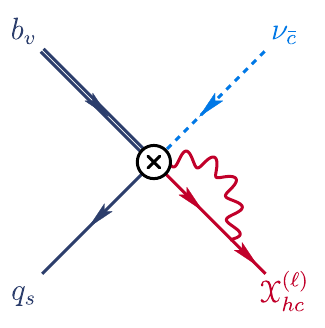}\qquad\qquad
\includegraphics[scale=0.65]{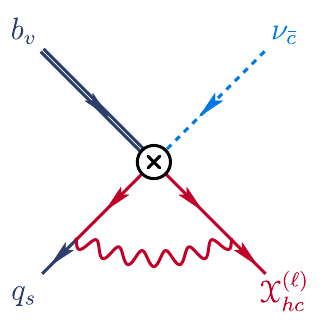}
\caption{Left: One-loop diagrams leading to a mixing of the operators $O_i^B$ and $O_i^A$ (with $i=1,2$). The lepton mass insertion is not shown explicitly. Right: One-loop diagram leading to a mixing of the operators $O_i^C$ and $O_i^A$ (with $i=1,2$). Here and in later figures, red lines refer to hard-collinear fields, gray lines to soft fields, and light blue lines to collinear or anti-collinear fields. The soft heavy-quark is represented by a double line.}
\label{fig:O1Aren_mixing_diags}
\end{figure}

In general, the SCET-1 operators with the same quantum numbers and mass dimension can mix with each other under renormalization. We define renormalization factors $Z_{ij}^{XX'}$ via the renormalization condition
\begin{equation}\label{eq:RGops}
   O_i^X(\mu) 
   = \sum_{j,X'}\spac Z_{ij}^{XX'}(\mu)\otimes O_j^{X'} \,,
\end{equation}
which relates the renormalized (and scale-dependent) operators $O_i^X(\mu)$ to the bare operators $O_j^{X'}$. The labels $X$ and $X'$ refer to the operator type ($A$, $B$ or $C$), and the symbol $\otimes$ means an integration over the momentum variables shared by the quantities $Z_{ij}^{XX'}$ and $O_j^{X'}$. Since we are working with a basis of subleading-power operators, there can be mixing of operators with different scaling dimensions. Specifically, the time-ordered products of the type-$B$ operators ($\sim\lambda^4$) with the power-suppressed lepton mass term ($\sim\lambda_\ell/\lambda^{\frac12}$) in the SCET-1 Lagrangian can mix with the type-$A$ operators ($\sim\lambda_\ell\,\lambda^{\frac72}$). Calculating the UV divergences of the one-loop diagram on the left in Figure~\ref{fig:O1Aren_mixing_diags}, we find that the corresponding off-diagonal renormalization factors are
\begin{equation}
   Z_{11}^{BA}(y,\mu) = Z_{22}^{BA}(y,\mu) 
   = Q_\ell\,\frac{\alpha}{2\pi\epsilon}\,y \,,
\end{equation}
while the remaining entries $Z_{ij}^{BA}$ vanish. The scale dependence of the renormalization factors arises via dimensional transmutation, since the dimensionless bare coupling $\alpha\propto\mu^{-2\epsilon}$. Similarly, the time-ordered products of some of the type-$C$ operators with the power-suppressed Lagrangian ${\cal L}_{\xi q}^{(1/2)}$ \cite{Beneke:2002ph}, which converts a hard-collinear to a soft quark, can mix with the type-$A$ operators. Calculating the UV divergences of the diagram shown on the right in Figure~\ref{fig:O1Aren_mixing_diags}, we obtain for the off-diagonal renormalization factors at one-loop order
\begin{equation}\label{eq:Z11CA}
\begin{aligned}
   Z_{11}^{CA}(y,\mu) 
   &= Q_\ell\spac Q_u\,\frac{\alpha}{2\pi}\,(1-y)
    \left[ \frac{1}{y^{1+\epsilon}} 
    \left( \frac{1}{\epsilon} + \ln\frac{\mu^2}{\nb\cdot\P_\hc\,n\cdot l_s} + 1 \right)
    + \frac{1}{\epsilon} \right]_{\rm poles} \,, \\ 
   Z_{22}^{CA}(y,\mu)  
   &= - Q_\ell\spac Q_u\,\frac{\alpha}{2\pi}\,y\spac(1-y)^2
    \left[ \frac{1}{y^{1+\epsilon}} 
    \left( \frac{1}{\epsilon} + \ln\frac{\mu^2}{\nb\cdot\P_\hc\,n\cdot l_s} \right)
    \right]_{\rm poles} \,.
\end{aligned}
\end{equation}
The remaining entries $Z_{ij}^{CA}$ vanish. Due to the appearance of a singularity for $y\to 0$, we have kept some subleading terms. One may be tempted to replace 
\begin{equation}\label{eq:distrib}
   \frac{1}{y^{1+\epsilon}} 
   = - \frac{1}{\epsilon}\,\delta(y) + \left[ \frac{1}{y} \right]_+
    + \mathcal{O}(\epsilon) \,,
\end{equation}
in which case all of the terms shown above would give rise to $1/\epsilon^n$ poles. However, as we will soon see, things are actually a bit more subtle, because these renormalization factors will be multiplied by hard functions which are themselves singular at $y=0$. The appearance of the incoming soft momentum of the $\bar u$ quark in the expression for $Z_{11}^{CA}$ is a bit worrisome, since this is an IR scale in SCET-1. We will see later how this puzzle is resolved. At $\mathcal{O}(\alpha)$ there is also a mixing of the type-$C$ operators with the type-$B$ operators, but the corresponding contributions to our process of interest would only appear at $\mathcal{O}(\alpha^2)$, so we can ignore them.

\begin{figure}
\centering
\includegraphics[scale=0.6]{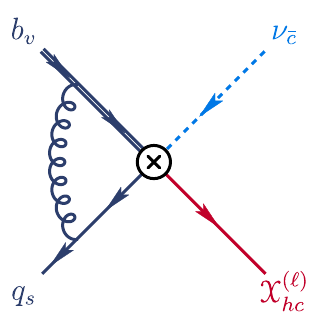} \quad
\includegraphics[scale=0.6]{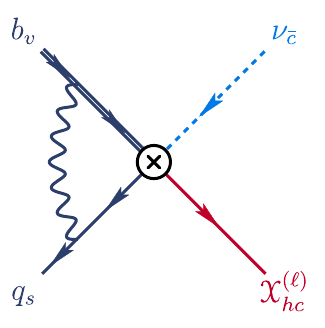} \quad
\includegraphics[scale=0.6]{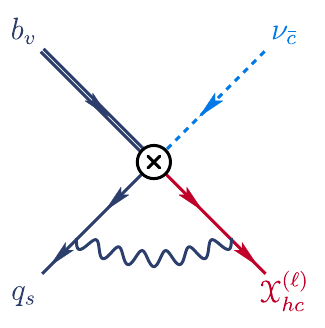} \\
\includegraphics[scale=0.6]{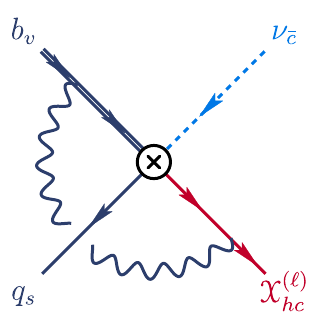} \quad
\includegraphics[scale=0.6]{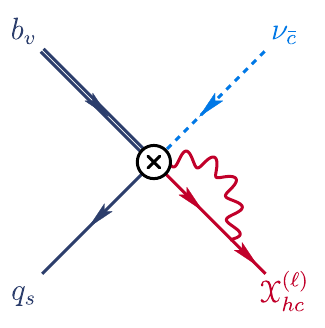}
\caption{One-loop diagrams needed for the calculation of the diagonal renormalization factors $Z_{ii}^{AA}$. These graphs need to be supplemented by wave-function renormalization.}
\label{fig:O1Aren_diags}
\end{figure}

Renormalization is also required within each class of operators. For the operators $O_{1,2}^A$, the corresponding renormalization factors are obtained from the UV poles of the diagrams shown in Figure~\ref{fig:O1Aren_diags}, supplemented by wave-function renormalization and the renormalization of the charged-lepton mass. To regularize IR divergences, we keep the soft external momenta of the quarks off shell and assign an off-shell collinear momentum to the charged lepton. The scale $(-p_\ell^2)$ enters the expressions for the last three diagrams as an infrared (IR) regulator, but it cancels out in their sum. We find that (for $i=1,2$)
\begin{align}\label{eq:ZiiAAfinal}
   Z_{ii}^{AA}(\mu)
   &=  1 + \frac{C_F\spac\alpha_s}{4\pi} \left( - \frac{3}{2\epsilon} \right) \\
   &\quad + \frac{\alpha}{4\pi}\,\bigg[\! 
    - Q_\ell\spac Q_b \left( \frac{1}{\epsilon^2} + \frac{L}{\epsilon} 
    + \frac{1}{\epsilon} \right) 
    + \spac Q_\ell\spac Q_u \left( \frac{2}{\epsilon^2} 
    + \frac{2}{\epsilon}\,\ln\frac{\mu^2}{\nb\cdot\P_\hc\,n\cdot l_s} \right) 
    + \frac{3}{2\epsilon} \left( Q_\ell^2 - Q_u^2 \right) \!\bigg] . \notag
\end{align}
Note that a dependence on the incoming soft momentum $l_s$ of the $\bar u$ quark remains in this expression.

From \eqref{eq:RGops} it is straightforward to derive the renormalization condition for the hard functions. At one-loop order, we obtain for the coefficients of the type-$A$ operators  (with $i=1,2$)
\begin{equation}\label{eq:Hiren}
   H_i^A(\mu) 
   = Z_{\rm EW}(\mu)\spac \big[ Z_{ii}^{AA}(\mu) \big]^{-1}\,H_i^A
    - \int_0^1\!dy\,Z_{ii}^{BA}(y,\mu)\,H_i^B(y)
    - \int_0^1\!dy\,Z_{ii}^{CA}(y,\mu)\,H_i^C(y) \,, 
\end{equation}
where
\begin{equation}
   Z_{\rm EW}(\mu) = 1 + 3\spac Q_\ell\spac Q_u\frac{\alpha}{4\pi\epsilon} \,.
\end{equation}
While the first two terms on the right-hand side of \eqref{eq:Hiren} are straightforward to evaluate, the last term is problematic, since both functions in the integrand are singular in the limit $y\to 0$. For $i=1$, the renormalization factor $Z_{11}^{CA}(y)\propto y^{-1-n\epsilon}$ at $n$-loop order, while the bare coefficient $H_1^C(y)\propto y^{-n'\epsilon}$ at $n'$ loops. For $i=2$, one finds $Z_{22}^{CA}(y)\propto y^{-n\epsilon}$ and $H_2^C(y)\propto y^{-1-n'\epsilon}$. In either case, the endpoint behavior of the product is $y^{-1-(n+n')\epsilon}$, where the strength of the singularity is determined by the loop order of the product of the two functions. We therefore cannot interpret $Z_{11}^{CA}$ in the distribution sense, as indicated in \eqref{eq:distrib}, and the same is true for $Z_{22}^{CA}$. Instead, we perform a plus-type subtraction of the integrand and rewrite the integral in the form
\begin{equation}\label{eq:3.75}
\begin{aligned}
   \int_0^1\!dy\,Z_{ii}^{CA}(y)\,H_i^C(y) 
   &= \int_0^1\!dy\,\Big[ Z_{ii}^{CA}(y)\,H_i^C(y) 
    - \llbracket Z_{ii}^{CA}(y) \rrbracket\,\llbracket H_i^C(y) \rrbracket \Big] \\
   &\quad + \int_0^1\!dy\,\llbracket Z_{ii}^{CA}(y) \rrbracket\,
    \llbracket H_i^C(y) \rrbracket \,,
\end{aligned}
\end{equation}
where the notation $\llbracket f(y)\rrbracket$ indicates that one must only keep the leading terms of the function $f(y)$ in the limit $y\to 0$. In the first integral on the right-hand side, the leading terms for $y\to 0$ cancel out, so that the integral is convergent and can be evaluated using the naive $\epsilon$ expansions 
\begin{equation}
\begin{aligned}
   Z_{11}^{CA} 
   &= Q_\ell\spac Q_u\,\frac{\alpha}{2\pi\epsilon}\,\frac{1-y^2}{y} \,, & \quad
   \llbracket Z_{11}^{CA} \rrbracket 
   &= Q_\ell\spac Q_u\,\frac{\alpha}{2\pi\epsilon}\,\frac{1}{y} \\ 
   Z_{22}^{CA} 
   &= - Q_\ell\spac Q_u\,\frac{\alpha}{2\pi\epsilon}\,(1-y)^2 \,, & \quad
   \llbracket Z_{22}^{CA} \rrbracket 
   &= - Q_\ell\spac Q_u\,\frac{\alpha}{2\pi\epsilon} \,,
\end{aligned}
\end{equation}
along with 
\begin{equation}
   \llbracket H_1^C(y) \rrbracket = 1 + \mathcal{O}(\alpha_s) \,, \qquad
   \llbracket H_2^C(y) \rrbracket = \mathcal{O}(\alpha_s) \,.
\end{equation}
For the last term, it is necessary to keep the full expressions in \eqref{eq:Z11CA}, consider the singularities of the product of the two functions {\em together}, and evaluate the integral in dimensional regularization. In order for the result to fit the structure of a renormalization condition, it must be possible to factorize the integral into one of the hard functions already present in our operator basis and a renormalization factor \cite{Liu:2019oav,Liu:2020wbn}. This is ensured by the exact $d$-dimensional refactorization conditions 
\begin{equation}\label{eq:refact}
\begin{aligned}
   \llbracket H_1^C(y) \rrbracket 
   &= H_1^A\,S_1^C\bigg(\frac{y\spac\nb\cdot\P_\hc}{\nb\cdot v} \bigg) \,, \\
   \llbracket H_2^C(y) \rrbracket 
   &= - \frac{H_2^A}{y}\,S_2^C\bigg(\frac{y\spac\nb\cdot\P_\hc}{\nb\cdot v} \bigg) \,,
\end{aligned}
\end{equation}
which must hold to all orders of perturbation theory. Here $H_{1,2}^A$ are the bare hard matching coefficient of the type-$A$ operators $O_{1,2}^A$, and $S_{1,2}^C=1+{\cal O}(\alpha_s)$ are functions of the indicated combination of momentum variables, which becomes soft in the region $y\to 0$. For example, performing a region analysis of the function $\llbracket H_1^C\rrbracket$ at one-loop order in QCD, we find 
\begin{equation}\label{eq:S1C}
   S_1^C(\bar\omega)
   = 1 - \frac{C_F\spac\alpha_s}{4\pi} \left( \frac{\mu^2}{\bar\omega^2} \right)^\epsilon 
    \left[ \frac{1}{\epsilon^2} + \frac{1}{\epsilon} + 2 + \frac{5\pi^2}{12} 
    + \mathcal{O}(\epsilon) \right] .
\end{equation}
Evaluating the last integral in \eqref{eq:3.75} using the expressions \eqref{eq:Z11CA} and 
\eqref{eq:refact}, we obtain
\begin{equation}
\begin{aligned}
   \int_0^1\!dy\,\llbracket Z_{11}^{CA}(y) \rrbracket\,\llbracket H_1^C(y) \rrbracket 
   &= - H_1^A\,Q_\ell\spac Q_u\,\frac{\alpha}{2\pi} \left( \frac{1}{\epsilon^2} 
    + \frac{1}{\epsilon} \ln\frac{\mu^2}{\nb\cdot\P_\hc\,n\cdot l_s}
    + \frac{1}{\epsilon} \right) + \mathcal{O}(\alpha\spac\alpha_s) \,, \\
   \int_0^1\!dy\,\llbracket Z_{22}^{CA}(y) \rrbracket\,\llbracket H_2^C(y) \rrbracket 
   &= - H_2^A\,Q_\ell\spac Q_u\,\frac{\alpha}{2\pi} \left( \frac{1}{\epsilon^2} 
    + \frac{1}{\epsilon} \ln\frac{\mu^2}{\nb\cdot\P_\hc\,n\cdot l_s}
    \right) + \mathcal{O}(\alpha\spac\alpha_s) \,.
\end{aligned}
\end{equation}
This contribution can be combined with the first term on the right-hand side of \eqref{eq:Hiren} and has the effect of canceling a corresponding term in $[Z_{ii}^A]^{-1}$. In the refactorization-based subtraction (RBS) scheme \cite{Liu:2019oav,Liu:2020wbn,Beneke:2020ibj,Beneke:2022obx} employed here, we find at one-loop order\footnote{This result is analogous to expression (4.29) in \cite{Liu:2020wbn}, where the RBS scheme was first applied to a subleading-power observable in SCET.}
\begin{equation}\label{eq:H1Arenorm}
\begin{aligned}
   H_1^A(\mu) 
   &= \bigg\{ 1 + \frac{C_F\spac\alpha_s}{4\pi} \left( \frac{3}{2\epsilon} \right) \\
   &\hspace{1.08cm} + \frac{\alpha}{4\pi} \left[ Q_\ell\spac Q_b 
    \left( \frac{1}{\epsilon^2} + \frac{L}{\epsilon} + \frac{1}{\epsilon} \right) 
    - \frac{3}{\epsilon}\,Q_\ell^2 + \frac{3}{2\epsilon}\,Q_b^2
    + \frac{2}{\epsilon}\,Q_\ell\spac Q_u 
    \right] \!\bigg\}\,H_1^A \\
   &\quad - \int_0^1\!dy\,Z_{11}^{BA}(y)\,H_1^B(y) 
    - \int_0^1\!dy\,\Big[ Z_{11}^{CA}(y)\,H_1^C(y) 
    - \llbracket Z_{11}^{CA}(y) \rrbracket\,\llbracket H_1^C(y) \rrbracket \Big] \\
   &= H_1^A + \frac{C_F\spac\alpha_s}{4\pi} \left( \frac{3}{2\epsilon} \right) 
    + \frac{\alpha}{4\pi} \left[ Q_\ell\spac Q_b 
    \left( \frac{1}{\epsilon^2} + \frac{L}{\epsilon} - \frac{1}{\epsilon} \right) 
    + \frac{3}{2\epsilon}\,Q_b^2 \right] ,
\end{aligned}
\end{equation}
and
\begin{equation}\label{eq:H2Arenorm}
\begin{aligned}
   H_2^A(\mu) 
   &= \bigg\{ 1 + \frac{C_F\spac\alpha_s}{4\pi} \left( \frac{3}{2\epsilon} \right) 
    + \frac{\alpha}{4\pi} \left[ Q_\ell\spac Q_b 
    \left( \frac{1}{\epsilon^2} + \frac{L}{\epsilon} + \frac{1}{\epsilon} \right) 
    - \frac{3}{\epsilon}\,Q_\ell^2 + \frac{3}{2\epsilon}\,Q_b^2 \right]
    \!\bigg\}\,H_2^A \\
   &\quad - \int_0^1\!dy\,Z_{22}^{BA}(y)\,H_2^B(y) 
    - \int_0^1\!dy\,\Big[ Z_{22}^{CA}(y)\,H_2^C(y) 
    - \llbracket Z_{22}^{CA}(y) \rrbracket\,\llbracket H_2^C(y) \rrbracket \Big] \\
   &= H_2^A - z\spac Q_\ell\spac Q_b\,\frac{\alpha}{4\pi\epsilon} \,.
\end{aligned}
\end{equation}
These are precisely the correct counterterms to remove the divergences in the bare functions $H_i^A$ in \eqref{eq:HAibare}. Note that the troublesome terms involving the IR scale $n\cdot l_s$ have disappeared.  

The appearance of the soft momentum $n\cdot l_s$ in the expressions for the renormalization factors in \eqref{eq:Z11CA} and \eqref{eq:ZiiAAfinal} suggests that one should redefine the operator basis. This will be discussed in more detail in Section~\ref{sec:SCET-2}.

\subsection{RG evolution equations}
\label{subsec:3.7}

We now discuss the scale evolution of the renormalized hard functions and the resummation of the leading large logarithmic corrections in both QCD and QED. The evolution equations governing the scale dependence of the hard functions take the form
\begin{equation}\label{eq:RGmaster}
\begin{aligned}
   \frac{d}{d\ln\mu}\,H_i^X(\mu)
   = - \gamma_\mathrm{EW}\,H_i^X(\mu) 
    + \sum_{j,X'} \Gamma_{ij}^{XX'}(\mu)
    \otimes H_j^{X'}(\mu) \,,
\end{aligned}
\end{equation}
where the anomalous dimensions are related to the renormalization factors by \cite{Becher:2009qa}
\begin{equation}\label{eq:gammadef}
   \Gamma_{ij}^{XX'} 
   = 2 \left( \alpha_s\,\frac{\partial}{\partial\alpha_s} 
    + \alpha\,\frac{\partial}{\partial\alpha} \right) Z_{ij}^{XX'\spac [1]} \,,
\end{equation}
where $Z_{ij}^{XX'\spac[1]}$ denotes the coefficient of the single $1/\epsilon$ pole term. The quantity $\gamma_\mathrm{EW}$ has been given in \eqref{eq:gammaEW}. The symbol $\otimes$ means an integration over the momentum variables shared by $\Gamma_{ij}^{XX'}$ and $H_j^{X'}$. From \eqref{eq:H1Arenorm}, we obtain at one-loop order in QED 
\begin{equation}\label{eq:H1Afin}
\begin{aligned}
   \frac{d}{d\ln\mu}\,H_1^A(\mu) 
   &= \left\{ \gamma_{\rm hl}(\alpha_s) 
    - \frac{\alpha}{2\pi} \left[ Q_\ell\spac Q_b \left( \ln\frac{\mu^2}{m_B^2} + 1 \right) 
    - 3\spac Q_\ell^2 + \frac{3}{2}\,Q_b^2 + 2\spac Q_\ell\spac Q_u 
    \right] \right\} H_1^A(\mu) \\
   &\quad + Q_\ell\,\frac{\alpha}{\pi} \int_0^1\!dy\,y\spac H_1^B(y,\mu) 
    + Q_\ell\spac Q_u\,\frac{\alpha}{\pi} \int_0^1\!\frac{dy}{y}\,
    \Big[ (1-y^2)\,H_1^C(y,\mu) - \llbracket H_1^C(y,\mu) \rrbracket \Big] \,, \\
   \frac{d}{d\ln\mu}\,H_2^A(\mu) 
   &= \left\{ \gamma_{\rm hl}(\alpha_s) 
    - \frac{\alpha}{2\pi}\,\bigg[ Q_\ell\spac Q_b \left( \ln\frac{\mu^2}{m_B^2} + 1 \right) 
    - 3\spac Q_\ell^2 + \frac{3}{2}\,Q_b^2 
    \bigg] \right\} H_2^A(\mu) \\
   &\quad + Q_\ell\,\frac{\alpha}{\pi} \int_0^1\!dy\,y\spac H_2^B(y,\mu) 
    - Q_\ell\spac Q_u\,\frac{\alpha}{\pi} \int_0^1\!dy\,
    \Big[ (1-y)^2\,H_2^C(y,\mu) - \llbracket H_2^C(y,\mu) \rrbracket \Big] \,,
\end{aligned}
\end{equation}
where we have used that $\nb\cdot\P_\hc/\nb\cdot v=m_B$ in our case to simplify the notation. The evolution in pure QCD is governed by the universal anomalous dimension of heavy-light current operators in HQET. Its two-loop expression reads \cite{Ji:1991pr}
\begin{equation}\label{eq:gammahl}
   \gamma_{\rm hl}(\alpha_s) 
   = - 3\spac C_F\,\frac{\alpha_s}{4\pi} 
    - \left( \frac{254}{9} + \frac{56\spac\pi^2}{27} - \frac{20}{9}\,n_f \right) 
    \left( \frac{\alpha_s}{4\pi} \right)^2 
    + \mathcal{O}(\alpha_s^3) \,,
\end{equation}
where $n_f=4$ is the appropriate number of active quark flavors below the scale $m_B$, and the two-loop coefficient has been evaluated for $N_c=3$. 

The remaining hard functions give contributions to the \blnu\ decay amplitude that are suppressed by a power of either $\alpha_s$ or $\alpha$. For these cases, an approximate treatment of QED evolution effects will be sufficient, especially since we will evolve the hard functions over a rather narrow scale interval, from $\mu_h\sim m_B$ to $\mu_\hc\sim\sqrt{m_B\spac\Lambda_\mathrm{QCD}}$. It is then a good approximation to treat QCD evolution effects exactly (to the order in which they are needed), while for QED we only keep the terms in the anomalous dimension giving rise to Sudakov double logarithms, which can be derived using the results of \cite{Becher:2009kw}. In this approximation, the relevant RG equations are diagonal and given by
\begin{equation}\label{eq:HBCfin}
\begin{aligned}
   \frac{d}{d\ln\mu}\,H_i^B(y,\mu) 
   &= \left[ \gamma_{\rm hl}(\alpha_s) 
    - \frac{\alpha}{2\pi} \left( Q_\ell\spac Q_b \ln\frac{\mu^2}{m_B^2} 
    + \dots \right) \right] H_i^B(y,\mu) \,;
    \quad i=1,2 \,, \\
   \frac{d}{d\ln\mu}\,H_i^C(y,\mu) 
   &= \left[ \gamma^A(y,\alpha_s) 
    - \frac{\alpha}{2\pi} \left( Q_\ell\spac Q_b \ln\frac{\mu^2}{(1-y)^2\spac m_B^2} 
    + Q_b\spac Q_u \ln\frac{\mu^2}{y^2\spac m_B^2} + \dots \right) \right] H_i^C(y,\mu) \,,
\end{aligned}
\end{equation}
where the dots indicate the neglected, non-logarithmic QED contributions. The QCD evolution of the type-$C$ hard functions is governed by the universal anomalous dimension $\gamma^A(\alpha_s)$ of heavy-light current operators in SCET \cite{Bauer:2003pi,Bosch:2004th,Beneke:2004rc,Hill:2004if},\footnote{Specifically, the quark current appearing in the operators $O_{1,4}^C$ can be related to $\nb_\mu\spac J_{V1}^{A\spac\mu}$, the quark current appearing in the operators $O_{2,5}^C$ can be related to $J_S^A$ and $\nb_\mu\spac J_{V2}^{A\spac\mu}$, and the quark current appearing in the operator $O_3^C$ can be related to the operator $g_{\perp\mu}^\alpha\spac J_{V3}^{A\spac\mu}$ in the notation of \cite{Hill:2004if}.} 
which takes the general form
\begin{equation}
   \gamma^A(y,\alpha_s)
   = - \frac{C_F}{2}\,\gamma_{\rm cusp}(\alpha_s)\,\ln\frac{\mu^2}{y^2\spac m_B^2} 
    + \tilde\gamma(\alpha_s) \,.
\end{equation}
Here 
\begin{equation}\label{eq:gammacusp}
   \gamma_{\rm cusp}(\alpha_s) 
    = \frac{\alpha_s}{\pi} 
     + \left( \frac{268}{3} - 4\pi^2 - \frac{9}{40}\,n_f \right) 
     \left( \frac{\alpha_s}{4\pi} \right)^2 
     + \mathcal{O}(\alpha_s^3) 
\end{equation}
denotes the light-like cusp anomalous dimension \cite{Korchemsky:1987wg,Korchemskaya:1992je}, where again we have set $N_c=3$ in the two-loop coefficient, and 
\begin{equation}\label{eq:gammatilde}
   \tilde\gamma(\alpha_s)
   = - 5\spac C_F\,\frac{\alpha_s}{4\pi} + \mathcal{O}(\alpha_s^2) \,.
\end{equation}
Note that the QED Sudakov evolution of the type-$C$ hard functions is different from that of the type-$A$ and type-$B$ hard functions. 

The solutions to the (approximate) evolution equations for the type-$B$ coefficients in \eqref{eq:HBCfin} reads 
\begin{equation}\label{eq:HiABres}
\begin{aligned}
   H_1^B(y,\mu) 
   &= U_\mathrm{hl}(\mu,m_B)\,\exp\left[ - Q_\ell\spac Q_b\,\frac{\alpha}{8\pi}\spac
    \ln^2\frac{\mu^2}{m_B^2} \right] \left( \frac{Q_u}{y} - Q_\ell \right) , \\   
   H_2^B(y,\mu) 
   &= U_\mathrm{hl}(\mu,m_B)\,\exp\left[ - Q_\ell\spac Q_b\,\frac{\alpha}{8\pi}\spac 
    \ln^2\frac{\mu^2}{m_B^2} \right] z\spac Q_b \,,
\end{aligned}
\end{equation}
where
\begin{equation}\label{eq:Uhl}
   U_\mathrm{hl}(\mu,m_B)
   = \left( \frac{\alpha_s(\mu)}{\alpha_s(m_B)} 
    \right)^{\!\!-\frac{\gamma_0^{\rm hl}}{2\beta_0}} \left[ 1 
    - \frac{\gamma_1^{\rm hl}\spac\beta_0-\gamma_0^{\rm hl}\spac\beta_1}{2\beta_0^2}\,
    \frac{\alpha_s(\mu)-\alpha_s(m_B)}{4\pi} + \dots \right] 
\end{equation}
encodes the QCD evolution at NLO in RG-improved perturbation theory. The one- and two-loop coefficients $\gamma_0^{\rm hl}$, $\gamma_1^{\rm hl}$ of the anomalous dimension (expanded in powers of $\alpha_s/4\pi$) can be read off from \eqref{eq:gammahl}, and $\beta_0$, $\beta_1$ denote the one- and two-loop coefficients of the QCD $\beta$-function. Corrections to the above solutions arise first at order $\alpha_s$ or $\alpha\ln(\mu^2/m_B^2)$ times the terms of order $[\alpha_s\ln(\mu^2/m_B^2)]^n$ and $[\alpha\ln^2(\mu^2/m_B^2)]^n$, which we have resummed to all orders, and hence it is consistent to drop the two-loop terms in \eqref{eq:Uhl}.

For the solutions to the (approximate) evolution equations for the relevant type-$C$ coefficients, we find
\begin{equation}\label{eq:HiCresRG}
\begin{aligned}
   H_1^C(y,\mu) 
   &= \frac{U_C(\mu,m_B,y)}{1-y}\,\exp\left[ - Q_b^2\,\frac{\alpha}{8\pi}\spac 
    \ln^2\frac{\mu^2}{m_B^2} \right] , \\   
   H_3^C(y,\mu) 
   &= \frac{U_C(\mu,m_B,y)}{y}\,\exp\left[ - Q_b^2\,\frac{\alpha}{8\pi}\spac 
    \ln^2\frac{\mu^2}{m_B^2} \right] , \\   
   H_4^C(y,\mu) 
   &= - \frac{U_C(\mu,m_B,y)}{1-y}\,\exp\left[ - Q_b^2\,\frac{\alpha}{8\pi}\spac 
    \ln^2\frac{\mu^2}{m_B^2} \right] ,  
\end{aligned}
\end{equation}
where
\begin{equation}\label{eq:U_C}
   U_C(\mu,m_B,y)
   = \exp\left[ C_F\spac S_{\gamma_{\rm cusp}}(m_B,\mu) 
    - \frac{C_F\spac\gamma_0\ln y + \tilde\gamma_0}{2\beta_0}\,
    \ln\frac{\alpha_s(\mu)}{\alpha_s(m_B)} \right] .
\end{equation}
The Sudakov exponent \cite{Neubert:2004dd} 
\begin{equation}\label{eq:Sudakovexp}
   S_{\gamma_{\rm cusp}}(m_B,\mu)
   = \frac{\gamma_0}{4\beta_0^2} \left[ \frac{4\pi}{\alpha_s(m_B)} 
    \left( 1 - \frac{1}{r} - \ln r \right)
    + \left( \frac{\gamma_1}{\gamma_0} - \frac{\beta_1}{\beta_0} \right) 
    \left( 1 - r + \ln r \right)
    + \frac{\beta_1}{2\beta_0^2} \ln^2 r + \dots \right] 
\end{equation}
with $r=\alpha_s(\mu)/\alpha_s(m_B)$ encodes the QCD evolution at LO in RG-improved perturbation theory. The one- and two-loop coefficients $\gamma_0$ and $\gamma_1$ of the cusp anomalous dimension can be read off from \eqref{eq:gammacusp} and the one-loop coefficient $\tilde{\gamma}_0$ from \eqref{eq:gammatilde}. We have neglected the $y$ dependence of the QED Sudakov logarithms, which is an effect beyond our accuracy. Note the interesting fact that, for $\mu<m_B$, QCD resummation leads to a power-like divergence of the evolution function for $y\to 0$, such that \begin{equation}
   U_C(\mu,m_B,y) = U_C(\mu,m_B)\,y^{-\delta(\mu)} \,,
    \quad \text{with} \quad
   \delta(\mu) = \frac{C_F\spac\gamma_0}{2\beta_0}\,
    \ln\frac{\alpha_s(\mu)}{\alpha_s(m_B)} \,,
\end{equation}
where $U_C(\mu,m_B)\equiv U_C(\mu,m_B,1)$. Once again, corrections to the above solutions arise first at order $\alpha_s$ or $\alpha\ln(\mu^2/m_B^2)$ times the terms resummed in the exponentials in \eqref{eq:U_C} and \eqref{eq:Sudakovexp}. 

As indicated in Table~\ref{tab:scalings}, the contributions of the type-$B$ and type-$C$ operators to the \blnu\ decay amplitude receive an additional $\mathcal{O}(\alpha)$ suppression from (hard-)collinear photon loops, implying that these contributions are determined up to higher-order terms scaling as $\alpha\spac\alpha_s$ or $\alpha^2\ln(\mu^2/m_B^2)$ times the exponentiated terms. We emphasize that the solutions in \eqref{eq:HiCresRG} are derived under the assumption that the dimensionless variable $y$ can be treated as an $\mathcal{O}(1)$ quantity. In cases where the functions $H_i^C(y,\mu)$ enter convolution integrals in which the region $y\ll 1$ is unsuppressed, it will be necessary to generalize these solutions (see Section~\ref{subsec:2-scales} below). 

We now turn to the solution of the evolution equations \eqref{eq:H1Afin}, which with the explicit results for the RG-evolved hard functions $H_i^B$ and $H_i^C$ take the form
\begin{equation}\label{eq:RGEsHiA}
\begin{aligned}
   \frac{d}{d\ln\mu}\,H_1^A(\mu) 
   &= \gamma_{\rm hl}(\alpha_s)\,H_1^A(\mu) 
    - \frac{\alpha}{2\pi}\,\bigg[ Q_\ell\spac Q_b\spac\ln\frac{\mu^2}{m_B^2} 
    - 2\spac Q_\ell^2 + \frac{3}{2}\,Q_b^2 
    + 3\spac Q_\ell\spac Q_u \bigg]\,H_1^A(\mu) \\
   &\quad + Q_\ell\spac(2\spac Q_u - Q_\ell)\,\frac{\alpha}{2\pi}\,
   U_\mathrm{hl}(\mu,m_B)\,\exp\left[ - Q_\ell\spac Q_b\,\frac{\alpha}{8\pi}\spac
    \ln^2\frac{\mu^2}{m_B^2} \right] \\
   &\quad + Q_\ell\spac Q_u\,\frac{\alpha}{\pi}\,\frac{U_C(\mu,m_B)}{1-\delta(\mu)}\,
    \exp\left[ - Q_b^2\,\frac{\alpha}{8\pi}\spac\ln^2\frac{\mu^2}{m_B^2} \right] , \\
   \frac{d}{d\ln\mu}\,H_2^A(\mu) 
   &= \gamma_{\rm hl}(\alpha_s)\,H_2^A(\mu) 
    - \frac{\alpha}{2\pi}\,\bigg[ Q_\ell\spac Q_b\spac\ln\frac{\mu^2}{m_B^2} \bigg]\,
    H_2^A(\mu) \\
   &\quad + z\spac Q_\ell\spac Q_b\,\frac{\alpha}{2\pi}\,
    U_\mathrm{hl}(\mu,m_B)\,\exp\left[ - Q_\ell\spac Q_b\,\frac{\alpha}{8\pi}\spac
    \ln^2\frac{\mu^2}{m_B^2} \right] .
\end{aligned}
\end{equation}
In the inhomogeneous terms of these equations, we are able to include the leading logarithmic QCD corrections to the terms multiplying $\alpha$, as encoded in the evolution functions $U_\mathrm{hl}(\mu,m_B)$ and $U_C(\mu,m_B)$. Unfortunately, the corresponding corrections to the homogeneous terms in the first line of each equation are not yet known. They would require a two-loop calculation of the contributions to the anomalous dimensions $\Gamma_{ii}^{AA}$ of $\mathcal{O}(\alpha\spac\alpha_s)$. In Section~\ref{sec:vir_ampl}, we will be able to bootstrap the leading higher-order terms from the requirement of RG invariance. 

Given the accuracy of the expressions for the type-$B$ and $C$ hard functions indicated above, the solutions for $H_i^A$ obtained from these equations will (at most) be accurate up to, but not including, corrections of order $\alpha_s^2$, $\alpha\spac\alpha_s\ln(\mu^2/m_B^2)$, or $\alpha^2\ln^2(\mu^2/m_B^2)$ times the exponentiated terms. Explicitly, the solutions are
\begin{equation}\label{eq:H12Asol}
\begin{aligned}
   H_1^A(\mu) 
   &= U_\mathrm{hl}(\mu,m_B)\,\exp\left[ 
    - Q_\ell\spac Q_b\,\frac{\alpha}{8\pi}\spac\ln^2\frac{\mu^2}{m_B^2} \right] \\
   &\quad \times \bigg[ H_1^A(m_B) + \frac{\alpha}{4\pi} \left( Q_\ell^2 - \frac{3}{2}\,Q_b^2 
    - Q_\ell\spac Q_u \right) \ln\frac{\mu^2}{m_B^2} 
    + Q_\ell\spac Q_u\,\frac{\alpha}{\pi} \int_{m_B}^\mu\!\frac{d\mu'}{\mu'}\,
    \frac{\widetilde U_C(\mu',m_B)}{1-\delta(\mu')} \bigg] \,, \\ 
   H_2^A(\mu) 
   &= U_\mathrm{hl}(\mu,m_B)\,\exp\left[ 
    - Q_\ell\spac Q_b\,\frac{\alpha}{8\pi}\spac\ln^2\frac{\mu^2}{m_B^2} \right] 
    \left[ H_2^A(m_B) 
    + z\spac Q_\ell\spac Q_b\,\frac{\alpha}{4\pi}\spac\ln\frac{\mu^2}{m_B^2} \right] ,
\end{aligned}
\end{equation}
where 
\begin{equation}
   \widetilde U_C(\mu,m_B) 
   \equiv \frac{U_C(\mu,m_B)}{U_{\rm hl}(\mu,m_B)}\spac 
    \exp\left[ - Q_b\spac Q_u\,\frac{\alpha}{8\pi}\spac\ln^2\frac{\mu^2}{m_B^2} \right] ,
\end{equation}
and we have expanded out the single-logarithmic QED corrections in the exponent of the first result. The initial conditions at the scale $m_B$ are given by
\begin{equation}\label{eq:HiAsolutions}
\begin{aligned}
   H_1^A(m_B) &= 1 - 3\spac C_F\,\frac{\alpha_s(m_B)}{4\pi} \left( 1 + \ln z \right) \\
   &\quad - \frac{Q_b\spac\alpha}{4\pi}\,\bigg\{\spac 
    Q_\ell\,\bigg[ - \frac{z\ln z}{z-1} + 2\spac\text{Li}_2(1-z) + 1 + \frac{\pi^2}{12} 
    \bigg] + Q_b \left( 2 + 3\ln z \right) + Q_u \bigg\} \\
   &\quad - Q_\ell\spac Q_u\,\frac{\alpha}{\pi} 
    - Q_\ell\spac(Q_b+Q_u)\,\frac{\alpha}{16\pi}\,\kappa \,, \\
   H_2^A(m_B) &= C_F\,\frac{\alpha_s(m_B)}{4\pi} 
    + \frac{Q_b\spac\alpha}{4\pi} \left[ Q_u
    + z\spac Q_\ell \left( \ln z +\frac{z-1}{z} \right) \right] .
\end{aligned}   
\end{equation}
In \eqref{eq:H12Asol}, we neglect the running of $\alpha$, which is a tiny effect in the scale window between $m_B$ and a typical hard-collinear scale. Since the hard function $H_1^A$ starts at tree level, we must keep the two-loop coefficient in \eqref{eq:Uhl} for consistency in the first equation.

\newpage
\section{Matching to SCET-2}
\label{sec:SCET-2}

At a hard-collinear scale $\mu_\hc\sim \sqrt{m_B\spac\Lambda_\mathrm{QCD}}$, the SCET-1 basis operators listed in Section~\ref{subsec:SCET1basis} are matched onto operators in SCET-2, consisting of soft and collinear fields with virtualities of order $\Lambda_\mathrm{QCD}$ and $m_\ell$, respectively, and momenta scaling as
\begin{equation} 
   p_s^\mu \sim m_B\,(\lambda,\lambda,\lambda) \,, \qquad
   p_c^\mu \sim m_B\,(\lambda_\ell^2,1,\lambda_\ell) \,.
\end{equation}
Soft and collinear fields do not interact with one another in SCET-2, since such interactions would violate momentum conservation. The sum of a soft momentum and a collinear momentum produces an off-shell momentum with hard-collinear scaling $(p_s^\mu+p_c^\mu)\sim(\lambda,1,\dots)$, where the perpendicular component scales like $\lambda$ or $\lambda_\ell$, whichever is larger. Such hard-collinear fluctuations are integrated out in the matching from SCET-1 onto SCET-2. This argument holds irrespectively of whether the charged lepton is a muon, for which $\lambda_\ell\sim\lambda$, or an electron, for which $\lambda_\ell\ll\lambda$.

The operators in SCET-2 can be conveniently written in terms of gauge-invariant building blocks of fields, which are invariant under soft and collinear gauge transformations \cite{Bauer:2002nz,Hill:2002vw}. The soft quark and gluon fields can be grouped into hadronic current operators, whereas the collinear lepton and photon fields, along with the anti-collinear neutrino field, can be grouped into leptonic currents. The two currents are decoupled from each other and do not interact. 

Already in SCET-1, the interactions coupling hard-collinear to soft fields can be removed from the leading-order effective Lagrangian by means of the field redefinitions \cite{Bauer:2001yt}
\begin{equation}\label{eq:softdecoupling}
\begin{aligned}
   \bar\X_\hc^{(\ell)}(x+t\spac\nb) 
   &\to \bar\X_\hc^{(\ell),(0)}(x+t\spac\nb)\,Y_n^{(\ell)\dagger}(x_-) \,, \\
   \bar\X_\hc^{(u)}(x+t\spac\nb)
   &\to \bar\X_\hc^{(u),(0)}(x+t\spac\nb)\,\overline{Y}_n^{(u)\dagger}(x_-) \,, \\
   \Asl_\hc^{(f)\perp}(x+t\spac\nb)
   &\to \Asl_\hc^{(f)\perp,(0)}(x+t\spac\nb) \,, \\
   \Gsl_\hc^\perp(x+t\spac\nb)
   &\to \overline{Y}_n(x_-)\,\Gsl_\hc^{\perp,(0)}(x+t\spac\nb)\,
    \overline{Y}_n^\dagger(x_-) \,,
\end{aligned}
\end{equation}
with $f=\ell,u,b$ denoting the charged fermions. Following \cite{Chay:2004zn,Feige:2013zla,Goerke:2017ioi,Beneke:2019slt}, we define the QCD soft Wilson lines for initial-state hard-collinear particles as
\begin{equation}
\begin{aligned}
   \overline{Y}_n(x)
   &= \bm{P}\exp\left[ + i\spac g_s\spac t^a\!\int_{-\infty}^0\!ds\,n\cdot G_s^a(x+sn)\,
    e^{-\epsilon|s|} \right] , \\
   \overline{Y}_n^\dagger(x)
   &= \overline{\bm{P}}\exp\left[ - i\spac g_s\spac t^a\!\int_{-\infty}^0\!ds\,
    n\cdot G_s^a(x+sn)\,e^{-\epsilon|s|} \right] ,
\end{aligned}
\end{equation}
where the path ordering $\bm{P}$ is such that fields are ordered from left to right according to decreasing values of $s$, and $\overline{\bm{P}}$ denotes the opposite ordering. The regulator $\epsilon>0$ ensures that the integrals converge at infinity. The soft Wilson lines including QED effects for the initial-state anti-quark and the final-state charged lepton are given by
\begin{equation}\label{eq:WLdef}
\begin{aligned}
   \overline{Y}_n^{(u)\dagger}(x)
   &= \exp\left[ - i\spac Q_u\spac e\!\int_{-\infty}^0\!ds\,n\cdot A_s(x+sn)\,
    e^{-\epsilon|s|} \right] \overline{Y}_n^\dagger(x) \,, \\
   Y_n^{(\ell)\dagger}(x)
   &= \exp\left[ + i\spac Q_\ell\spac e\!\int_0^\infty\!ds\,n\cdot A_s(x+sn)\,
    e^{-\epsilon|s|} \right] .
\end{aligned}
\end{equation}
In SCET-1, soft fields and Wilson lines interacting with hard-collinear fields need to be multipole-expanded in such a way that only the $n\cdot k_s$ component of soft momenta enter the interaction vertices \cite{Beneke:2002ph,Beneke:2002ni}. The soft Wilson lines and gauge fields must therefore be evaluated at the spacetime point $x_-^\mu\equiv\frac{\nb\cdot x}{2}\spac{n^\mu}$ rather than at $x^\mu$. Note that the displacement vector $t\spac\nb$ in the argument of the hard-collinear fields does not contribute to $x_-$. The superscript ``(0)'' on the fields appearing on the right-hand side of \eqref{eq:softdecoupling} indicates that the new hard-collinear fields are decoupled from leading-order soft interactions. Importantly, interactions among hard-collinear and soft fields are still allowed at subleading order in power counting. In the matching to SCET-2, the off-shell hard-collinear fields are integrated out, leaving effective operators built out of decoupled soft and collinear fields \cite{Bauer:2002aj,Beneke:2003pa,Becher:2005fg}.

For the operators $O_{3,4}^C$, one further needs the relations
\begin{equation}
\begin{aligned}
   \bar\X_\hc^{(u)}(x+t\spac\nb)\spac(-i\overleftarrow{D}_{\!\!s\perp}^\alpha)
   &\to \bar\X_\hc^{(u),(0)}(x+t\spac\nb)\,
    \big( \!-\!i\!\overleftarrow{\partial}_{\!\!s\perp}^\alpha
    + \G_{s\perp}^\alpha(x_-) + Q_u\,\A_{s\perp}^\alpha(x_-) \big)\,
    \overline{Y}_n^{(u)\dagger}(x_-) \,, \\
   \bar\X_\hc^{(\ell)}(x+t\spac\nb)\spac(-i\overleftarrow{\Dsl}_{\!\!s\perp})
   &\to \bar\X_\hc^{(\ell),(0)}(x+t\spac\nb)\,
    \big( \!-\!i\!\overleftarrow{\delsl}_{\!\!s}^\perp
    + Q_\ell\,\Asl_s^\perp(x_-) \big)\,Y_n^{(\ell)\dagger}(x_-) \,,
\end{aligned}
\end{equation}
where the gauge-invariant building blocks for the soft gauge fields are defined as \cite{Hill:2002vw} 
\begin{equation}
\begin{aligned}
   \G_s^\mu(x) 
   &= \overline{Y}_n^\dagger(x)\spac\big[ iD_s^\mu\spac\overline{Y}_n(x) \big] 
    = g_s \int_{-\infty}^0\!ds\,n_\alpha\spac
     \big[ \overline{Y}_n^\dagger\spac G_s^{\alpha\mu}\spac\overline{Y}_n \big](x+sn) 
    \equiv t^a\,\G_s^{\mu,a}(x) \,, \\
   \A_s^\mu(x) 
   &= \frac{1}{Q_\ell}\,Y_n^{(\ell)\dagger}(x)\spac\big[ iD_s^\mu\spac Y_n^{(\ell)}(x) \big] 
    = e \int_{-\infty}^0\!ds\,n_\alpha\spac F_s^{\alpha\mu}(x+sn) \,.
\end{aligned}
\end{equation}
For the abelian gauge field, the soft Wilson lines commute with $F_s^{\alpha\mu}$ and cancel out. 

\subsection{\texorpdfstring{Systematics of the SCET-1$\,\to\,$SCET-2 matching}{Systematics of the SCET-1 -> SCET-2 matching}}
\label{sec:SCET1toSCET2matching}

At tree level, the SCET-1 to SCET-2 matching relations for hard-collinear fields have been studied in detail in \cite{Beneke:2003pa,Becher:2005fg}. Besides the trivial possibility, in which the decoupled hard-collinear fields with superscript ``(0)'' in \eqref{eq:softdecoupling} are simply replaced by the corresponding collinear fields, we note the relation
\begin{equation}\label{eq:Xhctoqs}
   \bar\X_\hc^{(u),(0)}(x+t\spac\nb) 
   \to - \frac{Q_u}{2}\,\big( \bar u_s\spac\overline{Y}_n^{(u)}\big)(x_-)\,
    \frac{\nsl}{in\cdot\!\overleftarrow{\partial}_{\!\!s}}\,
    \big( \Gsl_c^\perp(x+t\spac\nb) + Q_u\,\Asl_c^\perp(x+t\spac\nb) \big) + \dots \,,
\end{equation}
in which a hard-collinear fermion splits into a soft quark and a collinear gauge field. The dots represent terms with higher power suppression. These tree-level matching relations are summarized in Table~\ref{tab:SCET-2}. The virtuality of collinear fields in SCET-2 is set by the mass of the charged lepton, and hence the power-counting rules for the collinear fields are
\begin{equation}
   \X_c^{(\ell)} \sim \lambda_\ell \,, \qquad
   \A_c^\perp \sim \lambda_\ell \,,
\end{equation}
with $\lambda_\ell\sim m_\ell/m_B$. Integrating out hard-collinear propagators can introduce inverse soft derivatives $(in\cdot\partial_s)^{-1}\sim\lambda^{-1}$ in the SCET-2 operators, as shown in \eqref{eq:Xhctoqs}. This has the effect of delocalizing the soft fields along the $n$ light-cone, i.e., the direction of motion of the charged lepton. The inverse derivative lifts the power-suppression of the soft quark field $\bar u_s\sim\lambda^{\frac32}$ relative to the hard-collinear quark field $\bar\X_\hc^{(u)}\sim\lambda^{\frac12}$. At first sight, it would seem that multiple inverse soft derivatives could upset the power counting in SCET-2. However, reparameterization invariance forces every inverse factor of $in\cdot\partial_s$ to be compensated by a power of $n$ in the numerator. The fact that $n^2=0$ then restricts the number of operators and ensures that SCET-2 operators are always power-suppressed relative to the SCET-1 operators they descend from \cite{Becher:2005fg}. 

\begin{table}
\centering
\renewcommand{\arraystretch}{1.5}
\begin{tabular}{cc|cc|c}
\hline
\rowcolor{\shadecolor{20}}
SCET-1 field & power counting & SCET-2 field & power counting & cost factor \\
\hline
$\bar\X_{hc}^{(\ell)}$ & $\lambda^{\frac12}$ & $\bar\X_c^{(\ell)}$ & $\lambda_\ell$ & $\lambda_\ell\,\lambda^{-\frac12}$ \\
\hline
$\A_{hc}^\perp$ & $\lambda^{\frac12}$ & $\A_c^\perp$ & $\lambda_\ell$ & $\lambda_\ell\,\lambda^{-\frac12}$ \\
\hline
$\bar\X_{hc}^{(u)}$ & $\lambda^{\frac12}$ & $\bar u_s\,\A_c^\perp$ & $\lambda_\ell\,\lambda^{\frac12}$ & $\lambda_\ell$ \\
\hline
$\bar\X_{hc}^{(\ell)}$ & $\lambda^{\frac12}$ & $\bar\X_c^{(\ell)}\spac\A_s^\perp$ & $\lambda_\ell^2$ & $\lambda_\ell^2\,\lambda^{-\frac12}$ \\
\hline
\end{tabular}
\caption{\label{tab:SCET-2} 
Relevant tree-level matching relations for hard-collinear fields splitting into collinear and soft fields in SCET-2. Here $\A_\hc^\perp$ and $\A_c^\perp$ refer to generic gauge bosons (gluons or photons). Additional soft gauge fields are contained in the SCET-2 Wilson lines (not shown). In all cases, adding extra fields implies a higher-order scaling with $\lambda$ and $\lambda_\ell$.}
\end{table}

\begin{figure}
\centering
\includegraphics[scale=0.65]{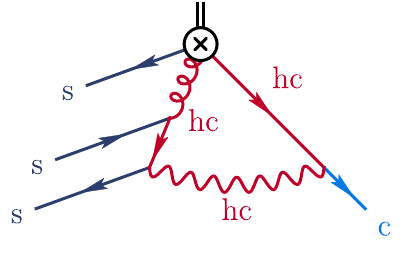} \quad
\includegraphics[scale=0.65]{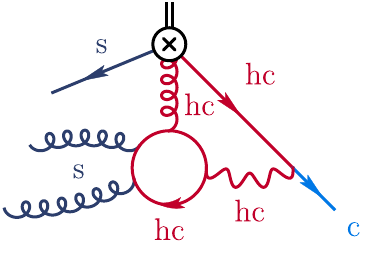} 
\caption{Representative hard-collinear loop diagrams for the matching of type-$B$ SCET-1 operators onto SCET-2. Red (dark blue) lines represent hard-collinear (soft) propagators and external states. Light blue lines represent collinear external states. The double line at the vertex represents the $b$-quark and neutrino fields.}
\label{fig:graphs_btype}
\end{figure}

We will now study in detail which SCET-2 operators can descend from the operators contained in the SCET-1 operator basis constructed in Section~\ref{subsec:SCET1basis}. For the operators of type-$A$ and the type-$B$ operators $O_{1,2}^B$ in the SCET-1 basis, there is no hard-collinear momentum flow between the quark and lepton currents. This implies that all hard-collinear loop diagrams are scaleless and vanish in dimensional regularization. The tree-level matching relations discussed above are then all that is needed to derive the corresponding expressions for the SCET-2 operators, in which all hard-collinear fields are simply replaced by collinear fields times the appropriate soft Wilson lines. Using the rules given in Table~\ref{tab:SCET-2}, one finds that these operators scale like\footnote{Recall that the external $B$-meson and charged-lepton states have power counting $\lambda^{-\frac32}$ and $\lambda_\ell^{-1}$, respectively, so that the decay amplitude scales like $\lambda_\ell\,\lambda^{\frac32}$, which is indeed the scaling of the product $m_\ell\spac f_{B}$.} 
$\lambda_\ell^2\,\lambda^3$. Adding additional fields or derivatives in the SCET-2 operators would give rise to additional power suppression. In particular, the transition $\bar\X_\hc^{(\ell)}\to\bar\X_c^{(\ell)}\spac\A_s^\perp$ produces a pair of SCET-2 fields with power counting $\lambda_\ell^2$, which implies a higher cost factor (see Table~\ref{tab:SCET-2}). The emission of the additional soft photon is thus a power-suppressed effect, which can be neglected. The descendants of the operators $O_{3,4}^A$ involve the leptonic current
\begin{equation}
   \bar\X_\hc^{(\ell)}\spac(-i\!\overleftarrow{\delsl}_{\!\!\perp})\spac P_L\spac\nu_\cb
   = \bar\X_\hc^{(\ell)}\spac(-i\!\overleftarrow{\partial}_{\!\!\rho}^{\!\perp})\,
    \gamma_\perp^\rho P_L\spac\nu_\cb \,,
\end{equation}
whose matrix element vanishes to all orders of perturbation theory, because we work in a reference frame where $p_\ell^\perp=0$. Hence, these two operators can be omitted from the SCET-2 basis. The operators $O_{3,4}^B$ contain a hard-collinear gluon field, which needs to be converted into a hard-collinear photon field before the hard-collinear momentum can be transferred from the quark current to the lepton current. As shown in Figure~\ref{fig:graphs_btype}, this conversion involves at least two power-suppressed interactions, yielding SCET-2 operators which are suppressed by (at least) a factor $\lambda$ with respect to $O_{1,2}^B$. In other words, the descendants of the operators $O_{3,4}^B$ do not match onto the leading-power SCET-2 operator basis.

For the remaining operators in the SCET-1 basis, it is straightforward to check that using the tree-level matching relations one obtains SCET-2 operators that are either power suppressed beyond $\mathcal{O}(\lambda_\ell^2\,\lambda^3)$ or have a vanishing projection onto the $B$ meson. The only non-trivial case is the operator $O_3^C$, for which the application of the tree-level matching relation \eqref{eq:Xhctoqs} yields the SCET-2 operator (omitting spacetime arguments)
\begin{equation}
\begin{aligned}
   O_3^C \to & - \frac{Q_u}{2}\,\frac{1}{\nb\cdot\P_c}\,
    \big( \bar u_s\spac\overline{Y}_n^{(u)} \big)\,
    \frac{\nsl}{in\cdot\!\overleftarrow{\partial}_{\!\!s}}\,\gamma_\perp^\beta\spac
    \big( \!-\!i\overleftarrow{\partial}_{\!\!s\perp}^\alpha
    + \G_{s\perp}^\alpha + Q_u\,\A_{s\perp}^\alpha \big)\,
    \nbsl P_L\spac Y_n^{(\ell)\dagger}\spac\overline{Y}_n^{(u)\dagger}\spac b_v \\
   &\times \bar\X_c^{(\ell)}\spac\A_{c,\beta[y]}^\perp\spac\gamma_\alpha^\perp\spac
    P_L\spac\nu_\cb \,, 
\end{aligned}
\end{equation}
which indeed exhibits the correct scaling $\lambda_\ell^2\,\lambda^3$. We can now use the fact that the external $B$-meson and lepton states have vanishing transverse momenta. Lorentz invariance then requires that the $B$-meson matrix element of the quark operator in the first line of this expression, and the lepton matrix element of the fermion bilinear in the second line, each contain two components multiplying $g_\perp^{\alpha\beta}$ and $i\epsilon_\perp^{\alpha\beta}$. Performing appropriate contractions of these matrix elements and using the relations \eqref{eq:nicerelations2}, it is straightforward to show that the hadronic matrix element is proportional to $\Pi_+^{\alpha\beta}$ and the leptonic one to $\Pi_{-\beta\alpha}$, where 
\begin{equation}
   \Pi_\pm^{\alpha\beta} 
   \equiv \frac{g_\perp^{\alpha\beta}\pm i\epsilon_\perp^{\alpha\beta}}{2} \,.
\end{equation}
The fact that $\Pi_+^{\alpha\beta}\,\Pi_-^{\beta\alpha}=0$ implies that the tree-level matching relation for the operator $O_3^C$ indeed vanishes when projected onto the $B$ meson. 

\begin{figure}
\centering
\includegraphics[scale=0.65]{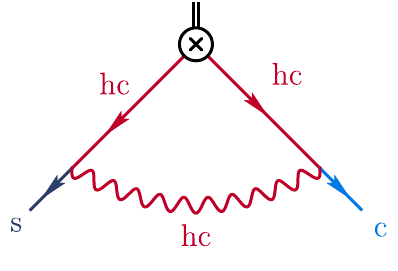} \qquad
\includegraphics[scale=0.65]{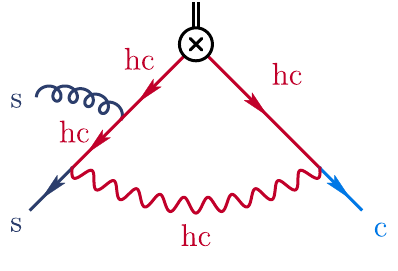} 
\caption{One-loop exchange of a hard-collinear photon, converting a hard-collinear to a soft quark. Red (dark blue) lines represent hard-collinear (soft) propagators and external states. Light blue lines represent collinear external states.} 
\label{fig:hcphoton}
\end{figure}

It is therefore necessary to extend the matching beyond the tree approximation. The present work is one of the first where the SCET-1$\,\to\,$SCET-2 matching relations are calculated at one-loop order. In this case, one needs to consider the matching relations for products of hard-collinear fields rather than individual fields. These relations involve loop diagrams, in which the exchange of a hard-collinear photon transfers the hard-collinear momentum from the quark current to the lepton current, as illustrated in Figure~\ref{fig:hcphoton}. This matching can be formalized by introducing an intermediate effective theory containing (off-shell) hard-collinear fields along with (on-shell) collinear fields and soft fields, and then integrating out the off-shell collinear fields \cite{Beneke:2003pa,Becher:2005fg}. Here we use symmetry arguments to constrain the form of the resulting operators, based on the following observations:
\begin{enumerate}
\item
Every SCET-1$\,\to\,$SCET-2 matching relation must contain at least one collinear and one soft field on the right-hand side to conserve momentum, where the soft field can also appear in the form of a soft Wilson line.
\item
The matching relation for a hard-collinear fermion field must vanish when multiplied with $\nsl$.
\item
The matching relation for a massless hard-collinear fermion field must preserve chirality.
\item
The matching relation for a collection of hard-collinear fields must preserve the total mass dimension of the fields.
\item
The $B$-meson 4-velocity $v^\mu$ does not appear in SCET-1$\,\to\,$SCET-2 matching relations. 
\end{enumerate}
The latter condition follows from the fact that the effective $b$-quark field $b_v$ solely participates in soft interactions, while only hard-collinear interactions contribute to the matching. It ensures that there can be at most one inverse soft derivative, since the reparameterization-invariant combination $n\cdot v/(in\cdot\!\overleftarrow\partial_{\!\!s})$ cannot appear. 

Conditions~2 and~3 imply that, to all orders of perturbation theory, a matching relation for a massless hard-collinear quark field onto a soft quark field must be of the form\footnote{Beyond one-loop order additional pairs of transverse Dirac matrices can appear, but they can be reduced to the expressions shown above using the relations \eqref{eq:nicerelations2} and \eqref{eq:nicerelations1}, apart from evanescent structures that do not appear in the physical operator basis.} 
\begin{equation}\label{eq:quarkmatching}
   \bar\X_{hc}^{(u),(0)} \ldots
   \to \big(\bar u_s\spac\overline{Y}_n^{(u)}\big)\,\frac{\nsl}{in\cdot\!\overleftarrow\partial_{\!\!s}}\,
    \gamma_\perp^\rho\spac\ldots \quad \text{or} \quad
   \bar\X_{hc}^{(u),(0)} \ldots
   \to \big(\bar u_s\spac\overline{Y}_n^{(u)}\big)\,\frac{\nbsl\spac\nsl}{4}\spac\ldots \,,
\end{equation}
where in the first case the mass dimension is lowered by one unit and an open transverse Lorentz index appears. In the SCET-1 basis, operators involving hard-collinear quark and leptons fields can have up to two additional transverse hard-collinear objects (derivatives or gauge fields), which also carry mass dimension and transverse Lorentz indices. The 17 basis operators involving hard-collinear quark and leptons fields can be written as collections of hard-collinear fields times one of the following four Dirac structures: 
\begin{equation}\label{eq:4structures}
\begin{aligned}
   S_1: && \nbsl\spac P_L\spac b_v 
   &\otimes P_L\spac\nu_\cb && (\text{operator}~O_1^C) \\
   S_2: && \gamma_\perp^\alpha\,\nbsl\spac P_R\,b_v 
   &\otimes P_L\spac\nu_\cb && (\text{operators}~O_2^C,\, O_{1,2}^D) \\
   S_3: && \nbsl\spac P_L\spac b_v 
   &\otimes \gamma_\perp^\alpha\spac P_L\spac\nu_\cb && (
   \text{operators}~O_{3,4}^C,\, O_{3,4}^D) \\
   S_4: && \gamma_\perp^\alpha\,\nbsl\spac P_R\,b_v 
   &\otimes \gamma_\perp^\beta\spac P_L\spac\nu_\cb
    &\quad& (\text{operators}~O_5^C,\, O_{5,6,7,8}^D,\, O_{1,2,3,4}^E)
\end{aligned}
\end{equation}
where the indices $\alpha$, $\beta$ are those of hard-collinear derivatives or gauge fields entering the loop function. Since neither the $b$ quark nor the neutrino participate in the matching, these structures can only be modified from their left-hand sides in each current. Matching the hard-collinear quark field to a soft quark field yields one of the two structures in \eqref{eq:quarkmatching}. For the hard-collinear lepton field, the possible matching relations are
\begin{equation}\label{eq:leptonmatching}
   \bar\X_{hc}^{(\ell),(0)} \ldots
   \to \bar\X_c^{(\ell)} \ldots \quad \text{or} \quad
   \bar\X_{hc}^{(\ell),(0)} \ldots
   \to m_\ell\,\bar\X_c^{(\ell)}\spac\gamma_\perp^\sigma \ldots \,,
\end{equation}
where the factor of the lepton mass in the second relation requires a chirality flip and hence an extra Dirac matrix. The hard-collinear loop functions carry the open transverse Lorentz indices in such a way that in the end all indices are contracted. 

\begin{table}[t]
\centering
\begin{tabular}{c|c|c}
\hline
\rowcolor{\shadecolor{20}}
 & $\displaystyle\frac{\nsl}{in\cdot\!\overleftarrow\partial_{\!\!s}}\,
 \gamma_\perp^\rho\spac\ldots$
 & $\displaystyle\frac{\nbsl\spac\nsl}{4}\spac\ldots$ \\ 
\hline
$S_1$ & ${\color{\shadecolor{100}} \big( i\!\overleftarrow\partial_{\!\!s\perp}^\sigma, 
 \G_{s\perp}^\sigma \big)}\,
 \displaystyle\frac{\nsl}{in\cdot\!\overleftarrow\partial_{\!\!s}}\,
 \gamma_\perp^\rho\spac\nbsl\spac P_L \otimes P_L$
 & $\nbsl\spac P_L \otimes P_L$ \\[3mm]
 & $\to \displaystyle - \spac\Pi_-^{\rho\sigma}\,
 \frac{\nsl\spac\nbsl}{in\cdot\!\overleftarrow\partial_{\!\!s}}\,
 \big( i\!\overleftarrow\delsl_{\!\!s}^{\!\perp}, \Gsl_s^\perp \big)\,P_L 
 \otimes P_L$ & \\
\hline
$S_2$ & $\displaystyle{\color{\shadecolor{100}} \nb\cdot\P_c\,in\cdot\!\overleftarrow\partial_{\!\!s}}\,
 \frac{\nsl}{in\cdot\!\overleftarrow\partial_{\!\!s}}\,\gamma_\perp^\rho\spac
 \gamma_\perp^\alpha\,\nbsl\spac P_R \otimes P_L$ & -- \\
 & $\to \displaystyle \Pi_-^{\rho\alpha}\,  
  \nb\cdot\P_c\,\nsl\spac\nbsl\,P_R \otimes P_L$ & \\
\hline
$S_3$ & ${\color{\shadecolor{100}} \big( i\!\overleftarrow\partial_{\!\!s\perp}^\sigma, 
 \G_{s\perp}^\sigma \big)}\,
 \displaystyle\frac{\nsl}{in\cdot\!\overleftarrow\partial_{\!\!s}}\,
 \gamma_\perp^\rho\spac\nbsl\spac P_L \otimes 
 {\color{\shadecolor{100}} \big( m_\ell\,\gamma_\perp^\gamma, \A_{c\perp}^\gamma \big)}\,
 \gamma_\perp^\alpha\spac P_L$ 
 & $\nbsl\spac P_L \otimes {\color{\shadecolor{100}} \big( m_\ell\,\gamma_\perp^\gamma, 
 \A_{c\perp}^\gamma \big)}\,\gamma_\perp^\alpha\spac P_L$ \\[3mm]
& $\to - \displaystyle \Pi_-^{\rho\sigma}\,\Pi_+^{\alpha\gamma}\,  
  \frac{\nsl\spac\nbsl}{in\cdot\!\overleftarrow\partial_{\!\!s}}\,
  \big( i\!\overleftarrow\delsl_{\!\!s}^{\!\perp}, \Gsl_s^\perp \big)\spac P_L 
  \otimes \big( m_\ell, \Asl_c^\perp \big)\spac P_L$ 
 & $\to \displaystyle \Pi_+^{\alpha\gamma}\,  
  \nbsl\spac P_L \otimes \big( m_\ell, \Asl_c^\perp \big) \spac P_L$ \\
\hline
$S_4$ & $\displaystyle{\color{\shadecolor{100}} \nb\cdot\P_c\,in\cdot\!\overleftarrow\partial_{\!\!s}}\,
 \frac{\nsl}{in\cdot\!\overleftarrow\partial_{\!\!s}}\,\gamma_\perp^\rho\spac
 \gamma_\perp^\alpha\,\nbsl\spac P_R \otimes 
 {\color{\shadecolor{100}} \big( m_\ell\,\gamma_\perp^\gamma, \A_{c\perp}^\gamma \big)}\,
 \gamma_\perp^\beta\spac P_L$ & -- \\
& $\to \displaystyle \Pi_-^{\rho\alpha}\,\Pi_+^{\beta\gamma}\,
  \nb\cdot\P_c\,\nsl\spac\nbsl\,P_R 
 \otimes \spac \big( m_\ell, \Asl_c^\perp \big)\spac P_L$ & \\
\hline
$S_5$ & $\displaystyle \frac{1}{\nb\cdot\P_c}\,
 {\color{\shadecolor{100}} \big( i\!\overleftarrow\partial_{\!\!s\perp}^\sigma, 
 \G_{s\perp}^\sigma \big)}\,
 \frac{\nsl}{in\cdot\!\overleftarrow\partial_{\!\!s}}\,
 \gamma_\perp^\rho\spac\nbsl\spac P_L \otimes 
 m_\ell\spac P_L$
 & $\displaystyle\frac{1}{\nb\cdot\P_c}\,\nbsl\spac P_L \otimes 
 m_\ell\spac P_L$ \\[3mm]
 & $\to \displaystyle - \frac{\Pi_-^{\rho\sigma}}{\nb\cdot\P_c}\,
 \frac{\nsl\spac\nbsl}{in\cdot\!\overleftarrow\partial_{\!\!s}}\,
 \big( i\!\overleftarrow\delsl_{\!\!s}^{\!\perp}, \Gsl_s^\perp \big)\,P_L 
 \otimes m_\ell\spac P_L$ & \\
\hline
\end{tabular}
\caption{\label{tab:structures}
Structures arising in the matching of the SCET-1 basis operators onto SCET-2. The two columns correspond to the two options for the matching of the hard-collinear quark field shown in \eqref{eq:quarkmatching}. The fermion fields and soft Wilson lines are not shown for brevity. A gluon field can also be replaced by a photon field. Open Lorentz indices are contracted with those of the hard-collinear loop functions.}
\end{table}

It is now a simple matter to combine the various possibilities in \eqref{eq:quarkmatching} and \eqref{eq:leptonmatching} with the four structures in \eqref{eq:4structures} in such a way that the overall mass dimension is preserved and the quark current does not contain an odd number of open transverse indices, since otherwise it would have a vanishing projection onto the $B$ meson. The resulting structures are summarized in Table~\ref{tab:structures}, omitting the fermion fields and soft Wilson lines. Extra derivatives, gauge fields and factors of the lepton mass needed to obtain the correct mass dimension and a non-zero projection onto the $B$-meson and lepton states are colored in blue. When there are two options $a$ and $b$, they are given in the form of a list as ${\color{\shadecolor{100}} (a,b)}$. In the second line of each expression, we show the results obtained after projection onto the $B$-meson and lepton matrix elements. A few comments are in order:
\begin{itemize}
\item 
The transverse index $\rho$ in the first structure for $S_1$ can in principle be contracted with a transverse derivative, a transverse soft or collinear gauge field, or an extra Dirac matrix in the lepton current, which changes the chirality and comes along with a factor $m_\ell$. However, in order to obtain a non-zero projection onto the $B$ meson, only a soft derivative or a soft gauge field are allowed. 
\item 
In an analogous way, the first structure for $S_3$ requires an extra soft derivative or a soft gauge field for a non-zero projection on the $B$ meson, In addition, one needs an additional mass insertion or a collinear gauge field to ensure a non-zero leptonic matrix element.  
\item 
In the case of the structures $S_2$ and $S_4$, it is possible to add a factor of the hard-collinear virtuality $\nb\cdot\P_c\,in\cdot\!\overleftarrow\partial_{\!\!s}$ with mass dimension~2 in the numerator and still obtain a non-vanishing projection onto the $B$ meson. Compared to this, insertions of two transverse objects would yield power-suppressed corrections.
\item 
The open transverse Lorentz indices in the obtained structures are contracted with open indices in the loop functions. In some cases, like in \eqref{eq:Xhctoqs}, this can give rise to vanishing results, because
\begin{equation}
   \Pi_\pm^{\alpha\beta}\,g_{\beta\gamma}^\perp\,\Pi_\mp^{\gamma\delta} = 0 \,.
\end{equation}
\end{itemize}

We finally consider the case of the soft-lepton operator $O_1^F$ in \eqref{typeF}, which comes with the Dirac structure:
\begin{equation}
\begin{aligned}
   S_5: && \nbsl\spac P_L\,b_v 
   &\otimes \nsl\spac P_L\spac\nu_\cb 
    &\quad& (\text{operator}~O_1^F)
\end{aligned}
\end{equation}
For the hard-collinear quark field, the two options for the matching are those shown in \eqref{eq:quarkmatching}. For the soft lepton field, on the other hand, the only option is
\begin{equation}
   \bar\ell_s \ldots
   \to m_\ell\,Y_n^{(\ell)\dagger}\,\bar\X_c^{(\ell)}\,
    \frac{\nbsl}{\nb\cdot\P_c} \ldots \,.
\end{equation}
The presence of $\nbsl$ is required in order to get a non-vanishing result, given the structure of $S_5$. The lepton mass is generated from soft dynamics. Its presence requires a chirality flip and hence a single additional Dirac matrix. Combining the two bilinears, and adding gauge fields or derivatives to ensure the correct mass dimension and non-zero $B$ meson and lepton matrix elements, we find the options shown in the lowest row of Table~\ref{tab:structures}.

\subsection{Construction of the SCET-2 operator basis}
\label{sec:SCET2op}

We are now in a position to write down the complete SCET-2 operator basis for our problem, consisting of operators scaling as $\lambda_\ell^2\,\lambda^3$ with a non-zero projection onto the $B$ meson. The type-$A$ and type-$B$ operators $O_{1,2}^A$ and $O_{1,2}^B$ match trivially onto the corresponding SCET-2 operators
\begin{equation}\label{eq:SCET2basis_local}
\begin{aligned}
   Q_1^A &= \frac{m_\ell\,\nb\cdot v}{\nb\cdot\P_c}\,
    \big( \bar u_s\,\frac{\nbsl}{\nb\cdot v}\spac P_L\spac b_v\spac Y_n^{(\ell)\dagger} \big)\,
    \big(\bar\X_c^{(\ell)} P_L\spac\nu_\cb\big) \,, \\
   Q_2^A &= \frac{m_\ell\,\nb\cdot v}{\nb\cdot\P_c}\,
    \big( \bar u_s\,\frac{\nsl}{n\cdot v}\spac P_L\spac b_v\spac Y_n^{(\ell)\dagger} \big)\,
    \big( \bar\X_c^{(\ell)} P_L\spac\nu_\cb \big) \,, \\
   Q_1^B &= \frac{\nb\cdot v}{\nb\cdot\P_c}\,
    \big( \bar u_s\,\frac{\nbsl}{\nb\cdot v}\spac P_L\spac b_v\spac Y_n^{(\ell)\dagger} \big)\,
    \big( \bar\X_c^{(\ell)}\spac\Asl_{c\spac[x]}^\perp\spac P_L\spac\nu_\cb \big) \,, \\
   Q_2^B &= \frac{\nb\cdot v}{\nb\cdot\P_c}\,
    \big( \bar u_s\,\frac{\nsl}{n\cdot v}\spac P_L\spac b_v\spac Y_n^{(\ell)\dagger} \big)\,
    \big( \bar\X_c^{(\ell)}\spac\Asl_{c\spac[x]}^\perp\spac P_L\spac\nu_\cb \big) \,.
\end{aligned}
\end{equation}
No hard-collinear momentum transfer from the quark to the lepton side is involved in this matching. We refer to these operators as ``local'', since all fermion fields are evaluated at the same spacetime point. Nonetheless, these operators still contain soft and collinear Wilson lines, which are non-local objects.

Before writing down the remaining basis operators, which result from the matching of the SCET-1 operators of type-$C$, $D$, $E$, and $F$ containing a hard-collinear quark field, we need to discuss one more subtlety. When a SCET-1 operator is matched onto SCET-2 and hard-collinear propagators are integrated out, the presence of inverse soft derivatives implies that the soft fields (but not the heavy-quark field $b_v$) in the SCET-2 operators can be smeared out along the light-like direction $n$. According to Table~\ref{tab:structures}, we find operators containing a soft anti-quark field $\bar u_s(x+s\spac n)$ and up to one soft gauge field $\A_s(x+s'n)$. In analogy to the discussion in Section~\ref{subsec:ingredients_SCET1}, we can write for a generic soft field
\begin{equation}
   \phi_s^{(i)}(x+s_i\spac n) = e^{-i s_i\spac n\cdot\P_s^{(i)}}\,\phi_s^{(i)}(x)
   = \int\!d\omega_i\,e^{-i s_i\spac n\cdot v\spac\omega_i}\,\phi_{s\spac[\omega_i]}^{(i)}(x) \,.
\end{equation}
where
\begin{equation}
   \phi_{s\spac[\omega_i]}^{(i)}(x)
   \equiv \delta\Big(\omega_i-\frac{n\cdot\P_s^{(i)}}{n\cdot v} \Big)\,\phi_s^{(i)}(x) \,.
\end{equation}
The operator $\P_s^{(i)}$ projects out the incoming momentum of the initial-state soft field. The auxiliary integration over $\omega_i$ is introduced to turn this operator into a light-cone momentum variable. The position-space jet functions arising as Wilson coefficients in the matching onto SCET-2 depend on the variables $s_i$, in the same way in which the position-space hard functions in \eqref{eq:3.14} depend on the variables $t_i$. Performing the Fourier integrals over the displacements $s_i$, one obtains the momentum-space jet functions $J_{O_i^X\to Q_j^{X'}}$ in the matching relations
\begin{equation}
   O_i^X(\{\underline{y}\},\mu)
   = \sum_{j,X'} J_{O_i^X\to Q_j^{X'}}(\{\underline{y}\},\{\underline{x}\},
    \{\underline{\omega}\},\mu) 
    \otimes Q_j^{X'}(\{\underline{x}\},\{\underline{\omega}\},\mu) \,,
\end{equation}
which now depend on the momentum variables $\omega_i$ conjugate to $s_i$. We use the notation $\{\underline{y}\}=\{y_1,\dots,y_n\}$ etc.\ for the lists of relevant momentum fractions $y_i$, $x_i$ and momentum variables $\omega_i$. The symbol $\otimes$ implies an integration over the common variables $\{\underline{x}\}$ and $\{\underline{\omega}\}$ shared by the jet functions and the SCET-2 operators, where the variables $x_i$ denote the collinear momentum fractions in case a SCET-2 operator contains more than one collinear field.

To construct the ``non-local'' SCET-2 basis operators for our problem, in which the various soft fields live at different spacetime points, we introduce gauge-invariant building blocks for the soft quark fields, defined as \cite{Beneke:2020vnb,Beneke:2021jhp} \begin{equation}\label{eq:softblocks}
   \bar\Q_s(x) = \big( \bar u_s\spac\overline{Y}_n^{(u)} \big)(x) \,, \qquad
   \H_v(x) = \big( Y_n^{(\ell)\dagger}\,\overline{Y}_n^{(u)\dagger}\spac b_v\big)(x) \,.
\end{equation}
Note that the gauge-invariant building block for the heavy-quark field contains the soft Wilson lines arising from the field redefinitions of the hard-collinear fermion fields in \eqref{eq:softdecoupling}. This ensures that $\H_v$ is indeed invariant under QCD and QED gauge transformations, since $Q_\ell+Q_u=Q_b$. With these definitions, we find that the remaining SCET-1 operators of type-$C$, $D$, $E$ and $F$ match onto the operators
\begin{equation}\label{eq:SCET2basis_nonlocal}
\begin{aligned}
   Q_1^C 
   &= \frac{m_\ell\,\nb\cdot v}{\nb\cdot\P_c}\,
    \big( \bar\Q_{s\spac[\omega]}\,\frac{\nbsl}{\nb\cdot v}\spac P_L\spac\H_v \big)\,
    \big( \bar\X_c^{(\ell)} P_L\spac\nu_\cb \big) \,, \\
   Q_2^C 
   &= \frac{m_\ell\,\nb\cdot v}{\nb\cdot\P_c}\,
    \big( \bar\Q_{s\spac[\omega]}\,\frac{\nsl}{n\cdot v}\spac P_L\,\H_v \big)\,
    \big( \bar\X_c^{(\ell)} P_L\spac\nu_\cb \big) \,, \\
   Q_1^D 
   &= \frac{\nb\cdot v}{\nb\cdot\P_c}\,
    \big( \bar\Q_{s\spac[\omega]}\,\frac{\nbsl}{\nb\cdot v}\spac P_L\spac\H_v \big)\,
    \big( \bar\X_c^{(\ell)}\spac\Asl_{c\spac[x]}^\perp\spac P_L\spac\nu_\cb \big) \,, \\
   Q_2^D 
   &= \frac{\nb\cdot v}{\nb\cdot\P_c}\,
    \big( \bar\Q_{s\spac[\omega]}\,\frac{\nsl}{n\cdot v}\spac P_L\,\H_v \big)\,
    \big( \bar\X_c^{(\ell)}\spac\Asl_{c\spac[x]}^\perp\spac P_L\spac\nu_\cb \big) \,, \\
   Q_1^E 
   &= \frac{m_\ell\,\nb\cdot v}{\omega\,\nb\cdot\P_c}\,
    \big( \bar\Q_{s\spac[\omega]}\,\Asl_{s\spac[\omega_g]}^\perp\,\frac{\nsl}{n\cdot v}\spac 
     P_R\,\H_v \big)\,
    \big( \bar\X_c^{(\ell)} P_L\spac\nu_\cb \big) \,, \\
   Q_2^E 
   &= \frac{m_\ell\,\nb\cdot v}{\omega\,\nb\cdot\P_c}\,
    \big( \bar\Q_{s\spac[\omega]}\,\Gsl_{s\spac[\omega_g]}^\perp\,\frac{\nsl}{n\cdot v}\spac 
     P_R\,\H_v \big)\,
    \big( \bar\X_c^{(\ell)} P_L\spac\nu_\cb \big) \,, \\
   Q_1^F 
   &= \frac{\nb\cdot v}{\omega\,\nb\cdot\P_c}\,
    \big( \bar\Q_{s\spac[\omega]}\,\Asl_{s\spac[\omega_g]}^\perp\,\frac{\nsl}{n\cdot v}\spac 
     P_R\,\H_v \big)\,
    \big( \bar\X_c^{(\ell)}\spac\Asl_{c\spac[x]}^\perp\spac P_L\spac\nu_\cb \big) \,, \\
   Q_2^F 
   &= \frac{\nb\cdot v}{\omega\,\nb\cdot\P_c}\,
    \big( \bar\Q_{s\spac[\omega]}\,\Gsl_{s\spac[\omega_g]}^\perp\,\frac{\nsl}{n\cdot v}\spac 
     P_R\,\H_v \big)\,
    \big( \bar\X_c^{(\ell)}\spac\Asl_{c\spac[x]}^\perp\spac P_L\spac\nu_\cb \big) \,,
\end{aligned}
\end{equation}
whose structure can be readily inferred from Table~\ref{tab:structures}. The non-locality of the operators is reflected in the fact that they depend on the light-cone momentum variables $\omega$ and $\omega_g$ corresponding to the (incoming) components $n\cdot l_s/n\cdot v$ and $n\cdot q_s/n\cdot v$ of the soft spectator anti-quark and a soft gluon or photon in the $B$ meson, respectively. In the last four operators we have pulled out an inverse power of $\omega$ to obtain the correct mass dimension. We see that the SCET-2 basis operators depend on at most two $\omega_i$ variables and at most one $x$ variable.

The results shown in Table~\ref{tab:structures} suggest that one needs additional two operators containing a soft derivative $i\!\overleftarrow{\delsl}_{\!\!s}^{\!\perp}$ acting on the soft quark field $\bar\Q_{s\spac[\omega]}$, but they turn out to be redundant and can be eliminated using the equation of motion $i\Dsl_s\,u_s=0$ of the massless quark field. When converted into gauge-invariant building blocks, this relation implies
\begin{equation}
   \frac{\nbsl\spac\nsl}{4}\,\frac{i\delsl_s^\perp}{in\cdot\partial_s}\spac\Q_{s\spac[\omega]}
   = - \frac{\nbsl}{2}\,Q_{s\spac[\omega]} 
    - \frac{\nb\cdot v}{\omega}\,\frac{\nbsl\spac\nsl}{4} 
    \int\!d\omega'\!\int\!d\omega_g\,\delta(\omega-\omega'-\omega_g)
    \left( \Gsl_{s\spac[\omega_g]}^\perp + Q_u\spac\Asl_{s\spac[\omega_g]}^\perp \right)
    Q_{s\spac[\omega']} \,.
\end{equation}
Using the hermitian conjugate of this identity, we find
\begin{equation}\label{eq:redundantops}
\begin{aligned}
   & \frac{m_\ell\,\nb\cdot v}{\nb\cdot\P_c}\,\bigg( \bar\Q_{s\spac[\omega]}\,
    \frac{i\!\overleftarrow{\delsl}_{\!\!s}^{\!\perp}}{in\cdot\!\overleftarrow{\partial}_{\!\!s}}\,\nsl\spac P_R\,\H_v \bigg)\,
    \big( \bar\X_c^{(\ell)} P_L\spac\nu_\cb \big) \\
   &= - Q_1^C(\omega) 
    - \int\!d\omega'\!\int\!d\omega_g\,\delta(\omega-\omega'-\omega_g)\,
    \Big[ Q_u\,Q_1^E(\omega',\omega_g) + Q_2^E(\omega',\omega_g) \Big] \,, \\ 
   & \frac{\nb\cdot v}{\nb\cdot\P_c}\,\bigg( \bar\Q_{s\spac[\omega]}\,
    \frac{i\!\overleftarrow{\delsl}_{\!\!s}^{\!\perp}}{in\cdot\!\overleftarrow{\partial}_{\!\!s}}\,\nsl\spac P_R\,\H_v \bigg)\,
    \big( \bar\X_c^{(\ell)}\spac\Asl_{c\spac[x]}^\perp P_L\spac\nu_\cb \big) \\
   &= - Q_1^D(x,\omega) 
    - \int\!d\omega'\!\int\!d\omega_g\,\delta(\omega-\omega'-\omega_g)\,
    \Big[ Q_u\,Q_1^F(x,\omega',\omega_g) + Q_2^F(x,\omega',\omega_g) \Big] \,,
\end{aligned}
\end{equation}
which shows that these two operators can be omitted from the basis.

\begin{table}[tp]
\centering
\begin{tabular}{ccc}
\hline
\rowcolor{\shadecolor{25}}
SCET-1 operator & SCET-2 operators & Jet function$^*$ \\ 
\hline
$O_1^A$ & $Q_1^A$ & 1 \\
$O_2^A$ & $Q_2^A$ & 1 \\
$O_{3,4}^A$ & -- & -- \\
\hline
$O_1^B$ & $Q_1^B$ & 1 \\
$O_2^B$ & $Q_2^B$ & 1 \\
$O_{3,4}^B$ & -- & -- \\
\hline
$O_1^C$ & $Q_1^C$, $Q_1^E$, $Q_2^E$ & $\alpha$ \\
$O_2^C$ & $Q_2^C$ & $\alpha$ \\
$O_{3,4}^C$ & $Q_1^C$, $Q_1^D$, $Q_1^E$, $Q_2^E$, $Q_1^F$, $Q_2^F$ & $\alpha$ \\
$O_5^C$ & $Q_2^C$, $Q_2^D$ & $\alpha^\dagger$ \\
\hline
$O_1^D$ & $Q_2^C$ & $\alpha^2$ \\
$O_2^D$ & $Q_2^C$ & $\alpha\spac\alpha_s$ \\
$O_3^D$ & $Q_1^C$, $Q_1^D$, $Q_1^E$, $Q_2^E$, $Q_1^F$, $Q_2^F$ & $\alpha^2$ \\
$O_4^D$ & $Q_1^C$, $Q_1^D$, $Q_1^E$, $Q_2^E$, $Q_1^F$, $Q_2^F$ & $\alpha\spac\alpha_s$ \\
$O_{5,7}^D$ & $Q_2^C$, $Q_2^D$ & $\alpha^2$ \\
$O_{6,8}^D$ & $Q_2^C$, $Q_2^D$ & $\alpha\spac\alpha_s$ \\
\hline
$O_1^E$ & $Q_2^C$, $Q_2^D$ & $\alpha^3$ \\
$O_2^E$ & $Q_2^C$, $Q_2^D$ & $\alpha^2\spac\alpha_s$ \\
$O_{3,4}^E$ & $Q_2^C$, $Q_2^D$ & $\alpha\spac\alpha_s^2$ \\
\hline
$O_1^F$ & $Q_1^C$, $Q_1^E$, $Q_2^E$ & $\alpha^\dagger$ \\
\hline
\end{tabular}
\caption{\label{tab:SCET2match}
Pattern of SCET-1$\,\to\,$SCET-2 matching relations, showing the set of SCET-2 operators onto which a given SCET-1 basis operator can be matched. The last column indicates at which order in perturbation theory this matching can first occur. This is the minimum perturbative suppression of the corresponding jet functions (times the low-energy matrix elements, if they contain a photon loop). It is not excluded that jet functions vanish upon explicit calculations. Indeed, for the cases marked with a dagger, we have explicitly shown that the jet functions vanish at the indicated order.}
\end{table}

Table~\ref{tab:SCET2match} shows the SCET-2 basis operators onto which the various SCET-1 basis operators can be matched and at which order of perturbation theory the corresponding jet functions can first arise. We do not exclude the possibility that some matching coefficients can be zero. As indicated by the symbol $\alpha^\dagger$, the one-loop jet functions in fact disappear in the cases of $O_5^C$ and $O_1^F$. For the SCET-2 operators of type-$B$, $D$ and $F$, the presence of the collinear photon field implies that their matrix elements involve a collinear photon loop and hence will be suppressed by a power of $\alpha$. Consequently, the type-$D$ and type-$F$ SCET-2 operators are only needed to calculate the decay amplitude $B^-\to\ell^-\spac\bar\nu_\ell$ in $\mathcal{O}(\alpha^2)$, which is beyond the scope of the present work.

\subsection{Results for the bare jet functions}
\label{sec:jet_func}

We now present the relevant jet functions arising in the matching of the SCET-1 basis operators $O_i^X$ onto the SCET-2 basis operators defined in \eqref{eq:SCET2basis_local} and \eqref{eq:SCET2basis_nonlocal}. As always in this paper, we restrict ourselves to terms contributing at $\mathcal{O}(\alpha)$ to the \blnu\ decay amplitude.

\subsubsection*{\boldmath Type-$A$ and type-$B$ SCET-1 operators}

For these operators we find the non-zero matching relations 
\begin{equation}\label{eq:4.25}
\begin{aligned}
   J_{O_1^A\to Q_1^A} &= J_{O_2^A\to Q_2^A} = 1 \,, \\
   J_{O_1^B\to Q_1^B}(y,x) & = J_{O_2^B\to Q_2^B}(y,x) = \delta(x-y) \,.
\end{aligned}
\end{equation}
Both relations hold to all orders of perturbation theory.

\subsubsection*{\boldmath Type-$C$ SCET-1 operators}

\begin{figure}
\centering
\includegraphics[scale=0.55]{feyn_ctype_1.pdf}\quad
\includegraphics[scale=0.55]{feyn_ctype_2.pdf}\quad
\includegraphics[scale=0.55]{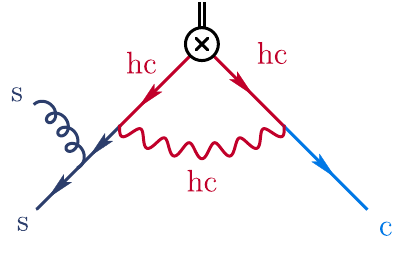}\quad
\includegraphics[scale=0.55]{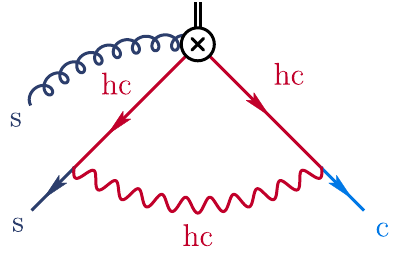} 
\caption{Representative hard-collinear loop diagrams for the matching of the type-$C$ SCET-1 operators onto SCET-2. Red (dark blue) lines represent hard-collinear (soft) propagators and external states. Light blue lines represent collinear external states. The double line at the vertex represents the $b$-quark and neutrino fields.}
\label{fig:graphs_ctype}
\end{figure}

For the type-$C$ operators, the jet functions start at one-loop order in QED and are obtained by calculating the diagrams shown in Figure~\ref{fig:graphs_ctype}. Starting with the matching onto type-$C$ SCET-2 operators (first diagram), we find for the non-vanishing jet functions in $d=4-2\epsilon$ spacetime dimensions (with $0\le y\le 1$)
\begin{equation}\label{eq:2pjet}
\begin{aligned}
   J_{O_1^C\to Q_1^C}(y,\omega) 
   &= - Q_\ell\spac Q_u\,\frac{\alpha}{2\pi}\,
    \frac{e^{\epsilon\gamma_E}\,\Gamma(\epsilon)}{1-\epsilon}
    \left( \frac{\mu^2\,\nb\cdot v}{y\spac(1-y)\,\nb\cdot\P_c\,\omega} \right)^\epsilon 
    (1-y) \left( \frac{1}{y}  + 1 - \epsilon \right) , \\
   J_{O_2^C\to Q_2^C}(y,\omega)
   &= Q_\ell\spac Q_u\,\frac{\alpha}{2\pi}\,e^{\epsilon\gamma_E}\,\Gamma(\epsilon) 
    \left( \frac{\mu^2\,\nb\cdot v}{y\spac(1-y)\,\nb\cdot\P_c\,\omega} \right)^\epsilon 
    (1-y)^2 \,, \\
   J_{O_3^C\to Q_1^C}(y,\omega)  
   &= Q_\ell\spac Q_u\,\frac{\alpha}{2\pi}\,\frac{e^{\epsilon\gamma_E}\,
    \Gamma(1+\epsilon)}{1-\epsilon}
    \left( \frac{\mu^2\,\nb\cdot v}{y\spac(1-y)\,\nb\cdot\P_c\,\omega} \right)^\epsilon 
    y\spac(1-y) \,, \\
   J_{O_4^C\to Q_1^C}(y,\omega)  
   &= - Q_\ell\spac Q_u\,\frac{\alpha}{2\pi}\,\frac{e^{\epsilon\gamma_E}\,
    \Gamma(1+\epsilon)}{1-\epsilon}
    \left( \frac{\mu^2\,\nb\cdot v}{y\spac(1-y)\,\nb\cdot\P_c\,\omega} \right)^\epsilon 
    y\spac(1-y) \,.
\end{aligned}
\end{equation}
Despite appearance, these bare expressions are independent of the scale $\mu$, since the dimensionless bare coupling $\alpha\propto\mu^{-2\epsilon}$. In the SCET-friendly projection scheme defined in Section~\ref{subsec:evanescent}, there are no evanescent contributions to these expressions. Note that the SCET-1 operator $O_5^C$ does not match onto the SCET-2 basis operators at one-loop order. Its contributions vanish by virtue of the magic identity \eqref{eq:magic}. We do not compute the jet functions arising in the matching relations onto the type-$D$ SCET-2 operators, because their matrix elements bring in a second factor of $\alpha$ and hence are beyond our accuracy. 

The calculation of the jet functions for the matching onto SCET-2 operators of type-$E$ (remaining diagrams in Figure~\ref{fig:graphs_ctype}) is subtle and deserves a more detailed discussion. If we denote the incoming soft gluon momentum by $q_s$ and its polarization vector by $\varepsilon^a(q_s)$, the calculation gives rise to the three structures
\begin{equation}\label{eq:4.28}
   g_s\spac t^a\spac\varepsilon_\mu^a(q_s) \left[ A_1(n\cdot q_s)\,\gamma_\perp^\mu
    + A_2(n\cdot q_s)\,\frac{\qsl_s^\perp}{n\cdot q_s}\,n^\mu 
    + A_3(n\cdot q_s)\,\nbsl\spac n^\mu \right]
\end{equation}
sandwiched between the quark spinors, where the coefficient functions can also depend on $n\cdot l_s$, with $l_s$ the incoming momentum of the anti-up quark. Gauge invariance requires that $A_2(n\cdot q_s)=-A_1(n\cdot q_s)$, so that the first two terms combine to give the Feynman rule for the gauge-invariant field $\Gsl_s^\perp$ contained in the operator $Q_2^E$ in \eqref{eq:SCET2basis_nonlocal}.\footnote{The third structure corresponds to soft gluon emission from one of the soft Wilson lines contained in the operators $Q_{1,2}^C$ and can be ignored for this matching calculation.} 
In a general projection scheme, it is essential to include the last diagram in Figure~\ref{fig:graphs_ctype} in order for this relation to be satisfied. It is also important to consider the third diagram, in which the soft gluon is emitted off the soft $\bar u$ quark. While naively this graph involves a soft propagator and thus would not contribute to the matching, closer inspection shows that it contains a ``short-distance''  contribution, in which the soft propagator $1/(\rlap/l_s+\rlap/q_s)$ is canceled by a factor of $(\rlap/l_s+\rlap/q_s)$ in the numerator.\footnote{Alternatively, one could work with a Green's basis containing the redundant operator shown in the first relation of \eqref{eq:redundantops}, match its coefficient, and then use the given relation to convert the result into a contribution to the SCET-2 operator $Q_2^E$.}   
Only this short-distance piece contributes to the matching calculation. In our SCET-friendly projection scheme, only the SCET-1 operator $O_1^C$ has a non-trivial matching onto SCET-2 operators of type-$E$. We consider the case of the operator $Q_2^E$ here, which contains a soft gluon field inside the $B$ meson. The analogous operator $Q_1^E$ containing a soft photon in the $B$ meson is suppressed, relative to the former one, by a factor of $\alpha/\alpha_s$. Also, we do not compute the matching onto type-$F$ operators containing a collinear photon field, because their matrix elements bring in a second factor of $\alpha$. Evaluating the last three diagrams in Figure~\ref{fig:graphs_ctype}, we obtain the bare jet function
\begin{equation}\label{eq:3pjet}
\begin{aligned}
   J_{O_1^C\to Q_2^E}(y,\omega,\omega_g) 
   &= Q_\ell\spac Q_u\,\frac{\alpha}{2\pi}\,\frac{e^{\epsilon\gamma_E}\,
    \Gamma(\epsilon)}{1-\epsilon}\,\frac{\omega}{\omega_g} 
    \left( \frac{\mu^2\,\nb\cdot v}{y\spac(1-y)\,\nb\cdot\P_c\,\omega} \right)^\epsilon 
    \frac{(1-y)^2}{y} \\
   &\quad\times \left[ \left( 1 + \epsilon\,\frac{\omega_g}{\omega+\omega_g} \right) 
    \left( \frac{\omega+\omega_g}{\omega} \right)^{-\epsilon} - 1 \right] .
\end{aligned}
\end{equation}
Note that this expression is regular in the limit $\omega_g\to 0$.

\subsubsection*{\boldmath Type-$D$ and  type-$E$ SCET-1 operators}

\begin{figure}[t]
\centering
\includegraphics[scale=0.55]{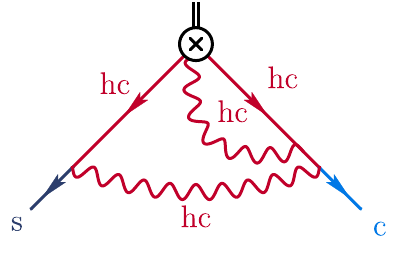} \quad
\includegraphics[scale=0.55]{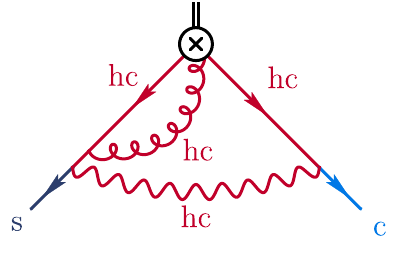} \quad
\includegraphics[scale=0.55]{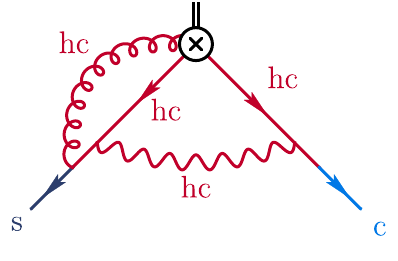} \quad
\includegraphics[scale=0.55]{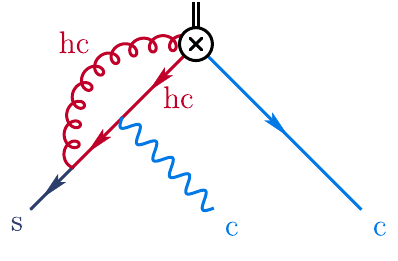} \\
\includegraphics[scale=0.55]{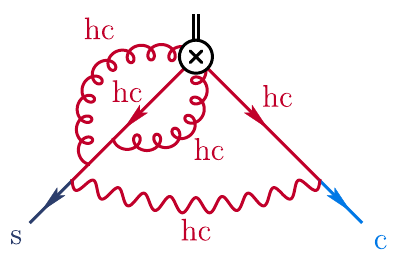} \quad
\includegraphics[scale=0.55]{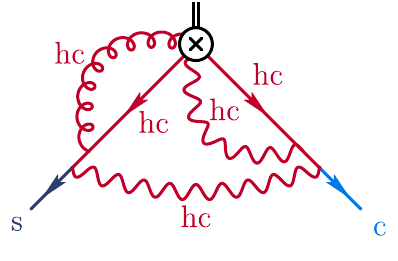} \quad
\includegraphics[scale=0.55]{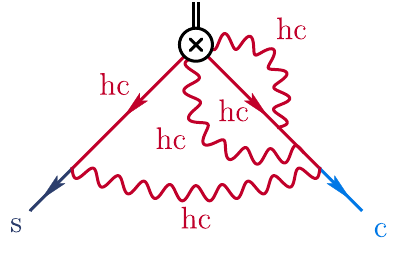} \quad
\includegraphics[scale=0.55]{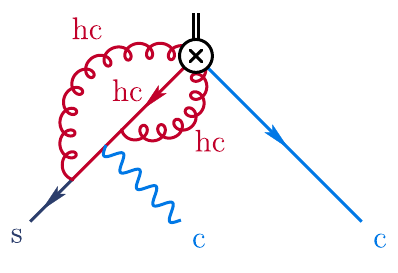} 
\caption{Representative hard-collinear loop diagrams for the matching of type-$D$ SCET-1 operators (upper row) and type-$E$ SCET-1 operators (lower row) onto SCET-2. Red (dark blue) lines represent hard-collinear (soft) propagators and external states. Light blue lines represent collinear external states.}
\label{fig:graphs_dtype}
\end{figure}

The jet functions arising from the SCET-1 operators of type-$D$ and type-$E$ are not needed for our purposes. As shown in Figure~\ref{fig:graphs_dtype}, they require at least two loops for type-$D$ (upper graphs) and three loops for type-$E$ operators (lower graphs). If the resulting SCET-2 operator contains a collinear photon field, the jet function starts one loop lower, but the matrix element of the operator contains a photon loop. Note that in each case the hard-collinear photon exchange between the up quark and the charged lepton is needed to transfer the hard-collinear momentum from the quark to the lepton side. For type-$D$ operators, we find that the $\mathcal{O}(\alpha\spac\alpha_s)$ jet functions for $O_{6,8}^D$ vanish in our projection scheme. 

\subsubsection*{\boldmath Type-$F$ SCET-1 operator}

\begin{figure}
\centering
\includegraphics[scale=0.65]{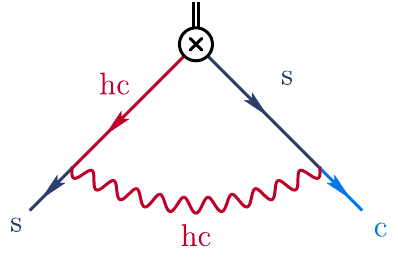} \quad
\includegraphics[scale=0.65]{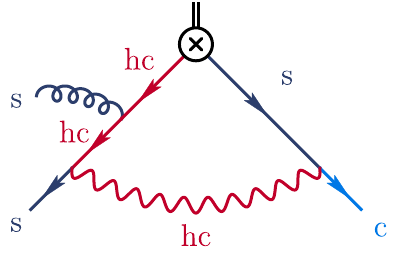} \quad
\includegraphics[scale=0.65]{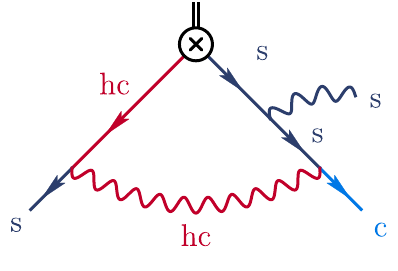} 
\caption{Representative hard-collinear loop diagrams for the matching of type-$F$ SCET-1 operators onto SCET-2. Red (dark blue) lines represent hard-collinear (soft) propagators and external states. Light blue lines represent collinear external states.}
\label{fig:graphs_ftype}
\end{figure}

The one-loop diagrams relevant for the matching of the type-$F$ operator $O_1^F$ onto our SCET-2 operator basis are depicted in Figure~\ref{fig:graphs_ftype}. As before, we only focus on the SCET-2 operators $Q_i^C$ and $Q_i^E$, because the matrix elements of operators involving a collinear photon field come with an extra factor of $\alpha$. The one-loop diagram for the matching onto type-$C$ operators (first graph) requires an analytic regulator to be well defined but then turns out to be scaleless. Hence this contribution vanishes. The one-loop diagrams for the matching onto type-$E$ operators have a non-vanishing loop integral, but their Dirac structure vanishes by virtue of the magic identity \eqref{eq:magic}. As a result, the operator $O_1^F$ does not contribute to the \blnu\ decay amplitude at $\mathcal{O}(\alpha)$.

\subsection{Endpoint-divergent convolution integrals}
\label{sec:endpoint}

Two of the jet functions calculated in \eqref{eq:2pjet} and \eqref{eq:3pjet}, namely $J_{O_1^C\to Q_1^C}$ and $J_{O_1^C\to Q_2^E}$, exhibit a singular behavior $\sim y^{-1-\epsilon}$ in the limit where $y\to 0$. One might be tempted to interpret this singularity in the sense of distributions, using the familiar relation 
\begin{equation}
   y^{-1-\epsilon}
   = - \frac{1}{\epsilon}\,\delta(y) + \left[ \frac{1}{y} \right]_+ + \mathcal{O}(\epsilon) \,,
\end{equation}
which would be valid if the two jet functions were convoluted with a function that is regular near $y=0$. But this condition is not satisfied in the case at hand, because the hard matching coefficient $H_1^C(y,\mu)$ in \eqref{eq:HiCres} is logarithmically divergent for $y\to 0$. The corresponding bare coefficient contains terms proportional to $y^{-\epsilon}$ at one-loop order. This causes a major problem, because the $1/\epsilon$ poles of the convolutions
\begin{equation}
   \int_0^1\!dy\,H_1^C(y)\,J_{O_1^C\to Q_1^C}(y,\omega) \quad\text{and}\quad
   \int_0^1\!dy\,H_1^C(y)\,J_{O_1^C\to Q_2^E}(y,\omega)
\end{equation}
cannot be removed by individually removing the poles of the hard and jet functions.

The appearance of such endpoint-divergent convolution integrals is a generic feature of SCET factorization theorems at subleading power in $\lambda$ \cite{Ebert:2018gsn,Moult:2019mog,Beneke:2019kgv,Moult:2019uhz,Beneke:2019oqx,Moult:2019vou,Liu:2019oav,Beneke:2020ibj,Liu:2020tzd,Liu:2020wbn,Beneke:2022obx,Bell:2022ott,Hurth:2023paz,Liu:2022ajh}. The ``refactorization-based subtraction (RBS) scheme'' introduced in \cite{Liu:2019oav,Liu:2020wbn} offers a systematic method for dealing with this problem. It consists of two steps: One first removes the singularities of the integrals by performing plus-type subtractions of the integrands, e.g.\
\begin{equation}\label{eq:RBFsubtractions}
   \int_0^1\!dy\,H_1^C(y)\,J_{O_1^C\to Q_1^C}(y,\omega) 
   \to \!\int_0^1\!dy\spac\Big[ H_1^C(y)\,J_{O_1^C\to Q_1^C}(y,\omega) 
    - \theta(\eta-y)\spac\llbracket H_1^C(y) \rrbracket\spac
    \llbracket J_{O_1^C\to Q_1^C}(y,\omega) \rrbracket \Big] 
\end{equation}
with $0<\eta\le 1$, and analogously for the second integral. This subtraction is similar to \eqref{eq:3.75}, but following \cite{Beneke:2022obx} we have introduced an auxiliary parameter $\eta$ in its definition. As before, the notation $\llbracket f(y)\rrbracket$ means that one retains only the leading terms of the function $f(y)$ in the limit $y\to 0$. In the second step, the subtraction terms need to be added back and combined with other terms in the factorization theorem for the process of interest. Exact $d$-dimensional refactorization conditions ensure that this is always possible, i.e., that the subtraction terms have the required form.

What is the physical meaning of the subtraction performed above? Note that $y\spac\nb\cdot p_\ell$ denotes the large momentum component of the hard-collinear anti-quark field, which for generic $y$ is of order the hard scale $m_B$. In the limit $y\to 0$, specifically in the region where $y\sim\lambda$ or smaller, this momentum component becomes soft, and the hard-collinear field in the operator $O_1^C$ should more appropriately be replaced by a soft field. To cut away the soft region from the integral, the parameter $\eta$ should scale like a soft parameter, $\eta=\mathcal{O}(\lambda)$, but it will be necessary to choose it sufficiently large that the soft scale $\Lambda\equiv\eta\spac m_B$ is in the perturbative domain. In other words, we require that
\begin{equation}\label{eq:etawindow}
   \lambda \ll \eta \ll 1 \,, 
\end{equation}
or equivalently $\Lambda_{\rm QCD}\ll\Lambda\ll m_B$. Based on the argument just presented, we expect that the subtraction term is structurally connected with the operator $O_1^A$, which is the analogue of $O_1^C$ containing a soft quark field. Indeed, the first relation in \eqref{eq:refact} shows that
\begin{equation}
   \llbracket H_1^C(y) \rrbracket
   = H_1^A\,S_1^C(\bar\omega) \,,
\end{equation}
where $H_1^A$ is the (bare) hard matching coefficient of the operator $O_1^A$, and we have defined $\bar\omega\equiv y\spac\nb\cdot p_\ell/\nb\cdot v$. The new (bare) soft function $S_1^C(\bar\omega)$ has been given at one-loop order in \eqref{eq:S1C}. Moreover, from \eqref{eq:2pjet} and \eqref{eq:3pjet} we obtain
\begin{equation}
   \llbracket J_{O_1^C\to Q_1^C}(y,\omega) \rrbracket\,dy
   = S_{O_1^C\to Q_1^C}(\bar\omega,\omega)\,\frac{d\bar\omega}{\bar\omega} 
\end{equation}
and analogously for $J_{O_1^C\to Q_2^E}$, with the (bare) soft functions
\begin{equation}
\begin{aligned}
   S_{O_1^C\to Q_1^C}(\bar\omega,\omega) 
   &= - Q_\ell\spac Q_u\,\frac{\alpha}{2\pi}\,
    \frac{e^{\epsilon\gamma_E}\,\Gamma(\epsilon)}{1-\epsilon}
    \left( \frac{\bar\omega\spac\omega}{\mu^2} \right)^{-\epsilon} 
    + \mathcal{O}(\alpha\spac\alpha_s) \,, \\
   S_{O_1^C\to Q_2^E}(\bar\omega,\omega) 
   &= Q_\ell\spac Q_u\,\frac{\alpha}{2\pi}\,
    \frac{e^{\epsilon\gamma_E}\,\Gamma(\epsilon)}{1-\epsilon}\,\frac{\omega}{\omega_g}
    \left( \frac{\bar\omega\spac\omega}{\mu^2} \right)^{-\epsilon} \\
   &\quad\times \left[ \left( 1 + \epsilon\,\frac{\omega_g}{\omega+\omega_g} \right) 
    \left( \frac{\omega+\omega_g}{\omega} \right)^{-\epsilon} - 1 \right] 
    + \mathcal{O}(\alpha\spac\alpha_s) \,.
\end{aligned}
\end{equation}
This allows us to rewrite the subtraction terms in the form
\begin{equation}
\begin{aligned}
   \int_0^\eta\!dy\,\llbracket H_1^C(y) \rrbracket\,
    \llbracket J_{O_1^C\to Q_1^C}(y,\omega) \rrbracket 
   &= H_1^A \int_0^\Lambda\frac{d\bar\omega}{\bar\omega}\,S_1^C(\bar\omega)\,
    S_{O_1^C\to Q_1^C}(\bar\omega,\omega) \\
   &= - H_1^A \int_\Lambda^\infty\frac{d\bar\omega}{\bar\omega}\,S_1^C(\bar\omega)\,
    S_{O_1^C\to Q_1^C}(\bar\omega,\omega) \,,
\end{aligned}
\end{equation}
and similarly for the second term, where we have defined $\Lambda\equiv\eta\,\nb\cdot\P_\hc/\nb\cdot v$. In the last step, we have used that the difference between the last two expressions is a scaleless integral, which vanishes in dimensional regularization. Performing the integral over $\bar\omega$, we find
\begin{equation}\label{eq:intsresu}
\begin{aligned}
   \int_\Lambda^\infty\!\frac{d\bar\omega}{\bar\omega}\,S_1^C(\bar\omega)\,
    S_{O_1^C\to Q_1^C}(\bar\omega,\omega)
   &= - Q_\ell\spac Q_u\,\frac{\alpha}{2\pi}\,
    \frac{e^{\epsilon\gamma_E}\,\Gamma(\epsilon)}{\epsilon\spac(1-\epsilon)}
    \left( \frac{\Lambda\spac\omega}{\mu^2} \right)^{-\epsilon} 
    + \mathcal{O}(\alpha\spac\alpha_s) \,, \\
   \int_\Lambda^\infty\!\frac{d\bar\omega}{\bar\omega}\,S_1^C(\bar\omega)\,
    S_{O_1^C\to Q_2^E}(\bar\omega,\omega,\omega_g) 
   &= Q_\ell\spac Q_u\,\frac{\alpha}{2\pi}\,
    \frac{e^{\epsilon\gamma_E}\,\Gamma(\epsilon)}{\epsilon\spac(1-\epsilon)}\,
    \frac{\omega}{\omega_g} \left( \frac{\Lambda\spac\omega}{\mu^2} \right)^{-\epsilon} \\
   &\quad\times \left[ \left( 1 + \epsilon\,\frac{\omega_g}{\omega+\omega_g} \right) 
    \left( \frac{\omega+\omega_g}{\omega} \right)^{-\epsilon} - 1 \right] 
    + \mathcal{O}(\alpha\spac\alpha_s) \,.
\end{aligned}
\end{equation}

In previous applications of the RBS scheme, in which the matrix elements of the operators containing soft fields were calculable in perturbation theory, adding back the subtraction term had the effect of imposing a cutoff on a convolution integral multiplying the hard function remaining after taking the endpoint limit, i.e.\ $H_1^A$ in our case (see e.g.\ \cite{Liu:2019oav,Liu:2020wbn,Liu:2022ajh,Beneke:2022obx,Hurth:2023paz}). We find that something analogous happens also in the present case, in which the soft matrix elements are non-perturbative hadronic functions (see Section~\ref{subsec:SCET2_matrix}). Concretely, the convolutions of the above expressions with the corresponding SCET-2 operators yield a new operator related closely to the operator $Q_1^A$, namely
\begin{equation}\label{eq:miracle}
\begin{aligned}
   & \int_\Lambda^\infty\!\frac{d\bar\omega}{\bar\omega}\,S_1^C(\bar\omega)\,
    \bigg[ \int\!d\omega\,S_{O_1^C\to Q_1^C}(\bar\omega,\omega)\,Q_1^C(\omega) 
    + \int\!d\omega\int\!d\omega_g\,
    S_{O_1^C\to Q_2^E}(\bar\omega,\omega,\omega_g)\,Q_2^E(\omega,\omega_g) \bigg] \\
   &= \frac{m_\ell}{\nb\cdot\P_c}\,
    \bigg[ \bar u_s\,\theta_T\bigg( 
    \frac{-i\nb\cdot\overleftarrow{D}_{\!\!s}}{\nb\cdot v}-\Lambda\bigg)\,
    \nbsl\spac P_L\spac b_v\,Y_n^{(\ell)\dagger} \bigg]\,
    \big( \bar\X_c^{(\ell)} P_L\spac\nu_\cb \big) \\
   &= \frac{m_\ell}{\nb\cdot\P_c}\,
    \bigg[ \big( \bar u_s\spac\overline{Y}_\nb^{(u)} \big)\,
    \theta_T\bigg( \frac{-i\nb\cdot\overleftarrow{\partial}_{\!\!s}}{\nb\cdot v}
    -\Lambda\bigg)\,\nbsl\spac P_L\spac 
    \big( Y_n^{(\ell)\dagger}\,\overline{Y}_\nb^{(u)\dagger}\spac b_v \big) \bigg]\,
    \big( \bar\X_c^{(\ell)} P_L\spac\nu_\cb \big) \,.
\end{aligned}
\end{equation}
In the last step, we have converted the covariant derivative into a regular one using soft Wilson lines along the direction $\nb$. We then encounter gauge-invariant building blocks for the quark fields, which are defined in analogy with the definitions in \eqref{eq:softblocks}, but with soft Wilson lines $\overline{Y}_\nb^{(u)}$ instead of $\overline{Y}_n^{(u)}$. The $\theta_T$ function restricts the component $\nb\cdot l_s/\nb\cdot v$ of the incoming anti-quark momentum to be negative and less than $-\Lambda$, which is only possible if QED effects are taken into account \cite{Beneke:2022msp}. The subscript ``$T$'' indicates that in evaluating the matrix elements of this operator one needs to perform a Taylor expansion, treating $\Lambda\gg\Lambda_\mathrm{QCD}$ as parametrically larger than all soft QCD scales. 

\begin{figure}
\centering
\includegraphics[height=2.5cm]{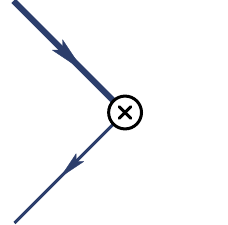} 
\includegraphics[height=2.5cm]{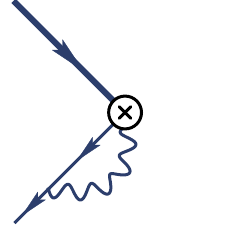} \quad
\includegraphics[height=2.5cm]{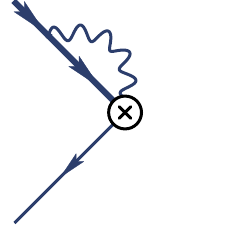} 
\includegraphics[height=2.5cm]{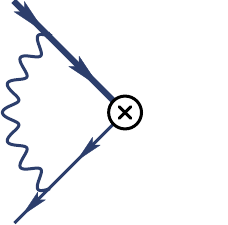} \\[2ex]
\includegraphics[height=2.5cm]{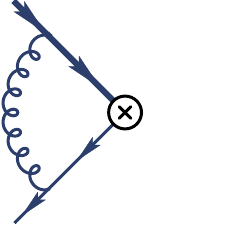} 
\includegraphics[height=2.5cm]{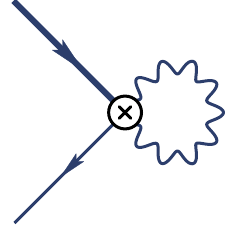} \quad
\includegraphics[height=2.5cm]{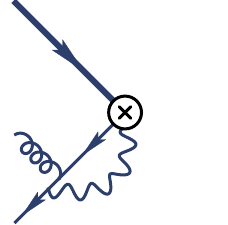} 
\includegraphics[height=2.5cm]{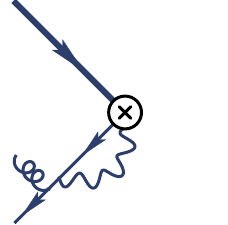}
\caption{Tree-level and one-loop diagrams contributing to the perturbative calculation of the matrix element in \eqref{eq:softloopwithLambda} in the region where $\Lambda\gg\Lambda_\mathrm{QCD}$. The tree diagram vanishes in this region. Diagrams in which a gluon is emitted from the current (not shown) give vanishing contributions.}
\label{fig:thetaops}
\end{figure}

To prove the above relation at one-loop order, we first rewrite the $\theta_T$ functions as an integral over a $\delta$-distribution (with the same Taylor expansion prescription)
\begin{equation}
   \theta_T\bigg( \frac{-i\nb\cdot\overleftarrow{\partial}_{\!\!s}}{\nb\cdot v}
    -\Lambda \bigg)
   = \int_\Lambda^\infty\!d\bar\omega\,
    \delta_T\bigg( \frac{-i\nb\cdot\overleftarrow{\partial}_{\!\!s}}{\nb\cdot v}
    -\bar\omega \bigg) \,.
\end{equation}
We then evaluate the one-loop diagrams shown in Figure~\ref{fig:thetaops} for external quark and gluon states with incoming soft momentum $l_s$ for the $\bar u$ quark, incoming residual momentum $k_s$ for the $b$ quark, and incoming momentum $q_s$ for the gluon (if present). Since $\bar\omega\ge\Lambda\gg\Lambda_\mathrm{QCD}$, we perform a systematic expansion in the ratios of the soft scales over $\bar\omega$ and keep the leading terms only. Photon fields in the operator originate from the three Wilson lines. Focusing first on the diagrams without external gluons, we find that the tree diagrams vanish, because the argument of the $\delta_T$ distribution is always negative for $\Lambda\gg\Lambda_\mathrm{QCD}$. The two diagrams in which the photon (or a gluon) is attached to the heavy quark vanish by residues. The photon tadpole graph is scaleless and vanishes in dimensional regularization. This leaves the second diagram, which gives rise to a non-vanishing contribution. Focusing next on the three-particle contributions, we find again that only diagrams with a photon emitted from the operator and attached to the $\bar u$ quark are non-vanishing. Allowing for an arbitrary Dirac structure $\Gamma$, we obtain
\begin{equation}\label{eq:softloopwithLambda}
\begin{aligned}
   & \left\langle \big( \bar u_s\spac\overline{Y}_\nb^{(u)} \big)\,
    \theta_T\bigg( \frac{-i\nb\cdot\overleftarrow{\partial}_{\!\!s}}{\nb\cdot v}
    -\Lambda \bigg)\,
    \Gamma\spac\big( Y_n^{(\ell)\dagger}\,\overline{Y}_\nb^{(u)\dagger}\spac b_v \big) \right\rangle \\
   &= - Q_\ell\spac Q_u\,\frac{\alpha}{2\pi}\,
    \frac{e^{\epsilon\gamma_E}\,\Gamma(\epsilon)}{\epsilon} 
    \int\!d\omega \left( \frac{\Lambda\spac\omega}{\mu^2} \right)^{-\epsilon} 
    \left\langle \bar\Q_{s\spac[\omega]} \left( 1 + \frac{\epsilon}{1-\epsilon}\,\frac{\nbsl\spac\nsl}{4} \right)
    \Gamma\,\H_v \right\rangle \\
   &\quad + Q_\ell\spac Q_u\,\frac{\alpha}{4\pi}\,
    \frac{e^{\epsilon\gamma_E}\,\Gamma(\epsilon)}{\epsilon\spac(1-\epsilon)}
    \int\!d\omega \int\!d\omega_g\,\frac{1}{\omega_g} 
    \left( \frac{\Lambda\spac\omega}{\mu^2} \right)^{-\epsilon} 
    \left[ \left( 1 + \epsilon\,\frac{\omega_g}{\omega+\omega_g} \right) 
    \left( \frac{\omega+\omega_g}{\omega} \right)^{-\epsilon} - 1 \right] \\
   &\quad\times \left\langle \bar\Q_{s\spac[\omega]}\,\Gsl_{s\spac[\omega_g]}^\perp\,
    \frac{\nsl}{n\cdot v}\,\Gamma\,\H_v \right\rangle .
\end{aligned}
\end{equation}
Setting $\Gamma=\nbsl\spac P_L$, we recover relations \eqref{eq:intsresu} and  \eqref{eq:miracle}. We stress that, while hadronic matrix elements of quark and gluon operators are non-perturbative quantities, which cannot be calculated in perturbation theory, the dependence of the matrix elements of the $\theta_T$ operator on the scale $\Lambda\gg\Lambda_\mathrm{QCD}$ {\em can\/} be calculated as long as $\Lambda$ is in the perturbative domain. In the following, we assume that the identification of the subtraction term with the matrix element of the $\theta_T$ operator holds to all orders of perturbation theory.

The outcome of this analysis is that removing the endpoint divergences as shown in \eqref{eq:RBFsubtractions} leads to a redefinition of the operator $O_1^A$ in the SCET-2, to which we add back the subtraction term to obtain the new operator
\begin{equation}\label{eq:thetaop}
   Q_{1,\theta}^A 
   = \frac{m_\ell}{\nb\cdot\P_c}\,
    \bigg( \bar u_s \bigg[ 1 
    - \theta_T\bigg( \frac{-i\nb\cdot\overleftarrow{D}_{\!\!s}}{\nb\cdot v}
    -\Lambda \bigg) \bigg]\spac
    \nbsl\spac P_L\spac b_v\,Y_n^{(\ell)\dagger} \bigg)\,
    \big(\bar\X_c^{(\ell)} P_L\spac\nu_\cb\big) \,.
\end{equation}
Due to the $T$ symbol in the Heaviside function, it is not legitimate to replace the terms inside the square brackets by a Heaviside function with the opposite-sign argument. In higher orders of perturbation theory, an analogous redefinition would need to be performed for the operators $Q_2^A$, $Q_1^B$ and $Q_2^B$. The result \eqref{eq:thetaop} has a nice physical interpretation: In the same way that the subtraction in \eqref{eq:RBFsubtractions} has removed the soft component in the integral over $y$, the Heaviside function in \eqref{eq:thetaop} removes the momentum modes of the soft $\bar u$ quark that would correspond to a {\em hard-collinear\/} momentum, with $\nb\cdot(- l_s)>\Lambda\gg\Lambda_{\rm QCD}$.

\subsection{Renormalized jet functions}

Once the endpoint divergences have been removed from the convolution integrals as shown in \eqref{eq:RBFsubtractions}, the hard and jet functions can be renormalized in the $\overline{\mathrm{MS}}$ scheme without paying special attention to the singularities at $y=0$. The non-vanishing jet functions that are non-trivial (i.e. different from unity) are those shown in \eqref{eq:2pjet} and \eqref{eq:3pjet}, of which only $J_{O_1^C\to Q_1^C}$ and $J_{O_2^C\to Q_2^C}$ require renormalization at one-loop order. From \eqref{eq:RGops}, it follows that after matching to SCET-2 the renormalized operators $O_i^C(\mu)$ with $i=1,2$ can be written as
\begin{equation}
   O_i(y,\mu) 
   = \int\!dx \int\!d\omega\,Z_{ii}^{CC}(y,x,\mu)\,J_{O_i^C\to Q_i^C}(x,\omega)\,
    Q_i^C(\omega) + Z_{ii}^{CA}(y,\mu)\,J_{O_i^A\to Q_i^A}\,Q_i^A \,,
\end{equation}
where the bare jet functions $J_{O_i^C\to Q_i^C}(x)$ and the renormalization factors $Z_{ii}^{CA}(y,\mu)$ start at $\mathcal{O}(\alpha)$ and have been given in \eqref{eq:2pjet} and \eqref{eq:3.75}, respectively, while $J_{O_i^A\to Q_i^A}=1$ and $Z_{ii}^{CC}(y,x,\mu)=\delta(y-x)+\mathcal{O}(\alpha)$. We now split up the bare functions in their $1/\epsilon$ pole terms, which are independent of the variable $\omega$, and finite remainders,
\begin{equation}
   J_{O_i^C\to Q_i^C}(x,\omega)
   \equiv J_{O_i^C\to Q_i^C}^{\mathrm{poles}}(x) + J_{O_i^C\to Q_i^C}^{ \mathrm{fin}}(x,\omega) \,.
\end{equation} 
Next, we use the fact that the bare operators satisfy the exact relations 
\begin{equation}
   \int\!d\omega\,Q_i^C(\omega) = Q_i^A \,.
\end{equation} 
At first order in $\alpha$, this leads to
\begin{equation}
   O_i(y,\mu) 
   = \left[ J_{O_i^C\to Q_i^C}^{\mathrm{poles}}(y) + Z_{ii}^{CA}(y,\mu) \right] Q_i^A 
    + \int\!d\omega\,J_{O_i^C\to Q_i^C}^{\mathrm{fin}}(y,\omega)\,Q_i^C(\omega) \,.
\end{equation}
It is easy to check from the explicit expressions that the two terms inside the rectangular brackets cancel each other, and only the second term remains. In the last step, we define renormalization factors for the SCET-2 operators via the condition
\begin{equation}\label{eq:RGopsSCET2}
   Q_i^X(\mu) 
   = \sum_{j,X'}\spac \mathcal{Z}_{ij}^{XX'}(\mu)\otimes Q_j^{X'} \,,
\end{equation}
which relates the renormalized (and scale-dependent) operators $Q_i^X(\mu)$ to the bare operators $O_j^{X'}$. As before, the labels $X$ and $X'$ refer to the operator type ($A$ through $F$), and the symbol $\otimes$ means an integration over the momentum variables shared by the quantities $\mathcal{Z}_{ij}^{XX'}$ and $Q_j^{X'}$. To the order we are working, we only need this relation at tree level, where $\mathcal{Z}_{ij}^{XX'}(\mu)$ is the unity operator and hence diagonal in $i,j$ and $X,X'$. This leads to the final result
\begin{equation}
   O_i(y,\mu) 
   = \int\!d\omega\,J_{O_i^C\to Q_i^C}^{\rm fin}(y,\omega)\,Q_i^C(\omega,\mu) 
    + \mathcal{O}(\alpha\spac\alpha_s) \,.
\end{equation}
Using again that $\nb\cdot\P_\hc/\nb\cdot v=m_B$ in our case to simplify the notation, we thus obtain the renormalized jet functions 
\begin{equation}\label{eq:Jrenset1}
\begin{aligned}
   J_{O_1^C\to Q_1^C}(y,\omega,\mu) 
   &= - Q_\ell\spac Q_u\,\frac{\alpha}{2\pi}\,\frac{1-y}{y} 
    \left[ (1+y) \ln\frac{\mu^2}{y\spac(1-y)\,m_B\,\omega} + 1 \right] , \\
   J_{O_2^C\to Q_2^C}(y,\omega)
   &= Q_\ell\spac Q_u\,\frac{\alpha}{2\pi}\,(1-y)^2\spac
    \ln\frac{\mu^2}{y\spac(1-y)\,m_B\,\omega} \,, \\
   J_{O_3^C\to Q_1^C}(y,\omega)  
   &= Q_\ell\spac Q_u\,\frac{\alpha}{2\pi}\,y\spac(1-y) \,, \\
   J_{O_4^C\to Q_1^C}(y,\omega)  
   &= - Q_\ell\spac Q_u\,\frac{\alpha}{2\pi}\,y\spac(1-y) \,,
\end{aligned}
\end{equation}
and
\begin{equation}\label{eq:Jrenset2}
   J_{O_1^C\to Q_2^E}(y,\omega,\omega_g) 
   = - Q_\ell\spac Q_u\,\frac{\alpha}{2\pi}\,\frac{(1-y)^2}{y}
    \left( \frac{\omega}{\omega_g} \ln\frac{\omega+\omega_g}{\omega} 
    - \frac{\omega}{\omega+\omega_g} \right) .
\end{equation}
For performing the subtracted convolution integrals in \eqref{eq:RBFsubtractions}, we also need the expressions
\begin{equation}
\begin{aligned}
   \llbracket J_{O_1^C\to Q_1^C}(y,\omega,\mu) \rrbracket 
   &= - Q_\ell\spac Q_u\,\frac{\alpha}{2\pi}\,\frac{1}{y} 
    \left( \ln\frac{\mu^2}{y\spac m_B\,\omega} + 1 \right) , \\
   \llbracket J_{O_1^C\to Q_2^E}(y,\omega,\omega_g) \rrbracket
   &= - Q_\ell\spac Q_u\,\frac{\alpha}{2\pi}\,\frac{1}{y}
    \left( \frac{\omega}{\omega_g} \ln\frac{\omega+\omega_g}{\omega} 
    - \frac{\omega}{\omega+\omega_g} \right) . 
\end{aligned}
\end{equation}

\subsection[\texorpdfstring{Two-scale RG evolution of the hard function $H_1^C$}{Two-scale RG evolution of the hard function H1C}]{\boldmath Two-scale RG evolution of the hard function $H_1^C$}
\label{subsec:2-scales}

Let us return to the subtracted convolution integral in \eqref{eq:RBFsubtractions}, now expressed in terms of renormalized functions, i.e.\footnote{In general, moving the cutoff $\eta$ from an integral over bare functions over to a corresponding integral of renormalized functions can give rise to additional correction terms, which must be calculated and properly accounted for \cite{Liu:2020wbn,Liu:2022ajh}. In our case, these terms could only arise the from the renormalization of the jet functions, which as discussed in the previous section is a higher-order effect.} 
\begin{equation}\label{eq:refact_ren}
\begin{aligned}
   & \int_0^1\!dy\spac\Big[ H_1^C(y,\mu)\,J_{O_1^C\to Q_1^C}(y,\omega,\mu) 
    - \theta(\eta-y)\spac\llbracket H_1^C(y,\mu) \rrbracket\spac
    \llbracket J_{O_1^C\to Q_1^C}(y,\omega,\mu) \rrbracket \Big] \\
   &= \int_0^1\!dy\spac\Big[ H_1^C(y,\mu)\,J_{O_1^C\to Q_1^C}(y,\omega,\mu) 
    - \llbracket H_1^C(y,\mu) \rrbracket\spac
    \llbracket J_{O_1^C\to Q_1^C}(y,\omega,\mu) \rrbracket \Big] \\
   &\quad + \int_\eta^1\!dy\,\llbracket H_1^C(y,\mu) \rrbracket\spac
    \llbracket J_{O_1^C\to Q_1^C}(y,\omega,\mu) \rrbracket \,.
\end{aligned}
\end{equation}
As mentioned earlier, from the point of view of subtracting the soft contribution from the integral, which involves hard and hard-collinear functions, one should require that $\lambda\ll\eta\ll 1$ is a parametrically small quantity separating the $\nb\cdot p$ components of soft and hard-collinear modes. In this case the convolution integral leads to large logarithms, which need to be resummed.\footnote{Alternatively, we could choose $\eta\sim 1$ in order to avoid these large logarithms, but then the matrix element of the soft operator contains large logarithms involving the hard scale $m_B$, which need to be resummed. This option will be discussed in Section~\ref{subsec:SCET2_matrix}.} 

The solutions \eqref{eq:HiCresRG} have been derived under the implicit assumption that the dimensionless variable $y$ can be treated as an $\mathcal{O}(1)$ quantity. This assumption is justified for the first integral on the right-hand side of \eqref{eq:refact_ren}, in which the $1/y$-enhanced terms of the integrand are removed and hence the region $y\ll 1$ gives a power-suppressed contribution. However, the assumption is invalidated for the second integral, which receives a leading contribution from the region where $y$ takes parametrically small values. We therefore need to construct an RG-improved expression for the hard function $H_1^C(y,\mu)$ that is simultaneously valid for both large and small $y$ values.\footnote{In principle the same is true for the jet functions. However, as mentioned earlier, in our case the jet functions are free of large logarithms for any reasonable choice of $\eta$.} 
Our explicit expression for the hard function $H_1^C$ in \eqref{eq:HiCres} shows that it depends on two different hard scales, $m_b\sim m_B=\nb\cdot\P_{\rm hc}/\nb\cdot v$ and $y\spac m_B$. For $y=\mathcal{O}(1)$ these scales are of the same order, but they are hierarchically separated for $y\ll 1$. The first refactorization condition in \eqref{eq:refact} suggests that we define a new function $\widetilde H_1^C$ via
\begin{equation}
   H_1^C(y,\mu) \equiv \frac{H_1^A(\mu)}{1-y}\,\widetilde H_1^C(y,\mu) \,.
\end{equation}
At one-loop order in QCD, we obtain (with $\nb\cdot\P_\hc/\nb\cdot v=m_B$)
\begin{equation}
\begin{aligned}
   \widetilde H_1^C(y,\mu) 
   &= 1 + \frac{C_F\spac\alpha_s(\mu)}{4\pi}\,\bigg[ 
    - \frac12\spac\ln^2\frac{\mu^2}{(y\spac m_B)^2} - \ln\frac{\mu^2}{(y\spac m_B)^2} 
    - 2 - \frac{5\pi^2}{12} \\
   &\hspace{3.25cm} + 2\spac\text{Li}_2(yz) + 2\ln yz\spac\ln(1-yz) 
    + \frac{yz\spac\ln yz}{1-yz} \bigg] \,,
\end{aligned}
\end{equation}
where the terms in the second line are of $\mathcal{O}(y)$. From the RG equations of the hard functions $H_1^C$ and $H_1^A$ in \eqref{eq:H1Afin} and \eqref{eq:HBCfin}, it follows that in our approximation
\begin{equation}
\begin{aligned}
   \frac{d}{d\ln\mu}\,\widetilde H_1^C(y,\mu) 
   &= \left\{ \gamma^A(y,\alpha_s) - \gamma_{\rm hl}(\alpha_s) 
    - \frac{\alpha}{2\pi}\!\left[ Q_b\spac Q_u\spac\ln\frac{\mu^2}{(y\spac m_B)^2} 
    - 2\spac Q_l\spac Q_b\spac\ln(1-y) \right] \right\}
    \widetilde H_1^C(y,\mu) \\
   &= \left[ - \frac{C_F}{2}\,\gamma_{\rm cusp}(\alpha_s)\spac\ln\frac{\mu^2}{(y\spac m_B)^2}
    - \frac{C_F\spac\alpha_s}{2\pi} 
    - Q_b\spac Q_u\,\frac{\alpha}{2\pi}\spac\ln\frac{\mu^2}{(y\spac m_B)^2} + \dots \right] \widetilde H_1^C(y,\mu) \spac ,
\end{aligned}
\end{equation}
where in the second line we have neglected the QED correction proportional to $\ln(1-y)$, which is beyond our approximation. Solving this equation yields
\begin{equation}\label{eq:tildeH1Csol}
\begin{aligned}
   \widetilde H_1^C(y,\mu) 
   &= \exp\left[ C_F\spac S_{\rm cusp}(y\spac m_B,\mu)
    + \frac{C_F}{\beta_0}\,\ln\frac{\alpha_s(\mu)}{\alpha_s(y\spac m_B)} 
    - Q_b\spac Q_u\,\frac{\alpha}{8\pi}\spac\ln^2\frac{\mu^2}{(y\spac m_B)^2} \right] 
    \widetilde H_1^C(y,y\spac m_B) \\
   &= \widetilde U_C(\mu,y\spac m_B)\,\widetilde H_1^C(y,y\spac m_B) \,,
\end{aligned}
\end{equation}
where
\begin{equation}
   \widetilde U_C(\mu,y\spac m_B) 
   \equiv \frac{U_C(\mu,y\spac m_B)}{U_{\rm hl}(\mu,y\spac m_B)}\spac 
    \exp\left[ - Q_b\spac Q_u\,\frac{\alpha}{8\pi}\spac\ln^2\frac{\mu^2}{(y\spac m_B)^2} \right] ,
\end{equation}
and the matching condition
\begin{equation}
   \widetilde H_1^C(y,y\spac m_B) 
   = 1 + \frac{C_F\spac\alpha_s(y\spac m_B)}{4\pi} \left[ - 2 - \frac{5\pi^2}{12} 
    + 2\spac\text{Li}_2(yz) + 2\ln yz\spac\ln(1-yz) 
    + \frac{yz\spac\ln yz}{1-yz} \right] 
\end{equation}
is free of large logarithms. With these results, the RG-improved expression for $H_1^C$ reads
\begin{equation}\label{eq:9}
   H_1^C(y,\mu) 
   = U_{\rm hl}(\mu,m_B)\,H_1^A(m_B)\spac
    \exp\left[ - Q_l\spac Q_b\,\frac{\alpha}{8\pi}\spac\ln^2\frac{\mu^2}{m_B^2} \right] 
    \widetilde U_C(\mu,y\spac m_B)\,\frac{\widetilde H_1^C(y,y\spac m_B)}{1-y} \,.
\end{equation}
It is valid for any relation between the hard scales $m_B$ and $y\spac m_B$ and thus can be used to evaluate the last convolution integral in \eqref{eq:refact_ren} in a consistent way. In particular, we find that $H_1^C(y,\mu)$ smoothly approaches the function
\begin{equation}
\begin{aligned}
   \llbracket H_1^C(y,\mu) \rrbracket 
   &= U_{\rm hl}(\mu,m_B)\,H_1^A(m_B)\spac
    \exp\left[ - Q_l\spac Q_b\,\frac{\alpha}{8\pi}\spac\ln^2\frac{\mu^2}{m_B^2} \right] \\
   &\quad\times \widetilde U_C(\mu,y\spac m_B)\,\bigg\{
    1 + \frac{C_F\spac\alpha_s(y\spac m_B)}{4\pi} \left[ - 2 - \frac{5\pi^2}{12} 
    \right] \!\bigg\} 
\end{aligned}
\end{equation}
in the limit where $y$ approaches small values. The fact that the solution \eqref{eq:9} involves the running coupling $\alpha_s(y\spac m_B)$ does not pose any conceptual problem, because the last integral in \eqref{eq:refact_ren} does not run below the scale $\eta\spac m_B=\Lambda$, which by assumption is above the Landau pole of the running coupling $\alpha_s(\mu)$. 

It is instructive to consider a further generalization of the result \eqref{eq:9}, obtained by introducing flexible matching scales $\mu_h\sim m_B$ and $\bar\mu_h\sim y\spac m_B$. This leads to (neglecting single-logarithmic QED effects)
\begin{equation}
   H_1^C(y,\mu) 
   = U_{\rm hl}(\mu,\mu_h)\,H_1^A(\mu_h)\spac
    \exp\left[ - Q_l\spac Q_b\,\frac{\alpha}{8\pi}\spac\ln^2\frac{\mu^2}{\mu_h^2} \right] 
    \widetilde U_C(\mu,\bar\mu_h)\,\frac{\widetilde H_1^C(y,\bar\mu_h)}{1-y}
    \left( \frac{y\spac m_B}{\bar\mu_h} \right)^{-\delta(\mu,\bar\mu_h)} ,
\end{equation}
where
\begin{equation}
   \delta(\mu,\bar\mu_h) 
   \equiv \frac{C_F\gamma_0}{2\beta_0}\spac\ln\frac{\alpha_s(\mu)}{\alpha_s(\bar\mu_h)} \,.
\end{equation}
Setting $\mu_h=\bar\mu_h=m_B$, one would reproduce the result for $H_1^C$ given in \eqref{eq:HiCresRG}. However, in this case the matching condition $\widetilde H_1^C(y,m_B)$ contains terms proportional to $\ln^n y$ with $n=1,2$, which become large for $y\ll 1$. This approximation is therefore only justified for $y$ values of $\mathcal{O}(1)$.

\subsection{RG evolution in SCET-2}

The SCET-2 basis operators in \eqref{eq:SCET2basis_local} and \eqref{eq:SCET2basis_nonlocal} consist of products of a hadronic current, composed of soft fields, and a leptonic current, composed of collinear fields and the (sterile) neutrino. Importantly, soft and collinear fields do not interact with each other, so the two currents can be treated separately. The hadronic currents include the two-particle operators 
\begin{equation}\label{eq:jhad1}
\begin{aligned}
   j_1^{\mathrm{had}}(\Lambda,\mu) 
   &= \bar u_s\spac\bigg[ 1 
    - \theta_T\bigg( \frac{-i\nb\cdot\overleftarrow{D}_{\!\!s}}{\nb\cdot v}
    -\Lambda \bigg) \bigg]\,
    \frac{\nbsl}{\nb\cdot v}\spac P_L\spac b_v\,Y_n^{(\ell)\dagger} \,, \\
   j_2^{\mathrm{had}}(\Lambda,\mu)
   &= \bar u_s\spac\bigg[ 1
    - \theta_T\bigg( \frac{-i\nb\cdot\overleftarrow{D}_{\!\!s}}{\nb\cdot v}
    -\Lambda \bigg) \bigg]\,
    \frac{\nsl}{n\cdot v}\spac P_L\spac b_v\,Y_n^{(\ell)\dagger} \,, \\
   j_3^{\mathrm{had}}(\omega,\mu)
   &= \bar\Q_{s\spac[\omega]}\,\frac{\nbsl}{\nb\cdot v}\spac P_L\,\H_v \,, \\
   j_4^{\mathrm{had}}(\omega,\mu)
   &= \bar\Q_{s\spac[\omega]}\,\frac{\nsl}{n\cdot v}\spac P_L\,\H_v \,,       
\end{aligned}
\end{equation}
and the three-particle operators 
\begin{equation}\label{eq:jhad2}
\begin{aligned}
   j_5^{\mathrm{had}}(\omega,\omega_g,\mu) 
   &= \frac{1}{\omega}\,\bar\Q_{s\spac[\omega]}\,\Gsl_{s\spac[\omega_g]}^\perp\, 
    \frac{\nsl}{n\cdot v}\spac P_R\,\H_v \,, \\
   j_6^{\mathrm{had}}(\omega,\omega_g,\mu)
   &= \frac{1}{\omega}\,\bar\Q_{s\spac[\omega]}\,\Asl_{s\spac[\omega_g]}^\perp\,
    \frac{\nsl}{n\cdot v}\spac P_R\,\H_v \,.
\end{aligned}
\end{equation}
The leptonic currents are 
\begin{equation}\label{eq:jlep}
\begin{aligned}
   j_1^{\mathrm{lep}}(\mu)
   &= m_\ell\,\bar\X_c^{(\ell)} P_L\spac\nu_\cb \,, \\
   j_2^{\mathrm{lep}}(x,\mu)
   &= \bar\X_c^{(\ell)}\spac\Asl_{c\spac[x]}^\perp\spac P_L\spac\nu_\cb \,, 
\end{aligned}
\end{equation}
where $m_\ell$ still is the running lepton mass in the $\overline{\mathrm{MS}}$ scheme. 

The scale dependence of these currents is controlled by RG evolution equations. Matrix elements of the hadronic currents can be expressed in terms of HQET decay constants and LCDAs of the $B$-meson, as will be discussed in Section~\ref{subsec:SCET2_matrix}. These objects should be evolved from the hard-collinear matching scale $\mu_\hc\sim\sqrt{m_B\spac\Lambda_{\rm QCD}}$ to a hadronic scale $\mu_0\sim\text{several times $\Lambda_\mathrm{QCD}$}$, which should still be in the perturbative domain. With $\Lambda_\mathrm{QCD}\sim 500$\,MeV, a typical value of the hard-collinear scale is $\mu_\hc\approx 1.6$\,GeV, which is approximately  $3\times\Lambda_\mathrm{QCD}$. Hence, in practice there is little room for this scale evolution, and hence no need to resum logarithms in the jet functions. We will simply perform the matching to SCET-2 at the scale $\mu_0\approx\mu_\hc$. Let us mention an interesting observation at this point. While the ``local'' current operators $\bar u_s\spac\Gamma\,b_v\,Y_n^{(\ell)\dagger}$ with $\Gamma=\nbsl\spac P_L$ or $\nsl\spac P_L$ are gauge invariant, they exhibit a peculiar behavior under scale evolution when QED effects are taken into account. As we will show in Section~\ref{subsec:SCET2_matrix}, these currents mix with the ``non-local'' currents $j_i^{\mathrm{had}}$ with $i=3,4,5,6$ under renormalization. This is an unexpected phenomenon, which does not occur when QED is switched off. Interestingly, we find that with the inclusion of the $\theta_T$ operator in the definitions of $j_1^{\mathrm{had}}$ and $j_2^{\mathrm{had}}$, this mixing disappears and the two currents evolve multiplicatively (at least to one-loop order). In essence, the $\theta_T$ operator removes contributions to the matrix elements of the currents from large negative values of the light-cone momentum $n\cdot l_s$ of the light spectator quark. These large negative momenta seem to be responsible for the mixing with the non-local operators.

Matrix elements of the leptonic currents should be evolved from the scale of the charged-lepton mass up to the matching scale $\mu_0$, thereby resumming large QED logarithms of the ratio $(\mu_0^2/m_\ell^2)$, which is larger than 200 in the muon case. At one-loop order, the RG equations satisfied by the leptonic currents $j_1^{\mathrm{lep}}$ and $j_2^{\mathrm{lep}}$ are 
\begin{equation}
\begin{aligned}
   \frac{d}{d\ln\mu}\,j_1^{\mathrm{lep}}(\mu) 
   &= Q_\ell^2\,\frac{\alpha}{\pi} \left( \ln\frac{\mu}{m_\ell} - \frac54 \right)
    j_1^{\mathrm{lep}}(\mu) \,, \\
   \frac{d}{d\ln\mu}\,j_2^{\mathrm{lep}}(x,\mu) 
   &= - Q_\ell\,\frac{\alpha}{\pi}\,x\,j_1^{\mathrm{lep}}(\mu) 
    + Q_\ell^2\,\frac{\alpha}{\pi} \int_0^1\!dx'\!
    \left[ \ln\frac{\mu}{m_\ell}\,\delta(x-x') + \gamma_{22}(x,x') \right]
    j_2^{\mathrm{lep}}(x',\mu) \,,
\end{aligned}
\label{eq:running_leptonic_currents}
\end{equation}
where the form of $\gamma_{22}$ is unknown.\footnote{Analogous equations for leptonic currents formed out of two charged muons were derived in \cite{Beneke:2019slt}. In equation~(A.43) of this paper, the quantity analogous to our $\gamma_{22}$ has been calculated at one-loop order.} 
The first equation can be solved to give (again neglecting the scale dependence of $\alpha$)
\begin{equation}\label{eq:j1lepRG}
   j_1^{\mathrm{lep}}(\mu) 
   = \exp\left[ Q_\ell^2\,\frac{\alpha}{2\pi} \left( \ln^2\frac{\mu}{m_\ell}
    - \frac52\,\ln\frac{\mu}{m_\ell} \right) \right]
    j_1^{\mathrm{lep}}(m_\ell) \,.
\end{equation}
For the second current, it will be sufficient to have the solution in the leading double-logarithmic approximation, since its matrix element starts at $\mathcal{O}(\alpha)$. This leads to 
\begin{equation}\label{eq:j2lepRG}
   j_2^{\mathrm{lep}}(x,\mu) 
   = \exp\left( Q_\ell^2\,\frac{\alpha}{2\pi}\spac\ln^2\frac{\mu}{m_\ell}
    \right) \left[ j_2^{\mathrm{lep}}(x,m_\ell) - Q_\ell\,\frac{\alpha}{\pi}\,x\,\ln\frac{\mu}{m_\ell}\,j_1^{\mathrm{lep}}(m_\ell) \right] + \dots \,,
\end{equation}
where the dots represent terms of subleading logarithmic order. 

\subsection{SCET-2 matrix elements}
\label{subsec:SCET2_matrix}

We now study the matrix elements of the SCET-2 operators between on-shell meson and lepton states and in the absence of photons emitted into the final state. In Section~\ref{sec:HHChiPT} below, we will construct a low-energy effective theory for the emission of very soft photons with energies much below the scale $\Lambda_\mathrm{QCD}$. In such a theory, the $B$ meson can be described as a point-like meson, and the charged lepton as a highly boosted heavy particle. We will find that the soft and collinear matrix elements considered below will play the role of the Wilson coefficients arising in the (non-perturbative) matching of SCET-2 onto this low-energy effective theory.  

We begin with the matrix elements of the leptonic currents defined in \eqref{eq:jlep} and define collinear functions $K_i(\mu)$ via ($i=1,2$)
\begin{equation}\label{eq:Kidef}
   \langle\ell(p_\ell)\spac\bar\nu(p_\nu)|\,j_i^{ \mathrm{\mathrm{lep}}}(x,\mu)\,|\spac 0\spac\rangle
   = m_\ell^{\mathrm{\mathrm{phys}}}\spac K_i(x,\mu)\,\,\bar u(v_\ell)\spac P_L\,v(p_\nu) \,, 
\end{equation}
where the on-shell lepton spinors satisfy $\slashed{v}_\ell\,u(v_\ell)=u(v_\ell)$ and $\nbsl\,v(p_\nu)=0$. The variable $x$ is only present for the case $i=2$. We express these matrix elements in terms of the physical (pole) mass of the charged lepton, which is related to the running mass in the $\overline{\mathrm{MS}}$ scheme by
\begin{equation}
   m_\ell^{ \mathrm{phys}} = m_\ell(\mu) \left[ 1 + Q_\ell^2\,\frac{\alpha}{2\pi} 
    \left( 3 \ln\frac{\mu}{m_\ell} + 2 \right) \right] ,
\end{equation}
where the superscript ``phys'' is dropped in the following. After a straightforward calculation, including the contribution from on-shell wave-function renormalization, we obtain at one-loop order
\begin{equation}
\begin{aligned}
   K_1(\mu)
   &= 1 + Q_\ell^2\,\frac{\alpha}{2\pi} 
    \left( \ln^2\frac{\mu}{m_\ell} - \frac52 \ln\frac{\mu}{m_\ell}
    + \frac{\pi^2}{24} - 1 \right) , \\
   K_2(x,\mu)
   &= - Q_\ell\,\frac{\alpha}{\pi}\,x
    \left( \ln\frac{\mu}{m_\ell} - \ln x - \frac12 \right) .
\end{aligned}
\end{equation}
For $\mu\gg m_\ell$, we can use the relations \eqref{eq:j1lepRG} and \eqref{eq:j2lepRG} to resum the large logarithmic corrections of these expressions, obtaining
\begin{equation}
\begin{aligned}
   K_1(\mu) 
   = \exp\left[ Q_\ell^2\,\frac{\alpha}{2\pi} \left( \ln^2\frac{\mu}{m_\ell}
    - \frac52\,\ln\frac{\mu}{m_\ell} \right) \right]
    \left[ 1 + Q_\ell^2\,\frac{\alpha}{2\pi} 
    \left( \frac{\pi^2}{24} - 1 \right) \right] , \\
   K_2(x,\mu) 
   =  - Q_\ell\,\frac{\alpha}{\pi}\,x\,
    \exp\left( Q_\ell^2\,\frac{\alpha}{2\pi}\spac\ln^2\frac{\mu}{m_\ell}
    \right) \left( \ln\frac{\mu}{m_\ell} - \ln x - \frac12 \right) ,
\end{aligned}
\end{equation}
where in the second relation only the double-logarithmic corrections are resummed. 

We now turn to the matrix elements of the hadronic currents defined in \eqref{eq:jhad1} and \eqref{eq:jhad2}. It has been shown in \cite{Beneke:2019slt,Beneke:2020vnb} that, in the presence of QED effects, these matrix elements contain an overlap contribution from the soft and collinear regions, which is in conflict with factorization. It is necessary to eliminate this contribution by dividing the hadronic currents by the vacuum matrix element of two soft Wilson lines corresponding to the $B^-$ meson and the charged lepton, 
\begin{equation}\label{eq:Rdef}
   \myR \equiv \langle\spac 0\spac|\,Y_n^{(\ell)\dagger}\,\overline{Y}_v^{(B)}\spac 
    |\spac 0\spac\rangle \,.
\end{equation}
Only when this is done, the hadronic matrix elements become independent of unphysical IR regulators (see Appendix~\ref{app:F_anodim} for a detailed discussion). 

\subsubsection*{Preliminary definitions}

We will define reduced matrix elements using the trace formalism of HQET \cite{Falk:1990yz,Neubert:1993mb}. Starting with the ``local'' current operators without the $\theta_T$ operator included, we define
\begin{equation}\label{eq:Fdefnaive}
\begin{aligned}
   & \frac{1}{R^{(\ell,B)}}\,
    \langle\spac 0\spac|\,\bar u_s\spac\Gamma\,b_v\,Y_n^{(\ell)\dagger}\spac 
    |B^-(v)\rangle \\
   &= - \frac{i}{2}\spac\sqrt{m_B}\,\, 
    \mathrm{Tr}\left\{ \left[ F_+(\mu) - \frac{\nsl}{2\spac n\cdot v}\,\Big( F_-(\mu) - F_+(\mu) \Big) \right] 
    \Gamma\spac\frac{1+\slashed{v}}{2}\spac\gamma_5\spac \right\} \\
   &= \frac{i}{2}\spac\sqrt{m_B}\,\,
    \mathrm{Tr}\left\{ \left[ \frac{\nbsl}{2\spac\nb\cdot v}\,F_+(\mu)
    + \frac{\nsl}{2\spac n\cdot v}\,F_-(\mu) \right] 
    \Gamma\spac\frac{1+\slashed{v}}{2}\spac\gamma_5\spac \right\} ,
\end{aligned}
\end{equation}
where $F_\pm$ are referred to as HQET decay constants, and $\Gamma$ can be an arbitrary Dirac structure. Due to the presence of the soft Wilson line $Y_n^{(\ell)\dagger}$, the matrix element ``knows'' about the direction $n$ of the charged lepton, and this allows for the presence of a second structure proportional to $(F_--F_+)$, which is absent in pure QCD. It follows that 
\begin{equation}
    F_\pm(\mu) = F_\mathrm{QCD}(\mu) + \mathcal{O}(\alpha) \,, 
\end{equation}
where $F_\mathrm{QCD}$ is the decay constant defined in QCD without electromagnetic effects. In an analogous way, we can express the matrix elements of the ``non-local'' currents $j_3^{\mathrm{had}}$ and $j_4^{\mathrm{had}}$ in terms of functions $\phi_\pm^B(\omega,\mu)$ defined via \cite{Grozin:1996pq}
\begin{equation}\label{eq:phidefnaive}
\begin{aligned}
   & \frac{1}{R^{(\ell,B)}}\,
    \langle\spac 0\spac|\,\bar\Q_s(z)\spac\Gamma\spac\H_v(0)\,|B^-(v)\rangle \\
   &= \frac{i}{2}\spac\sqrt{m_B}\,\int\!d\omega\,e^{-i\omega\tau}\,\,
    \mathrm{Tr}\left\{ \left[ \frac{\nbsl}{2\spac\nb\cdot v}\,F_+(\mu)\,\phi_+^B(\omega,\mu)
    + \frac{\nsl}{2\spac n\cdot v}\,F_-(\mu)\,\phi_-^B(\omega,\mu) \right] 
    \Gamma\spac\frac{1+\slashed{v}}{2}\spac\gamma_5\spac \right\} ,
\end{aligned}
\end{equation}
where $\tau=v\cdot z$, and $z\parallel n$ with $z^2=0$. Here $\phi_+^B$ and $\phi_-^B$ are the leading-twist and subleading-twist two-particle LCDAs of the $B$ meson. This definition implies
\begin{equation}\label{eq:norm}
   \int\!d\omega\,\phi_\pm^B(\omega,\mu) = 1 \,.
\end{equation}
Strictly speaking, this relation holds only for the bare LCDAs prior to the subtraction of UV pole terms. When corrections of $\mathcal{O}(\alpha_s)$ are included, the LCDAs develop a radiative tail, so that these integrals no longer converge at infinity \cite{Grozin:1996pq,Lange:2003ff,Lee:2005gza}. One then needs to refactorize these integrals in the region where $\Lambda_\mathrm{QCD}\ll\omega\ll m_b$ and match them onto the $B$-meson LCDAs defined in full QCD \cite{Lee:2005gza,Beneke:2023nmj}. For the purposes of our discussion in this work, where corrections of $\mathcal{O}(\alpha\spac\alpha_s)$ are neglected, we can ignore these subtleties and assume that relation \eqref{eq:norm} holds for the renormalized LCDAs, too.

\subsubsection*{Final definitions}

After refactorization, we have seen that the ``local'' currents get modified by the presence of the $\theta_T$ operator, which introduces the scale $\Lambda$. We thus need to generalize the definition \eqref{eq:Fdefnaive} to account for this effect, which we do by writing
\begin{equation}\label{eq:Fdeffinal}
\begin{aligned}
   & \frac{1}{R^{(\ell,B)}}\,
    \langle\spac 0\spac|\,\bar u_s\spac\bigg[ 1 
    - \theta_T\bigg( \frac{-i\nb\cdot\overleftarrow{D}_{\!\!s}}{\nb\cdot v}
    -\Lambda \bigg) \bigg]\,\Gamma\spac b_v\,Y_n^{(\ell)\dagger}\spac 
    |B^-(v)\rangle \\
   &= \frac{i}{2}\spac\sqrt{m_B}\,\,
    \mathrm{Tr}\left\{ \left[ \frac{\nbsl}{2\spac\nb\cdot v}\,F_+(\Lambda,\mu)
    + \frac{\nsl}{2\spac n\cdot v}\,F_-(\Lambda,\mu) \right] 
    \Gamma\spac\frac{1+\slashed{v}}{2}\spac\gamma_5\spac \right\} ,
\end{aligned}
\end{equation}
where as before
\begin{equation}
    F_\pm(\Lambda,\mu) = F_\mathrm{QCD}(\mu) + \mathcal{O}(\alpha) \,.
\end{equation}
But now the definitions \eqref{eq:Fdeffinal} and \eqref{eq:phidefnaive} involve different $F_\pm$ parameters, and hence the normalization condition \eqref{eq:norm} no longer holds. One may therefore adopt a different definition, employed in \cite{Beneke:2022msp}, in which all QED effects are absorbed into the LCDAs, i.e.\
\begin{equation}\label{eq:phideffinal}
\begin{aligned}
   & \frac{1}{R^{(\ell,B)}}\,
    \langle\spac 0\spac|\,\bar\Q_s(z)\spac\Gamma\spac\H_v(0)\,|B^-(v)\rangle \\
   &= \frac{i}{2}\spac\sqrt{m_B}\,F_\mathrm{QCD}(\mu)
    \int\!d\omega\,e^{-i\omega\tau}\,\,
    \mathrm{Tr}\left\{ \left[ \frac{\nbsl}{2\spac\nb\cdot v}\,\phi_+^B(\omega,\mu)
    + \frac{\nsl}{2\spac n\cdot v}\,\phi_-^B(\omega,\mu) \right] 
    \Gamma\spac\frac{1+\slashed{v}}{2}\spac\gamma_5\spac \right\} ,
\end{aligned}
\end{equation}
with
\begin{equation}
   \int\!d\omega\,\phi_\pm^B(\omega,\mu) = 1 + \mathcal{O}(\alpha) \,.
\end{equation}

With these definitions, we obtain
\begin{equation}\label{eq:Sidef}
\begin{aligned}
   S_1 = \frac{\langle\spac 0\spac|\,j_1^{ \mathrm{had}}(\Lambda,\mu)\,|B^-\rangle}{\myR}
   &= - \frac{i}{2} \sqrt{m_B}\,F_-(\Lambda,\mu) \,, \\
   S_2 = \frac{\langle\spac 0\spac|\,j_2^{ \mathrm{had}}(\Lambda,\mu)\,|B^-\rangle}{\myR}
   &= - \frac{i}{2} \sqrt{m_B}\,F_+(\Lambda,\mu) \,, \\
   S_3 = \frac{\langle\spac 0\spac|\,j_3^{ \mathrm{had}}(\omega,\mu)\,|B^-\rangle}{\myR}
   &= - \frac{i}{2} \sqrt{m_B}\,F_\mathrm{QCD}(\mu)\,
    \phi_-^B(\omega,\mu) \,, \\
   S_4 = \frac{\langle\spac 0\spac|\,j_4^{ \mathrm{had}}(\omega,\mu)\,|B^-\rangle}{\myR}
   &= - \frac{i}{2} \sqrt{m_B}\,F_\mathrm{QCD}(\mu)\,
    \phi_+^B(\omega,\mu) \,.
\end{aligned}
\end{equation}
The remaining hadronic matrix elements can be expressed in terms of three-particle LCDAs of the $B$ meson, corresponding to the Fock state with an extra gluon or photon. Generalizing the definitions in \cite{Kawamura:2001jm,Braun:2017liq} to include QED effects, we define 
\begin{equation}
\begin{aligned}\label{eq:Sinldef}
   S_5 = \frac{\langle\spac 0\spac|\,j_5^{ \mathrm{had}}(\omega,\omega_g,\mu)\,|B^-\rangle}{\myR}
   &= i\spac\sqrt{m_B}\,F_\mathrm{QCD}(\mu)\,
    \frac{\phi_{3g}^B(\omega,\omega_g,\mu)}{\omega} \,, \\
   S_6 = \frac{\langle\spac 0\spac|\,j_6^{ \mathrm{had}}(\omega,\omega_g,\mu)\,|B^-\rangle}{\myR}
   &= i\spac\sqrt{m_B}\,F_\mathrm{QCD}(\mu)\,
    \frac{\phi_{3\gamma}^B(\omega,\omega_g,\mu)}{\omega} \,.
\end{aligned}
\end{equation}
Our 3-particle LCDA $\phi_{3g}^B$ is related to the definitions in used in these references by
\begin{equation}
   \phi_{3g}^B(\omega,\omega_g,\mu)
   = \frac{\psi_A(\omega,\omega_g,\mu)-\psi_V(\omega,\omega_g,\mu)}{\omega_g}
   = \frac{\phi_3(\omega,\omega_g,\mu)}{\omega_g} \,,
\end{equation}
and the 3-particle LCDA $\phi_{3\gamma}^B$ is the analogous object defined with an electromagnetic gauge field. This latter function is suppressed relatve to $\phi_{3g}^B$ by a factor $\alpha/\alpha_s$. It would contribute to the \blnu\ process starting at $\mathcal{O}(\alpha^2)$, but can be ignored for our purposes. In the limit of very large $\mu$, the asymptotic behavior of the LCDAs for small values of the momentum variables is dictated by conformal symmetry. One finds \cite{Braun:2017liq}
\begin{equation}
   \phi_+^B(\omega,\mu) \sim \omega \,, \qquad
   \phi_-^B(\omega,\mu) \sim 1 \,, \qquad
   \phi_{3g,3\gamma}^B(\omega,\omega_g,\mu) \sim \omega\,\omega_g \,. 
\end{equation}

\subsubsection*{\boldmath Evolution equations for the parameters $F_\pm(\Lambda,\mu)$}

By carefully computing the UV divergences of the hadronic currents $j_{1,2}^{\rm had}$ defined in \eqref{eq:jhad1}, we find that the parameters $F_\pm(\Lambda,\mu)$ satisfy the RG evolution equations
\begin{equation}\label{eq:RGEsforF}
   \frac{d}{d\ln\mu}\,F_\mp(\Lambda,\mu) 
   = - \gamma_{F_\mp}(\Lambda,\mu)\,F_\mp(\Lambda,\mu) \,,
\end{equation}
with 
\begin{equation}\label{eq:gamma_F}
\begin{aligned}
   \gamma_{F_-}(\Lambda,\mu) 
   &= \gamma_{\rm hl}(\alpha_s) 
    + \frac{\alpha}{4\pi} \left( 3\spac Q_\ell^2 - 3\spac Q_b^2 
    - 2\spac Q_\ell\spac Q_u\spac\ln\frac{\mu^2}{\Lambda^2} \right) 
    + \mathcal{O}(\alpha\spac\alpha_s) \,, \\
   \gamma_{F_+}(\Lambda,\mu) 
   &= \gamma_{\rm hl}(\alpha_s) 
    + \frac{\alpha}{4\pi} \left( - Q_\ell^2 - 3\spac Q_b^2 + 4\spac Q_\ell\spac Q_b
    - 2\spac Q_\ell\spac Q_u\spac\ln\frac{\mu^2}{\Lambda^2} \right) 
    + \mathcal{O}(\alpha\spac\alpha_s) \,,
\end{aligned}   
\end{equation}
where the QCD anomalous dimension has been given in \eqref{eq:gammahl}. The two results can be simplified using charge conservation and combined into
\begin{equation}\label{eq:4.90}
   \gamma_{F_\mp}(\Lambda,\mu) 
   = \gamma_{\rm hl}(\alpha_s) 
    + \frac{Q_u\spac\alpha}{4\pi} \left( - (4\pm 2)\spac Q_\ell - 3\spac Q_u 
    - 2\spac Q_\ell\spac\ln\frac{\mu^2}{\Lambda^2} \right) 
    + \mathcal{O}(\alpha\spac\alpha_s) \,.
\end{equation}
The technical details of this calculation are presented in Appendix~\ref{app:F_anodim}. For the case of $F_-$, the above result was first derived in \cite{Cornella:2022ubo}. 

The fact that the $\Lambda$-dependent decay constants satisfy multiplicative evolution equations is non-trivial. Without refactorization, the parameters $F_\pm(\mu)$ introduced in \eqref{eq:Fdefnaive} would exhibit an off-diagonal mixing with the matrix elements of non-local operators, such that 
\begin{equation}\label{eq:F_nonrefac_RG}
\begin{aligned}
   \frac{d}{d\ln\mu}\,F_\mp(\mu) 
   &= - \left[ \gamma_{\rm hl}(\alpha_s) 
    + \frac{Q_u\spac\alpha}{4\pi} \left( - 2\spac Q_\ell - 3\spac Q_u \right) 
    \right] F_\mp(\mu) \\
   &\quad - \frac{\alpha}{2\pi}\,Q_\ell\spac Q_u\,F_\mathrm{QCD}(\mu)
\int_0^\infty\!d\omega\,\phi_\mp^B(\omega,\mu)\,
    \ln\frac{\mu^2}{\omega^2} \,.
\end{aligned}
\end{equation}
The anomalous dimensions in this result follow directly from \eqref{eq:ZiiAAfinal}. 

The $\Lambda$ dependence of the parameters $F_\mp(\Lambda,\mu)$ is compensated by an equal and opposite cutoff dependence of the RBS-subtracted convolution integrals of the hard and jet functions to render the final result independent of the cutoff scale. It is entirely determined by the subtraction terms derived from \eqref{eq:softloopwithLambda}. Taking a derivative with respect to $\ln\Lambda$, and subtracting the $1/\epsilon$ pole terms in the $\overline{\rm MS}$ scheme, we find the $\Lambda$ evolution equations\footnote{The right-hand sides of these relations can be related to the negative tails of the QED-corrected LCDAs of the $B$ meson, as defined in \cite{Beneke:2022msp}. This connection will be studied elsewhere.}
\begin{equation}\label{Lambda_evolution}
\begin{aligned}
   \frac{d}{d\ln\Lambda}\,\frac{F_-(\Lambda,\mu)}{F_{\rm QCD}(\mu)}
   &= - Q_\ell\spac Q_u\,\frac{\alpha}{2\pi}\,\bigg[ 
    \int_0^\infty\!d\omega\,\phi_-^B(\omega,\mu) 
    \left( \ln\frac{\mu^2}{\Lambda\spac\omega} + 1 \right) \\
   &\quad - 2 \int_0^\infty\!d\omega \int_0^\infty\!d\omega_g\,\phi_{3g}^B(\omega,\omega_g,\mu) 
    \left( \frac{1}{\omega_g}\,\ln\frac{\omega+\omega_g}{\omega} 
    - \frac{1}{\omega+\omega_g} \right) \bigg] \,, \\
   \frac{d}{d\ln\Lambda}\,\,\frac{F_+(\Lambda,\mu)}{F_{\rm QCD}(\mu)}
   &= - Q_\ell\spac Q_u\,\frac{\alpha}{2\pi} \int_0^\infty\!d\omega\,
    \phi_+^B(\omega,\mu)\spac\ln\frac{\mu^2}{\Lambda\spac\omega} \,,
\end{aligned}
\end{equation} 
which are valid at one-loop order. 

It is possible to include the leading-logarithmic QCD corrections to these results, which become important if $\Lambda\gg\mu_0$ (whereas $\sqrt{\Lambda\Lambda_{\rm QCD}}\sim\mu_0$ even for the extreme choice $\Lambda=m_B$). We will discuss the case of the parameter $F_-$ for concreteness. So far, we have derived the dependence of $F_-$ on $\Lambda$ by starting from the bare expressions in \eqref{eq:softloopwithLambda} and extracting the one-loop anomalous dimension $\gamma_{F_-}$ from the $1/\epsilon$ pole terms. We can alternatively derive the $\Lambda$ dependence of the renormalized function $F_-(\Lambda,\mu)$ using that the renormalized SCET-2 operator $Q_{1,\theta}^A(\mu)$ contains the integral (only the upper integration limit matters)
\begin{equation}
\begin{aligned}
   Q_{1,\theta}^A(\mu) &\ni \int^\eta\!dy\,\llbracket \widetilde H_1^C(y,\mu) \rrbracket\,
    \bigg[ \int\!d\omega\,\llbracket J_{O_1^C\to Q_1^C}(y,\omega,\mu) \rrbracket\,
    Q_1^C(\omega,\mu) \\
   &\hspace{3.92cm} + \int\!d\omega \int\!d\omega_g\,
    \llbracket J_{O_1^C\to Q_2^E}(y,\omega,\mu) \rrbracket\, 
    Q_2^E(\omega,\omega_g,\mu) \bigg] \,,
\end{aligned}
\end{equation}
where
\begin{equation}
\begin{aligned}
   \llbracket \widetilde H_1^C(y,\mu) \rrbracket
   &= \widetilde U_C(\mu,\bar\omega)\,\Big[ 1 + \mathcal{O}(\alpha_s) \Big] \,, \\
   \llbracket J_{O_1^C\to Q_1^C}(y,\omega,\mu) \rrbracket
   &= - Q_\ell\spac Q_u\,\frac{\alpha}{2\pi}\,\frac{1}{y}\,
    \bigg[ \left( \ln\frac{\mu^2}{\bar\omega\spac\omega} + 1 \right) 
    + \mathcal{O}(\alpha_s) \bigg] \,, \\
   \llbracket J_{O_1^C\to Q_2^E}(y,\omega,\omega_g,\mu) \rrbracket
   &= - Q_\ell\spac Q_u\,\frac{\alpha}{2\pi}\,\frac{1}{y}\,
    \bigg[ \left( \frac{\omega}{\omega_g} \ln\frac{\omega+\omega_g}{\omega} 
    - \frac{\omega}{\omega+\omega_g} \right) 
    + \mathcal{O}(\alpha_s) \bigg] \,,
\end{aligned}
\end{equation}
with $\bar\omega\equiv y\spac m_B$, and $\mathcal{O}(\alpha_s)$ corrections that are free of large logarithms even for $\Lambda\lesssim m_B$. The first relation follows from \eqref{eq:tildeH1Csol} by taking the limit of small $y$, and the one-loop expressions for the renormalized jet functions have been given in \eqref{eq:Jrenset1} and \eqref{eq:Jrenset2}. Taking the matrix elements of the operators involved, we then obtain 
\begin{equation}\label{eq:new_Lambda_evolution}
\begin{aligned}
   \frac{d}{d\ln\Lambda}\,\frac{F_-(\Lambda,\mu)}{F_{\rm QCD}(\mu)} 
   &= - Q_l\spac Q_u\,\frac{\alpha}{2\pi}\,\widetilde U_C(\mu,\Lambda)\,\bigg[
    \int_0^\infty\!d\omega\,\phi_-^B(\omega,\mu) 
    \left( \ln\frac{\mu^2}{\Lambda\spac\omega} + 1 \right) \\
   &\qquad - 2 \int_0^\infty\!d\omega \int_0^\infty\!d\omega_g
    \left( \frac{1}{\omega_g} \ln\frac{\omega+\omega_g}{\omega} 
    - \frac{1}{\omega+\omega_g} \right) \phi_{3g}^B(\omega,\omega_g,\mu) 
    + \mathcal{O}(\alpha_s) \bigg] \,,
\end{aligned}
\end{equation}
and similarly
\begin{equation}
   \frac{d}{d\ln\Lambda}\,\frac{F_+(\Lambda,\mu)}{F_{\rm QCD}(\mu)} 
   = - Q_l\spac Q_u\,\frac{\alpha}{2\pi}\,\widetilde U_C(\mu,\Lambda)
    \int_0^\infty\!d\omega\,\phi_+^B(\omega,\mu)\spac
    \ln\frac{\mu^2}{\Lambda\spac\omega} \,.
\end{equation}
These result differ from those in \eqref{Lambda_evolution} by the Sudakov factor $\widetilde U_C(\mu,\Lambda)$ on the right-hand side, which becomes relevant if $\Lambda$ is significantly different from the scale $\mu$. The first equation can be solved to give 
\begin{equation}\label{eq:Lambdaevolution}
\begin{aligned}
   \frac{F_-(m_B,\mu)}{F_{\rm QCD}(\mu)}
   &= \frac{F_-(\Lambda,\mu)}{F_{\rm QCD}(\mu)}
    - Q_l\spac Q_u\,\frac{\alpha}{2\pi} 
    \int_\Lambda^{m_B}\!\frac{d\bar\omega}{\bar\omega}\, 
    \widetilde U_C(\mu,\bar\omega)\,\Bigg[ \int_0^\infty\!d\omega\,\phi_-^B(\omega,\mu) 
    \left( \ln\frac{\mu^2}{\bar\omega\spac\omega} + 1 \right) \\
   &\quad - 2 \int_0^\infty\!d\omega \int_0^\infty\!d\omega_g
    \left( \frac{1}{\omega_g} \ln\frac{\omega+\omega_g}{\omega} 
    - \frac{1}{\omega+\omega_g} \right) \phi_{3g}^B(\omega,\omega_g,\mu) 
    + \mathcal{O}(\alpha_s) \Bigg] \,,
\end{aligned}
\end{equation} 
which can be used to calculate the parameter $F_-(m_B,\mu)$ in terms of $F_-(\Lambda,\mu)$ in leading logarithmic approximation.

Given these results, it is straightforward to show that\footnote{When QCD corrections to the jet functions are included, there will be additional contributions to the right-hand side, which we expect to be numerically subdominant.} 
\begin{equation}
   \frac{d}{d\ln\mu}\,\frac{d}{d\ln\Lambda}\,\frac{F_\mp(\Lambda,\mu)}{F_{\rm QCD}(\mu)} 
   = - Q_l\spac Q_u\,\frac{\alpha}{\pi}\,\widetilde U_C(\mu,\Lambda)\,
    \bigg[ 1 + \mathcal{O}\bigg(\alpha_s\spac\ln\frac{\mu^2}{\Lambda^2} \bigg) \bigg] \,,
\end{equation}
where the leading, single-logarithmic corrections arise from the $\mu$-derivative of $\widetilde U_C(\mu,\Lambda)$. It follows that the anomalous dimensions $\gamma_{F_\mp}$ in \eqref{eq:4.90} must be generalized to 
\begin{equation}\label{eq:gammaFnew}
   \gamma_{F_\mp}(\Lambda,\mu) 
   = \gamma_{\rm hl}(\alpha_s) 
    + \frac{Q_u\spac\alpha}{4\pi} \left[ - (4\pm 2)\spac Q_\ell - 3\spac Q_u 
    - 4\spac Q_\ell \int_\Lambda^\mu\!\frac{d\bar\omega}{\bar\omega}\,
    \widetilde U_C(\mu,\bar\omega) \right] 
    + \mathcal{O}\bigg(\alpha\spac\alpha_s\spac\ln^2\frac{\mu^2}{\Lambda^2} \bigg) \,.
\end{equation}
The fact that this anomalous dimension is beyond the Sudakov type, as it involves a non-linear function of $\ln(\mu^2/\Lambda^2)$ and needs to be resummed if the two scales $\Lambda$ and $\mu$ are of different order, appears to be a generic feature of SCET factorization problems at next-to-leading power (see e.g.\ \cite{Vogt:2010cv,Almasy:2010wn} and the Introduction of \cite{Beneke:2020ibj}, as well as \cite{Liu:2020wbn,Liu:2022ajh}). It deserves further exploration.

\subsection[\texorpdfstring{The virtual \blnu\ amplitude}{The virtual B->lnu amplitude}]{\boldmath The virtual \blnu\ amplitude}
\label{sec:vir_ampl}

We are now in a position to combine our results to obtain the factorized and RG-improved expression for the \blnu\ decay amplitude without additional soft photon emission. 
We stress that this is an unphysical, scale-dependent quantity, which is nevertheless a well-defined in our framework. We eliminate the HQET parameter $F_\mathrm{QCD}$ in terms of the $B$-meson decay constant $f_B$ defined in pure QCD, which can be calculated with high accuracy using lattice QCD. The relevant relation is 
\begin{equation}\label{eq:Fhqet_fB} 
   \sqrt{m_B}\spac f_B
   = \left[ H_1^A(\mu) + H_2^A(\mu) \right]_\mathrm{QCD} F_\mathrm{QCD}(\mu) \,, 
\end{equation}
which holds up to power corrections of $\mathcal{O}(\Lambda_\mathrm{QCD}/m_b)$ \cite{Neubert:1992fk}. The subscript ``QCD'' on the hard functions indicates that they must be evaluated setting $\alpha\to 0$. We then write the decay amplitude in the form 
\begin{equation}\label{eq:4.98}
   \mathcal{M}(B^-\to\ell^-\spac\bar\nu_\ell)_{\mu}
   = i\spac\sqrt2\,G_F^{(\mu)}\,V_{ub}\,m_\ell\,f_B\,
    \bar u(v_\ell)\spac P_L\spac v(p_\nu)\,\sum_{i,X}\,T_i^X{(\mu)} \,,
\end{equation}
where we have used that the vacuum matrix element $\langle\spac 0\spac|\spac R^{(\ell,B)}\spac|\spac 0\spac\rangle=1$ is scaleless without emitted photons and hence equals unity. The contributions $T_i^X$ result from the matrix elements of the relevant SCET-1 operators $O_i^X$, matched onto SCET-2 as described in this section. In full generality, we obtain 
\begin{equation}\label{eq:T12AB}
\begin{aligned}
   T_{1,2}^A{(\mu)}
   &= K_\mathrm{EW}(\mu)\,\frac{H_{1,2}^A(\mu)}{\left[H_1^A(\mu)+H_2^A(\mu)\right]_\mathrm{QCD}}\, 
    K_1(\mu)\,\frac{F_{-,+}(\Lambda,\mu)}{F_\mathrm{QCD}(\mu)} \,, \\
   T_{1,2}^B{(\mu)} 
   &= K_\mathrm{EW}(\mu) \int_0^1\!dy\,
    \frac{H_{1,2}^B(y,\mu)}{\left[H_1^A(\mu)+H_2^A(\mu)\right]_\mathrm{QCD}}\,
    K_2(y,\mu)\,\frac{F_{-,+}(\Lambda,\mu)}{F_\mathrm{QCD}(\mu)} \,, 
\end{aligned}
\end{equation}
and
\begin{equation}\label{eq:T134C}
\begin{aligned}
   T_1^C{(\mu)} 
   &= K_\mathrm{EW}(\mu)\,K_1(\mu) \int_0^1\!dy\,
    \frac{1}{\left[H_1^A(\mu)+H_2^A(\mu)\right]_\mathrm{QCD}} \\
   &\times\!\int_0^\infty\!\!d\omega\spac\bigg\{\!\! 
    \left[ H_1^C(y,\mu)\,J_{O_1^C\to Q_1^C}(y,\omega,\mu)
    - \theta(\eta-y)\,\llbracket H_1^C(y,\mu) \rrbracket\,
    \llbracket J_{O_1^C\to Q_1^C}(y,\omega,\mu) \rrbracket \right]\!
    \phi_-^B(\omega,\mu) \\
   &\quad - 2 \int_0^\infty\frac{d\omega_g}{\omega}\,
    \Big[ H_1^C(y,\mu)\,J_{O_1^C\to Q_2^E}(y,\omega,\omega_g,\mu) \\
   &\hspace{2.9cm} - \theta(\eta-y)\,\llbracket H_1^C(y,\mu) \rrbracket
    \llbracket J_{O_1^C\to Q_2^E}(y,\omega,\omega_g,\mu) \rrbracket \Big]\,
    \phi_{3g}^B(\omega,\omega_g\mu) \bigg\} \,, \\
   T_{3,4}^C{(\mu)} 
   &= K_\mathrm{EW}(\mu)\,K_1(\mu) \int_0^1\!dy \int_0^\infty\!d\omega\,
    \frac{H_{3,4}^C(y,\mu)}{\left[H_1^A(\mu)+H_2^A(\mu)\right]_\mathrm{QCD}}\,
    J_{O_{3,4}^C\to Q_1^C}(y,\omega,\mu)\,\phi_-^B(\omega,\mu) \,.
\end{aligned}
\end{equation}
The dependence on the cutoff $\Lambda=\eta\spac m_B$ cancels between the parameters $F_\mp(\Lambda,\mu)$ in \eqref{eq:T12AB} and the $\eta$-dependent terms in \eqref{eq:T134C}. As mentioned earlier, we will evaluate the scale-dependent quantities $T_i^X$ at a hadronic scale $\mu=\mu_0$, which is still in the perturbative domain. In our numerical work, we will take $\mu_0=1.5$\,GeV as default value, and estimate scale uncertainties by varying $\mu_0$ up and down by a factor $\sqrt{2}$.

For the sum of the type-$A$ and type-$B$ contributions, we find in our approximation scheme
\begin{equation}\label{eq:T12ABsum}
\begin{aligned}
   T_{1+2}^{A+B}{(\mu_0)}
   &= U_\mathrm{EW}(m_B,m_Z)\spac 
    \exp\left[ \frac{\alpha}{8\pi} \left( Q_\ell^2\spac\ln^2\frac{\mu_0^2}{m_\ell^2}
    - Q_\ell\spac Q_b\spac\ln^2\frac{\mu_0^2}{m_B^2} \right) \right]
    \frac{F_-(\Lambda,\mu_0)}{F_\mathrm{QCD}(\mu_0)} \\
   &\quad \times \bigg\{ \left[ H_1^A(m_B) + H_2^A(m_B) \right]_\mathrm{QED}^{\kappa=0} \\
   &\hspace{1.2cm} + \frac{\alpha}{4\pi}\,\bigg[
    \left( Q_\ell\spac Q_u 
     + (1+z)\,Q_\ell\spac Q_b - \frac32\,Q_b^2 \right) \ln\frac{\mu_0^2}{m_B^2} \\
   &\hspace{2.cm} - \left( 2\spac Q_\ell\spac Q_u + \frac32\,Q_\ell^2 
    + z\spac Q_\ell\spac Q_b \right) \ln\frac{\mu_0^2}{m_\ell^2} 
    + \frac92\,Q_\ell\spac Q_u + Q_\ell\spac Q_b 
    + \left( \frac{\pi^2}{12} - 2 \right) Q_\ell^2 \\
   &\hspace{2.cm} + 4\spac Q_\ell\spac Q_u
    \int_{m_B}^{\mu_0}\!\frac{d\mu'}{\mu'}\,
    \frac{\widetilde U_C(\mu',m_B)}{1-\delta(\mu')} \bigg] \bigg\} \,,
\end{aligned}
\end{equation}
where $U_\mathrm{EW}(m_B,m_Z)$ has been given in \eqref{eq:electroweak_ev}. We have used that the quantity $F_+$ appears only in the one-loop QED corrections, so it is sufficient to keeps its double-logarithmic terms, which are the same as those of $F_-$. In the matching conditions for the hard functions $H_{1,2}^A(m_B)$, given explicitly in \eqref{eq:HiAsolutions}, we only keep the QED contributions and set $\kappa=0$. At one-loop order, the $\kappa$-dependent corrections cancel between  $K_{\rm EW}$ and $H_1^A$. We further obtain for the sum of the type-$C$ contributions 
\begin{equation}\label{eq:T123C}
\begin{aligned}
   T_{1+3+4}^C{(\mu_0)}
   &= - U_\mathrm{EW}(m_B,m_Z)\spac
    \exp\left[ \frac{\alpha}{8\pi} \left( Q_\ell^2\spac\ln^2\frac{\mu_0^2}{m_\ell^2}
    - Q_\ell\spac Q_b\spac\ln^2\frac{\mu_0^2}{m_B^2} \right) \right] \\
   &\quad\times Q_\ell\spac Q_u\,\frac{\alpha}{2\pi}\,\Bigg\{\!
    \left[ \frac{1}{1-\delta(\mu_0)}\,\ln\frac{\mu_0^2}{m_B\spac\omega_-(\mu_0)} 
    - h_1\big(\delta(\mu_0)\big) \right] \widetilde U_C(\mu_0,m_B) \\
   &\hspace{3.cm} + \int_\Lambda^{m_B}\!\frac{d\bar\omega}{\bar\omega}
    \left( \ln\frac{\mu_0^2}{\bar\omega\spac\omega_-(\mu_0)} + 1 \right) 
    \widetilde U_C(\mu_0,\bar\omega) \\ 
   &\hspace{3.0cm} + 2 \left[ \frac{\widetilde U_C(\mu_0,m_B)}{1-\delta(\mu_0)} 
    - \int_\Lambda^{m_B}\!\frac{d\bar\omega}{\bar\omega}\,\widetilde U_C(\mu_0,\bar\omega)
    \right] \\
   &\hspace{3.4cm} \times  
    \int_0^\infty\!d\omega\! \int_0^\infty\!d\omega_g\,\phi_{3g}^B(\omega,\omega_g,\mu_0)
    \left[ \frac{1}{\omega_g}\,\ln\frac{\omega+\omega_g}{\omega} - \frac{1}{\omega+\omega_g} \right] \!\Bigg\} \,,
\end{aligned}
\end{equation}
where 
\begin{equation}
   h_1(\delta) = \frac{H(-\delta)}{\delta} - \frac{H(1-\delta)}{1-\delta}
    - \frac{\delta}{(1-\delta)^2} \,, 
\end{equation}
and $H(x)=\psi(1+x)+\gamma_E$ is the harmonic-number function. The parameter $\omega_-$ is defined as the logarithmic moment of the $B$-meson LCDA,
\begin{equation}\label{eq:omegaminusdef}
   \ln\frac{\omega_-(\mu)}{\nu} 
   = \int_0^\infty\!d\omega\,\phi_-^B(\omega,\mu)\,\ln\frac{\omega}{\nu} \,.
\end{equation}
where $\nu$ is an arbitrary reference scale introduced to render the arguments of the logarithms dimensionless. Note that $\omega_-(\mu)$ is scale dependent, but this dependence is beyond our accuracy. The same is true for $\delta(\mu)$. 

Given the above results, it is straightforward to show that, at one-loop order,
\begin{equation}\label{eq:Tmudep}
\begin{aligned}
   \frac{d}{d\ln\mu_0}\,\ln\sum_{i,X}\,T_i^X(\mu_0)
   &= \gamma_\mathrm{soft}(\alpha) + Q_\ell\spac Q_u\,\frac{\alpha}{2\pi}
    \left( 2 \int_{m_B}^{\mu_0}\!\frac{d\bar\omega}{\bar\omega}\,
    \widetilde U_C(\mu_0,\bar\omega)
    - \ln\frac{\mu_0^2}{m_B^2} \right) \\
   & + \frac{d}{d\ln\mu_0}\spac
    \ln\frac{F_-(\Lambda,\mu_0)}{F_\mathrm{QCD}(\mu_0)}
    + \frac{\alpha}{4\pi} \left( 3\spac Q_\ell^2 - 3\spac Q_b^2 
    - 4\spac Q_\ell\spac Q_u\!\int_\Lambda^{\mu_0}\!\frac{d\bar\omega}{\bar\omega}\,
    \widetilde U_C(\mu_0,\bar\omega) \right) ,
\end{aligned}
\end{equation}
where the first term on the right-hand side is the soft anomalous dimension
\begin{equation}\label{eq:gammasoft}
   \gamma_\mathrm{soft} 
   = Q_\ell^2\,\frac{\alpha}{2\pi} \left( \ln\frac{m_B^2}{m_\ell^2} - 2 \right) ,
\end{equation}
which cancels the scale dependence of the real-photon emission contributions calculated in a low-energy effective theory in Section~\ref{sec:HHChiPT}. The quantity $\gamma_\mathrm{soft}$ is closely related to the velocity-dependent anomalous dimension of a QED current in heavy-particle effective theory, consisting of a charged $B$-meson field with 4-velocity $v\equiv v_B$ and a soft lepton field with 4-velocity $v_\ell$, which reads \cite{Falk:1990yz} 
\begin{equation}\label{eq:gamma_soft_full}
   \gamma_\mathrm{soft}(v_\ell\cdot v_B,\alpha)
   = Q_\ell\spac Q_B\,\frac{\alpha}{\pi} \left[
    \frac{w}{\sqrt{w^2-1}}\,\ln\left( w + \sqrt{w^2-1} \right) - 1 \right] , \quad
   w = v_\ell\cdot v_B \,.
\end{equation}
With $w=(m_B^2+m_\ell^2)/(2\spac m_B\spac m_\ell)$, cf.~\eqref{eq:hardscales}, one finds
\begin{equation}
   \gamma_\mathrm{soft}(v_\ell\cdot v_B,\alpha)
   = Q_\ell\spac Q_B\,\frac{\alpha}{\pi} \left(
    \frac{m_B^2+m_\ell^2}{m_B^2-m_\ell^2}\,\ln\frac{m_B}{m_\ell} - 1 \right) ,
\end{equation}
which reduces to the result \eqref{eq:gammasoft} in the limit $m_\ell\ll m_B$.

The term in brackets in the second line of \eqref{eq:Tmudep} vanishes due to the scale dependence of the ratio $F_-(\Lambda,\mu)/F_{\rm QCD}(\mu)$ obtained from \eqref{eq:gammaFnew}. This leaves the second term in the first line, which differs from~1 by terms of order $\alpha\ln(\mu_0^2/m_B^2)\big[\alpha_s\ln^2(\mu_0^2/m_B^2)\big]^n$ with $n\in\mathbb{N}$. These leading-logarithmic QCD corrections are, however, within our accuracy goal. The fact that they do not cancel out is related to the missing two-loop terms of $\mathcal{O}(\alpha\spac\alpha_s)$ in the RG evolution equations for the hard functions $H_{1,2}^A$ in \eqref{eq:RGEsHiA}. To fix the problem, we must replace
\begin{equation}
   Q_\ell\spac Q_b\spac\ln\frac{\mu^2}{m_B^2}  
   \to Q_\ell^2\spac\ln\frac{\mu^2}{m_B^2}
    + 2\spac Q_\ell\spac Q_u \int_{m_B}^\mu\!\frac{d\bar\omega}{\bar\omega}\,
    \widetilde U_C(\mu,\bar\omega)
\end{equation}
in the homogeneous terms shown in the first line of each equation. When this is done, the QED Sudakov exponent in \eqref{eq:T12ABsum} and \eqref{eq:T123C} is replaced by
\begin{equation}
   Q_\ell^2\,\frac{\alpha}{8\pi} \left( \ln^2\frac{\mu_0^2}{m_\ell^2}
    - \ln^2\frac{\mu_0^2}{m_B^2} \right)
   - Q_\ell\spac Q_u\,\frac{\alpha}{\pi} \int_{m_B}^{\mu_0}\!\frac{d\mu'}{\mu'}
    \int_{m_B}^{\mu'}\!\frac{d\bar\omega}{\bar\omega}\,
    \widetilde U_C(\mu',\bar\omega) \,, 
\end{equation}
and the extra terms in the first line of \eqref{eq:Tmudep} vanish. 

We now rearrange the QED double and single logarithms in such a way that we factor out
\begin{equation}
   \exp\left[ Q_\ell^2\,\frac{\alpha}{8\pi} \left( 
    \ln^2\frac{\mu^2}{m_\ell^2} - \ln^2\frac{\mu^2}{m_B^2} 
    - 2 \ln\frac{\mu^2}{m_\ell^2} - 2 \ln\frac{\mu^2}{m_B^2}\right) \right] 
   = \left( \frac{\mu^2}{m_B\spac m_\ell} \right)^{\gamma_\mathrm{soft}/2} .
\end{equation}
We then define a scale-invariant quantity $\mathcal{R_{\mathrm{virt}}}$ by \begin{equation}\label{eq:total_T}
   \sum_{i,X}\,T_i^X{ (\mu_0)} 
   = \left( \frac{\alpha(m_Z)}{\alpha(m_B)} \right)^{\frac{9}{40}} 
    \left( \frac{\mu_0^2}{m_B\spac m_\ell} \right)^{\gamma_\mathrm{soft}/2}\, 
    \mathcal{R}_\mathrm{virt} \,.
\end{equation}
Within our approximations, it is given by
\begin{equation}\label{eq:4.114}
\begin{aligned}
   \mathcal{R}_{\rm virt}
   &= \exp\bigg[ - Q_\ell\spac Q_u\,\frac{\alpha}{\pi}
    \int_{m_B}^{\mu_0}\!\frac{d\mu'}{\mu'} \int_{m_B}^{\mu'}\!\frac{d\bar\omega}{\bar\omega}\,\widetilde U_C(\mu',\bar\omega) \bigg]\,
    \frac{F_-(\Lambda,\mu_0)}{F_{\rm QCD}(\mu_0)} \\
   &\quad \times \Bigg\{ 1 + \frac{\alpha}{4\pi}\,\bigg[
    \frac32\,Q_\ell^2\spac\ln\frac{\mu_0^2}{m_\ell^2} 
    - \frac32\,Q_b^2\spac\ln\frac{\mu_0^2}{m_B^2} 
    - (2+z)\,Q_\ell\spac Q_b\spac\ln\frac{m_B^2}{m_\ell^2} 
    + \left( \frac{\pi^2}{12} - \frac52 \right) Q_\ell^2 \\
   &\hspace{2.67cm} 
    + \left( - \frac12 + \frac{z^2\ln z}{z-1} + z  
    - 2\spac\text{Li}_2(1-z) - \frac{\pi^2}{12} \right) Q_\ell\spac Q_b 
    - \left( 2 + 3\ln z \right) Q_b^2 \bigg] \\
   &\hspace{1.15cm} + Q_\ell\spac Q_u\,\frac{\alpha}{\pi}\,
    \int_{m_B}^{\mu_0}\!\frac{d\mu'}{\mu'}\,\frac{\widetilde U_C(\mu',m_B)}{1-\delta(\mu')} \\
   &\hspace{1.15cm} - Q_\ell\spac Q_u\,\frac{\alpha}{2\pi}\,\Bigg[ 
    \left[ \frac{1}{1-\delta(\mu_0)}\,\ln\frac{\mu_0^2}{m_B\spac\omega_-(\mu_0)} 
    - h_1\big(\delta(\mu_0)\big) \right] \widetilde U_C(\mu_0,m_B) \\
   &\hspace{3.6cm} + \int_\Lambda^{m_B}\!\frac{d\bar\omega}{\bar\omega}
    \left( \ln\frac{\mu_0^2}{\bar\omega\spac\omega_-(\mu_0)} + 1 \right) 
    \widetilde U_C(\mu_0,\bar\omega) \Bigg] \\ 
   &\hspace{1.15cm} - Q_\ell\spac Q_u\,\frac{\alpha}{\pi} 
    \left[ \frac{\widetilde U_C(\mu_0,m_B)}{1-\delta(\mu_0)} 
    - \int_\Lambda^{m_B}\!\frac{d\bar\omega}{\bar\omega}\,\widetilde U_C(\mu_0,\bar\omega)
    \right] \\
   &\hspace{1.6cm} \times  
    \int_0^\infty\!d\omega\! \int_0^\infty\!d\omega_g\,\phi_{3g}^B(\omega,\omega_g,\mu_0)
    \left[ \frac{1}{\omega_g}\,\ln\frac{\omega+\omega_g}{\omega} - \frac{1}{\omega+\omega_g} \right]
    \!\Bigg\} \,.
\end{aligned}
\end{equation}
{ The remaining scale dependence in \eqref{eq:total_T} will cancel against a corresponding scale dependence of the real soft-photon emission contribution to the decay amplitude, see \eqref{eq:rad_rate} in Section~\ref{subsec:decayrate} below.}

Expression \eqref{eq:4.114} is the main result of our paper up to now, and it constitutes the RG-improved expression of a result presented in \cite{Cornella:2022ubo}. Using the result \eqref{eq:new_Lambda_evolution}, we obtain
\begin{equation}
   \frac{d\spac\mathcal{R}_\mathrm{virt}}{d\ln\Lambda} = 0 \,,
\end{equation}
up to terms that are beyond our approximation. A particularly simple form of the result is obtained by setting $\Lambda=m_B$, in which case 
\begin{equation}\label{eq:Rvirt_final}
\begin{aligned}
   \mathcal{R}_{\rm virt} 
   &= \exp\bigg[ - Q_\ell\spac Q_u\,\frac{\alpha}{\pi}
    \int_{m_B}^{\mu_0}\!\frac{d\mu'}{\mu'} \int_{m_B}^{\mu'}\!\frac{d\bar\omega}{\bar\omega}\,\widetilde U_C(\mu',\bar\omega) \bigg]\,
    \frac{F_-(m_B,\mu_0)}{F_{\rm QCD}(\mu_0)} \\
   &\quad \times \Bigg\{ 1 + \frac{\alpha}{4\pi}\,\bigg[
    \frac32\,Q_\ell^2\spac\ln\frac{\mu_0^2}{m_\ell^2} 
    - \frac32\,Q_b^2\spac\ln\frac{\mu_0^2}{m_B^2} 
    - (2+z)\,Q_\ell\spac Q_b\spac\ln\frac{m_B^2}{m_\ell^2} 
    + \left( \frac{\pi^2}{12} - \frac52 \right) Q_\ell^2 \\
   &\hspace{2.67cm} 
    + \left( - \frac12 + \frac{z^2\ln z}{z-1} + z  
    - 2\spac\text{Li}_2(1-z) - \frac{\pi^2}{12} \right) Q_\ell\spac Q_b 
    - \left( 2 + 3\ln z \right) Q_b^2 \bigg] \\
   &\hspace{1.15cm} + Q_\ell\spac Q_u\,\frac{\alpha}{\pi}\,
    \int_{m_B}^{\mu_0}\!\frac{d\mu'}{\mu'}\,\frac{\widetilde U_C(\mu',m_B)}{1-\delta(\mu')} \\
   &\hspace{1.15cm} - Q_\ell\spac Q_u\,\frac{\alpha}{2\pi} 
    \left[ \frac{1}{1-\delta(\mu_0)}\,\ln\frac{\mu_0^2}{m_B\spac\omega_-(\mu_0)} 
    - h_1\big(\delta(\mu_0)\big) \right] \widetilde U_C(\mu_0,m_B) \\ 
   &\hspace{1.15cm} - Q_\ell\spac Q_u\,\frac{\alpha}{\pi}\, 
    \frac{\widetilde U_C(\mu_0,m_B)}{1-\delta(\mu_0)} 
    \int_0^\infty\!d\omega\! \int_0^\infty\!d\omega_g\,\phi_{3g}^B(\omega,\omega_g,\mu_0)
    \left[ \frac{1}{\omega_g}\,\ln\frac{\omega+\omega_g}{\omega} - \frac{1}{\omega+\omega_g} \right]
    \!\Bigg\} \,.
\end{aligned}
\end{equation}
The choice $\Lambda=m_B$ (corresponding to $\eta=1$) lies outside the window indicated in \eqref{eq:etawindow}, but since our result is explicitly $\Lambda$-independent, we are free to take this choice. The large logarithms associated with the parameter $\eta$ are then entirely contained in the parameter $F_-(m_B,\mu_0)$, which now depends on the hadronic $B$-meson mass. With the help of relation \eqref{eq:Lambdaevolution}, we can related this parameter to a parameter $F_-(\Lambda,\mu_0)$ with a different choice of $\Lambda$. 

Let us finally estimate the perturbative uncertainty of the above result based on its residual scale dependence. We find that
\begin{equation}
   \frac{d}{d\ln\mu_0}\,\ln\mathcal{R}_{\rm virt}
   = \mathcal{O}\bigg(\alpha\spac\alpha_s\spac\ln\frac{\mu_0^2}{m_B^2},
    \alpha^2\spac\ln^2\frac{\mu_0^2}{m_B^2}\bigg) \,,
\end{equation}
where the former arise from the residual scale dependence of the terms involving $\widetilde U_c(\mu_0,m_B)$, while the latter are due to our restriction to the leading QED double logarithms at two-loop order and beyond. The discussion preceding the results \eqref{eq:H12Asol} shows that these ambiguities are consistent with our approximation scheme.
\footnote{Relation \eqref{eq:gammaFnew} suggests that the scale ambiguity of the ratio $F_-(m_B,\mu_0)/F_{\rm QCD}(\mu_0)$ starts at order $\alpha\spac\alpha_s\ln^2(\mu_0^2/m_B^2)$, which is one power of logarithm higher than what we found here. Since the remaining terms in the expression for $\mathcal{R}_{\rm virt}$ do not exhibit a corresponding scale ambiguity, we conclude that the true scale ambiguity in \eqref{eq:gammaFnew} is reduced to $\mathcal{O}\big(\alpha\spac\alpha_s\ln(\mu^2/m_B^2)\big)$ when higher-order QCD corrections are included.}

\subsection{Power-enhanced QED corrections from new physics}
\label{subsec:pow-enhanced-QED}

The decay $B_s\to\ell^+\spac\ell^-$ shares several similarities with the $B^-\to\ell^-\spac\bar\nu_\ell$ process considered here, in particular it is also chirally suppressed, i.e., its decay amplitude is proportional to the charged-lepton mass $m_\ell$. It has been shown in \cite{Beneke:2017vpq} that some of the QED corrections to the $B_s\to\ell^+\spac\ell^-$ decay rate are enhanced by a factor of $\mathcal{O}(m_B/\Lambda_\mathrm{QCD})$ relative to the leading-order rate. For $B^-\to\ell^-\spac\bar\nu_\ell$ in the SM, on the other hand, we have shown in Section \ref{subsec:SCET1basis} that such power-enhanced QED corrections are absent -- a fact that was also noted in \cite{Beneke:2017vpq}. This is a direct consequence of the $(V-A)\otimes(V-A)$ structure of the effective four-fermion operator mediating this decay. 

In the presence of new physics, this conclusion can be changed. Consider as an example the operator with structure $(V+A)\otimes(V-A)$ multiplying the Wilson coefficient $L_\ell^{V,LR}$ in the effective Lagrangian \eqref{eq:Lleft}. When matching this operator onto SCET-1, one encounters the operator
\begin{equation}
   \big(\bar\X_\hc^{(u)}\spac\gamma_\perp^\alpha P_R\,b_v\big)\,
   \big(\bar\X_\hc^{(\ell)}\spac\gamma_\alpha^\perp\spac P_L\spac\nu_\cb\big) 
   \sim \lambda^{\frac52} \,.
\end{equation}
Matching this operator onto SCET-2, using the same rules as in the previous section, one finds the operators
\begin{equation}
\begin{aligned}
   & \bigg(\bar u_s\,\frac{\nsl}{in\cdot\!\overleftarrow\partial_{\!\!s}}\,
    \gamma_\perp^\rho\spac\gamma_\perp^\alpha P_R\,b_v\bigg)\,
    \big(\bar\X_c^{(\ell)}\,\{ m_\ell\,\gamma_\rho^\perp, \A_{c\rho}^\perp \}\,
    \gamma_\alpha^\perp\spac P_L\spac\nu_\cb\big) \\
   &\to \bigg(\bar u_s\,\frac{\nsl}{in\cdot\!\overleftarrow\partial_{\!\!s}}\,P_R\,b_v\bigg)\,
    \big(\bar\X_c^{(\ell)}\,\{ m_\ell, \Asl_c^\perp \}\,P_L\spac\nu_\cb\big) 
    \sim \lambda_\ell^2\,\lambda^2 \,,
\end{aligned}
\end{equation}
which are enhanced by a factor $1/\lambda$ compared with the leading-order type-$A$ and type-$B$ operators 
\begin{equation}
   \frac{1}{\nb\cdot\P_c}\,\big(\bar u_s\spac\nbsl\spac P_R\,b_v\big)\,
    \big(\bar\X_c^{(\ell)}\,\{ m_\ell, \Asl_c^\perp \}\,P_L\spac\nu_\cb\big)
    \sim \lambda_\ell^2\,\lambda^3
\end{equation}
encountered in the absence of QED corrections. This is completely analogous to what happens in the case of the $B_s\to\ell^+\spac\ell^-$ process, which in the SM is mediated by operators with equal or opposite chirality. In the presence of new physics, the naive magnitude $\sim\alpha/\pi$ of QED corrections may thus be boosted by a factor $m_B/m_\ell$, which is about 50 for the case of the muon. Such an enhancement could bring QED corrections to similar size as perturbative QCD effects.

\newpage
\section{Low-energy description and real corrections}
\label{sec:HHChiPT}

We now outline the construction of the effective theory valid below the QCD confinement scale $\OurLambda\approx 500$\,MeV. In this regime, the $B$ meson can be described as a point-like object. Our primary focus is the dynamics of photons with total energy below an experimentally imposed cut $E_\gamma^{\rm tot}\le E_\mathrm{cut}\ll\OurLambda$. These very soft photons see the $B$ meson and the charged lepton as static sources of electric charge, allowing for an HQET-like description of both particles. However, it is important to include the possibility of the photon-induced excitation of the $B$ meson into the $B^*$ vector meson, whose leptonic weak decay is not chirally suppressed. The dynamics of other light degrees of freedom, specifically the light pseudoscalar mesons, can be included within the framework of heavy-hadron chiral perturbation theory (HH$\chi$PT) \cite{Wise:1992hn,Yan:1992gz,Burdman:1992gh}. For the lepton, the appropriate theory is boosted heavy-lepton effective theory (bHLET) \cite{Fleming:2007xt,Fleming:2007qr}. The ingredients of these EFT constructions will be presented below. 

\subsection{Heavy-particle effective theory} 
\label{subsec:HMET}

 Which mass and energy scales should an effective theory valid below the scale of QCD confinement capture? A first relevant scale is set by $E_{\rm cut}$, the maximal energy allowed for final-state radiation. Another important scale is the mass splitting between the $B$ meson and its first excited state, the $B^*$, which vanishes in the limit $m_b\to\infty$ due to heavy-quark spin symmetry. The associated parameter $(m_{B^*}-m_B)\sim\OurLambda^2/m_b$ is parametrically smaller than $\OurLambda$. Finally, the light pseudoscalar mesons $\pi,K,\eta$ associated with spontaneous chiral symmetry breaking in QCD are parametrically lighter than all other QCD resonances. Their masses squared are proportional to the light-quark masses, $m_{\pi,K,\eta}^2\sim m_q\spac\OurLambda$, with $m_u,m_d,m_s\ll\OurLambda$. We begin with a discussion of the first two scales and will later include the dynamics of the light pseudoscalar mesons.

\begin{figure}[t]
\centering
\includegraphics[scale=0.5]{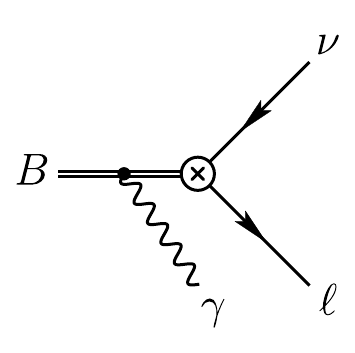} \quad
\includegraphics[scale=0.5]{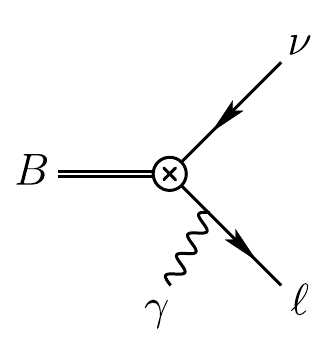} \quad
\includegraphics[scale=0.5]{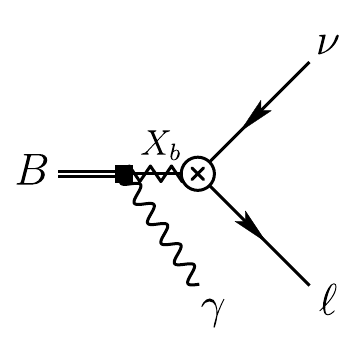} \quad
\includegraphics[scale=0.5]{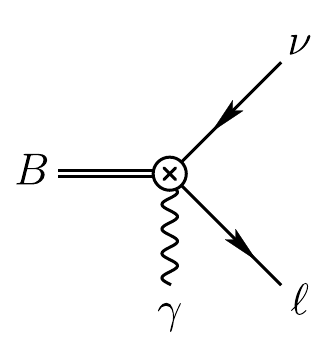}
\caption{Decay topologies describing the $B^-\to\ell^-\spac\bar\nu_\ell\spac\gamma$ process at energies far below $\OurLambda$. The photon energy is restricted to be less than $E_{\rm cut}\ll\OurLambda$ in the $B$-meson rest frame. The crossed circle indicates the weak interaction vertex. In the third graph, the $B$ meson transitions to an excited meson $X_b$ via the emission of a photon. The black square indicates that the corresponding vertex is a power-suppressed interaction.}
\label{fig:graphs_mesons}
\end{figure}

Figure~\ref{fig:graphs_mesons} shows the different topologies for the real emission of low-energy photons in the effective description of point-like particles. A region analysis reveals that the relevant momentum modes are either ``ultrasoft" (us) or ``ultrasoft-collinear" (usc), scaling as 
\begin{equation}\label{eq:us_usc_scaling}
\begin{aligned}
    p_{us} &\sim E_\mathrm{cut} 
     \equiv \OurLambda \left( \zeta, \zeta, \zeta \right) , \\
    p_{usc} &\sim E_\mathrm{cut} 
     \left( \frac{m_\ell^2}{m_B^2}, 1, \frac{m_\ell}{m_B} \right) 
     = \OurLambda\spac\zeta \left( \lambda_\ell^2, 1, \lambda_\ell \right) ,
\end{aligned}
\end{equation}
where $\zeta\equiv E_\mathrm{cut}/\OurLambda$ is the expansion parameter of the low-energy theory. The ultrasoft-collinear photons obey the homogeneous scaling $p^\prime\sim\frac{E_\mathrm{cut}\spac m_\ell}{m_B}\,(1,1,1)$ when boosted to the rest frame of the charged lepton. Hence, in their respective rest frames, both modes are a factor $E_\mathrm{cut}/m_B=\lambda\spac\zeta$ softer than the masses of the $B$~meson and the charged lepton. The physical reason for the appearance of the ultrasoft-collinear mode is that its energy is of order $E_{\rm cut}$ in the $B$-meson rest frame, where the measurement is performed.

The real-emission corrections come from two different mechanisms: one involving the emission of a photon off the charged external particles, and one where the $B$ meson emits a photon by transitioning into a virtual excited state $X_b$ containing a $b$ quark, which subsequently decays into the lepton pair. In the literature, these two topologies are commonly referred to as ``inner bremsstrahlung" and ``structure-dependent" contributions, respectively.  This nomenclature, introduced for leptonic decays of light mesons \cite{Bryman:1982et,Bijnens:1992en} and then adopted also for $B$-meson decays \cite{Becirevic:2009aq,Carrasco:2015xwa}, is somewhat misleading, since, as we have shown in the previous sections, also the inner bremsstrahlung contribution contains structure dependence through the weak vertex. We therefore choose to refer to the two contributions as the ``direct" and ``indirect" ones, where in the latter case the weak decay occurs through an excited state of the $B$ meson. In principle, an additional topology exists (last graph in Figure~\ref{fig:graphs_mesons}), in which the photon is emitted from a local effective $B\spac\ell\spac\bar\nu_\ell\spac\gamma$ interaction vertex \cite{Boyd:1994pa}, which can arise e.g.\ from $\rho$-meson exchange \cite{Chiladze:1998ny}. We will show below that this contribution is further power suppressed by a factor of $\order{\zeta^2}$, and will therefore be neglected in our analysis.

For the indirect contribution, heavy-quark spin symmetry implies that only the state $X_b=B^\ast$ needs to be included as an intermediate state as long as $E_\mathrm{cut}\ll\OurLambda$. To see this, it is instructive to examine the propagator of the intermediate state $X_b$. In the rest frame of the $B$ meson, we have
\begin{equation}
   \frac{1}{(p_B-k_\gamma)^2-m_X^2} 
   = - \frac{1}{2\spac m_B\spac(E_\gamma+\delta_{X_b})} \,; 
    \quad \text{with} \quad
   \delta_{X_b} = \frac{m_X^2-m_B^2}{2\spac m_B} \,,
\end{equation}
indicating that only states satisfying $\delta_{X_b}\lesssim E_\mathrm{cut}$ contribute at leading power in $\zeta=E_{\rm cut}/\OurLambda$. In the heavy-quark limit, one finds that $m_B^*-m_B\to 0$ due to heavy-quark symmetry, whereas $m_{X_b}-m_B=\mathcal{O}(\OurLambda)$ \cite{Neubert:1993mb} for all other excited states, such that
\begin{equation}
   \delta_{X_b}
   \sim \left\{ \begin{array}{cc} \displaystyle
    \frac{\OurLambda^2}{m_B} \,; &\quad \text{for} \quad X_b=B^* \,, \\[4mm]
    \OurLambda \,; &\quad \text{for} \quad X_b\neq B^* \,.
    \end{array} \right.
\end{equation}
Numerically, one finds $\delta_{B^*}\simeq 45.4$\,MeV for the $B^{*-}$ meson, and $\delta_{B_1}\simeq 465.5$\,MeV for the next excited state $B_1(5721)$. A significant number of excited $B$ states with masses below 6\,GeV are known \cite{ParticleDataGroup:2024cfk}, for which $\delta_{X_b}<770$\,MeV. For the validity of the low-energy effective theory it is therefore essential to impose the condition $E_\mathrm{cut}\ll\OurLambda$. Choosing a larger value $E_\mathrm{cut}\gtrsim \OurLambda$ would require including the full tower of excited hadronic states or, equivalently, knowing the two form factors parameterizing the $B^-\to\ell^-\spac\bar\nu_\ell\spac\gamma$ transition for arbitrary photon energy (see e.g.\ \cite{Kurten:2022zuy}). It would then not be possible to integrate out the scale $\OurLambda$, and both real and virtual photon corrections would contribute in the low-energy theory at the QCD scale. For $E_\mathrm{cut}\ll\OurLambda$, these form factors simplify and are controlled by the product of two non-perturbative parameters, the $BB^*\gamma$ coupling and the decay constant $f_{B^*}$.

\subsubsection*{Leptonic sector}

We begin with the construction of the effective description of the charged lepton (the neutrino is trivially described by $\nu_\cb$ at all scales). For $\mu\ll m_\ell$, the charged lepton can be treated as a heavy field in the spirit of HQET. Ultrasoft radiation in the lepton rest frame appears boosted in the rest frame of the decaying $B$ meson, giving rise to the ultrasoft-collinear scaling in \eqref{eq:us_usc_scaling}. To describe these fluctuations we adopt the bHLET framework \cite{Fleming:2007xt,Fleming:2007qr,Beneke:2023nmj}. The lepton momentum is decomposed as
\begin{equation}
   p_\ell^\mu = m_\ell\spac v_\ell^\mu + k_\ell^\mu \,,
\end{equation}
where $v_{\ell}$ is the 4-velocity of the lepton, and $k_\ell$ denotes its residual (ultrasoft-collinear) momentum. In the $B$-meson rest frame, the components of $v_\ell$ are given by
\begin{equation}
   (n \cdot v_{\ell}, \barn \cdot v_{\ell}, v^{\mu}_{\ell \perp}) 
   = \bigg( \frac{m_\ell}{m_B}, \frac{m_B}{m_\ell}, 0 \bigg) 
   \sim (\lambda_\ell, \lambda_\ell^{-1}, 0) \,.
\end{equation}
Following the standard HQET construction, we define a rephased lepton field 
\begin{equation}\label{eq:HQETproj}
   \ell(x) = e^{-i m_\ell\spac v_\ell\cdot x}\,\big[
    h_{v_\ell}(x) + H_{v_\ell}(x) \big] \,,
\end{equation}
where $\slashed{v}_\ell\spac h_{v_\ell}=h_{v_\ell}$ and $\slashed{v}_\ell\spac H_{v_\ell}=-H_{v_\ell}$. The field $h_{v_\ell}$ describes ultrasoft-collinear fluctuations with virtuality of order $\frac{E_{\rm cut}\spac m_\ell}{m_B}\sim\lambda\spac\zeta\spac m_\ell$ of the lepton about its mass shell, whereas $H_{v_\ell}$ describes fluctuations with virtuality of order $m_\ell$. The latter are integrated out in bHLET and give rise to power-suppressed terms. This leads to the effective Lagrangian
\begin{equation}\label{eq:Ll}
    \mathcal{L}_\ell = \mathcal{L}_{\rm bHLET}
    = \bar h_{v_\ell}\spac i v_\ell\cdot D\spac h_{v_\ell} + \dots \,,
\end{equation}
where the terms not shown are multiplied by powers of $1/m_\ell$ and are suppressed, relative to the leading term, by powers of $\lambda\spac\zeta\sim E_{\rm cut}/m_B$. 

In the low-energy theory, the charged lepton couples only to ultrasoft-collinear radiations, because soft radiations have already been decoupled in SCET-2. These interactions can be decoupled in the bHLET Lagrangian via the field redefinition
\begin{equation}\label{eq:leptondecoupling}
   h_{v_\ell}(x) = C_{v_\ell}^{(\ell)}(x)\,h^{(0)}_{v_\ell}(x) \,, 
\end{equation}
where we have defined the ultrasoft-collinear Wilson line
\begin{equation}
   C_{v_\ell}^{(\ell)}(x)
   = \exp\left[ - i\spac Q_\ell\spac e \int_0^\infty\!ds\,n\cdot A_{usc}(x+s v_\ell)\,
    e^{-\epsilon|s|} \right] .
\end{equation}
In terms of the decoupled field $h^{(0)}_{v_\ell}$, the leading-order effective Lagrangian \eqref{eq:Ll} becomes that of a free theory. 

For the collinear charged-lepton field in SCET-2, the above field redefinition implies
\begin{equation}
   \bar\X_c^{(\ell)}(x) 
   = e^{i m_\ell\spac v_\ell\cdot x}\,C_{v_\ell}^{(\ell)\dagger}(x)\,
    \bar h_{v_\ell}^{(0)}(x)\,\frac{\nbsl\spac\nsl}{4} + \dots \,.
\end{equation}
In the leptonic current, the SCET projector can be absorbed in the neutrino field. At leading power in bHLET, we thus obtain for the $(V-A)$ weak lepton current in the low-energy theory
\begin{equation}\label{eq:Jmulepton}
   J_{\rm lep}^\mu(x) 
   = \bar\ell\spac\gamma^\mu\spac P_L\spac\nu_\ell
   \,\to\, e^{i m_\ell\spac v_\ell\cdot x}\,C_{v_\ell}^{(\ell)\dagger}(x) 
    \left[ \bar h_{v_\ell}^{(0)}\spac\gamma_\perp^\mu\spac P_L\spac\nu_\cb
    + \frac{m_\ell}{\nb\cdot p_\ell}\,\nb^\mu\,
    \bar h_{v_\ell}^{(0)}\spac P_L\spac\nu_\cb \right] ,
\end{equation}
up to $\mathcal{O}(\lambda\spac\zeta)$ power corrections. 

\subsubsection*{Hadronic sector}

In the heavy-quark limit, the ground-state mesons containing a heavy quark $Q$ and a light antiquark $\bar{q}$ form a doublet under the heavy-quark spin symmetry and a triplet under $SU(3)_V$, containing pseudoscalar meson field $\varphi_v$ and vector meson field $\rho_v$ (with $v\cdot\rho_v=0$), which are degenerate in mass. For the case $Q=b$ and $q=u$ these are the $B^-$ and $B^{\ast -}$  mesons. To construct the heavy-meson effective theory (HMET) describing the interactions of these particles with photons, it is convenient to define the superfield 
\begin{equation}\label{eq:Hdef}
    \Phi_B = \frac{e^{-i\overline{m}_B\spac v\cdot x}}{\sqrt{2\spac\overline{m}_B}}\spac H \,, 
     \quad \text{with} \quad
    H = \frac{1+\vsl}{2}\,\big( \slashed{\rho}_v - \varphi_v \spac \gamma_5 \big) \,,
\end{equation}
where $\vsl\spac H=H=-H\spac\vsl$.\footnote{The meson field $H$ is not to be confused with the field $H_{v_\ell}$ introduced earlier.} 
We denote by $\overline{m}_B=m_b+\OurLambda$ the common mass of the $B$ and $B^*$ states in the heavy-quark limit. The field $H$ describes ultrasoft fluctuations about the mass shell of these states. 

The leading-order Lagrangian of the $H$ field can be written in the compact form 
\begin{equation}\label{eq:L_leading}
   \mathcal{L}_{\rm HMET}
   = -\frac{1}{2}\,\mathrm{Tr}\!\left[ \bar H\spac iv\cdot D\spac H \right] 
   = \varphi_v^\dagger\,iv\cdot D\spac\varphi_v 
    - \rho_{v,\mu}^\dagger\spac iv\cdot D\spac\rho_v^\mu \,,
\end{equation}
where $\bar H=\gamma^0 H^\dagger\gamma^0$. Momentum conservation allows these fields to interact with both ultrasoft and ultrasoft-collinear photons. The corresponding covariant derivative reads \begin{equation}
   i D^\mu = i\partial^\mu + Q_B\spac e\spac A_{us}^\mu(x) 
    + Q_B\spac e\,\frac{n^\mu}{2}\,\nb\cdot A_{usc}(x_+) \,,
\end{equation}
where $Q_B=-1$ is the charge of the $B^-$ meson, and we have multipole expanded the ultrasoft-collinear gauge field about the point $x_+$. This ensures that only the large momentum component $\nb\cdot k_{usc}\sim E_{\rm cut}$ is kept at the interaction vertices and smaller components are expanded out. In analogy with the discussion of the lepton sector, the leading-power interactions of the $H$ field can be decoupled at the Lagrangian level via the field redefinition
\begin{equation}\label{eq:5.15}
   H(x)= \overline{Y}_v^{(B)}(x)\,\overline{C}_\nb^{(B)}(x_+)\,H^{(0)}(x) \,,
\end{equation}
where we define the ultrasoft and ultrasoft-collinear Wilson lines
\begin{equation}
\begin{aligned}
   \overline{Y}_v^{(B)}(x) 
   &= \exp\left[ i\spac Q_B\spac e \int_{-\infty}^0\!ds\,v\cdot A_{us}(x+sv)\,
    e^{-\epsilon|s|} \right] , \\
   \overline{C}^{(B)}_\nb(x_+) 
   &= \exp\left[ i\spac Q_B\spac e \int_{-\infty}^0\!ds\,
    \nb\cdot A_{usc}(x_+ +s\nb)\,e^{-\epsilon|s|} \right] .
\end{aligned} 
\end{equation}
Written in terms of the field $H^{(0)}$, the leading-order Lagrangian \eqref{eq:L_leading} describes a free theory. 

The effective field $H$ has mass dimension $D=\frac32$ and EFT power counting  $(\OurLambda\spac\zeta)^\frac32$, hence the leading-order Lagrangian \eqref{eq:L_leading} has scaling $(\OurLambda\spac\zeta)^4\sim E_{\rm cut}^4$, which compensates the scaling of the measure $d^4x_{us}\sim E_{\rm cut}^{-4}$ to give unsuppressed contributions to the action. It will be necessary for our analysis to include some subleading terms in the hadronic sector. At mass dimension $D=5$, the only relevant subleading operator is \cite{Cho:1992nt}
\begin{equation}\label{eq:photonBs}
   c_{\rm dip}^{\rm HMET}\,\frac{Q_u\spac e}{8\OurLambda}\,
   \mathrm{Tr}\!\left[\sigma_{\mu\nu} \bar H^{(0)} H^{(0)} \right] F_{us}^{\mu\nu} \,,
\end{equation} 
with a non-perturbative low-energy constant $c_{\rm dip}^{\rm HMET}=\mathcal{O}(1)$. This operator conserves $SU(2)_v$ and yields contributions to the action suppressed by one power of $\zeta=E_{\rm cut}/\OurLambda$. It can be understood as arising from dipole interactions of the light spectator quark inside the heavy mesons \cite{Colangelo:1993zq}, which is why we have included the charge $Q_u$ in the normalization of this operator. The analogous operator with ultrasoft-collinear photons is further power-suppressed by a factor $m_\ell/m_B$ and can be safely neglected. In order to account for the small mass splitting between the $B$ and $B^*$ mesons, it will also be necessary to include the leading corrections to the heavy-quark limit $m_b\to\infty$, which are best identified at the parton level and incorporated into the EFT via spurions. In the HQET Lagrangian, the leading power corrections are given by \cite{Neubert:1993mb}
\begin{equation}
   C_{\mathrm{kin}}\,\frac{1}{2 m_b}\,\bar b_v\spac(iD_{\perp_v})^2\spac b_v
   + C_{\mathrm{mag}}(\mu)\,\frac{g_s}{4 m_b}\,
    \bar b_v\spac\sigma_{\mu\nu}\spac G_{us}^{\mu\nu}\spac b_v + \dots \,,
\end{equation}
where the symbol $\perp_v$ projects out the components of a vector orthogonal to $v$ (i.e.\ $v_\mu\spac D_{\perp_v}^\mu=0$), $G_{us}^{\mu\nu}$ denotes the ultrasoft gluon field-strength tensor, and the dots refer to terms suppressed by at least two powers of $1/m_b$. The Wilson coefficient of the kinetic operator satisfies $C_{\rm kin}=1$ to all orders of perturbation theory \cite{Luke:1992cs}, while the expression for the coefficient $C_{\rm mag}$ at NLO in RG-improved perturbation theory has been derived in \cite{Amoros:1997rx}. The terms shown here are suppressed relative to the leading-order terms by a factor $\zeta\spac\OurLambda/m_b\sim\lambda\spac\zeta$. From this we can read off the spurions
\begin{equation}\label{eq:spurion}
   \Sigma_{\rm kin} = C_{\rm kin} = 1 \,, \qquad 
   \Sigma_{\rm mag}^{\mu\nu} 
   = C_{\mathrm{mag}}(\mu)\,\sigma^{\mu\nu} \,,
\end{equation}
both of which transform as triplets under $SU(2)_v$. It is now a straightforward matter to construct the relevant power-suppressed corrections to the effective Lagrangian \eqref{eq:L_leading}. We obtain 
\begin{equation}\label{eq:5.20}
\begin{aligned} 
   \mathcal{L}_H
   &= - \frac{1}{2}\,\mathrm{Tr}\!\left[ 
    \bar H^{(0)}\spac iv\cdot\partial\spac H^{(0)} \right] 
    - \frac{1}{4 m_B}\,\mathrm{Tr}\!\left[ 
    \bar H^{(0)} \spac(iD_{\perp_v})^2\spac H^{(0)} \right] 
    + c_{\rm dip}^{\rm HMET}\,\frac{Q_u\spac e}{8\OurLambda}\,
    \mathrm{Tr}\!\left[ \sigma_{\mu\nu} \bar H^{(0)} H^{(0)} \right] 
    F_{us}^{\mu\nu} \\
   &\quad - \frac{\lambda_1}{4 m_b}\,\mathrm{Tr}\!\left[ \bar H^{(0)} H^{(0)} \right] 
    - \frac{\lambda_2(\mu)}{8 m_b}\,\mathrm{Tr}\!\left[ 
    \bar H^{(0)}\spac\Sigma_{\rm mag}^{\mu\nu} H^{(0)}\sigma_{\mu\nu} \right] 
    + \dots \,,
\end{aligned}
\end{equation}
where the second term in the first line arises as a consequence of the field redefinition in \eqref{eq:Hdef} and plays no role for our analysis. It is suppressed relative to the leading term by $E_{\rm cut}/m_B\sim\lambda\spac\zeta$. As mentioned earlier, the third operator is suppressed by $\zeta$. The operators shown in the second line scale like $\OurLambda^2/(m_B\spac E_{\rm cut})\sim\lambda/\zeta$ relative to the leading term. They are suppressed for $E_{\rm cut}\gg\OurLambda^2/m_B$ but become of leading order for $E_{\rm cut}\sim\delta_{B^*}$. Further operators not shown are suppressed by at least two powers of $\zeta$ or $\lambda^2/\zeta$. The quantities $\lambda_{1,2}\sim\OurLambda^2$ are well-known hadronic matrix elements defined in HQET, and the product $C_{\rm mag}\spac\lambda_2\approx 0.12\,\text{GeV}^2$ is scale independent \cite{Falk:1992wt}. Written out in component fields, we find 
\begin{equation}\label{eq:5.21}
\begin{aligned} 
   \mathcal{L}_H
   &= \varphi_v^{\dagger\spac(0)}\spac iv\cdot\partial\spac\varphi_v^{(0)}
    - \delta m_B\,\varphi_v^{\dagger\spac(0)}\spac\varphi_v^{(0)}
    - \rho_{v,\mu}^{\dagger\spac(0)}\spac iv\cdot\partial\spac\rho_v^{\mu\spac(0)} 
    + \delta m_{B^*}\,\rho_{v,\mu}^{\dagger\spac(0)}\spac\rho_v^{\mu\spac(0)} \\
  &\quad + \frac{e\spac c_{B B^*\gamma}}{2\OurLambda} 
   \left( v^\mu \rho_v^{\dagger\spac\nu\spac(0)}\spac\varphi_v^{(0)}\spac
    \tilde{F}_{\mu\nu}^{us} + \mathrm{h.c.} \right) 
   - \frac{i e\spac c_{B^* B^*\gamma}}{2\OurLambda}\,
   \rho_v^{\dagger\spac\mu\spac(0)}\spac\rho_v^{\nu\spac(0)}\spac 
   F_{\mu\nu}^{us} + \dots \,,
\end{aligned}
\end{equation}
with $\tilde{F}_{\mu\nu}^{us}=\frac{1}{2}\spac\epsilon_{\mu\nu\rho\sigma}\spac F^{\rho\sigma}_{us}$ being the dual field-strength tensor, and the dipole couplings 
\begin{equation}\label{eq:5.22}
   c_{B B^*\gamma} = c_{B^* B^*\gamma} = Q_u\spac c_{\rm dip}^{\rm HMET} \,. 
\end{equation}
Our coupling $c_{B B^\ast\gamma}$ is related to the parameter $g_{B B^\ast\gamma}$ defined in \cite{Becirevic:2009aq} via $c_{B B^\ast\gamma}/\OurLambda=g_{B B^\ast\gamma}$. The corrections to the meson masses arising at first order in $1/m_b$ are
\cite{Falk:1992wt}
\begin{equation}\label{deltamBBst}
   \delta m_B 
   = - \frac{\lambda_1+3\spac C_{\rm mag}(\mu)\spac\lambda_2(\mu)}{2m_b} \,, \qquad 
   \delta m_{B^*}
   = - \frac{\lambda_1-C_{\rm mag}(\mu)\spac\lambda_2(\mu)}{2m_b} \,, 
\end{equation}
from which we obtain 
\begin{equation}\label{eq:deltaB*}
   \delta_{B^*} = \frac{m_{B^*}^2-m_B^2}{2\spac m_B} 
   = C_{\rm mag}(\mu)\,\frac{2\spac \lambda_2(\mu)}{m_b} + \dots \,, 
\end{equation}
up to terms of order $1/m_b^2$ and higher. The physical masses of the $B$ and $B^*$ mesons are given by $m_{B^{(*)}}=\overline{m}_B+\delta m_{B^{(*)}}$. Since we have used the parameter $\overline{m}_B$ in the field redefinition  \eqref{eq:Hdef}, the propagators of $B^{(*)}$ mesons with residual momentum $q$ in HMET are given by $i/(v\cdot q-\delta m_{B^{(*)}}+i0)$. External states for on-shell $B^{(*)}$ mesons carry the residual momentum $q^\mu=\delta m_{B^{(*)}}\spac v^\mu$.

We now discuss the representation of the flavor-changing quark current $\bar u\spac\gamma^\mu\spac P_L\spac b$ in the low-energy theory. In this discussion we neglect QED corrections, because we will consider real-photon emission topologies, whose rate is already suppressed by a power of $\alpha$. Including short-distance QCD corrections, the relevant weak current is of the form \cite{Neubert:1993mb}
\begin{equation}
   J_{\rm had}^\mu(x)
   = e^{-i m_B\spac v\cdot x}\,Y_n^{(\ell)\dagger}(x_-) 
    \left[ \mathcal{C}_1^{\rm HQET}(\mu)\,\bar u\spac\gamma^\mu P_L\spac b_v
    + \mathcal{C}_2^{\rm HQET}(\mu)\,\bar u\spac v^\mu P_R\,b_v \right] ,
\end{equation}
where the soft Wilson line $Y_n^{(\ell)\dagger}$ is inherited from the lepton current, see relation \eqref{eq:softdecoupling}, and the Wilson coefficients have been given at one-loop order in \eqref{eq:HQETcoefs}. These currents introduce sources of $SU(2)_v$-breaking, which can be captured in HMET via the spurion
\begin{equation}
    \Gamma_{\rm weak}^\mu 
    = \mathcal{C}_1^{\rm HQET}(\mu)\,\gamma^\mu P_L
     + \mathcal{C}_2^{\rm HQET}(\mu)\,v^\mu P_R \,.
\end{equation}
There is no need to introduce two spurions, because heavy-quark spin ensures that the combination of the two terms remains unchanged in the matching to the low-energy theory. The representation of the weak current in HMET is obtained as
\begin{equation}\label{eq:Jhadronic}
   J_{\rm had}^\mu(x) 
   \to e^{-i m_B\spac v\cdot x}\,\overline{Y}_v^{(B)}(x)\,
    \overline{C}_\nb^{(B)}(x_+)\,Y_n^{(\ell)\dagger}(x_-)\,
    \frac{i\spac\F_{\rm HMET}(\mu)}{2\spac\sqrt2}\,
    \mathrm{Tr}\!\left[ \Gamma_{\rm weak}^\mu\spac H^{(0)} \right] ,
\end{equation}
where $\F_{\rm HMET}$ is a non-perturbative parameter. In components, we find 
\begin{equation}
   \mathrm{Tr}\!\left[ \Gamma_{\rm weak}^\mu\spac H^{(0)} \right] 
   = - \left[ \mathcal{C}_1^{\rm HQET}(\mu) + \mathcal{C}_2^{\rm HQET}(\mu) \right]
    v^\mu\spac\varphi_v^{(0)} 
    + \mathcal{C}_1^{\rm HQET}(\mu)\,\rho_v^{\mu\spac(0)} \,.
\end{equation}
The normalization of the fields $\varphi_v$ and $\rho_v^\mu$ is such that \begin{equation}\label{eq:norms}
   \langle\spac 0\spac|\spac\varphi_v|B(v)\rangle 
   = \sqrt{2\spac m_B} \,, \qquad
   \langle\spac 0\spac|\spac\rho_v^\mu|B^*(v)\rangle 
   = \sqrt{2\spac m_{B^*}}\,\epsilon^\mu(v) \,.
\end{equation}
where for convenience we normalize the states to the physical masses of the $B$ and $B^*$ mesons. Matching \eqref{eq:Jhadronic} to the usual definition of the meson decay constants, 
\begin{equation}
   \langle\spac 0\spac|\spac\bar u\spac\gamma^\mu\spac P_L\spac b\spac|B(v)\rangle 
   = - \frac{i}{2}\,m_B\spac f_B\,v^\mu \,, \qquad
   \langle\spac 0\spac|\spac\bar u\spac\gamma^\mu\spac P_L\spac b\spac|B^*(v)\rangle 
   = \frac{i}{2}\,m_{B^*} f_{B^*}\,\epsilon^\mu \,, 
\end{equation}
we obtain
\begin{equation}
\begin{aligned}
   m_B\spac f_B^{\rm QCD}
   &= \left[ \mathcal{C}_1^{\rm HQET}(\mu) + \mathcal{C}_2^{\rm HQET}(\mu) \right]
    \sqrt{m_B}\,\spac\F_{\rm HMET}(\mu) \,, \\
   m_{B^*} f_{B^*}^{\rm QCD}
   &= \mathcal{C}_1^{\rm HQET}(\mu)\, 
    \sqrt{m_{B^*}}\,\spac\F_{\rm HMET}(\mu) \,,
\end{aligned}
\end{equation}
valid at leading power in $1/m_b$. From these relations, we can identify $\F_{\rm HMET}(\mu)$ with the HQET decay constant $F_{\rm QCD}(\mu)$ introduced in \eqref{eq:Fhqet_fB}, taking into account that $\mathcal{C}_1^{\rm HQET}+\mathcal{C}_2^{\rm HQET}=H_1^A+H_2^A$.

\subsubsection*{Effective weak interactions}

Contracting the representation of the weak flavor-changing quark current with the representation of the lepton current in \eqref{eq:Jmulepton}, the effective weak interaction in the low-energy theory (cf.\ \eqref{eq:Lleft}) takes the form (we set $x=0$ for simplicity)
\begin{equation}\label{eq:OVLL_low_energy}
\begin{aligned}
   \mathcal{L}_{H\ell}
   &= i\sqrt2\spac G_F^{(\mu)}\spac V_{ub}\,
    \overline{Y}_v^{(B)}\,\overline{C}_\nb^{(B)}\,
    Y_n^{(\ell)\dagger}\,C_{v_\ell}^{(\ell)\dagger} \\
   &\quad \times \left\{
    \frac{f_B^{\rm QCD}\spac m_\ell}{\sqrt{2\spac m_B}}\,
    \varphi_v^{(0)}\,\bar h_{v_\ell}^{(0)}\spac P_L\spac\nu_\cb
    - \frac{f_{B^*}^{\rm QCD}\spac m_{B^*}}{\sqrt{2\spac m_{B^*}}}\,
    \bigg[ \bar h_{v_\ell}^{(0)}\spac\rlap{\spac/}{\rho}_v^{\perp(0)} 
    P_L\spac\nu_\cb
    + \frac{m_\ell}{m_{B^*}}\,\frac{\nb\cdot\rho_v^{(0)}}{\nb\cdot v}\,
    \bar h_{v_\ell}^{(0)} P_L\spac\nu_\cb \bigg] \right\} .
\end{aligned}
\end{equation}
As a last step, we need to account for QED radiative corrections.   This is accomplished by multiplying the terms in \eqref{eq:OVLL_low_energy} with factors $y_B$ (for $\varphi_v$) and $y_{B^*}^{\perp,\parallel}$ (for $\rho_v^\perp$ and $\nb\cdot\rho_v$). For the case of the $B^*$ meson, it will be sufficient to work with the lowest-order approximation $y_{B^*}^{\perp,\parallel}=1+\order{\alpha}$, since a real photon is involved in converting the $B$ meson into a $B^*$ meson, and the rate for this process is already suppressed by one power of $\alpha$. For the $B$ meson, however, we need to perform a consistent matching of SCET-2 onto the low-energy effective theory to determine $y_B$ including perturbative and non-perturbative QED effects. This will be discussed in Section~\ref{subsec:hhchipt_matching}.

\begin{figure}[t]
\centering
\includegraphics[scale=0.5]{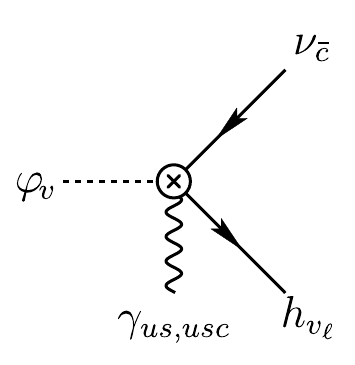} \quad
\includegraphics[scale=0.5]{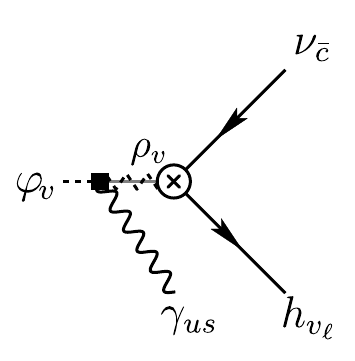} \quad
\includegraphics[scale=0.5]{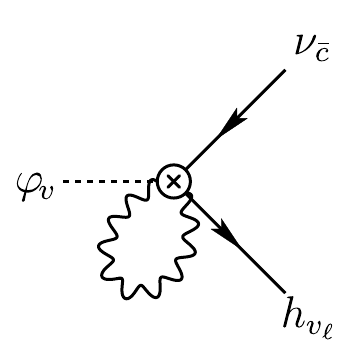} \quad
\includegraphics[scale=0.5]{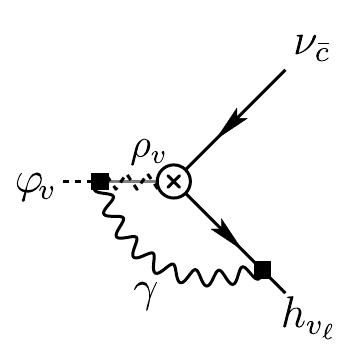} 
\caption{Feynman diagrams describing QCD corrections to the \blnu\ process at $\order{\alpha}$ in the heavy-particle effective theory. For clarity, we label the lines by the corresponding EFT fields. A crossed circle indicates the weak interaction vertex, while a black square denotes a power-suppressed interaction. In the first and third graphs the photons are emitted from the Wilson lines contained in the effective weak Lagrangian \eqref{eq:OVLL_low_energy}.}
\label{fig:graphs_mesons_eft}
\end{figure}

According to Low's theorem \cite{Low:1958sn}, the emissions of very soft photons -- i.e. with energies below all relevant masses and excitation thresholds -- can be obtained, for our process, from the squared matrix element of the product of soft QED Wilson lines $S_{v_\ell}^{(\ell)\dagger}(x)\,\overline{S}_v^{(B)}(x)$, where $v\equiv v_B$ denotes the 4-velocity of the $B$ meson. It is interesting to ask how this product of two Wilson lines is related to the product of four Wilson lines appearing in \eqref{eq:OVLL_low_energy}. The answer is simple: after integrating over the photon phase space, the squared matrix element of $S_{v_\ell}^{(\ell)\dagger}\,\overline{S}_v^{(B)}$ can only depend on the Lorentz invariants $v^2=v_\ell^2=1$ and $v\cdot v_\ell$. For the \blnu\ process with $\ell=\mu,e$, the latter is parametrically large,
\begin{equation}
   v\cdot v_\ell = \frac{m_B^2+m_\ell^2}{2\spac m_B\spac m_\ell}
   \approx \frac{m_B}{2\spac m_\ell}\gg 1 \,.
\end{equation}
A region analysis shows that, in this limit, two different momentum modes contribute to the squared matrix element arising from Low's theorem, corresponding to the ultrasoft and ultrasoft-collinear regions identified earlier. In the spirit of asymptotic expansions, one finds that 
\begin{equation}
   S_{v_\ell}^{(\ell)\dagger}\,\overline{S}_v^{(B)}
   = \left[ Y_n^{(\ell)\dagger}\,\overline{Y}_v^{(B)} \right]
    \left[ C_{v_\ell}^{(\ell)\dagger}\,\overline{C}_\nb^{(B)} \right]
    + \order{\frac{m_\ell}{m_B}} \,.
\end{equation}
Combining all ingredients, at this stage the low-energy effective Lagrangian reads 
\begin{equation}\label{eq:full_low_energy_lagrangian}
   \mathcal L_{\spac\text{low-$E$}} 
   = \mathcal{L}_\ell + \mathcal{L}_{H} + \mathcal{L}_{H\ell} \,. 
\end{equation}
From this construction, the direct and indirect real-emission contributions introduced earlier can be clearly identified. They are shown in the first two graphs in Figure~\ref{fig:graphs_mesons_eft}. For the direct contribution, the $B$ meson couples to the leptons via the $m_\ell/m_B$-suppressed weak interaction of the pseudoscalar meson in $\mathcal{L}_{H\ell}$, see \eqref{eq:OVLL_low_energy}, with photons emitted from the Wilson lines (first diagram). Effective $B\spac\ell\spac\bar\nu_\ell\spac\gamma$ interactions beyond those induced by photon emission from the Wilson lines, represented by the last graph in Figure~\ref{fig:graphs_mesons}, would require the insertion of an ultrasoft (or ultrasoft-collinear) field-strength tensor $F_{us}^{\mu\nu}$ (or $F_{usc}^{\mu\nu}$) in the weak-interaction operators shown in \eqref{eq:OVLL_low_energy}, which costs an additional power suppression of order $\zeta^2\sim(E_{\rm cut}/\OurLambda)^2$ (or $(\lambda\spac\zeta)^2\sim(E_{\rm cut}/m_B)^2$). In the indirect contribution, the $B$ meson transitions to a $B^*$ through the emission of a photon via the $E_{\rm cut}/\OurLambda$-suppressed dipole interaction in $\mathcal{L}_{H}$, see \eqref{eq:5.21}. The intermediate $B^*$ meson then couples to the leptonic current with the leading-power interaction in $\mathcal{L}_{H\ell}$ (second diagram). The relative importance of these two contributions depends on the experimental cut on the photon energy. Naively, the indirect contribution becomes relevant for $E_\mathrm{cut}/m_\ell\gtrsim\OurLambda/m_B$, which yields $E_\mathrm{cut}>10$\,MeV for the muon channel and $E_\mathrm{cut}>50$\,keV for the electron case (for $\OurLambda\approx 500$\,MeV).  However, as we will show later, the onset of the indirect contribution is pushed to larger values of $E_\mathrm{cut}$ by an additional suppression factor $\sim(E_{\rm cut}/\delta_{B^*})^2$, with $\delta_{B^*}\simeq 45$\,MeV.
 From these estimates, one can infer that for $\ell=e$ the decay is dominated by the $B\to B^\ast\spac\gamma\to e\spac\bar\nu_e\spac\gamma$ contribution, while for $\ell=\mu$ the direct and indirect channels compete for low values of $E_{\mathrm{cut}}$.

The last two graphs in Figure~\ref{fig:graphs_mesons_eft} show the virtual QED corrections to the \blnu\ process in the low-energy theory as discussed so far. In principle, they must be added to the virtual corrections arising from scales above $\Lambda_{\rm QCD}$, which, as we will show in Section \ref{subsec:hhchipt_matching},  are accounted for by the coupling $y_B$ discussed below \eqref{eq:OVLL_low_energy}. Virtual corrections arising from photons emitted from Wilson lines (third graph) are scaleless and vanish in dimensional regularization.\footnote{These virtual graphs have the effect of replacing the IR $1/\epsilon$ poles in the real-emission contributions by UV poles, which are removed by renormalization of the operators in the low-energy theory.} 
The last diagram is {\em a priori\/} not scaleless, since the $B^*$ propagator in the loop contains the mass parameter $\delta_{B^*}$ defined in \eqref{eq:deltaB*}. By angular momentum conservation, all these diagrams feature the chiral suppression $\sim m_\ell/m_B$ of the tree-level \blnu\ decay amplitude. Upon explicit calculation, we find that the fourth diagram vanishes when the leading-order couplings of the photon to the Wilson lines are used. A non-zero result is obtained if the photon couples to an effective $B\spac\ell\spac\bar\nu_\ell\spac\gamma$ interaction not induced by photon emission from the Wilson lines, which, as we have discussed above, costs an additional power suppression of order $\zeta^2\sim(E_{\rm cut}/\OurLambda)^2$. Alternatively, the photon can couple to the charged lepton through a power-suppressed dipole interaction
\begin{equation}
   \frac{Q_\ell\spac e}{4m_\ell}\,\bar h_{v_\ell}\spac\sigma_{\mu\nu}\spac F_{usc}^{\mu\nu}\,h_{v_\ell} \,,
\end{equation}
which can be added to the bHLET Lagrangian in \eqref{eq:Ll}. The corresponding contribution is suppressed, relative to the leading-order \blnu\ amplitude, by a factor of order $(\lambda\spac\zeta)^2=(E_{\rm cut}/m_B)^2$. We conclude that, at leading order in the heavy-particle effective theory discussed in this section, all virtual QED corrections are accounted for by the coupling $y_B$.

\subsection{Heavy-hadron chiral effective theory} 
\label{subsec:HHchiPT}

In our discussion so far we have ignored the fact that below the scale $\OurLambda$ the point-like mesons can interact not only with electromagnetic radiation, but also with the pseudo-Nambu-Goldstone bosons associated with the breaking of the chiral symmetry $SU(3)_L\times SU(3)_R\to SU(3)_V$, i.e., the light pseudoscalar mesons $\pi$, $K$ and $\eta$. To include these interactions, we need to generalize the HMET to heavy-hadron chiral perturbation theory (HH$\chi$PT) in the low-energy theory, following \cite{Wise:1992hn,Yan:1992gz,Burdman:1992gh}. To this end, we promote the effective ground-state heavy-hadron field $H$ to a triplet (with $a=1,2,3$) 
\begin{equation}   
    H_a = \frac{1+\vsl}{2}\,
     \big( \slashed{\rho}_{v,a}- \varphi_{v,a}\spac\gamma_5 \big)
\end{equation}
transforming as a $\overline{3}$ under flavor $SU(3)$. It describes the pseudoscalar mesons $(B^-,\bar B^0,\bar B_s^0)$ and their vector partners $(B^{*-},\bar B^{*0},\bar B_s^{*0})$. We then couple these fields to the octet of the light pseudoscalar mesons $(\pi,K,\eta)$ using a chiral Lagrangian. Below the scale of chiral symmetry breaking, $(\pi,K,\eta)$ are collected in the unitary field 
\begin{equation}
   \Sigma = \exp\left[ \frac{i\sqrt2\,\pi^a\spac\lambda^a}{f} \right] ,
    \quad \text{with} \quad
   \frac{\pi^a\spac\lambda^a}{\sqrt2}
   = \begin{pmatrix}
    \frac{1}{\sqrt2}\spac\pi^0+\frac{1}{\sqrt6}\spac\eta_8 & \pi^+ & K^+ \\
    \pi^- & -\frac{1}{\sqrt2}\spac\pi^0+\frac{1}{\sqrt6}\spac\eta_8 & K^0 \\
    K^- & \bar K^0 & -\sqrt{\frac23}\spac\eta_8 
    \end{pmatrix} ,   
\end{equation}
where the matrices $\lambda^a$ satisfy $\mathrm{tr}\spac[\spac\lambda^a\spac\lambda^b\spac]=2\spac\delta^{ab}$, and $f\simeq f_\pi=130.2$\,MeV is related to the pion decay constant. This field  transforms as $\Sigma\to L\spac\Sigma\spac R^\dagger$ under chiral $SU(3)_L\times SU(3)_R$ transformations. Next, we define a new unitary field $\xi$ via $\xi^2=\Sigma$. It transforms as $\xi\to L\,\xi\spac U^\dagger=U\spac\xi\spac R^\dagger$, where in general the special unitary matrix $U$ is a complicated nonlinear function of $L$, $R$ and the pseudo-Nambu-Goldstone boson fields \cite{Georgi:1984zwz}. The heavy-meson field $H$ transforms as $H\to H\spac U^\dagger$.

The leading-order HH$\chi$PT Lagrangian reads \cite{Wise:1992hn,Yan:1992gz,Burdman:1992gh} 
\begin{equation}\label{eq:LHHchiptLP} 
\begin{aligned}
   \mathcal{L}_{\mathrm{HH\chi PT}} 
   &= - \frac12\,\mathrm{Tr}\!\left[ \bar H\spac iv\cdot D H \right]
    + \frac12\,\mathrm{Tr}\!\left[ \bar H H\,v\cdot\V \right] 
    + \frac{g}{2}\,\mathrm{Tr}\!\left[ \bar H H\,\Asl\spac\gamma_5 \right] \\
   &\quad + \varrho_1\spac\mathrm{Tr}\big[ \bar H H\,m_q^\xi \big] 
    + \varrho_2\spac\mathrm{Tr}\!\left[ \bar H H \right] 
    \mathrm{tr}\!\left[ m_q^\xi \right] 
    + \varrho_3\spac Q_b\spac\alpha\spac
    \mathrm{Tr}\big[ \bar H H\spac Q^\xi \big] \\
   &\quad + \frac{f^2}{8}\,\mathrm{tr}\!\left[ (D^\mu\spac\Sigma)\spac 
    (D_\mu\spac\Sigma^\dagger) \right] +
    \frac{f^2}{4}\,B_0\,\mathrm{tr}\!\left[ m_q\spac\Sigma^\dagger 
    + \Sigma\spac m_q^\dagger \right] ,
\end{aligned}
\end{equation}
where here and below the symbol ``Tr'' implies a combined trace over Dirac and flavor indices, while ``tr'' refers to a flavor trace. Note that $(\bar H H)_{ab}=\bar H_a H_b$ are the components of a $3\times 3$ matrix in flavor space. The light pseudoscalar mesons couple to the heavy mesons through the vector and axial currents
\begin{equation}
   \V_\mu = \frac{i}{2} \left( \xi^\dagger D_\mu\spac\xi 
    + \xi\spac D_\mu\spac\xi^\dagger \right) , \qquad
   \A_\mu = \frac{i}{2} \left( \xi^\dagger D_\mu\spac\xi 
    - \xi\spac D_\mu\spac\xi^\dagger \right) .
\end{equation}
The covariant derivative acting on a field $\psi(x)$ is defined as 
\begin{equation}\label{eq:5.42}
   i D^\mu\spac\psi(x) = i\partial^\mu\spac\psi(x) 
    + e \left[ A_{us}^\mu(x) + \frac{n^\mu}{2}\,\nb\cdot A_{usc}^\mu(x_+) \right] 
    [Q,\psi(x)] \,,
\end{equation}
where $Q=\mathrm{diag}(Q_u,Q_d,Q_s)$ contains the light-quark charges. 
For the heavy-hadron field, we define
\begin{equation}
   [Q,H] \equiv Q' H - H Q = H\spac Q_B \,,
\end{equation}
where $Q'=Q_b\spac\mathbbm{1}$ gives the charge of the $b$ quark, and $Q_B=\mathrm{diag}(-1,0,0)$ contains the charges of the three mesons described by $H_a$. Under the chiral symmetry, the spurions $m_q=\mathrm{diag}(m_u,m_d,m_s)$ and $Q$ transform as $m_q\to L\spac m_q R^\dagger$ and $Q\to L\spac Q L^\dagger$ or $Q\to R\spac Q R^\dagger$, and we have defined related quantities \
\begin{equation}
   m^\xi = \frac12 \left( \xi^\dagger\spac m_q\spac\xi^\dagger 
    + \xi\spac m_q^\dagger\spac\xi \right) , \qquad  
   Q^\xi = \frac{1}{2} \left( \xi^\dagger Q\spac\xi 
    + \xi\spac Q\spac\xi^\dagger \right) ,
\end{equation}
which transform as $m_q^\xi\to U m_q^\xi\,U^\dagger$ and $Q^\xi\to U\spac Q^\xi\spac U^\dagger$. The first term in \eqref{eq:LHHchiptLP} contains $\mathrm{Tr}\big[\bar H_a H_b\spac Q_{ba}\big]$ from the covariant derivative acting on $H$, and this expression is not invariant under $SU(3)_L\times SU(3)_R$ transformations. The second term contains the same expression arising from the covariant derivatives inside $v\cdot \V$. The condition that the two terms cancel each other fixes the coefficient of the second term to be $+\frac12$. 

The coupling $g$ in \eqref{eq:LHHchiptLP} is proportional (at lowest order) to the $BB^*\pi$ coupling constant. The parameter $B_0$ is defined such that, at leading order in the chiral expansion, $m_\pi^2=B_0\spac(m_u+m_d)$, and similarly for the other mesons. The parameter $\varrho_1$ governs the mass splittings between the components of the triplet $H$, whereas $\varrho_2$ leads to an overall mass shift of the three meson masses proportional to the sum of the quark masses \cite{Wise:1992hn}. The parameter $\varrho_3$ yields an electromagnetic correction to the masses of the heavy mesons from the Coulomb interaction between their constituent quarks. At the lowest order in the chiral expansion, we can identify
\begin{equation}\label{eq:masssplittings}
\begin{aligned}
   m_{B_s} - m_{B_d} 
   &= 2\varrho_1(\mu) \left[ m_s(\mu) - m_d(\mu) \right] 
    \stackrel{!}{\simeq} 87.2 \,\text{MeV}\,, \\
   m_{B_d} - m_{B_u} 
   &= 2\varrho_1(\mu) \left[ m_d(\mu) - m_u(\mu) \right] 
    + \frac{2\alpha}{3}\,\varrho_3 
    \stackrel{!}{\simeq} 0.31\,\text{MeV} \,, 
\end{aligned}
\end{equation}
where the scale dependence cancels out. With $(m_s-m_d)\simeq 88.8$\,MeV and $(m_d-m_u)\simeq 2.54$\,MeV at $\mu=2$\,GeV \cite{ParticleDataGroup:2024cfk}, we obtain $\varrho_1(2\,\text{GeV})\simeq 0.49$ and $\varrho_3\simeq-436$\,MeV. It is natural to expect that the coefficient $\varrho_2$ takes an $\order{1}$ value, but this value cannot be determined from spectroscopy. In the presence of the parameters $\varrho_i$, the shifts of the meson masses in \eqref{deltamBBst} must be generalized to
\begin{equation}
\begin{aligned}
   \delta m_{B_q} 
   &= 2\varrho_1\spac m_q + 2\varrho_2\spac(m_u+m_d+m_s) 
    + 2\varrho_3\spac Q_b\spac Q_q\spac\alpha 
    - \frac{\lambda_1+3\spac C_{\rm mag}(\mu)\spac\lambda_2(\mu)}{2m_b} \,, \\ 
   \delta m_{B_q^*}
   &= 2\varrho_1\spac m_q + 2\varrho_2\spac(m_u+m_d+m_s) 
    + 2\varrho_3\spac Q_b\spac Q_q\spac\alpha 
    - \frac{\lambda_1-C_{\rm mag}(\mu)\spac\lambda_2(\mu)}{2m_b} \,. 
\end{aligned}
\end{equation}

The generalization of the remaining operators proceeds in an analogous way. In HH$\chi$PT, the electromagnetic dipole operator in \eqref{eq:photonBs} is generalized to \begin{equation}\label{eq:5.44}
   \frac{c_{\mathrm{dip}}\spac e}{8\OurLambda}\,
    \mathrm{Tr}\!\left[\sigma_{\mu\nu} \bar H H\spac Q^\xi \right] F_{us}^{\mu\nu} \,, 
\end{equation} 
while the two operators describing the $1/m_b$ corrections in \eqref{eq:5.20} remain unchanged. Finally, the representation of the left-handed, flavor-changing quark current \eqref{eq:Jhadronic} appearing in the weak interactions is generalized to \begin{equation}\label{eq:JhadChPT}
   J_{\mathrm{had},\spac a}^\mu(x) 
   = e^{-i m_B\spac v\cdot x}\,Y_n^{(\ell)\dagger}(x_-)\,
    \frac{i\spac\F}{2\spac\sqrt2}\,
    \mathrm{tr}\!\left[ \Gamma_{\rm weak}^\mu\spac(H\spac\xi^\dagger)_a \right] .
\end{equation}
Only the charged current (with $a=1$) enters the effective weak Lagrangian \eqref{eq:OVLL_low_energy} relevant to our process. Contracting it with the lepton current in \eqref{eq:Jmulepton}, we obtain for the effective weak Lagrangian including pseudoscalar mesons
\begin{equation}\label{eq:full_low_energy_lagrangian_pions}
   \mathcal{L}_{H\ell} 
   = - 2\sqrt2\spac G_F^{(\mu)}\spac V_{ub}\,g_{\mu\nu}\,
    J_{\mathrm{had},\spac 1}^\mu J_{\mathrm{lep}}^\nu \,.
\end{equation}

\begin{figure}[t]
\centering
\includegraphics[scale=0.5]{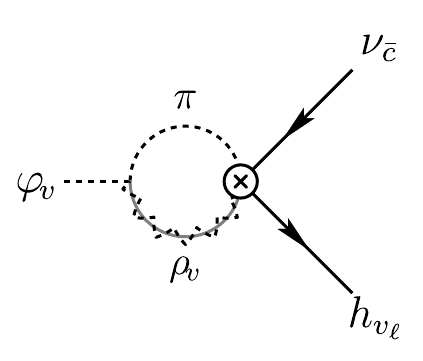} \quad
\includegraphics[scale=0.5]{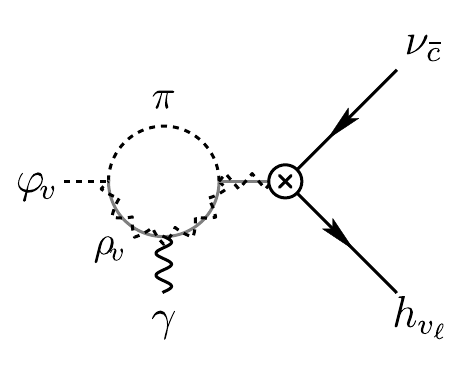} \quad
\includegraphics[scale=0.5]{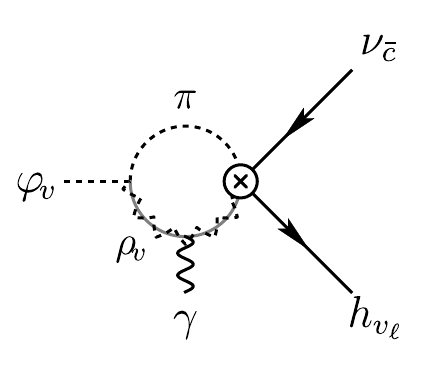} \quad 
\includegraphics[scale=0.5]{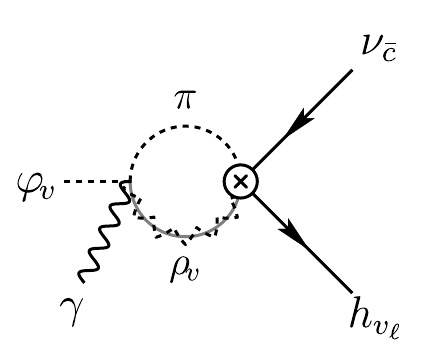}
\caption{Representative one-loop diagrams in HH$\chi$PT giving rise to higher-order corrections to the heavy-meson decay constant $F_{\rm QCD}$ (first graph) and the $BB^*\gamma$ coupling (second graph), and ``genuine'' one-loop corrections to the $B^-\to\ell^-\spac\bar\nu_\ell\spac\gamma$ decay amplitude (last two diagrams). Contributions from local higher-order operators in HH$\chi$PT, which are needed as counterterms for these diagrams, exist but are not shown. A crossed circle indicates the weak interaction vertex. Photons are not yet decoupled from the hadrons, and we do not distinguish between leading-order and power-suppressed vertices. Instead of pions, other pseudoscalar mesons can also propagate in the loops.}
\label{fig:chiral_loops}
\end{figure}

In \eqref{eq:JhadChPT}, the parameter $\F$ denotes the leading-order term in the chiral expansion of the HQET decay constant $F_{\rm QCD}(\mu)$. Higher-order corrections arise from chiral loops and local higher-order interactions in HH$\chi$PT. At one-loop order,  chiral logarithms in these corrections have been calculated in \cite{Goity:1992tp,Grinstein:1992qt}, and a more complete expression including also (some) finite terms has been derived in \cite{Boyd:1994pa}. A representative contribution is shown in the first graph of Figure~\ref{fig:chiral_loops}. Likewise, the dipole coupling $c_{\mathrm{dip}}$ in \eqref{eq:5.44}, which determines the effective $BB^*\gamma$ and $B^*B^*\gamma$ couplings according to \eqref{eq:5.22}, receives higher-order corrections, which have been studied at leading logarithmic order in \cite{Cheng:1993kp} and including finite terms in \cite{Stewart:1998ke}. A representative contribution is shown in the second graph. For our purposes, it is sufficient to simply substitute the ``physical'' values of $F_{\rm QCD}$ and $c_{BB^*\gamma}$ for the leading-order parameters. Finally, there are also ``genuine'' chiral loop corrections to the $B^-\to\ell^-\spac\bar\nu_\ell\spac\gamma$ decay amplitude, illustrated in the last two diagrams in Figure~\ref{fig:chiral_loops}. These diagrams scale like $m_P^2/\Lambda_\chi^2$ (with $P=\pi,K,\eta_8$) relative to the tree-level real-emission contributions, and they involve loop functions depending on dimensionless ratios of the light-meson masses, the parameter $\delta_{B^*}$, and the photon-energy cut $E_{\rm cut}$. {\it A priori}, one would expect a rather complicated result from the sum of all graphs. However, we will now show that the genuine chiral loop corrections to the $B^-\to\ell^-\spac\bar\nu_\ell\spac(\gamma)$ process vanish identically. 

\subsubsection*{\boldmath Reorganizing HH$\chi$PT}

We can further simplify the HH$\chi$PT Lagrangian by means of a non-linear field redefinition, using $H'\equiv H\spac\xi^\dagger$ rather than the original $H$ as the interpolating field for the heavy hadrons. The equation $H'=H \left[ 1+\order{\pi^a/f} \right]$ shows that this is a legitimate choice. The new field transforms as $H'\to H' L^\dagger$ under $SU(3)_L\times SU(3)_R$. We then obtain the leading-order Lagrangian (dropping the prime on the fields for convenience)
\begin{equation}\label{eq:LHHchiptLPnew}
\begin{aligned}
   \mathcal{L}_{\mathrm{HH\chi PT}} 
   &= - \frac12\,\mathrm{Tr}\!\left[ \bar H\spac iv\cdot D H \right]
    + \frac14\,\mathrm{Tr}\!\left[ \bar H H\spac v\cdot L \right] 
    - \frac{g}{4}\,\mathrm{Tr}\!\left[ 
    \bar H H \rlap{\,/}{L}\spac\gamma_5 \right] \\
   &\quad + \frac{\varrho_1}{2}\,
    \mathrm{Tr}\big[ \bar H H\spac S \big] 
    + \frac{\varrho_2}{2}\,\mathrm{Tr}\!\left[ \bar H H \right] 
    \mathrm{tr}\!\left[ S \right] 
    + \frac{\varrho_3}{2}\,Q_b\spac\alpha\,
    \mathrm{Tr}\!\left[ \bar H H \spac\big( Q
    + \Sigma\spac Q\spac\Sigma^\dagger \big) \right] \\
   &\quad + \frac{f^2}{8}\,\mathrm{tr}\!\left[ L^2 \right] 
    + \frac{f^2}{4}\,B_0\,\mathrm{tr}\!\left[ S \right] ,
\end{aligned}
\end{equation}
where we have used the standard definitions of the ``left-handed'' meson current $L^\mu$ and the scalar density $S$, defined as
\begin{equation}
   L^\mu = \Sigma\,iD^\mu\spac\Sigma^\dagger \,, \qquad
   S = m_q\spac\Sigma^\dagger + \Sigma\spac m_q^\dagger \,.
\end{equation}
In deriving the structure multiplying the coupling $g$, we have used the identity
\begin{equation}
   \xi \left( \xi^\dagger\spac iD_\mu\spac\xi - \xi\spac iD_\mu\spac\xi^\dagger \right) 
    \xi^\dagger
   = - \Sigma\,iD^\mu\spac\Sigma^\dagger \,,
\end{equation}
which follows from $(iD_\mu\spac\xi)\spac\xi^\dagger+\xi\spac(iD_\mu\spac\xi^\dagger)=0$. The power-suppressed terms relevant to our analysis include the dipole interaction and the $1/m_b$ corrections to the heavy-meson masses. They take the form
\begin{equation}\label{eq:LHHchiptNLPnew}
\begin{aligned}
   \mathcal{L}_{\mathrm{HH\chi PT}}^{\rm power}
   &= \frac{c_{\mathrm{dip}}\spac e}{16\OurLambda}\,
    \mathrm{Tr}\left[\sigma_{\mu\nu}\spac\bar H H\spac\big( Q
    + \Sigma\spac Q\spac\Sigma^\dagger \big) \right] F_{us}^{\mu\nu} \\
   &\quad - \frac{\lambda_1}{4 m_b}\,\mathrm{Tr}\!\left[ \bar H H \right] 
    - \frac{\lambda_2}{8 m_b}\,\mathrm{Tr}\!\left[ 
    \bar H\spac\Sigma_{\rm mag}^{\mu\nu}\spac H\sigma_{\mu\nu} \right] + \dots \,.
\end{aligned}
\end{equation} 
Finally, after the field redefinition, the flavor-changing quark current reads 
\begin{equation}
   J_{\mathrm{had}, a}^\mu(x) 
   = e^{-i m_B\spac v\cdot x}\,Y_n^{(\ell)\dagger}(x_-)\,
    \frac{i\F}{2\spac\sqrt2}\,
    \mathrm{Tr}\left[ \Gamma_{\rm weak}^\mu\spac H_a' \right] .
\end{equation}
Importantly, the pion fields have disappeared from the weak current, and they appear in the leading-order HH$\chi$PT Lagrangian through objects defined in terms of the field $\Sigma$. 

At this point, the photon interactions in the leading-order HH$\chi$PT Lagrangian can be decoupled via the field redefinition 
\begin{equation}\label{eq:5.51}
   H(x) = H^{(0)}(x)\,\EuScript{R}(x)  \,, 
    \quad \text{with} \quad
   \EuScript{R}(x) = \text{diag}\!\left[ 
    \overline{Y}_v^{(B)}(x)\,\overline{C}_\nb^{(B)}(x_+), 1, 1 \right] .
\end{equation}
This replaces the covariant derivative in the first term of \eqref{eq:LHHchiptLPnew} by an ordinary derivative, $iv\cdot\partial$. However, most of the terms in the effective Lagrangian containing the pseudoscalar fields are not invariant under this field redefinition. For instance, the second term in \eqref{eq:LHHchiptLPnew} transforms into
\begin{equation}
   \frac14\,\mathrm{Tr}\!\left[ \bar H^{(0)} H^{(0)}\spac
    \EuScript{R}\,v\cdot L\,\EuScript{R}^\dagger \right] ,
\end{equation}
and the same applies to the operators multiplying the couplings $g$, $\varrho_1$, $\varrho_3$, and $c_{\rm dip}$. After decoupling, the flavor-changing hadronic current takes the final form (for $a=1$)
\begin{equation}\label{eq:5.54}
   J_{\mathrm{had},\spac 1}^\mu(x) 
   = e^{-i m_B\spac v\cdot x}\,\overline{Y}_v^{(B)}(x)\,
    \overline{C}_\nb^{(B)}(x_+)\,Y_n^{(\ell)\dagger}(x_-)\,
    \frac{i\spac\F}{2\spac\sqrt2}\,
    \mathrm{Tr}\left[ \Gamma_{\rm weak}^\mu\spac H_1^{(0)} \right] .
\end{equation}
This result is identical to the corresponding expression obtained in heavy-particle effective theory, see \eqref{eq:Jhadronic}, and thus the effective weak Lagrangian $\mathcal{L}_{H\ell}$ in \eqref{eq:OVLL_low_energy} remains valid. Moreover, all terms in \eqref{eq:5.20} are still present in the HH$\chi$PT Lagrangians \eqref{eq:LHHchiptLPnew} and \eqref{eq:LHHchiptNLPnew} (where we have omitted the irrelevant kinetic operator proportional to $1/m_B$). The additional interactions of the heavy hadrons with the pseudoscalar mesons present in HH$\chi$PT renormalize the parameters and fields in the Lagrangian, but apart from the wave-function renormalization of the field $H^{(0)}$ they do not connect to the weak vertex. This implies that from the loop diagrams shown in Figure~\ref{fig:chiral_loops}, only the second one is present after the field redefinition $H\to H'$. In general, the only chiral loop corrections contributing to our process are those renormalizing the $B$-meson wave function and the $BB^*\gamma$ coupling. In particular, the only corrections to the coupling $\F$ come from wave-function renormalization.

\begin{figure}
\centering
\includegraphics[scale=0.55]{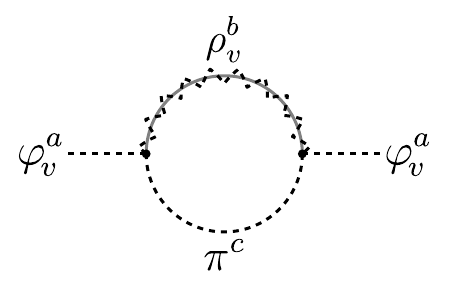}\qquad 
\includegraphics[scale=0.55]{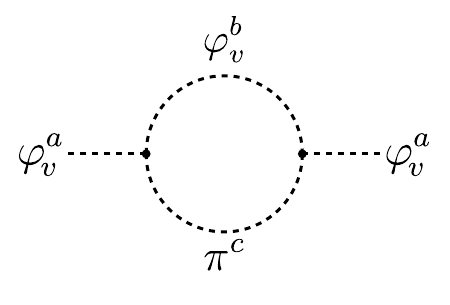}
\caption{One-loop diagrams contributing to the wave-function renormalization of the pseudo-scalar heavy-meson fields $\varphi_v^a$ (additional tadpole graphs contribute to the self-energy but not to the wave-function renormalization). The label $\pi^c$ represents the pseudoscalar mesons $\pi,K,\eta_8$, while $\rho_v^b$ and $\varphi_v^b$ represent $B^*$ and $B$ mesons of different flavors.}
\label{fig:WFR}
\end{figure}

Figure~\ref{fig:WFR} shows the Feynman diagrams contributing at one-loop order to the wave-function renormalization constant of the $B_q$ mesons in HH$\chi$PT, where several terms in the leading-order Lagrangian \eqref{eq:LHHchiptLPnew} contribute to the vertices. Working in the isospin limit $m_u=m_d$ and ignoring QED effects, we find 
\begin{equation}\label{eq:ZB}
\begin{aligned}
   Z_{B^-}
   &= 1 + \frac{1}{(4\pi f)^2}\,\Bigg\{ \frac{3\spac m_\pi^2}{2}\spac L_\pi 
    + \frac{m_{\eta_8}^2}{6}\spac L_{\eta_8} + m_K^2\spac L_K 
    \\
   &\hspace{2.9cm} + \frac{3\spac g^2}{2} \left[ m_\pi^2 \left( 3 L_\pi - 2 \right)
    - 2\spac\delta_{B^*}^2 \left( 3 L_\pi + 1 
    - 6 f_1\bigg(\frac{\delta_{B^*}}{m_\pi}\bigg) \right) \right] \\
   &\hspace{2.9cm} + \frac{g^2}{6} \left[ m_{\eta_8}^2 \left( 3 L_{\eta_8} - 2 \right)
    - 2\spac\delta_{B^*}^2 \left( 3 L_{\eta_8} + 1 
    - 6 f_1\bigg(\frac{\delta_{B^*}}{m_{\eta_8}}\bigg) \right) \right] \\
   &\hspace{2.9cm} + g^2 \left[ m_K^2 \left( 3 L_K - 2 \right)
    - 2\spac\delta_{B_s^*}^2 \left( 3 L_K + 1 
    - 6 f_1\bigg(\frac{\delta_{B_s^*}}{m_K}\bigg) \right) \right] 
     + \text{CTs} \Bigg\} \,,
\end{aligned}
\end{equation}
where 
\begin{equation}
   L_P = \left( \frac{1}{\epsilon} + 1 \right) + \ln\frac{\mu^2}{m_P^2} 
    \quad \stackrel{{\rm renorm}}{\longrightarrow} \quad 
    \ln\frac{\mu^2}{m_P^2}
\end{equation}
are the chiral logarithms including the divergent $1/\epsilon$ UV poles obtained in dimensional regularization. These poles (along with the $+1$, as is conventional in $\chi$PT \cite{Gasser:1983yg,Gasser:1984gg}) are removed by counterterms (CTs) in the HH$\chi$PT Lagrangian at NLO in the chiral expansion, whose explicit construction has been discussed in \cite{Goity:1992tp,Boyd:1994pa}, leaving behind a pure logarithm after renormalization, as indicated above. At leading order in the chiral Lagrangian, the mass of the $\eta_8$ meson is determined by the Gell-Mann--Okubo mass relation $m_{\eta_8}^2=(4\spac m_K^2-m_\pi^2)/3$, and for simplicity we neglected the effects of $\pi^0$--$\eta$--$\eta'$ mixing. The values of the relevant $\delta_{X_b}$ parameters are $\delta_{B^*}\simeq 45.4$\,MeV and $\delta_{B_s^*}\simeq 137.7$\,MeV. The loop functions $f_i(x)$ are defined as (for $0<x<1$)
\begin{equation}
   f_1(x) = \frac{\sqrt{1-x^2}}{x} \left[ \arctan\bigg(\frac{x}{\sqrt{1-x^2}}\bigg)
    - \frac{\pi}{2} \right] , \qquad
   f_2(x) = \frac{x^2}{1-x^2}\,f_1(x) \,.
\end{equation} 
The diagram with a pseudoscalar $B_s$ meson and a charged kaon in the loop (second graph in Figure~\ref{fig:WFR}) receives contributions involving the $B^-\to B_s\spac K^-$ vertices from the second and fourth operators in the effective Lagrangian \eqref{eq:LHHchiptLPnew}. Using the relation  $2\varrho_1\spac(m_s-m_u)\approx\delta_{B_s}$, which is valid up to power corrections, we find that the sum of these contributions vanishes.

The ``effective'' decay constant of the $B^-$ meson in HH$\chi$PT is now obtained as
\begin{equation}\label{eq:Frenorm}
   F_B = \sqrt{Z_{B^-}}\,\spac\F + \order{\frac{1}{m_b}} .
\end{equation}
It is interesting to compare our results \eqref{eq:ZB} and \eqref{eq:Frenorm} with corresponding expressions available in the literature. Setting the parameters $\delta_{X_b}$ and $\varrho_1$ to zero, we find that the chiral logarithms $L_P$ agree with the findings of \cite{Goity:1992tp}. In \cite{Boyd:1994pa} a more complete expression was derived for the chiral loop corrections to the decay constant of neutral $B$ and $B^*$ mesons. It includes finite terms but neglects corrections proportional to $m_\pi^2$. Also, for the mass of the $\eta_8$ meson the relation $m_{\eta_8}^2=\frac43\,m_K^2$ is used. In the isospin limit, we can compare our result \eqref{eq:Frenorm} with their expression for the decay constant of the $B^0$ meson. We find agreement with the terms proportional to $m_K^2$, $m_{\eta_8}^2$ and $\delta_{B_s^*}^2$ including all finite terms.\footnote{The only exception is the term involving $\ln(4/3)$ from $L_{\eta_8}$ in the first line of \eqref{eq:ZB}, which appears to be missing in \cite{Boyd:1994pa}, in conflict with the findings of \cite{Goity:1992tp}.} 
The terms proportional to $\delta_{B^*}^2$ in the second and third line of \eqref{eq:ZB} were dropped in \cite{Boyd:1994pa}, because $\delta_{B^*}^2$ is formally of $\order{\OurLambda^2/m_b^2}$ in the heavy-quark limit and the authors worked consistently to first order in $\OurLambda/m_b$. Our agreement of the NLO corrections to the $B$-meson decay constant in HH$\chi$PT with existing results provides a non-trivial cross check of our rewriting of the HH$\chi$PT Lagrangian. Using the original Lagrangian, the decay constant would receive loop corrections from wave-function renormalization and vertex corrections, the latter of which are absent in our scheme. The expression for the wave-function renormalization factor $Z_{B^-}$ obtained from the traditional HH$\chi$PT Lagrangian differs from our result \eqref{eq:ZB} \cite{Stewart:1998ke}, but the sum of all corrections agrees with our result \eqref{eq:Frenorm}.

\subsection{Integrating out the pseudoscalar mesons}

After decoupling the pseudoscalar mesons from the weak vertex using a field redefinition, we have seen that chiral loop corrections plus higher-order counterterms have the effect of correcting the parameters $\F$, $c_{\rm dip}$ and $g$ in the HH$\chi$PT Lagrangian to their ``physical'' values, e.g.\ 
\begin{equation}
\begin{aligned}
   F_{\rm QCD}(\mu_0) 
   &= \sqrt{Z_B}\,\F = \F + \order{\frac{m_P^2}{(4\pi f)^2}} , \\
   c_{BB^*\gamma}
   &= c_{\rm dip} + \order{\frac{m_P^2}{(4\pi f)^2}} , \\
   g_{BB^*\pi}
   &= g + \order{\frac{m_P^2}{(4\pi f)^2}} ,
\end{aligned}
\label{eq:def_hhchipt_couplings}
\end{equation}
and similarly for the meson decay constants and masses, but they do not give rise to genuinely new effects. The question then arises whether, for processes without pseudoscalar mesons in the external states, one can integrate out these modes to obtain a heavy-meson effective Lagrangian involving the $B$ and $B^*$ mesons, in analogy with the treatment in Section~\ref{subsec:HMET}. This can indeed be done, but as we will see later there is one important subtlety that needs to be taken into account. 

In the interactions of the $B$ and $B^*$ mesons with photons, the neutral pion and its di-photon decay can play a crucial role. The terms in the effective Lagrangian \eqref{eq:LHHchiptLPnew} producing a single $\pi^0$ are
\begin{equation}
   \mathcal{L}_{\mathrm{HH\chi PT}}^{(\pi^0)} 
   = \frac{1}{2\sqrt2 f_\pi}\,\mathrm{Tr}\!\left[ 
    \bar H H\spac t^3 \right] v\cdot\partial\pi^0
    - \frac{g_{BB^*\pi}}{2\sqrt2 f_\pi}\,\mathrm{Tr}\!\left[ 
    \bar H H\spac t^3\spac\gamma_\mu\spac\gamma_5 \right] \partial^\mu\pi^0 \,,
\end{equation}
where $t^3=\text{diag}(1,-1,0)$, and we neglect effects of $\mathcal{O}(\alpha)$.\footnote{As written, these couplings are only present for $B^{(*)-}$ and $B_d^{(*)}$ mesons. When $\pi^0$--$\eta$ mixing is taken into account, also the $B_s^{(*)}$ states exhibit a coupling to the neutral pion, which is suppressed by $\sin\theta_{\pi\eta}$.} 
The $\pi^0\to\gamma\gamma$ decay is mediated by the axial anomaly. It can be described in the context of the chiral Lagrangian by adding a Wess--Zumino--Witten term \cite{Wess:1971yu,Witten:1983tw,Gasser:1984gg}. The corresponding decay amplitude can be expressed as the matrix element of the operator 
\begin{equation}
   \mathcal{L}_{\rm eff}^{\rm WZW}
   = \frac{\alpha}{4\pi f_\pi}\,\pi^0\spac 
    F_{\alpha\beta}^{us}\spac\tilde F_{us}^{\alpha\beta} \,.
\end{equation}
Combining the two expressions above, we obtain the effective Lagrangian for the coupling of two heavy mesons to two photons,
\begin{equation}
   \mathcal{L}_{\rm eff}^{\gamma\gamma}(x)
   = T \left\{ \mathcal{L}_{\mathrm{HH\chi PT}}^{(\pi^0)}(x) ,\spac 
    i\!\int\!d^4y\,\mathcal{L}_{\rm eff}^{\rm WZW}(y) \right\} ,
\end{equation}
where $T$ stands for time ordering. Integrating out the pion fields amounts to calculating the full pion propagator
\begin{equation}
   \langle\spac0\spac|\,T \left\{ \pi^0(x), \pi^0(y) \right\} |\spac 0\spac\rangle
   = \int\!\frac{d^4p}{(2\pi)^4}\,
    \frac{i\,e^{ip\cdot(x-y)}}{p^2-m_\pi^2+i m_\pi\spac\Gamma_\pi}
   = - \frac{i}{\Box+m_\pi^2-i m_\pi\spac\Gamma_\pi}\,\delta^{(4)}(x-y) \,.
\end{equation}
Here $\Gamma_\pi\simeq 7.8$\,eV denotes the total width of the neutral pion, which to good approximation is given by
\begin{equation}
   \Gamma_\pi \approx \Gamma(\pi^0\to\gamma\gamma)
   \approx \frac{\alpha^2}{64\pi^3}\,\frac{m_\pi^3}{f_\pi^2} \,.
\end{equation}
Using this result, we obtain
\begin{equation}\label{eq:Leffgaga}
   \mathcal{L}_{\rm eff}^{\gamma\gamma}
   = \frac{1}{2\sqrt2 f_\pi}\,\frac{\alpha}{4\pi f_\pi}\,\mathrm{Tr}\!\left[ 
    \bar H H\spac t^3 \left( v_\mu - g_{BB^*\pi}\spac\gamma_\mu\spac\gamma_5 \right) 
    \right] \frac{\partial^\mu}{\Box+m_\pi^2-i m_\pi\spac\Gamma_\pi}\,
    F_{\alpha\beta}^{us}\spac\tilde F_{us}^{\alpha\beta} \,.
\end{equation}
Since $f_\pi$ and $m_\pi$ are scales of the low-energy theory that are parametrically smaller than $\OurLambda$, this Lagrangian describes leading-order interactions in the HMET, which should be added to the effective Lagrangian in \eqref{eq:5.20}. Despite appearances, these interactions are not suppressed by the QED coupling $\alpha$. The squared momentum-space amplitude for a process involving one of these effective interactions contains the Breit--Wigner function
\begin{equation}\label{eq:narrowBW}
   \frac{1}{\left( s_{\gamma\gamma} - m_\pi^2 \right)^2 
            + \left( m_\pi\spac\Gamma_\pi \right)^2}
   = \frac{\pi}{m_\pi\spac\Gamma_\pi}\,\delta(s_{\gamma\gamma}-m_\pi^2)
    + \order{(\Gamma_\pi)^0} ,
\end{equation}
where $s_{\gamma\gamma}$ is the di-photon invariant mass squared. Keeping only the leading term in this expansion is an excellent approximation, since the neutral-pion width is seven orders of magnitude smaller than the pion mass. The decay amplitude squared obtained from the effective Lagrangian \eqref{eq:5.20} scales like $\alpha^2$, but the factor $1/\Gamma_\pi$ scales like $1/\alpha^2$, so that the di-photon decay rate is in fact independent of the QED coupling (see Section~\ref{subsec:indirect} for details). 

The $\delta$-distribution on the right-hand side of \eqref{eq:narrowBW} allows us to factorize the phase-space integration in a decay process mediated by the effective Lagrangian \eqref{eq:Leffgaga}. Consider a decay of a particle with 4-momentum $P$ into $n$ particles with 4-momenta $p_i$, where the first two ($i=1,2$) refer to the photons produced by the effective interaction. When the $n$-particle phase-space integral is multiplied with $\delta\big((p_1+p_2)^2-m_\pi^2\big)$, we can insert a dummy integration over a 4-momentum $p_\pi\equiv p_1+p_2$ to obtain
\begin{align}\label{eq:phasespacefact}
   & \prod_{i=1}^n \int\!\frac{d^3\bm{p}_i}{(2\pi)^3\spac 2E_{\bm{p}_i}}\,
    \delta^{(4)}\Big(\sum_{i=1}^n p_i - P\Big)\,
    \frac{\pi}{m_\pi\spac\Gamma_\pi}\,\delta\big((p_1+p_2)^2-m_\pi^2\big)
    \int\!\frac{d^4 p_\pi}{(2\pi)^4}\,(2\pi)^4\,\delta^{(4)}(p_\pi-p_1-p_2) \notag\\
   &= \frac{\pi}{m_\pi\spac\Gamma_\pi} 
    \int\!\frac{d^4 p_\pi}{(2\pi)^4}\,\theta(p_\pi^0)\,2\pi\,\delta(p_\pi^2-m_\pi^2) 
    \spac\prod_{i=3}^n \int\!\frac{d^3\bm{p}_i}{(2\pi)^3\spac 2E_{\bm{p}_i}}\,
    \delta^{(4)}\bigg(p_\pi+\sum_{i=3}^n p_i - P\bigg) \notag\\
   &\quad \times (2\pi)^3 \int\!\frac{d^3\bm{p}_1}{(2\pi)^3\spac 2E_{\bm{p}_1}}
    \int\!\frac{d^3\bm{p}_2}{(2\pi)^3\spac 2E_{\bm{p}_2}}\,\delta^{(4)}(p_\pi-p_1-p_2) \\
   &= \prod_{i=\pi,3}^n \int\!
    \frac{d^3\bm{p}_i}{(2\pi)^3\spac 2E_{\bm{p}_i}}\,
    \delta^{(4)}\bigg(p_\pi+\sum_{i=3}^n p_i - P\bigg) \notag\\
   &\quad \times \frac{1}{\Gamma_\pi}\,\frac{1}{2\spac m_\pi}  
    \int\!\frac{d^3\bm{p}_1}{(2\pi)^3\spac 2E_{\bm{p}_1}}
    \int\!\frac{d^3\bm{p}_2}{(2\pi)^3\spac 2E_{\bm{p}_2}}\,
    (2\pi)^4\,\delta^{(4)}(p_\pi-p_1-p_2) \,. \notag
\end{align}
In the second step we have inserted $\theta(p_\pi^0)=1$, which holds since $p_1$ and $p_2$ are the 4-momenta of on-shell particles with positive energies. The product in the last expression includes the $(n-1)$ particles with momenta $p_\pi,p_3,\dots,p_n$ and masses $m_\pi,m_3,\dots,m_n$, whereas the expression in the last line contains the phase-space integral for the decay $\pi^0\to\gamma\gamma$ (including the proper normalization $1/2 m_\pi$, but not the symmetry factor 1/2). The di-photon matrix element of the fields $F_{\alpha\beta}^{us}\spac\tilde F_{us}^{\alpha\beta}$ in \eqref{eq:5.20}, when multiplied with $\frac{\alpha}{4\pi f_\pi}$, gives the $\pi^0\to\gamma\gamma$ decay amplitude. Integrating the square of this amplitude over the di-photon phase-space (last line in \eqref{eq:phasespacefact}), taking into account a symmetry factor 1/2,  yields the corresponding decay rate, which gives the $\pi^0\to\gamma\gamma$ branching ratio when divided by the total decay rate $\Gamma_\pi$. 

\subsection[\texorpdfstring{Matching SCET-2 to HH$\chi$PT}{Matching SCET-2 to HHchiPT}]{\boldmath Matching SCET-2 to HH$\chi$PT}
\label{subsec:hhchipt_matching}

Having constructed the low-energy effective theory, we must now perform the matching of the results we obtained in SCET-2 to HH$\chi$PT$\,\otimes\,$bHLET. Ignoring QED effects, we have already accomplished this task, since in Section~\ref{subsec:HHchiPT} we have related the non-perturbative coupling parameter $\F$ in \eqref{eq:5.54} to the ``physical'' decay constant of the $B$ meson. Since the heavy hadrons are treated in the heavy-quark limit in the effective theory below the scale $m_B$, this parameter equals the HQET parameter $F_{\rm QCD}(\mu_0)$ at the scale where the matching onto HH$\chi$PT is performed. This condition yields the non-perturbative matching condition 
\begin{equation}
   \F = \frac{F_{\rm QCD}(\mu_0)}{\sqrt{Z_{B^-}}} \,.
\end{equation}
On the other hand, we have seen earlier in Section~\ref{sec:SCET-2} that the calculation of QED effects involves additional non-perturbative hadronic quantities, namely $F_-$ and various $B$-meson LCDAs. We will now make contact between SCET-2 and the low-energy theory and study the non-perturbative matching of these theories in the presence of QED effects in detail.

In a first step, we need to address the fact that in our discussion of SCET-2 in Section~\ref{sec:SCET-2} we have not explored the possible relevance of soft-collinear modes, with momentum scaling $p_{sc}\sim\lambda\spac(\lambda_\ell^2,1,\lambda_\ell)$. These modes have one momentum component of order $\lambda\spac m_B=\Lambda_{\rm QCD}$, but their virtuality $p_{sc}^2\sim\lambda_\ell^2\spac\Lambda_{\rm QCD}^2$ is far below the QCD confinement scale $\OurLambda$, so one would expect that they are blind to the QCD dynamics inside hadrons. This is indeed the case, but it is nevertheless interesting to look at these modes in detail. The relevant quark currents in SCET-2 are of the form
\begin{equation}\label{eq:SCET2ops}
   \bar u_s(x)\spac\Gamma\,b_v(x)\spac Y_n^{(\ell)\dagger}(x_-)
    \quad \text{or} \quad
   \bar\Q_s(x+sn)\spac\Gamma\,\H_v(x)\spac Y_n^{(\ell)\dagger}(x_-) \,,
\end{equation}
where the first form applies to the ``local'' operators in \eqref{eq:SCET2basis_local}, and second one to the non-local operators in \eqref{eq:SCET2basis_nonlocal}, some of which contain an additional soft gauge field $\A(x+s_g n)$. The interactions of soft-collinear gluons and photons interacting with soft fields are of eikonal form, and due to their momentum scaling, only the minus components of the soft-collinear momenta are transferred to the soft particles. The corresponding interaction term $\psi_s(x)\spac i\nb\cdot A_{sc}(x_+)\spac\psi(x)$ can be decoupled via the field redefinition $\psi(x)\to\overline{C}_\nb(x_+)\,\psi^{(0)}(x)$ for an initial-state soft field $\psi$. For both types of operators in \eqref{eq:SCET2ops} this produces the Wilson line (this is also true for operators containing an additional soft gauge field)
\begin{equation}
   \overline{C}_\nb^{(u)\dagger}(x_+)\,\overline{C}_\nb^{(b)}(x_+)
   = \overline{C}_\nb^{(B)}(x_+) \,,
\end{equation}
where the QCD parts of the Wilson lines cancel each other, and the QED parts add up to a combined Wilson line with the electric charge of the $B$ meson, since $Q_b-Q_u=Q_B$. The result is the Wilson line for soft-collinear photon emission off a point-like $B$ meson. This Wilson line is transmitted to the Wilson-line operator $R^{(\ell,B)}$ in \eqref{eq:Rdef}, which we must then redefine as
\begin{equation}
   \myR \equiv \langle\spac 0\spac|\,Y_n^{(\ell)\dagger}\,\overline{Y}_v^{(B)}\,
    \overline{C}_\nb^{(B)}\spac|\spac 0\spac\rangle \,.
\end{equation}
In the definitions of the soft hadronic matrix elements $S_i$ in terms of hadronic parameters and LCDAs in \eqref{eq:Sidef} and \eqref{eq:Sinldef}, the Wilson line $ \overline{C}_\nb^{(B)}$ cancels out between the numerator and denominator, and the hadronic quantities remain unchanged.

Let us now reconsider the definition of the leptonic functions $K_i$ in \eqref{eq:Kidef}. For performing the calculation of these functions, we have considered the simplest leptonic state consisting of a charged lepton and an anti-neutrino. What would change if we had allowed for the presence of soft-collinear photons in this calculation? The spinor product on the right-hand of the equation can trivially be rewritten in bHLET as
\begin{equation}
   \bar u(v_\ell)\spac P_L\,v(p_\nu)
   = \langle\ell(p_\ell)\spac\bar\nu(p_\nu)|\,
    \bar h_{v_\ell}\spac P_L\,\nu_\cb\,|\spac 0\spac\rangle \,,
\end{equation}
where $\bar u(v_\ell)\spac\vsl_\ell=\bar u(v_\ell)$ obeys the same constraint as the field $\bar h_{v_\ell}$, see \eqref{eq:HQETproj}. We can thus rewrite \eqref{eq:Kidef} as a matching relation connecting leptonic matrix elements in SCET-2 and bHLET, namely (for $i=1,2$)
\begin{equation}
   \langle\ell(p_\ell)\spac\bar\nu(p_\nu)|\,j_i^{ \mathrm{\mathrm{lep}}}(x,\mu)\,
    |\spac 0\spac\rangle
   = m_\ell^{\mathrm{\mathrm{phys}}}\spac K_i(x,\mu)\,
    \langle\ell(p_\ell)\spac\bar\nu(p_\nu)|\,
    \bar h_{v_\ell}\spac P_L\,\nu_\cb\,|\spac 0\spac\rangle \,. 
\end{equation}
Since the matching of operators in two effective theories can be done using any choice of external states, we can allow for the presence of soft-collinear photons in the final state and generalize the relation to
\begin{equation}
\begin{aligned}
   \langle\ell\spac\bar\nu+n\spac\gamma_{sc}|\,
    j_i^{ \mathrm{\mathrm{lep}}}(x,\mu)\,|\spac 0\spac\rangle
   &= m_\ell^{\mathrm{\mathrm{phys}}}\spac K_i(x,\mu)\,
    \langle\ell\spac\bar\nu+n\spac\gamma_{sc}|\,
    \bar h_{v_\ell}\spac P_L\,\nu_\cb\,|\spac 0\spac\rangle \\  
   &= m_\ell^{\mathrm{\mathrm{phys}}}\spac K_i(x,\mu)\,
    \langle\ell\spac\bar\nu+ n\spac\gamma_{sc}|\,C_{v_\ell}^{(\ell)\dagger}\,
    \bar h_{v_\ell}^{(0)}\spac P_L\,\nu_\cb\,
    |\spac 0\spac\rangle \\ 
   &= m_\ell^{\mathrm{\mathrm{phys}}}\spac K_i(x,\mu)\,
    \langle n\spac\gamma_{sc}|\,C_{v_\ell}^{(\ell)\dagger}\,|\spac 0\spac\rangle\,
    \bar u(p_\ell)\spac P_L\,v(p_\nu) \,, 
\end{aligned}
\end{equation}
where the number $n$ of photons is arbitrary, and in the last step we have applied the decoupling transformation \eqref{eq:leptondecoupling} to the charged-lepton field. If bHLET is the correct low-energy theory in the leptonic sector, the functions $K_i$ must be the same for all values of $n$. They are the Wilson coefficients associated with the matching of the two theories. 

An analogous discussion applies for the hadronic sector, where we can rewrite the relations in \eqref{eq:Sidef} and \eqref{eq:Sinldef} in the form (for $i=1,\dots,6$)
\begin{equation}
\begin{aligned}
   \langle\spac 0\spac|\,j_i^{ \mathrm{had}}(\Lambda,\mu)\,|B^-\rangle
   &= S_i\,\myR 
    = S_i\,\langle\spac 0\spac|\,Y_n^{(\ell)\dagger}\,\overline{Y}_v^{(B)}\,
    \overline{C}_\nb^{(B)}\spac|\spac 0\spac\rangle \\
   &= \frac{S_i}{\sqrt{2m_B}}\,\langle\spac 0\spac|\,
    Y_n^{(\ell)\dagger}\,\overline{Y}_v^{(B)}\,
    \overline{C}_\nb^{(B)}\spac\varphi_v^{(0)}\spac|B^-\rangle \,,
\end{aligned}
\end{equation}
where in the last step we have reintroduced the decoupled initial-state $B$-meson field and used relation \eqref{eq:norms}. This result again takes the form of a matching relation between two effective theories, and we can generalize it by allowing for any number of soft or soft-collinear photons in the final state, i.e.\
\begin{equation}
\begin{aligned}
   \langle n_s\spac\gamma_s + n_{sc}\spac\gamma_{sc}|\,
    j_i^{ \mathrm{had}}(\Lambda,\mu)\,|B^-\rangle
   &= \frac{S_i}{\sqrt{2m_B}}\,\langle n_s\spac\gamma_s+n_{sc}\spac\gamma_{sc}|\, 
    Y_n^{(\ell)\dagger}\,\overline{Y}_v^{(B)}\,
    \overline{C}_\nb^{(B)}\spac\varphi_v^{(0)}\spac|B^-\rangle \,.
\end{aligned}
\end{equation}
Once again, the SCET-2 hadronic matrix elements $S_i$ appear as the Wilson coefficients of this matching relation, and they are the same for any numbers $n_s$ and $n_{sc}$ of photons if the low-energy theory is constructed in a consistent way.

It follows from these observations that the matrix elements of all SCET-2 operators can be written as a combination of a collinear function $K_i$ (with $i=1,2$), a soft function $S_i$ (with $i=1,\dots,6$), and the $B$-meson-to-leptons matrix element of the operator
\begin{equation}\label{eq:softstring}
   \overline{Y}_v^{(B)}\,\overline{C}_\nb^{(B)}\,
   Y_n^{(\ell)\dagger}\,C_{v_\ell}^{(\ell)\dagger}\,
   \frac{\varphi_v^{(0)}}{\sqrt{2m_B}}\,
   \bar h_{v_\ell}^{(0)}\spac P_L\,\nu_\cb \,,
\end{equation}
with any number of soft or soft-collinear photons in the final state. Due to the experimental constraint that the energy of these photons should be less than a threshold value $E_{\rm cut}\ll\OurLambda$, we need to lower the scale of the soft and soft-collinear photons by a factor of $\zeta=E_{\rm cut}/\OurLambda$, turning them into ultrasoft and ultrasoft-collinear modes. With this in mind, we see that the operator in \eqref{eq:softstring} is precisely the operator for the scalar heavy meson in \eqref{eq:OVLL_low_energy}. With this identification, and using the definitions in \eqref{eq:4.98} and \eqref{eq:total_T}, we now obtain the matching coefficient
\begin{equation}\label{eq:matching_yB}    
    y_B(\mu_0) = \left(\frac{\alpha(m_Z)}{\alpha(m_B)} \right)^\frac{9}{40}
     \left( \frac{\mu_0^2}{m_B\spac m_\ell} \right)^\frac{\gamma_\mathrm{soft}}{2}\, \mathcal{R}_{\mathrm{virt}} \,,
\end{equation}
with the matching scale $\mu_0=1.5$\,GeV. The soft anomalous dimension $\gamma_{\rm soft}$ has been defined in \eqref{eq:gammasoft}, and the remaining virtual corrections contained in $\mathcal{R}_{ \mathrm{virt}}$ have been given in \eqref{eq:Rvirt_final}.

\subsection{Direct contribution to the decay rate}
\label{subsec:decayrate} 

We now compute the $B^-\to\ell^-\spac\bar\nu_\ell\spac(\gamma)$ decay rate for a photon-energy cut $E_{\rm cut}\ll\OurLambda$ at $\order{\alpha}$ in the low-energy theory. Here and below, the symbol ``$(\gamma)$'' refers to any form of electromagnetic radiation, from one or several low-energetic photons. The total decay rate can, to good approximation, be written as 
\begin{equation}\label{eq:rad_rate}
   \Gamma(E_\mathrm{cut}) 
   = \Gamma_\mathrm{dir}(E_\mathrm{cut}) + \Gamma_\mathrm{indir}(E_\mathrm{cut}) \,.
\end{equation}
The second term corresponds to the indirect contributions with an excited $B^*$ resonance as an intermediate state. It will be discussed in detail in the next section, where we will also show that interference terms between the direct and indirect contributions are doubly power suppressed by factors of $(m_\ell/m_B)^2$ and $(E_{\rm cut}/\OurLambda)^2$, and thus can be safely neglected. The direct component factorizes into
\begin{equation}\label{eq:IB_rate}
   \Gamma_\mathrm{dir}(E_\mathrm{cut}) 
   = \Gamma_\mathrm{tree}\,\spac y_B^2(\mu_0)\,R(E_\mathrm{cut},\mu_0) \,,
\end{equation}
with the ``tree-level'' decay rate 
\begin{equation}\label{eq:treelevelrate}
   \Gamma_\mathrm{tree}
   = \frac{m_\ell^2\,m_B}{8\pi}
    \left( G_F^{(\mu)}\spac|V_{ub}|\spac f_B^{\rm QCD} \right)^2
    \left( 1 - \frac{m_\ell^2}{m_B^2} \right)^2 ,
\end{equation}
in which all QED corrections are omitted. The product $\Gamma_\mathrm{tree}\,\spac y_B^2$ is the non-radiative rate and $R(E_\mathrm{cut},\mu)$ is the radiation function. The former includes virtual QED corrections via the coupling $y_B(\mu)$ in \eqref{eq:matching_yB}, while the latter describes the real emission of ultrasoft or ultrasoft-collinear photons. At leading order in power counting, virtual corrections to the \blnu\ process in the low-energy theory arise solely from the tadpole diagram shown by the third graph in Figure~\ref{fig:graphs_mesons_eft}, which is scaleless and vanishes. 

The radiation function is given by the convolution of the squared matrix elements of the ultrasoft and ultrasoft-collinear Wilson lines, integrated over phase space. We have
\begin{equation}\label{eq:soft_function}
   R(E_\mathrm{cut},\mu) 
   = \int_0^\infty\!d\omega_{us} \int_0^\infty\!d\omega_{usc}\, 
    \theta\big( E_\mathrm{cut}-\omega_{us}-\omega_{usc} \big)\, 
    W_{us}(\omega_{us},\mu)\,W_{usc}(\omega_{usc},\mu) \,, 
\end{equation}
where $\omega_{us}$ and $\omega_{usc}$ are the total energies of ultrasoft and ultrasoft-collinear radiation (measured in the rest frame of the $B$ meson), and the Heaviside function implements the cut on the total energy of all photons in the final state. The (bare) ultrasoft and ultrasoft-collinear functions $W_{us}$ and $W_{usc}$ are given by 
\begin{equation}\label{eq:us_usc_definition}
\begin{aligned}
   W_{us}(\omega_{us}) 
   &= \sum_{n=0}^\infty \left[ \int\prod_{i=1}^n\, 
    \frac{d^{d-1}\boldsymbol{q}_i}{(2\pi)^{d-1}\spac 2 E_i}
    \left| \bra{\gamma_{us}^n} \overline{Y}_v^{(B)}\spac Y_n^{(\ell)\dagger} 
    \ket{0} \right|^2 \delta\Big( \omega_{us} - \sum_{j=1}^n E_j \Big) \right] , \\
   W_{usc}(\omega_{usc}) 
   &= \sum_{n=0}^{\infty} \left[ \int\prod_{i=1}^n\, 
    \frac{d^{d-1}\boldsymbol{q}_i}{(2\pi)^{d-1}\spac 2 E_i} 
    \left| \bra{\gamma_{usc}^n} \overline{C}_\nb^{(B)}\spac 
     C_{v_\ell}^{(\ell)\dagger} \ket{0} \right|^2 
     \delta \Big( \omega_{usc} - \sum_{j=1}^n E_j \Big) \right] ,
\end{aligned} 
\end{equation}
where $\bra{\gamma_{i}^n}=\bra{\gamma_{i}(q_1)\ldots\gamma_{i}(q_n)}$ with $i=us$ or $i=usc$ denotes the $n$-photon Fock state for ultrasoft or ultrasoft-collinear photons, respectively. At $\order{\alpha}$, we obtain in the $\overline{\rm MS}$ scheme (with $q_+\equiv\frac{n\cdot q}{n\cdot v}$ and $q_-\equiv\frac{\nb\cdot q}{\nb\cdot v}$) 
\begin{equation}\label{eq:2Ws}
\begin{aligned}
   W_{us}(\omega_{us}) 
   &= \delta(\omega_{us}) + \frac{Q_\ell^2\spac\alpha}{\pi}\,\mu^{2\epsilon}\,
    \frac{e^{\epsilon\gamma_E}}{\Gamma(1-\epsilon)} 
    \int_0^\infty\!dq_+ \int_0^\infty\!dq_-\,
    \frac{q_+^{-1-\epsilon}\spac q_-^{1-\epsilon}}{(q_+ + q_-)^2}\,\spac
    \delta\!\left( \omega_{us} - \frac{q_+ + q_-}{2} \right) , \\
   W_{usc}(\omega_{usc}) 
   &= \delta(\omega_{usc}) + \frac{Q_\ell^2\spac\alpha}{\pi}\,\mu^{2\epsilon}\,
    \frac{e^{\epsilon\gamma_E}}{\Gamma(1-\epsilon)} 
    \int_0^\infty\!dq_+ \int_0^\infty\!dq_-\,
    \frac{q_+^{1-\epsilon}\spac q_-^{-1-\epsilon}}%
         {\left(q_+ + \frac{m_\ell^2}{m_B^2}\,q_- \right)^2}\,\spac
    \delta\!\left( \omega_{usc} - \frac{q_-}{2} \right) ,
\end{aligned}
\end{equation}
where  we have used the relation
\begin{equation}
   \int d^dq\,\delta(q^2)\,\theta(q^0)
   = \frac12 \int_0^\infty\!dq_+ \int_0^\infty\!dq_-\,
    \int d^{d-2}q_\perp\,\delta(q_+ q_- + q_\perp^2) \,.
\end{equation}
While the integrands of the two expressions in \eqref{eq:2Ws} coincide in the limit $m_\ell=m_B$, the integrals differ in the arguments of the $\delta$-functions, since
$q_+\ll q_-$ for ultrasoft-collinear photons, and hence the photon energy is equal to $q_-/2$ up to power-suppressed terms. Performing the integrations, we obtain
\begin{equation}\label{eq:us_usc_implicit}
\begin{aligned}
   W_{us}(\omega_{us}) 
   &= \delta(\omega_{us}) + \frac{Q_\ell^2\spac\alpha}{\pi}\,\frac{1}{\omega_{us}} 
    \left( \frac{\mu^2}{(2\omega_{us})^2} \right)^\epsilon e^{\epsilon\gamma_E}\,
    \frac{(1-\epsilon)\spac\Gamma(-\epsilon)}{\Gamma(2-2\epsilon)} 
    + \order{\alpha^2} , \\
   W_{usc}(\omega_{usc}) 
   &= \delta(\omega_{usc}) + \frac{Q_\ell^2\spac\alpha}{\pi}\,\frac{1}{\omega_{usc}}
    \left(\frac{\mu^2\spac m_B^2}{(2\omega_ {usc})^2\spac m_\ell^2} \right)^\epsilon 
    e^{\epsilon\gamma_E}\,(1-\epsilon)\spac\Gamma(\epsilon) + \order{\alpha^2} .
\end{aligned}
\end{equation}
Inserting these expressions into \eqref{eq:soft_function}, performing the two energy integrals, and expanding the result in a Laurent series in $\epsilon$, we obtain for the (bare) radiation function 
\begin{equation}\label{eq:hhchipt_one-loop-soft-function}
   R(E_\mathrm{cut}) 
   = 1 + \frac{Q_\ell^2\spac\alpha}{2\pi} \left[
    - \left( \frac{1}{\epsilon} 
    + \ln\frac{\mu^2\spac m_B}{(2E_\mathrm{cut})^2\spac m_\ell} \right) 
    \left( \ln\frac{m_B^2}{m_\ell^2} - 2 \right) + 2 - \frac{\pi^2}{3} \right] 
    + \mathcal{O}(\alpha^2) .
\end{equation}

A problematic feature of this derivation is the fact that the integrations in \eqref{eq:soft_function}, when carried out with bare functions, produce additional $1/\epsilon$ poles not present in the functions $W_{us}$ and $W_{usc}$. The result \eqref{eq:hhchipt_one-loop-soft-function} thus does not provide a basis for a consistent scale separation and resummation of large logarithmic corrections. Instead, we should renormalize the ultrasoft and ultrasoft-collinear functions separately, and then perform the integrations over renormalized functions in \eqref{eq:soft_function}. Due to their singular behavior at the origin, the functions $W_{us}$ and $W_{usc}$ must be treated as distributions and renormalized using convolutions, such that \begin{equation}\label{eq:Wreneq}
   W_{us}(\omega,\mu)
   = \int_0^\Omega\!d\omega'\,Z_{us}(\omega,\omega',\mu)\,W_{us}(\omega') \,,
\end{equation}
and similarly for $W_{usc}$. Here $\Omega$ denotes the upper value of the interval on which the renormalized functions are defined, in our case $\Omega=E_{\rm cut}$. To derive the renormalization factors, we regularize the bare functions in \eqref{eq:us_usc_implicit} using star distributions \cite{DeFazio:1999ptt,Bosch:2004th}, which are defined as 
\begin{equation}\label{eq:stardists}
\begin{aligned}
   \int_0^\Omega\!d\omega\,f(\omega) 
    \left( \frac{1}{\omega} \right)_{\!*}^{\![\mu/2]}
   &= \int_0^\Omega\!d\omega\,\frac{f(\omega)-f(0)}{\omega}
   + f(0)\spac\ln\frac{2\Omega}{\mu} \,, \\
   \int_0^\Omega\!d\omega\,f(\omega) 
    \left( \frac{\ln\frac{2\omega}{\mu}}{\omega} \right)_{\!*}^{\![\mu/2]}
   &= \int_0^\Omega\!d\omega\,\frac{f(\omega)-f(0)}{\omega}\,\ln\frac{2\omega}{\mu}
   + \frac{f(0)}{2} \spac \ln^2\frac{2\Omega}{\mu} \,.
\end{aligned}
\end{equation}
Using the relation
\begin{equation}
   \frac{1}{\omega} \left( \frac{\mu^2}{(2\omega)^2} \right)^\epsilon
   = - \frac{1}{2\epsilon}\,\delta(\omega) + \left( \frac{1}{\omega} \right)_{\!*}^{\![\mu/2]}
    - 2\epsilon  \left( \frac{\ln\frac{2\omega}{\mu}}{\omega} \right)_{\!*}^{\![\mu/2]} + \order{\epsilon^2} ,
\end{equation}
we find that
\begin{equation}
\begin{aligned}
   W_{us}(\omega) 
   &= \delta(\omega) \left[ 1 + \frac{Q_\ell^2\spac\alpha}{\pi}
    \left( \frac{1}{2\epsilon^2} + \frac{1}{2\epsilon} + 1 - \frac{\pi^2}{8} \right) 
    \right] \\
   &\quad + \frac{Q_\ell^2\spac\alpha}{\pi} 
    \left[ - \left( \frac{1}{\epsilon} + 1 \right) 
    \left( \frac{1}{\omega} \right)_{\!*}^{\![\mu/2]}
    + 2 \left( \frac{\ln\frac{2\omega}{\mu}}{\omega} \right)_{\!*}^{\![\mu/2]} \right]
    + \order{\alpha^2} , \\
   W_{usc}(\omega) 
   &= \delta(\omega) \left[ 1 + \frac{Q_\ell^2\spac\alpha}{\pi}
    \left( - \frac{1}{2\epsilon^2} 
    + \frac{1}{2\epsilon} \left( 1 - \ln r_\ell \right) 
    - \frac{\pi^2}{24} + \frac{\ln r_\ell}{2} 
    - \frac{\ln^2 r_\ell}{4} \right) \right] \\
   &\quad + \frac{Q_\ell^2\spac\alpha}{\pi} 
    \left[ \left( \frac{1}{\epsilon} - 1 + \ln r_\ell \right) 
    \left( \frac{1}{\omega} \right)_{\!*}^{\![\mu/2]} 
    - 2 \left( \frac{\ln\frac{2\omega}{\mu}}{\omega} \right)_{\!*}^{\![\mu/2]} \right]
    + \order{\alpha^2} ,
\end{aligned}
\end{equation}
where $r_\ell=m_B^2/m_\ell^2$. From these expressions we read off the renormalization factors
\begin{equation}\label{eq:Zus_Zusc}
\begin{aligned}
   Z_{us}(\omega,\omega',\mu) 
   &= \delta(\omega-\omega') \left[ 1 - \frac{Q_\ell^2\spac\alpha}{2\pi}
    \left( \frac{1}{\epsilon^2} + \frac{1}{\epsilon} \right) \right] 
    + \frac{Q_\ell^2\spac\alpha}{\pi}\, 
    \frac{1}{\epsilon} \left( \frac{1}{\omega-\omega'} \right)_{\!*}^{\![\mu/2]}
    + \order{\alpha^2} , \\
   Z_{usc}(\omega,\omega',\mu) 
   &= \delta(\omega-\omega') \left[ 1 + \frac{Q_\ell^2\spac\alpha}{2\pi}
    \left( \frac{1}{\epsilon^2} 
    - \frac{1}{\epsilon} \left( 1 - \ln r_\ell \right) \right) \right] 
    - \frac{Q_\ell^2\spac\alpha}{\pi}\,\frac{1}{\epsilon} 
    \left( \frac{1}{\omega-\omega'} \right)_{\!*}^{\![\mu/2]} 
    + \order{\alpha^2} ,
\end{aligned}
\end{equation}
and the renormalized functions
\begin{equation}\label{eq:us_usc_renorm}
\begin{aligned}
   W_{us}(\omega,\mu) 
   &= \delta(\omega) \left[ 1 + \frac{Q_\ell^2\spac\alpha}{\pi}
    \left( 1 - \frac{\pi^2}{8} \right) \right] 
    + \frac{Q_\ell^2\spac\alpha}{\pi} 
    \left[ - \left( \frac{1}{\omega} \right)_{\!*}^{\![\mu/2]} 
    + 2 \left( \frac{\ln\frac{2\omega}{\mu}}{\omega} \right)_{\!*}^{\![\mu/2]} \right]
    + \order{\alpha^2} , \\
   W_{usc}(\omega,\mu) 
   &= \delta(\omega) \left[ 1 + \frac{Q_\ell^2\spac\alpha}{\pi}
    \left( - \frac{\pi^2}{24} + \frac{\ln r_\ell}{2} - \frac{\ln^2 r_\ell}{4} \right) 
    \right] \\
   &\quad + \frac{Q_\ell^2\spac\alpha}{\pi} \left[ \left( \ln r_\ell - 1 \right) 
    \left( \frac{1}{\omega} \right)_{\!*}^{\![\mu/2]} 
    - 2 \left( \frac{\ln\frac{2\omega}{\mu}}{\omega} \right)_{\!*}^{\![\mu/2]} \right]
    + \order{\alpha^2} .
\end{aligned}
\end{equation}
Using these distributions in \eqref{eq:soft_function}, we find the renormalized expression
\begin{equation}
   R(E_\mathrm{cut},\mu) 
   = 1 + \frac{Q_\ell^2\spac\alpha}{2\pi} \left[
    - \ln\frac{\mu^2\spac m_B}{(2E_\mathrm{cut})^2\spac m_\ell} 
    \left( \ln\frac{m_B^2}{m_\ell^2} - 2 \right) + 2 - \frac{\pi^2}{3} \right] 
    + \mathcal{O}(\alpha^2) ,
\end{equation}
in which the $1/\epsilon$ pole present in \eqref{eq:hhchipt_one-loop-soft-function} has been removed.

\subsubsection*{Resummation of large logarithmic corrections}

The renormalized functions in \eqref{eq:us_usc_renorm} contain logarithms of the form $\ln(\mu/2\omega)$ and $\ln(m_B^2/m_\ell^2)$. This gives rise to ultrasoft and ultrasoft-collinear logarithms of the type $\ln(\mu/2E_{\rm cut})$ and $\ln(\mu\spac m_B/2E_{\rm cut}\spac m_\ell)$ in the radiation function $R(E_\mathrm{cut},\mu)$, which cannot be simultaneously made small by an appropriate choice of the scale $\mu$. Both types of logarithms can be resummed by solving the RG equations satisfied by the various functions. It follows from \eqref{eq:Wreneq} that the renormalized ultrasoft and ultrasoft-collinear functions satisfy the evolution equations
\begin{equation}\label{eq:Wusevol}
   \frac{d}{d\ln\mu}\,W_{us}(\omega,\mu)
   = \int_0^\Omega\!d\omega'\,\Gamma_{us}(\omega,\omega',\mu)\,W_{us}(\omega',\mu) \,,
\end{equation}
and similarly for $W_{usc}$, where the anomalous dimensions are obtained from the single $1/\epsilon$ pole terms of the renormalization factors in \eqref{eq:Zus_Zusc}. They are
\begin{equation}
\begin{aligned}
   \Gamma_{us}(\omega,\omega',\mu) 
   &= \frac{Q_\ell^2\spac\alpha}{\pi} \left[ - \delta(\omega-\omega')  
    + 2 \left( \frac{1}{\omega-\omega'} \right)_{\!*}^{\![\mu/2]} \right]
    + \order{\alpha^2} , \\
   \Gamma_{usc}(\omega,\omega',\mu) 
   &= \frac{Q_\ell^2\spac\alpha}{\pi} \left[ \delta(\omega-\omega') 
    \left( \ln\frac{m_B^2}{m_\ell^2} - 1 \right)  
    - 2 \left( \frac{1}{\omega-\omega'} \right)_{\!*}^{\![\mu/2]} \right] 
    + \order{\alpha^2} .
\end{aligned}
\end{equation}
The integro-differential evolution equation \eqref{eq:Wusevol} has the same form as the RG equation for the $B$-meson shape function \cite{Neubert:1993um}, which was  obtained in \cite{Grozin:1994ni,Bosch:2004th}. Its closed-form solution has been derived in \cite{Neubert:2004dd} using a technique developed in \cite{Lange:2003ff}. For our purposes, it will however be more convenient to use a variant of the method formulated in Laplace space \cite{Neubert:2005nt,Becher:2006nr}, where the convolution in \eqref{eq:soft_function} turns into a product, i.e.\
\begin{equation}\label{eq:Rlaplace}
   \widetilde{R}(s,\mu) 
   = \int_0^\infty\!dE_\mathrm{cut}\,e^{-s E_\mathrm{cut}}\spac R(E_\mathrm{cut},\mu) 
   = \frac{\widetilde{W}_{us}(s,\mu)\,\widetilde{W}_{usc}(s,\mu)}{s} \,.
\end{equation}
The Laplace transform
\begin{equation}
   \widetilde{W}_{us}(s,\mu) 
   = \int_0^\infty\!d\omega\,e^{-s\spac\omega}\spac W_{us}(\omega,\mu) \,,
\end{equation}
and similarly $\widetilde{W}_{usc}(s,\mu)$, cannot be derived using the distribution-valued expressions in \eqref{eq:us_usc_renorm}, because the star distributions are defined for integrals with a finite upper cutoff $\omega\le\Omega$, see \eqref{eq:stardists}. We thus need to go back to the expressions for the bare functions given in \eqref{eq:us_usc_implicit}. Using the relation
\begin{equation}\label{eq:powerLaplace}
   \int_0^\infty\!d\omega\,e^{-s\spac\omega}\,\omega^{-1+a}
   = \Gamma(a)\,s^{-a} 
\end{equation}
for $a=-2\epsilon$, we find for the bare ultrasoft and ultrasoft-collinear functions in Laplace space
\begin{equation}
\begin{aligned}
   \widetilde{W}_{us}(s) 
   &= 1 + \frac{Q_\ell^2\spac\alpha}{2\pi} \left( \frac{1}{\epsilon^2} 
    + \frac{1+L_{us}}{\epsilon} + \frac{L_{us}^2}{2} + L_{us} + \frac{\pi^2}{12} 
    + 2 \right) + \order{\alpha^2} , \\
   \widetilde{W}_{usc}(s) 
   &= 1 + \frac{ Q_\ell^2\spac\alpha}{2\pi} \left( - \frac{1}{\epsilon^2} 
    + \frac{1-L_{usc}}{\epsilon} - \frac{L_{usc}^2}{2} + L_{usc} 
    - \frac{5\pi^2}{12} \right) + \order{\alpha^2} ,
\end{aligned}
\end{equation}
where all dependence on the Laplace variable $s$ is contained in the logarithms
\begin{equation}
   L_{us} = \ln\bigg( \frac{\mu^2\spac s^2\spac e^{2\gamma_E}}{4} \bigg) \,, \qquad  L_{usc} = \ln\bigg( \frac{\mu^2\spac s^2\spac e^{2\gamma_E}\spac m_B^2}%
                            {4\spac m_\ell^2} \bigg) \,.  
\end{equation}
The pole terms can now be subtracted locally, such that $\widetilde{W}_{us}(s,\mu)=Z_{us}(s,\mu)\,\widetilde{W}_{us}(s)$, and analogously for the ultrasoft-collinear function. The renormalized functions
\begin{equation}\label{eq:us_usc_explicit}
\begin{aligned}
   \widetilde{W}_{us}(s,\mu) 
   &= 1 + \frac{Q_\ell^2\spac\alpha}{2\pi} \left( \frac{L_{us}^2}{2} + L_{us} 
    + \frac{\pi^2}{12} + 2 \right) + \order{\alpha^2} , \\
   \widetilde{W}_{usc}(s,\mu) 
   &= 1 + \frac{ Q_\ell^2\spac\alpha}{2\pi} \left( - \frac{L_{usc}^2}{2} + L_{usc} 
    - \frac{5\pi^2}{12} \right) + \order{\alpha^2} 
\end{aligned}
\end{equation}
obey local RG equations, e.g.\
\begin{equation}\label{eq:5.104}
   \frac{d}{d\ln\mu}\,\widetilde{W}_{us}(s,\mu)
   = \gamma_{us}(s,\mu)\,\widetilde{W}_{us}(s,\mu) \,,
\end{equation}
with anomalous dimensions
\begin{equation}\label{eq:5.105}
\begin{aligned}    
   \gamma_{us}(s,\mu) 
   &= \frac{Q_\ell^2\spac\alpha}{\pi} \left( L_{us} + 1 \right) 
    + \order{\alpha^2} , \\
   \gamma_{usc}(s,\mu) 
   &= \frac{Q_\ell^2\spac\alpha}{\pi} \left( - L_{usc} + 1 \right) 
    + \order{\alpha^2} .
\end{aligned}
\end{equation}
It is important for the resummation of large logarithms that, to all orders in perturbation theory, the anomalous dimensions contain only a single power of logarithms, whose coefficient is proportional to the light-like cusp anomalous dimension of QED  \cite{Korchemsky:1987wg,Korchemskaya:1992je}. Note that 
\begin{equation}
   \gamma_{us} + \gamma_{usc} = - 2\spac\gamma_{\rm soft} \,,
\end{equation} 
where the soft anomalous dimension has been given in \eqref{eq:gammasoft}. This exact relation holds to all orders of perturbation theory. As we will see, it ensures that the direct contribution to the decay rate is RG invariant. 

In general, the renormalized functions in \eqref{eq:us_usc_explicit} depend on the scale $\mu$ via the logarithms $L_{us}$ and $L_{usc}$ and the renormalized QED coupling $\alpha(\mu)$. In the approximation where the running of $\alpha$ is neglected, we can write
\begin{equation}\label{eq:Wshortforms}
   \widetilde{W}_{us}(s,\mu) \equiv \widetilde{W}_{us}(L_{us}) \,, \qquad
   \widetilde{W}_{usc}(s,\mu) \equiv \widetilde{W}_{usc}(L_{usc}) \,.
\end{equation}
The RG equations are then solved by
\begin{equation}
\begin{aligned}
   \widetilde{W}_{us}(L_{us})
   &= \exp\left[ \frac{Q_\ell^2\spac\alpha}{2\pi} 
    \left( \frac{L_{us}^2}{2} + L_{us} \right) \right]
    \widetilde{W}_{us}(0) \,, \\
   \widetilde{W}_{usc}(L_{usc})
   &= \exp\left[ \frac{Q_\ell^2\spac\alpha}{2\pi} 
    \left( - \frac{L_{usc}^2}{2} + L_{usc} \right) \right]
    \widetilde{W}_{usc}(0) \,,
    \end{aligned}
\end{equation}
where the matching scales are chosen such that $L_{us}=0$ and $L_{usc}=0$, respectively. Combining these solutions, we obtain the RG-improved expression
\begin{equation}
   \widetilde{W}_{us}(s,\mu)\,\widetilde{W}_{usc}(s,\mu)
   = \left( \frac{\mu^2\spac s^2\spac e^{2\gamma_E}\spac m_B}{4\spac m_\ell} 
    \right)^{-\gamma_{\rm soft}}
    \left[ 1 + \frac{Q_\ell^2\spac\alpha}{2\pi} \left( 2 - \frac{\pi^2}{3} \right) 
    + \order{\alpha^2} \right] .
\end{equation}
Inserting this result in \eqref{eq:Rlaplace}, and performing the inverse Laplace transformation using \eqref{eq:powerLaplace}, we obtain for the RG-improved radiation function 
\begin{equation}\label{eq:REfinal}
   R(E_\mathrm{cut},\mu) 
   = \left( \frac{\mu^2\spac m_B}{(2 E_{\rm cut})^2\,m_\ell} 
    \right)^{-\gamma_{\rm soft}} \frac{e^{-2\gamma_E\spac\gamma_{\rm soft}}}{\Gamma(1+2\gamma_{\rm soft})}
    \left[ 1 + \frac{\alpha\spac Q_\ell^2}{2\pi} \left( 2 - \frac{\pi^2}{3} \right) 
    + \mathcal{O}(\alpha^2) \right] .
\end{equation}
In this result, the large rapidity logarithms contained in $\gamma_{\rm soft}$ are resummed to all orders. While in our derivation we have neglected the scale dependence of the QED coupling, the formalism developed in \cite{Becher:2006nr} allows one to include the effects of the running of $\alpha$ in a simple and systematic way. This is discussed in Appendix~\ref{app:running}.

From \eqref{eq:matching_yB} and \eqref{eq:IB_rate}, we now obtain for the direct contribution to the $B^-\to\ell^-\spac\bar\nu_\ell\spac(\gamma)$ decay rate 
\begin{equation}\label{eq:Gammadir}
\begin{aligned}   
   \Gamma_{\rm dir}(E_{\rm cut})
   &= \Gamma_{\rm tree}\,\spac y_B^2(\mu_0)\,R(E_\mathrm{cut},\mu_0) \\
   &= \Gamma_{\rm tree} \left(\frac{\alpha(m_Z)}{\alpha(m_B)} \right)^\frac{9}{20}\!
    \left( \frac{2 E_{\rm cut}}{m_B} \right)^{2\gamma_{\rm soft}}\! 
    \frac{e^{-2\gamma_E\spac\gamma_{\rm soft}}}{\Gamma(1+2\gamma_{\rm soft})}\,
    \mathcal{R}_{\mathrm{virt}}^2 
    \left[ 1 + \frac{Q_\ell^2\spac\alpha}{2\pi} \left( 2 - \frac{\pi^2}{3} \right) 
    + \mathcal{O}(\alpha^2) \right] .
\end{aligned}   
\end{equation}
Note that the matching scale $\mu_0$ has disappeared from the final result. Also, the lepton-mass dependence in the radiation function $R(E_\mathrm{cut},\mu_0)$ cancels against that of the prefactor in \eqref{eq:matching_yB}, ensuring that there are no double logarithms of the charged-lepton mass left in the physical decay rate.\footnote{The quantity $\mathcal{R}_{\mathrm{virt}}$ in \eqref{eq:Rvirt_final} contains single-logarithmic corrections involving $m_\ell$. However, since the virtual decay rate is proportional to $m_\ell^2$ due to its chiral suppression, the limit $m_\ell\to 0$ is smooth.}

\subsection{Indirect contributions to the decay rate}
\label{subsec:indirect}

We now turn to the discussion of the indirect contributions to the $B^-\to\ell^-\spac\bar\nu_\ell\spac(\gamma)$ decay rate. They originate from the production of an off-shell $B^{*-}$ resonance, which undergoes a weak $B^{*-}\to\ell^-\spac\bar\nu_\ell$ decay. Analogous contributions of higher resonances are power suppressed in the limit of small photon energy. We split up the indirect rate into two contributions,  
\begin{equation}\label{eq:Gamma2terms}
   \Gamma_\mathrm{indir}(E_\mathrm{cut}) 
   = \Gamma_{B^\ast\gamma}(E_\mathrm{cut})+ \Gamma_{B^\ast\pi}(E_\mathrm{cut}) \,,
\end{equation}
one in which a photon is produced in a $BB^*\gamma$ interaction, and one in which a neutral pion is produced in a $BB^*\pi^0$ interaction, which subsequently decays into electromagnetic radiation.

\begin{figure}
\centering
\includegraphics[scale=0.60]{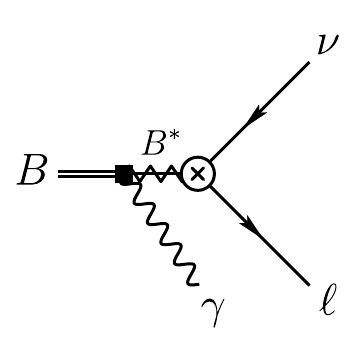} \qquad
\includegraphics[scale=0.60]{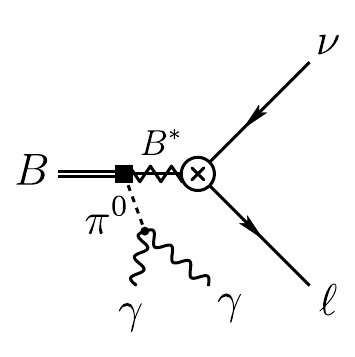}
\caption{Indirect decay topologies contributing to $B^-\to\ell^-\spac\bar\nu_\ell\spac(\gamma)$ decay rate, which arise at leading order in the low-energy theory. In the first graph, a photon is emitted through the effective $BB^\ast\gamma$ coupling. In the second graph an on-shell pion is first emitted, which subsequently decays into two photons.}
\label{fig:indirect}
\end{figure}

\subsubsection*{\boldmath Contribution involving the $BB^*\gamma$ vertex}

The low-energy effective theory allows for the process $B^-\to B^{*-}\spac\gamma$ followed by the weak vector-meson decay $B^{*-}\to\ell^-\spac\bar\nu_\ell$, which does not suffer the chiral suppression of the decay \blnu. The tree-level diagram for this process is shown in the first graph in Figure~\ref{fig:indirect}. The corresponding decay amplitude is given by
\begin{equation}
\begin{aligned}
   \mathcal{A}_{B^*}(B^-\to\ell^-\spac\bar\nu_\ell\,\gamma) 
   &= \frac{G_F^{(\mu)}\spac V_{ub}}{\sqrt2}\,f_{B^*}\spac m_{B^*}\,
    \sqrt{\frac{m_B}{m_{B^*}}}\,\frac{e\spac g_{BB^*\gamma}}{v\cdot q+\delta_{B^*}}\, \epsilon_{\mu\nu\rho\sigma}\,\varepsilon^{*\mu}(q)\spac q^\nu\spac v^\rho\,
    \bar u(v_\ell)\spac\gamma_\perp^\sigma P_L\spac v(p_\nu) \,,
\end{aligned}
\end{equation}
where $q$ denotes the ultrasoft photon 4-momentum.\footnote{One may wonder whether there is another contribution to the amplitude in which photon has ultrasoft-collinear momentum scaling. In this case the phase-space measure $d^3\bm{q}$ in \eqref{eq:indirect1} is suppressed by a factor of $\order{\lambda_\ell^2}$ relative to the measure for an ultrasoft photon, while the integrand obeys the same scaling on both cases. Hence the ultrasoft-collinear contribution is suppressed by a factor $\sim(m_\ell/m_B)^2$ and can be neglected.} 
The effective heavy-lepton spinor $u(v_\ell)$ is normalized to $\sqrt{2\spac m_\ell}$, such that 
\begin{equation}
   \sum_{\rm pol}\,u(v_\ell)\,\bar u(v_\ell)
   = 2\spac m_\ell\,\frac{1+\vsl_\ell}{2} = \psl_\ell + m_\ell \,.
\end{equation}
For the squared decay amplitude, summed over polarizations and normalized to the squared amplitude for the tree-level \blnu\ process, we obtain (up to power-suppressed terms $\sim m_\ell^2/m_B^2$) 
\begin{equation}\label{eq:5.99}
\begin{aligned}
   \frac{\sum_{\rm pol}\,\left|
         \mathcal{A}_{B^*}(B^-\to\ell^-\spac\bar\nu_\ell\,\gamma)\right|^2}%
        {\sum_{\rm pol}\,\left|
         \mathcal{A}_{\rm tree}(B^-\to\ell^-\spac\bar\nu_\ell)\right|^2}
   &= \frac{m_B^2}{2\spac m_\ell^2}\,\frac{f_{B^*}^2\spac m_{B^*}}{f_B^2\,m_B}
    \left( \frac{e\spac g_{BB^*\gamma}}{E_{\bm{q}}+\delta_{B^*}} \right)^2
    \bm{q}^2\,\frac{1+\cos^2\theta}{2} \,,
\end{aligned}
\end{equation}
where we have aligned the $z$-axis with the direction of the charged lepton, and $\theta=\sphericalangle(\bm{q},\bm{v}_\ell)$ denotes the angle between the directions of the ultrasoft photon and the charged lepton in the $B$-meson rest frame. In the ultrasoft-photon limit $q\to 0$, the three-particle phase space integral factorizes into a two-particle phase-space times a phase-space integral of the emitted photon,
\begin{equation}
\begin{aligned}
   & \int\!\frac{d^3\bm{p}_\ell}{(2\pi)^3\spac 2E_{\bm{p}_\ell}} 
    \int\!\frac{d^3\bm{p}_\nu}{(2\pi)^3\spac 2E_{\bm{p}_\nu}} 
    \int\!\frac{d^3\bm{q}}{(2\pi)^3\spac 2E_{\bm{q}}}\,
    (2\pi)^4\,\delta^{(4)}(m_B\,v-p_\ell-p_\nu-q) \\
   &\stackrel{q\to 0}{=}\,
    \int\!\frac{d^3\bm{p}_\ell}{(2\pi)^3\spac 2E_{\bm{p}_\ell}} 
    \int\!\frac{d^3\bm{p}_\nu}{(2\pi)^3\spac 2E_{\bm{p}_\nu}} 
    (2\pi)^4\,\delta^{(4)}(m_B\,v-p_\ell-p_\nu)
   \int\!\frac{d^3\bm{q}}{(2\pi)^3\spac 2E_{\bm{q}}} \,.
\end{aligned}
\end{equation}
It follows that (with $E_{\bm{q}}=|\bm{q}|$ for a real photon)
\begin{equation}\label{eq:indirect1}
\begin{aligned}
   \frac{\Gamma_{B^*\gamma}(E_{\rm cut})}{\Gamma_{\rm tree}}
   &= \frac{m_B^2}{2\spac m_\ell^2}\,\frac{f_{B^*}^2\spac m_{B^*}}{f_B^2\,m_B} 
    \left( e\spac g_{BB^*\gamma} \right)^2\!
    \int\limits_{E_{\bm{q}}\le E_{\rm cut}}\!\!
    \frac{d^3\bm{q}}{(2\pi)^3\spac 2E_{\bm{q}}}\,
    \frac{\bm{q}^2}{\left( E_{\bm{q}}+\delta_{B^*} \right)^2}\,
    \frac{1+\cos^2\theta}{2} \\
   &= \frac{m_B^2}{m_\ell^2}\,\frac{f_{B^*}^2\spac m_{B^*}}{f_B^2\,m_B}\,
    \frac{\alpha}{6\pi} \left( g_{BB^*\gamma}\spac E_{\rm cut} \right)^2 
    I\bigg(0,\frac{\delta_{B^*}}{E_{\rm cut}}\bigg) \,.
\end{aligned}
\end{equation}
The phase-space integral $I(0,z)$ is a special case of the function 
\begin{equation}
   I(y,z) = 2\int_y^1\!dx\,\frac{\left(x^2-y^2\right)^{\frac32}}{(x+z)^2} \,,
\end{equation}
which is normalized such that $I(0,0)=1$. For $y=0$, we find
\begin{equation}
   I(0,z) = 3 - 6z - \frac{2}{1+z} + 6 z^2\ln\frac{1+z}{z} 
   = 9 - 2 r - \frac{6}{r} - \frac{6\spac(1-r)^2}{r^2}\spac\ln(1-r) \,,
\end{equation}
where in the second form we have substituted $r=\frac{1}{1+z}=E_\mathrm{cut}/(E_\mathrm{cut}+\delta_{B^*})$. This function approaches~1 for $E_{\rm cut}\gg\delta_{B^*}\simeq 45$\,MeV, while for $E_{\rm cut}\lesssim\delta_{B^*}$ it is given by a power series in $E_{\rm cut}/\delta_{B^*}$ starting with $\frac12(E_{\rm cut}/\delta_{B^*})^2-\frac45(E_{\rm cut}/\delta_{B^*})^3\pm \dots$. The factor $m_{B^*}^2/m_\ell^2$ in \eqref{eq:indirect1} removes the chiral suppression of the tree-level rate and hence gives a strong enhancement of the indirect rate. However, the factor $(g_{BB^*\gamma}\spac E_{\rm cut})^2\sim (E_\mathrm{cut}/\OurLambda)^2$ leads to a power suppression. In the limit $E_{\rm cut}\to 0$, the indirect rate scales like $E_{\mathrm{cut}}^4$. 

The amplitudes for the direct and indirect contributions to the $B^-\to\ell^-\spac\bar\nu_\ell\spac\gamma$ process can interfere. However, this requires right-handed lepton polarization in the indirect amplitude, which leads to an additional power suppression $\sim m_\ell/m_B$. As a result, the interference contribution to the decay rate carries a chiral suppression $\sim(m_\ell/m_B)^2$ in addition to the suppression $\sim(E_{\rm cut}/\OurLambda)^2$ from the dipole operator contributing to the indirect rate. This makes the interference term negligible. 

In the literature of leptonic weak decays of pseudoscalar mesons, the indirect contributions are often expressed in terms of the meson-to-photon matrix element of the time-ordered product of the weak current and the electromagnetic current, which for an on-shell photon can be parametrized by a vector form factor $F_V(x_\gamma)$ and an axial-vector form factor $F_A(x_\gamma)$, with $x_\gamma=2 E_\gamma/m_B$, which we define as in \cite{Becirevic:2009aq}.\footnote{
Lattice QCD determinations of the form factors $F_{V,A}(x_\gamma)$ are available for the light pseudoscalars $\pi^\pm$ and $K^\pm$ (see e.g. \cite{Desiderio:2020oej} and \cite{DiPalma:2025iud}), and for the heavier mesons $D_s$~\cite{Frezzotti:2023ygt} and $B_s$~\cite{Frezzotti:2024kqk}.} After integration over the charged-lepton energy $x_\ell=2 E_\ell/m_B$ over the interval (with $\lambda_\ell=m_\ell/m_B$)
\begin{equation}
   1 - x_\gamma + \frac{\lambda_\ell^2}{1-x_\gamma} \le x_\ell \le 1 + \lambda_\ell^2 \,,
\end{equation}
one finds that the differential decay rate $d\spac\Gamma_{B^*}(B^-\to\ell^-\spac\bar\nu_\ell\spac\gamma)/dx_\gamma$ is proportional to the sum $\left|F_V(x_\gamma)\right|^2+\left|F_A(x_\gamma)\right|^2$. Our results imply that in the soft-photon limit $x_\gamma\ll 1$ 
\begin{equation}\label{eq:FVexp}
   F_V(x_\gamma)
   = \frac{f_{B^*}\spac m_{B^*}}{2}\,
    \frac{g_{B B^*\gamma}}{E_\gamma+\delta_{B^*}} \,, \qquad 
   F_A(x_\gamma) = 0 \,,  
\end{equation}
in agreement with \cite{Becirevic:2009aq}. Our HH$\chi$PT construction thus corresponds to the first-vector pole approximation for the form factors. Corrections to these relations have energy denominators in which $\delta_{B^*}$ is replaced by $\delta_{X_b}\sim\OurLambda$ and thus give contributions suppressed by $E_\gamma/\OurLambda$ for $E_\gamma>\delta_{B^*}$, and $\delta_{B^*}/\OurLambda\sim\OurLambda/m_B$ for $E_\gamma<\delta_{B^*}$. The first axial-vector resonance $B_1(5721)$, whose contribution was estimated in \cite{Becirevic:2009aq}, therefore gives a power-suppressed contribution to the $B^-\to\ell^-\spac\bar\nu_\ell\spac\gamma$ decay rate suppressed by $(E_\gamma/\OurLambda)^2$ or $(\OurLambda/m_B)^2$, respectively, in the soft-pion limit. The first vector resonance with $J^P=1^-$ and mass above the $B^*$ mass would give contributions suppressed by $E_\gamma/\OurLambda$ or $\OurLambda/m_B$, since it interferes with the $B^*$ contribution. However, for orbitally excited states one expects reduced couplings to the weak current, because in quark models their decay constants probe the wave function at the origin.

\begin{figure}[t]
\centering
\includegraphics[scale=1]{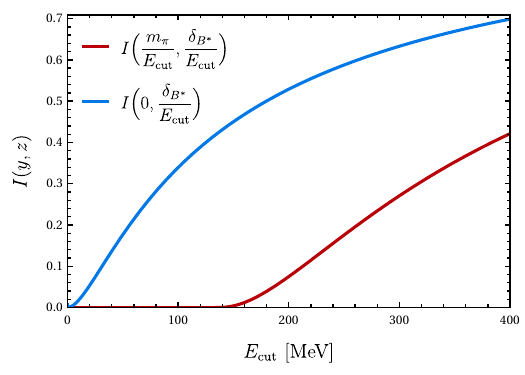} 
\caption{Dependence of the functions $I(0,\delta_{B^*}/E_{\rm cut})$ (blue) and $I(m_\pi/E_{\rm cut},\delta_{B^*}/E_{\rm cut})$ (red) on the photon energy cut. Both functions approach~1 from below in the limit $E_{\rm cut}\to\infty$.}
\label{fig:Ifunctions}
\end{figure}

\subsubsection*{\boldmath Contribution involving the $BB^*\pi^0$ vertex}

For values of $E_{\rm cut}$ below the pion mass, the $B^-\to B^{*-}\spac\gamma\to\ell^-\spac\bar\nu_\ell\,\gamma$ process is the dominant contribution to the indirect rate. However, experimental considerations might require raising the cutoff to values $E_{\rm cut}\gtrsim m_\pi$. In this case, a second indirect contribution becomes kinematically allowed, and it turns out to be the dominant contribution by far. The low-energy effective theory allows for the transition $B^-\to B^{*-}\spac\pi^0\to\ell^-\spac\bar\nu_\ell\,\pi^0$, shown in the right diagram in Figure~\ref{fig:indirect}. The ultrasoft $\pi^0$ meson decays promptly, within few $10^{-8}$\,m from the production point, with branching fractions $\text{Br}(\pi^0\to\gamma\spac\gamma)\simeq 98.82\%$ and $\text{Br}(\pi^0\to e^+\spac e^-\spac\gamma)\simeq 1.174\%$. Both signals are seen as ``soft photons'' in the electromagnetic calorimeter. We can thus reinterpret the associated final state as $B^-\to\ell^-\spac\bar\nu_\ell\,\gamma\spac\gamma$. Given that the width of the neutral pion is around 7.8\,eV \cite{ParticleDataGroup:2024cfk}, this decay can only occur for an on-shell pion, which implies that the energy of the emitted electromagnetic radiation is bounded by $E_{\rm rad}\ge m_{\pi^0}\simeq 135$\,MeV. The decay amplitude for the $B^-\to\ell^-\spac\bar\nu_\ell\,\pi^0$ process via an intermediate $B^{*-}$ resonance is obtained as
\begin{equation}
\begin{aligned}
   \mathcal{A}_{B^*}(B^-\to\ell^-\spac\bar\nu_\ell\,\pi^0) 
   &= - \frac{G_F^{(\mu)}\spac V_{ub}}{f_\pi}\,f_{B^*}\spac m_{B^*}\,
    \sqrt{\frac{m_B}{m_{B^*}}}\,\frac{g_{BB^*\pi}}{v\cdot q+\delta_{B^*}}\, 
    \bar u(v_\ell)\,\qsl_\perp P_L\spac v(p_\nu) \,,
\end{aligned}
\end{equation}
where now $q$ denotes the 4-momentum of the pion. For the squared amplitude summed over polarizations, we find in this case (again up to power-suppressed terms $\sim(m_\ell/m_B)^2$)
\begin{equation}
\begin{aligned}
   \frac{\sum_{\rm pol}\,\left|
         \mathcal{A}_{B^*}(B^-\to\ell^-\spac\bar\nu_\ell\,\pi^0)\right|^2}%
        {\sum_{\rm pol}\,\left|
         \mathcal{A}_{\rm tree}(B^-\to\ell^-\spac\bar\nu_\ell)\right|^2}
   &= \frac{m_B^2}{2\spac m_\ell^2}\,\frac{f_{B^*}^2\spac m_{B^*}}{f_B^2\,m_B}
    \left( \frac{g_{BB^*\pi}/f_\pi}{E_{\bm{q}}+\delta_{B^*}} \right)^2
    \bm{q}^2 \left( 1 - \cos^2\theta \right) .
\end{aligned}
\end{equation}
Compared with \eqref{eq:5.99}, we find $e\spac g_{BB^*\gamma}\approx 0.44\,\text{GeV}^{-1}$ and $g_{BB^*\pi}/f_\pi\approx 4.3\,\text{GeV}^{-1}$ (see Table~\ref{tab:inputs} in Section~\ref{sec:pheno} for a discussion of our input parameter choices), so at the level of the squared decay amplitude the pion mode is enhanced by two orders of magnitude in the region $E_q>m_\pi$, where it is kinematically allowed. In the ultrasoft-pion limit, the three-particle decay rate can be approximated as (with $E_{\bm{q}}=\sqrt{m_\pi^2+\bm{q}^2}$)
\begin{equation}\label{eq:indirect2}
\begin{aligned}
   \frac{\Gamma_{B^*}(B^-\to\ell^-\spac\bar\nu_\ell\,\pi^0)}{\Gamma_{\rm tree}}
   &= \frac{m_B^2}{2\spac m_\ell^2}\,\frac{f_{B^*}^2\spac m_{B^*}}{f_B^2\,m_B} 
    \left( \frac{g_{BB^*\pi}}{f_\pi} \right)^2\!
    \int\limits_{m_\pi\le E_{\bm{q}}\le E_{\rm cut}}\!\!
    \frac{d^3\bm{q}}{(2\pi)^3\spac 2 E_{\bm{q}}}\,
    \frac{\bm{q}^2}{\left(E_{\bm{q}}+\delta_{B^*}\right)^2} 
    \left( 1-\cos^2\theta \right) \\
   &= \frac{m_B^2}{m_\ell^2}\,\frac{f_{B^*}^2\spac m_{B^*}}{f_B^2\,m_B}\,
    \frac{g_{BB^*\pi}^2}{24\pi^2} \left( \frac{E_{\rm cut}}{f_\pi} \right)^2
    I\bigg(\frac{m_\pi}{E_{\rm cut}},\frac{\delta_{B^*}}{E_{\rm cut}}\bigg) \,,
\end{aligned}
\end{equation}
where for $z<y<1$
\begin{equation}
\begin{aligned}
   I(y,z) 
   &= \frac{\sqrt{1-y^2}}{1+z} \left[ 1+2 y^2-3 z\spac(1 + 2 z) \right]
    - 3 \left( y^2-2 z^2 \right)\,\text{arctanh}\big(\sqrt{1-y^2}\big) \\
   &\qquad + 6 z\sqrt{y^2-z^2} \,
    \arctan\bigg(\frac{\sqrt{1-y^2}\sqrt{y^2-z^2}}{y^2+z}\bigg) \,.
\end{aligned}
\end{equation}
Taking into account that the neutral pion decays electromagnetically and that energy is conserved in its decay, we obtain 
\begin{equation}
\begin{aligned}
   \Gamma_{B^*\pi}(E_{\rm cut})
   &= \Gamma_{B^*\pi}(B^-\to\ell^-\spac\bar\nu_\ell\,\gamma\gamma)\spac 
    \big|_{E_{\gamma\gamma}\le E_{\rm cut}}
    + \Gamma_{B^*\pi}(B^-\to\ell^-\spac\bar\nu_\ell\,\gamma\spac e^+ e^-)\spac 
    \big|_{E_{\gamma ee}\le E_{\rm cut}} \\
   &= \left[ \text{Br}(\pi^0\to\gamma\gamma) 
    + \text{Br}(\pi^0\to\gamma\spac e^+ e^-) \right]
    \Gamma(B^-\to\ell^-\spac\bar\nu_\ell\,\pi^0)\spac\big|_{E_\pi\le E_{\rm cut}}
\end{aligned}
\end{equation}
to excellent approximation, because the two branching fractions add up to 99.997\%. In Figure~\ref{fig:Ifunctions} we show the two phase-space integrals in \eqref{eq:indirect1} and \eqref{eq:indirect2} as functions of $E_{\rm cut}$. The $\pi^0$-induced decay mode opens up for $E_{\rm cut}>m_{\pi^0}\simeq 135$\,MeV, and due to the strong enhancement of the squared amplitude for this process, it quickly becomes the dominant indirect decay mode. For $E_{\rm cut}=150$\,MeV (200\,MeV), we find $\Gamma_{B^*\pi}/\Gamma_{B^*\gamma}\approx 0.8~(13)$. 

Away from the soft-pion limit, the $B^-\to\ell^-\spac\bar\nu_\ell\,\pi^0$ decay amplitude can be expressed in a model-independent way in terms of the scalar form factor $f_+^{B\to\pi}(q^2)$ (see e.g.\ \cite{Wirbel:1985ji}), where $q^2=(p_B-p_\pi)^2=m_B^2+m_\pi^2-2m_B\spac v\cdot p_\pi$, where $v$ denotes the 4-velocity of the $B$ meson, and $v\cdot p_\pi=E_\pi$ in the $B$-meson rest frame. In the limit of vanishing lepton mass, one obtains the differential decay rate 
\begin{equation}\label{eq:Blnupi}
   \frac{d\spac\Gamma(B^-\to\ell^-\spac\bar\nu_\ell\,\pi^0)}{d(v\cdot p_\pi)}
   = \frac{G_F^2\spac m_B}{24\pi^3}\,|V_{ub}|^2
    \left[ (v\cdot p_\pi)^2 - m_\pi^2 \right]^\frac32
    \left| f_+(q^2) \right|^2 .
\end{equation}
A systematic analysis of the $B\to\pi$ form factors in the heavy quark expansion was performed in \cite{Burdman:1993es}, where it was shown that in the soft-pion limit
\begin{equation}
   \lim_{v\cdot p_\pi\to m_\pi} f_+(q^2) 
   = \frac{f_{B^*}\spac m_B }{2 f_\pi}\,
    \frac{g_{BB^*\pi}}{v\cdot p_\pi+\delta_{B^*}} \,,
\end{equation}
in agreement with \cite{Wise:1992hn,Yan:1992gz,Burdman:1992gh}. Inserting this expression in \eqref{eq:Blnupi} and integrating over the pion energy, we recover our previous result \eqref{eq:indirect2} times a factor $m_B/m_{B^*}=1+\order{1/m_b^2}$. 

In the present work, we assume for simplicity that experimentally one measures the total energy of electromagnetic radiation, and that this energy is below the cutoff $E_{\rm cut}$. The phenomenological results presented in Section~\ref{sec:pheno} are obtained based on this assumption. Using a finely grained electromagnetic calorimeter, it is possible to measure further properties of the deposited electromagnetic showers. For example, in events with two (or more) final-state photons it may be possible to reconstruct the energies and directions of the photons individually. For a two-photon signature, one could then check whether the invariant mass of the photon pair is compatible with the pion mass, them being produced from the decay of a neutral pion (see e.g.\ \cite{Belle:2018jqd}). Such an analysis, which typically uses modern tools of machine learning and deep neural networks, may allow one to veto against the indirect contribution $\Gamma_{B^*\pi}(E_{\rm cut})$ which, as we will see below, turns out to be a significant background.

\subsection{Electromagnetic radiation spectrum}

Given the results derived in the previous two sections, it is possible to obtain the resummed differential decay rate for the $B^-\to\ell^-\spac\bar\nu_\ell\spac(\gamma)$ process with respect to the total energy of low-energetic electromagnetic radiation. In the phase-space region where $E_{\rm rad}\ll\OurLambda$, we obtain 
\begin{equation}
    \frac{d\spac\Gamma}{dE_{\rm rad}}
    = \frac{d\spac\Gamma_{\rm dir}({E_\mathrm{cut}})}{dE_\mathrm{cut}}
     +  \frac{d\spac\Gamma_{\rm indir}({E_\mathrm{cut}})}{dE_\mathrm{cut}}
    \bigg|_{E_\mathrm{cut}=E_{\rm rad}} \,.
\end{equation}  
From the resummed expression \eqref{eq:Gammadir} for the direct contribution, we deduce that  
\begin{equation}
   \frac{d\spac\Gamma_{\rm dir}}{dE_\mathrm{rad}} 
   =\frac{2\spac\gamma_{\rm soft}}{E_{\rm rad}}\,\Gamma_{\rm dir}(E_{\rm rad})
    \propto E_{\rm rad}^{-1+2\gamma_{\rm soft}} \,.
\end{equation} 
This contribution diverges for $E_{\rm rad}\to 0$ but is integrable after RG resummation. The indirect contribution to the radiation spectrum vanishes for $E_{\rm rad}\to 0$. It scales like $E_{\rm rad}^3$ near the origin and is monotonically rising, exhibiting a linear behavior for values $E_{\rm rad}\gg\delta_{B^*}$, with an abrupt change of the slope above the pion threshold.

\newpage
\section{Phenomenology}
\label{sec:pheno}

We are finally in a position to present our numerical predictions for the $B^-\to\mu^-\spac\bar\nu_\mu\spac(\gamma)$ decay rate. The input values for the parameters entering our calculation are collected in Table~\ref{tab:inputs}. The RG evolution of the strong coupling $\alpha_s(\mu)$ is provided by the \texttt{RunDec} package \cite{Chetyrkin:2000yt}. For the electromagnetic coupling, we perform the running between the weak scale and $m_B$ at one-loop order. 

\subsection{Non-perturbative input parameters}\label{subsec:Non_pert_input}

A number of non-perturbative hadronic parameters are needed for our analysis. We determine them as follows.

\subsubsection*{\boldmath Input for the $B B^\ast$ couplings}

For the coupling $g_{B B^\ast\gamma}$, neither experimental nor first-principle determinations are currently available. An experimental determination would require measuring the radiative decay $B^\ast\to B\spac\gamma$ or the total $B^\ast$ width, both of which are unlikely to be accessible in the foreseeable future. A first-principle result might eventually come from lattice QCD. Existing estimates based on quark models, HH$\chi$PT, and light-cone sum rules (LCSR) span a wide range (see e.g.\ Table~7 of \cite{Pullin:2021ebn}). The central values of the two most recent calculations based on LCSRs are in good agreement with each other \cite{Li:2020rcg, Pullin:2021ebn}. To be conservative, we adopt the larger of the two quoted uncertainties.\footnote{In principle, a theoretically clean determination of $g_{BB^*\gamma}$ could come from a measurement of the $B^-\to e^-\spac\bar\nu_e$ branching ratio with a cut $E_{\rm cut}\ll\OurLambda$, because this channel is completely dominated by the indirect contribution $B\to B^\ast\spac\gamma\to e^-\spac\bar\nu_e\spac\gamma$. The associated decay rate is thus proportional to $|g_{BB^\ast\gamma}|^2$.}

Results for the $g_{B B^\ast \pi}$ coupling exist both from lattice QCD and LCSRs. Here we take as input the most recent lattice determination of the heavy-quark coupling $g_b$ \cite{Flynn:2015xna}, which coincides with the coupling $g$ in the HH$\chi$PT Lagrangian up to $\mathcal{O}(1/m_b)$ corrections and thus with the physical coupling $g_{B B^\ast\pi}$ up to chiral corrections by virtue of \eqref{eq:def_hhchipt_couplings}. With this choice, the value of $g_{BB^\ast\pi}$ is approximately 30\% larger than the LCSR determination \cite{Belyaev:1994zk,Khodjamirian:2020mlb}. 

\begin{table}[t]
\centering
\renewcommand{\arraystretch}{1.15}
\setlength{\tabcolsep}{6pt}
\begin{tabular}{l|l|c}
\hline\rowcolor{\shadecolor{20}}
Parameter & Value & Reference \\
\hline
$G_F^{(\mu)}$ & $1.1663788(6)\cdot 10^{-5}\ \mathrm{GeV}^{-2}$ & \cite{ParticleDataGroup:2024cfk} \\
$\alpha(m_Z)^{-1}$ &$127.930(8)$  &\cite{ParticleDataGroup:2024cfk} \\ 
$\alpha_s(m_Z)$ &$0.1180(9)$  &\cite{ParticleDataGroup:2024cfk} \\ 
$f_\pi$ & $130.2 \,\mathrm{MeV}$  & \cite{FlavourLatticeAveragingGroupFLAG:2024oxs} \\
$f_B$ & $190.0(1.3) \,\mathrm{MeV}$ & \cite{FlavourLatticeAveragingGroupFLAG:2024oxs} \\
$f_{B^\ast}/f_B$ & $0.951 (17)$ & \cite{Colquhoun:2015oha,Lubicz:2017asp} \\
$\lambda_E^2$ & $0.03(2) \,\mathrm{GeV}^2$ & \cite{Nishikawa:2011qk}\\
$\lambda_H^2$ &  $0.06(3) \,\mathrm{GeV}^2$ & \cite{Nishikawa:2011qk}\\
$\lambda_B$ & $0.383(153) \, \mathrm{GeV}$ &\cite{Khodjamirian:2020hob} \\
$\abs{g_{BB^\ast\gamma}}$ &$1.45(27)\ \mathrm{GeV^{-1}}$ &\cite{Pullin:2021ebn} \\
$\abs{g_{BB^\ast \pi}}$ & $0.56(8)$ & \cite{Flynn:2015xna} \\
\hline
\end{tabular}
\caption{\label{tab:inputs}
Numerical input values for the relevant parameters. For the ratio $f_{B^\ast}/f_B$ we use the average of the available $N_f = 2+1+1$ results, HPQCD \cite{Colquhoun:2015oha} and ETMC \cite{Lubicz:2017asp}.}
\end{table}

\subsubsection*{Modeling of the LCDAs}

Another important set of non-perturbative ingredients are the light-cone distribution amplitudes (LCDAs) of the $B$ meson, which encode information about its inner structure as probed in decays with highly energetic light final-state particles. For our analysis, the relevant LCDAs are the subleading-twist distribution amplitudes $\phi_-^B$ and $\phi_{3g}^B$, which can be related to the leading-twist LCDA $\phi_+^B$ using equations of motion \cite{Kawamura:2001jm,Braun:2017liq}. In this way, one can derive the representations \cite{Braun:2015pha}
\begin{equation}\label{eq:LCDAs_representations}
\begin{aligned}
   \phi_-^B(\omega,\mu) 
   &= \int_\omega^\infty\!\frac{d\omega'}{\omega'}\,\phi_+^B(\omega',\mu) 
    + \int_0^\infty\!ds\,J_0(2\sqrt{\omega s})\,\eta_3^{(0)}(s,\mu) \,, \\
   \phi_{3g}^B(\omega,\omega_g,\mu) 
   &= \int_0^\infty\!ds \left[ \eta_3^{(0)}(s,\mu)\,
    Y_3^{(0)}(s\spac|\spac\omega,\omega_g) 
    + \frac12 \int_{-\infty}^\infty\!dx\,\eta_3(s,x,\mu)\,
    Y_3(s,x\spac|\spac\omega,\omega_g) \right] .
\end{aligned}
\end{equation}
In these expressions, the non-perturbative contributions are encoded in the functions $\eta_3^{(0)}$ and $\eta_3$. The first one captures the subleading-twist contribution to the two-particle LCDA $\phi_-^B$, while the second describes the genuine three-particle contribution to the three-particle LCDA $\phi_{3g}^B$. These functions multiply the Bessel function $J_0$ and the eigenfunctions $Y_3^{(0)}$, $Y_3$ of the associated evolution equation (see \cite{Braun:2017liq} for details). 

In our analysis, we assume the exponential model functions \cite{Grozin:1996pq,Braun:2017liq} 
\begin{equation}\label{eq:LCDAs_modeling}
\begin{aligned}
   \phi_+^B(\omega,\mu_0) 
   &= \frac{\omega}{\omega_0^2}\,e^{-\frac{\omega}{\omega_0}} \,, \\
   \eta_3^{(0)}(s,\mu_0) 
   &= - \frac{\lambda_E^2-\lambda_H^2}{18}\,s^2\,e^{-\omega_0\spac s} \,, \\
   \phi_{3g}^B(\omega,\omega_g,\mu_0) 
   &= \frac{\lambda_E^2-\lambda_H^2}{6\spac\omega_0^5}\,\omega\spac\omega_g\,
    e^{-\frac{\omega+\omega_g}{\omega_0}} \,,   
\end{aligned}
\end{equation}
from which we obtain
\begin{equation}
   \phi_-^B(\omega,\mu_0)
   = \frac{1}{\omega_0}\,e^{- \frac{\omega}{\omega_0}}
    \left[ 1 - \frac{\lambda_E^2-\lambda_H^2}{18\spac\omega_0^2}
    \left( 2 - \frac{4\spac\omega}{\omega_0} + \frac{\omega^2}{\omega_0^2} \right) 
    \right] .
\end{equation}
Using these model functions, the relevant convolution integrals over the LCDAs in \eqref{eq:omegaminusdef} and \eqref{eq:Rvirt_final} become
\begin{equation} 
   \ln\frac{\omega_-(\mu_0)}{\omega_0}
   = \int_0^\infty\!d\omega\,\phi_-^B(\omega,\mu_0)\spac
    \ln\frac{\omega}{\omega_0}  
   = \frac{\lambda_E^2-\lambda_H^2}{18\spac\omega_0^2} - \gamma_E \,, 
\end{equation}
and
\begin{equation} 
   \int_0^\infty\!d\omega \int_0^\infty\!d\omega_g\,
    \phi_{3g}^B(\omega,\omega_g,\mu_0)
    \left( \frac{1}{\omega_g}\,\ln\frac{\omega+\omega_g}{\omega} 
    - \frac{1}{\omega+\omega_g} \right) 
   = \frac{\lambda_E^2-\lambda_H^2}{36\spac\omega_0^2} \,.
\end{equation}
In practice, we fix $\omega_0$ using a LCSR determination of the first inverse moment ($\lambda_B$) of the LCDA $\phi_+^B$, which in our model is obtained as $\lambda_B=\omega_0$. For the remaining model parameters we use the values given in Table~\ref{tab:inputs}. Note that our model functions for the LCDAs neglect the perturbative radiative tails at large values of $\omega$ and $\omega_g$ \cite{Lange:2003ff,Lee:2005gza}. This is legitimate, since these tails would contribute to \blnu\ decay starting at $\order{\alpha\spac\alpha_s}$, which is beyond our accuracy goal in this work.  

\subsubsection*{\boldmath QED corrections to the $B$-meson decay constant}

A central ingredient to the structure-dependent QED corrections is the QED-corrected HQET decay constant $F_-(\Lambda,\mu)$ defined in \eqref{eq:Fdeffinal}. This object is genuinely non-perturbative, since it results from the interplay between soft photons and the non-perturbative dynamics inside the $B$ meson. Given that a reliable determination of $F_-$ is unavailable at present, we parameterize it as a relative $\order{\alpha}$ correction to the HQET decay constant,
\begin{equation}
   F_-(\Lambda, \mu)
   = F_\mathrm{QCD}(\mu) \left[ 1 + \frac{\alpha}{\pi}\,
    f^{(1)}(\Lambda,\mu) + \order{\alpha^2} \right] ,
\end{equation}
and treat $f^{(1)}$ evaluated at the low scales $\Lambda=\mu=\mu_0=1.5$\,GeV as an unknown $\order{1}$ parameter by including it in the theoretical uncertainty,
\begin{equation}
   f^{(1)}(\mu_0,\mu_0)=0\pm 1 \,.
\end{equation}
This scale choice avoids large logarithms inside $f^{(1)}$, so that this range seems reasonable. The evolution of $F_-$ in $\Lambda$ between $\mu_0$ and $m_B$ and is then accomplished using \eqref{eq:Lambdaevolution}. It resums large logarithms of the ratio $m_B/\mu_0$. With our model LCDAs, we obtain the shift
\begin{equation}
   \frac{F(m_B,\mu_0) - F(\mu_0,\mu_0)}{F_\mathrm{QCD}(\mu_0)} 
   = (2.32\pm 0.51_{\phi_B})\cdot 10^{-3} \,,
\end{equation}
where the subscript ``$\phi_B$'' indicates the uncertainty from the variation of the model parameters of the LCDAs.

\subsection{Numerical estimates}

We are now ready to present numerical estimates of our findings for the direct and indirect contributions to the QED-corrected $B^-\to\mu^-\spac\bar\nu_\mu\spac(\gamma)$ decay rate in \eqref{eq:rad_rate}. As mentioned earlier, the symbol ``$(\gamma)$'' stands for low-energetic electromagnetic radiation in the form of one or multiple photons. We begin with the direct contribution, whose explicit form is given in \eqref{eq:Gammadir}. In order to convey a sense of the numerical impact of the various ingredients in this formula, we consider them one by one.  For the factor encoding short-distance electroweak corrections from the evolution between $m_Z$ and $m_B$, we obtain
\begin{align}
    \left( \frac{\alpha(m_Z)}{\alpha(m_B)}\right)^{\frac{9}{20}} = 1.01406 \,,
\end{align}
where the leading uncertainty comes from the determination of $\alpha^{-1}(m_Z)$ and can be safely neglected. Next, $\mathcal{R}_\mathrm{virt}$ encodes QED and part of the mixed QED\,--\,QCD corrections from the evolution between $m_B$ and $\OurLambda$. As shown in Section~\ref{sec:vir_ampl}, this quantity is scale-independent up to $\order{\alpha\spac\alpha_s\spac \ln\mu_0^2/m_B^2}$ terms, which are only partially included in our calculation. We estimate them by varying the scale $\mu_0$ by a factor of $1/\sqrt{2}$ to $\sqrt{2}$ about its default value. Using the inputs in Table~\ref{tab:inputs}, we find 
\begin{equation}
   \mathcal{R}_\mathrm{virt}^2 
   = 1 + (15.8\pm 4.8_{f^{(1)}} \pm 1.8_{\phi_B}\pm 1.1_\mathrm{\mu_0})
    \cdot 10^{-3} \,,
\end{equation} 
with the dominant error stemming from our estimate of the unknown QED effects in $F_{-}$, followed by the scale uncertainty and that on the LCDA parameters, which is largely driven by the uncertainty on $\lambda_B$. The scale uncertainty coming from the RG evolution of the LCDAs is neglected. It is natural to expect that  $\mathcal{R}^2_\mathrm{virt}-1$ should be at the few percent level, since it contains the large logarithms $\frac{\alpha}{\pi} \ln(m_B/m_\mu)\approx 1.9\%$ and $\frac{\alpha}{\pi} \ln^2(m_B/\lambdaqcd)\sim(1.5\!-\!5.4)\%$ for $\Lambda_{\rm QCD}\sim(0.5\!-\!1.5)$\,GeV. This is in accordance with our numerical findings.

Moving to lower scales, the object
\begin{equation}
   \Omega(E_\mathrm{cut})
   \equiv \left( \frac{2 E_\mathrm{cut}}{m_B} \right)^{2\gamma_\mathrm{soft}} 
\end{equation} 
encodes the dependence of the direct contribution to the decay rate on the experimental cut. For $E_\mathrm{cut}=25, 100$ and $200$ MeV, we obtain 
\begin{equation}
\begin{aligned}
   \Omega(25\,\mathrm{MeV}) &= 1 - 62.6\cdot 10^{-3} \,, \\
   \Omega(100\,\mathrm{MeV}) &= 1 - 44.4\cdot 10^{-3} \,, \\
   \Omega(200\,\mathrm{MeV}) &= 1 - 35.2\cdot 10^{-3} \,.
\end{aligned}
\end{equation}
For the factor
\begin{equation}
   \mathcal{W}
   = \frac{e^{-2\gamma_E\spac\gamma_{\rm soft}}}{\Gamma(1+2\gamma_{\rm soft})} 
    \left[ 1 + \frac{Q_\ell^2\spac\alpha}{2\pi} 
    \left( 2 - \frac{\pi^2}{3} \right) \right] ,
\end{equation}
which resums single logarithms of the boost factor $m_B/m_\ell$, we find 
\begin{equation}
   \mathcal{W} = 1 - 1.69\cdot 10^{-3} \,.  
\end{equation}
The resummation of large logarithms affects $\mathcal{R}_\mathrm{virt}$, $\Omega$, and $\mathcal{W}$ numerically in the range from $(1\spac\text{--}\spac 10)\%$, with the largest impact in $\mathcal{W}$.

\begin{figure}[t]
\centering
\includegraphics[width=0.48\textwidth]{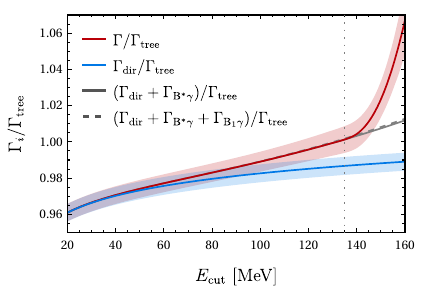} 
\includegraphics[width=0.48\textwidth]{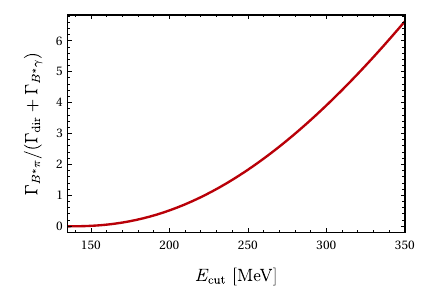} 
\caption{
Left panel: The \bmunu\ decay rate including QED corrections and resummation effects, normalized to the leading-order rate. The red curve shows the full rate including the indirect contributions from $B\to B^\ast\gamma$ and $B\to B^\ast\pi^0$ transitions, while the blue curve shows only the direct contribution for comparison. Shaded areas indicate the combined uncertainties. The gray lines show the rate excluding the pion contribution, with the dashed line including an estimate of the next higher $B$ resonance, the $B_1$. Right panel: Size of the $BB^\ast\pi$ contribution relative to all other contributions as a function of $E_\mathrm{cut}$.}
\label{fig:delta_BR}
\end{figure}

In order to better elucidate the QED corrections to the individual contributions to the decay rate and their dependence on the cut parameter, we define
\begin{equation}\label{eq:useless_splitup}
   \Gamma(E_{\rm cut})
   = \Gamma_\mathrm{tree}\,\Big[ 1 + \Delta\Gamma_\mathrm{dir}(E_\mathrm{cut})
    + \Delta\Gamma_{B^\ast\gamma}(E_\mathrm{cut}) 
    + \Delta\Gamma_{B^\ast\pi}(E_\mathrm{cut}) \Big] \,.
\end{equation}
Combining the various ingredients above according to \eqref{eq:Gammadir}, the direct contribution for the three values of the cut chosen above is
\begin{equation}
\begin{aligned}
   \Delta\Gamma_\mathrm{dir}(25\,\mathrm{MeV}) 
   &= (-36.1\pm 4.6_{f^{(1)}}\pm 1.7_{\phi_B}\pm 0.9_\mathrm{\mu_0})
    \cdot 10^{-3} \,, \\
   \Delta\Gamma_\mathrm{dir}(100\,\mathrm{MeV}) 
   &= (-17.4\pm 4.7_{f^{(1)}}\pm 1.7_{\phi_B}\pm 0.9_\mathrm{\mu_0}) 
    \cdot 10^{-3} \,, \\
   \Delta\Gamma_\mathrm{dir}(200\,\mathrm{MeV}) 
   &= (-\phantom{1}7.9\pm 4.7_{f^{(1)}}\pm 1.8_{\phi_B}\pm 1.0_\mathrm{\mu_0})
    \cdot 10^{-3} \,.
\end{aligned}
\end{equation}
The quoted uncertainties refer to the quantity $f^{(1)}$, the parameters of the LCDAs, and scale variations. In the left panel of Figure~\ref{fig:delta_BR}, the blue band shows the direct contribution to the decay rate as a function of $E_{\rm cut}$.

We now turn to the indirect contributions to the decay rate, given by the difference between the red and blue bands in the figure. The contribution arising from the $BB^\ast\gamma$ interaction is negligible for $E_\mathrm{cut}=25\,\mathrm{MeV}$. It grows with the radiation veto and becomes sizable for $E_\mathrm{cut}\gtrsim 50\,\mathrm{MeV}$. We find 
\begin{equation}
\begin{aligned}
   \Delta\Gamma_{B^\ast\gamma}(25\,\mathrm{MeV})
   &= (0.09\pm 0.03_{g_{BB^\ast\gamma}})\cdot 10^{-3} \,, \\
   \Delta\Gamma_{B^\ast\gamma}(100\,\mathrm{MeV})
   &= (6.4\pm 2.4_{g_{BB^\ast\gamma}}\pm 0.2_{f_{B^\ast}/f_B})\cdot 10^{-3} \,, \\
   \Delta\Gamma_{B^\ast\gamma}(200\,\mathrm{MeV}) 
   &= (40.2\pm 15.0_{g_{BB^\ast\gamma}}\pm 1.44_{f_{B^\ast}/f_B}) \cdot 10^{-3} \,.
\end{aligned}
\end{equation}
The growth continues for even larger values of the cut, but as discussed earlier the validity of our low-energy effective description requires $E_\mathrm{cut}\ll\OurLambda\approx 500$\,MeV, hence we do not provide values of $\Delta\Gamma_{B^\ast\gamma}$ for $E_\mathrm{cut}$ larger than 200\,MeV. The sum of this contribution plus the direct one is shown by the solid gray line in the left panel of Figure~\ref{fig:delta_BR}. For comparison, we also illustrate the approximate size of the contribution from the next excited $B$-meson state, the $B_1(5721)$. It can be obtained in a straightforward manner from \eqref{eq:indirect1} by replacing $(m_{B^\ast}, f_B^\ast, g_{BB^\ast\gamma})\to (m_{B_1}, f_{B_1}, g_{BB_1\gamma})$. With $m_{B_1}=5726$\,MeV, and using $f_{B_1}=248$\,MeV and $g_{BB_1\gamma}=0.73\,\mathrm{GeV}^{-1}$ from \cite{Pullin:2021ebn}, we obtain the dashed gray line for the sum of the direct contribution and the  indirect contributions induced by the effective $BB^*\gamma$ and $BB_1\gamma$ interactions. The very small difference between the dashed and solid lines is the $B_1$ contribution, and it is negligible for values $E_{\rm cut}<160$\,MeV.

\begin{figure}[t]
\centering
\includegraphics[scale=1.1]{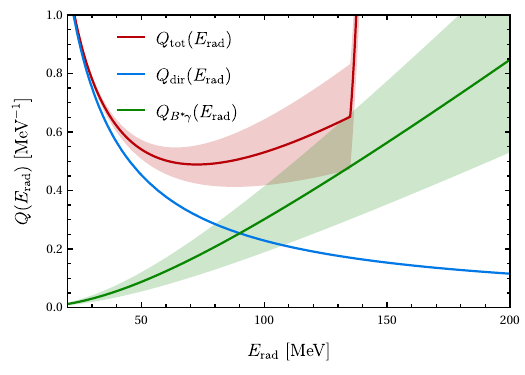} 
\caption{Different contributions to the radiation-energy spectrum $Q(E_{\rm rad})$ as a function of the radiation energy. The three curves show the direct contribution (blue), the indirect $B^\ast\gamma$ contribution (green), and the full spectrum (red) including also the $B^*\pi$ contribution. Shaded areas denote the uncertainty.}
\label{fig:diff_rate}
\end{figure}

The contribution from the $BB^\ast\pi$ interaction is absent for values of $E_\mathrm{cut}$ below the neutral pion mass, but then grows rapidly with $E_\mathrm{cut}$ above this threshold, quickly becoming the dominant contribution to the rate. 
For example, we find
\begin{equation}
\begin{aligned}
   \Delta\Gamma_{B^\ast\pi}(150\,\mathrm{MeV}) 
   &= (16.0\pm 4.6_{g_{BB^\ast\pi}}\pm 0.6_{f_{B^\ast}/f_B}\pm 0.2_{f_\pi})
    \cdot 10^{-3} \,, \\
   \Delta\Gamma_{B^\ast\pi}(200\,\mathrm{MeV}) 
   &= 0.53\pm 0.15_{g_{BB^\ast\pi}}\pm 0.02_{f_{B^\ast}/f_B}\pm 0.01_{f_\pi} \,.
\end{aligned}
\end{equation}
The sum of the direct contribution and the indirect contributions from the $BB^*\gamma$ and $BB^*\pi$ couplings is shown by the red band in the left panel of Figure~\ref{fig:delta_BR}. The pion-induced contribution is responsible for the sharp rise of the rate above $E_{\rm cut}=135$\,MeV. The right panel shows the size of just the pion contribution, $\Gamma_{B^\ast\pi}=\Gamma_\mathrm{tree}\,\Delta\Gamma_{B^\ast\pi}$, without uncertainty, normalized to the other components for values of $E_\mathrm{cut}>m_{\pi^0}$. We observe that the pion-induced contribution dominates all other contributions for $E_{\rm cut}\gtrsim 220$\,MeV.

For completeness, Figure~\ref{fig:diff_rate} displays the $B^-\to\mu^-\spac\bar\nu_\mu\spac(\gamma)$ radiation-energy spectrum, normalized as \cite{Carrasco:2015xwa,Frezzotti:2020bfa}
\begin{equation}\label{eq:blnugammaQ}
   Q(E_\mathrm{rad})
   \equiv \frac{4\pi}{\alpha}\,\frac{1}{\Gamma_\mathrm{tree}}\,
    \frac{d\spac\Gamma(E_\mathrm{rad})}{dE_\mathrm{rad}} \,.
\end{equation}
The widths of the bands reflect the theoretical uncertainty in each contribution. Note that the absolute uncertainty in the direct contribution is much smaller than that of the indirect contributions, which is also true for the bands in Figure~\ref{fig:delta_BR}. The pion-induced contribution results in a cusp at $E_\mathrm{rad}=m_{\pi^0}$, followed by a steep rise of the spectrum for larger values of $E_\mathrm{rad}$ (red band). 

A few observations follow from Figure \ref{fig:delta_BR}. First,  QED effects induce a shift in the rate ranging from $-3.9(5)\%$ to $0.1(7)\%$ for $E_\mathrm{cut}\in [20\,\mathrm{MeV}, m_{\pi^0}]$. The structure-dependent component of these corrections, encoded in the sum of its virtual piece, $\mathcal{R}^2_\mathrm{virt} -1$, and its real one, $\Delta\Gamma_{B^\ast\gamma}$, is always positive and varies between $+1.6(5)\%$ and $+3.0(8)\%$ in this range. While these QED effects are negligible at present, when only upper limits on the $B^-\to\mu^-\spac\bar\nu_\mu$ decay rate exist, but they will become relevant by the end of the Belle\,II data taking (with $50\,\mathrm{ab}^{-1}$ of integrated luminosity), when a 5\% precision on the $B^-\to\mu^-\spac\bar\nu_\mu$ branching ratio is expected -- and even more so at the FCC-ee, where a further four-fold improvement may be anticipated (based on statistics alone). Second, for $E_\mathrm{cut}\lesssim 100$\,MeV, the uncertainty on the QED correction is entirely dominated by the one in $\mathcal{R}_\mathrm{virt}^2$, which in turn is dominated by the unknown $\order{\alpha}$ correction $f^{(1)}$ entering $F_{-}$. For larger values of the cut it receives a comparable contribution from the uncertainty on $g_{BB^\ast\gamma}$. As soon as $E_\mathrm{cut}\gtrsim m_\pi$, the pion channel turns on and $\Delta \Gamma_{B^\ast\pi}$ rapidly becomes the dominant contribution to the rate. Including radiation in this energy range, while still measuring $B^-\to\mu^-\spac\bar\nu_\mu$ precisely, therefore requires a thorough understanding of the pion contribution. Its theory prediction is governed by $g_{BB^\ast\pi}$ or, more generally, the $B\to\pi$ form factor at small recoil, both of which are known with reasonable precision, so that this background can be quantified reliably. Indeed, the left panel of Figure~\ref{fig:delta_BR} shows that the theory uncertainty band does not increase significantly for $E_\mathrm{cut}>m_\pi$. The more critical limitation is instead the domain of validity of the HH$\chi$PT framework, which as discussed requires $E_\mathrm{cut}\ll\Lambda_c\approx 500$\,MeV. For this reason, we find it advisable to choose $E_\mathrm{cut}$ as low as possible, definitely not exceeding 200\,MeV to remain on the conservative side -- with the caveat that for $E_\mathrm{cut}>m_\pi$ the pion contribution must be included in the signal model or subtracted as an explicit background.

When constructing the full rate from \eqref{eq:useless_splitup}, one needs the tree-level rate $\Gamma_{\rm tree}$, a pseudo observable defined in \eqref{eq:treelevelrate} as the $B^-\to\mu^-\spac\bar\nu_\mu$ decay rate in a world where QED effects are absent. Note that this quantity is {\em not\/} equal to the rate in the limit $E_{\rm cut}\to 0$. Indeed, the physical decay rate vanishes in this limit. Numerically, we find
\begin{equation}
   \Gamma_\mathrm{tree} 
   = \left( \frac{\abs{V_{ub}}}{3.6\cdot 10^{-3}} \right)^2 
    \times (1.49 \pm 0.02_{f_B})\cdot 10^{-19}\,\mathrm{GeV} \,,
\end{equation}
where $\abs{V_{ub}}$, which currently is the leading source of uncertainty, is left explicit and normalized to the central value from the latest FLAG global fit \cite{FlavourLatticeAveragingGroupFLAG:2024oxs}. In the full rate, the QED-sourced uncertainty we have discussed above is subleading to the uncertainty in $f_B$ for  $E_\mathrm{cut}\le 150$\,MeV, and comparable between 150 and 200\,MeV. It would become dominant for larger cuts because of the growing indirect contributions, however as discussed in this region our description ceases to be reliable. We therefore conclude that the QED corrections (and the dominant mixed QED\,--\,QCD corrections) to the $B^-\to\mu^-\spac\bar\nu_\mu\spac(\gamma)$ decay rate are well under theoretical control in the range of validity of our calculations. 

The only result available in the literature for comparison was presented in \cite{Rowe:2024jml}, which missed the pion-induced contribution. A direct comparison of the remaining components is challenging due to the different methodology employed there. Indeed, without a strict separation of scales, the distinction between universal and structure-dependent correc tions appears ambiguous. One can, however, compare the overall magnitude of the QED effects. Reading off the total QED correction $\Delta\Gamma_\mathrm{QED}$ from Figure~5 of \cite{Rowe:2024jml} (corresponding to our $\Delta\Gamma_\mathrm{dir}+\Delta\Gamma_{B^\ast\gamma}$), we find a reasonably similar net effect at the level of central values over the range of $E_\mathrm{cut}$ considered. On the other hand, we disagree with the statement in this paper that ``virtual structure-dependent'' corrections amount to about a $+5\%$ relative effect. In our framework, the virtual structure-dependent contribution is unambiguously identified with $\mathcal{R}_\mathrm{virt}^2-1$, which we find to be much smaller, specifically $+1.5(5)\%$. This discrepancy can be traced to a different definition of ``virtual structure-dependent''. Indeed, we believe that the authors' definition of what classifies as such is misleading, since a distinction from universal contributions cannot be achieved by selecting graph topologies alone. Consequently, their ``virtual structure-dependent'' corrections are in large part universal, leading to a substantial overestimate of their magnitude. Furthermore, in \cite{Rowe:2024jml} the theoretical uncertainty on the QED corrections appears to be essentially saturated by the $B^\ast$ contribution. Our analysis instead shows that, for cuts $E_\mathrm{cut}\le 100~\mathrm{MeV}$, the dominant uncertainty originates from the virtual structure-dependent term $\mathcal{R}_\mathrm{virt}^2$ (through $f^{(1)}$ and the $B$-meson LCDAs). 

Finally, an important aspect to consider is that existing measurements of the $B^-\to\mu^-\spac\bar\nu_\mu\spac(\gamma)$ process (see e.g.\ the measurement by the Belle collaboration in \cite{Belle:2019iji}) do not impose a photon-energy veto. Instead, they select events where the muon momentum in the $B$-meson rest frame is peaked around the two-body value, $|\boldsymbol{p}_{\mu}|=\frac{m_B^2-m_\mu^2}{2m_B}$. This cut does not remove events with energetic photons (see e.g.\ the Dalitz plots in the ($E_\gamma,E_\mu$) plane in \cite{Becirevic:2009aq,Rowe:2024jml}). In order to cleanly isolate the $B^-\to\mu^-\spac\bar\nu_\mu\spac(\gamma)$ signal from the background, one should apply a cut on the photon energy as small as experimentally feasible. Relaxing the cut does not enhance the signal, but only the background. Moreover, if for experimental reasons it is not possible to lower the cut on the photon energy below 200\,MeV, a meaningful comparison with theory is not possible at present. Such a comparison would require either a first-principle calculation of the $B^-\to\mu^-\spac\bar\nu_\mu\spac\gamma$ decay rate over a large range in photon energy, which could only come from the lattice.

\newpage
\section{Conclusions}
\label{sec:conclusions}

We have derived a state-of-the-art prediction for the rate of the leptonic decay $B^-\to\mu^-\spac\bar\nu_\mu\spac(\gamma)$ at next-to-leading order in QED, including the resummation of the dominant logarithmic QED and QCD corrections. QED effects above the hadronic scale probe the internal structure of the $B$ meson, rendering a point-like treatment of the meson insufficient to reach the percent accuracy needed to match the precision of future measurements. The central result of this work is a systematic and explicit characterization of the structure-dependent contributions in terms of perturbatively calculable functions and a well-defined set of non-perturbative input parameters. Due to the large number of scales involved, the chiral suppression of the process, and the non-perturbative dynamics involved, this has proven to be a technically challenging endeavor. 

We have handled the multi-scale nature of the problem through a sequence of effective field theories, which we used to write the final rate in the form of a factorization theorem valid to all orders in QED$\times$QCD. Below the electroweak scale, the decay is described within the low-energy effective theory (LEFT). For scales below the $B$-meson mass, the relevant dynamics is captured by soft-collinear effective theory (SCET) and heavy-quark effective theory (HQET). We have constructed a complete basis of SCET operators at next-to-leading power for heavy-to-light quark currents coupled to weak leptonic currents. Corrections at the scale of the meson mass $\mu_h\sim m_B$ are encoded in hard functions, while interactions between the energetic charged lepton and the light spectator quark in the $B$ meson introduce an intermediate hard-collinear scale $\mu_\hc^2\sim m_B\spac\lambdaqcd$. This is encoded in an additional matching step within SCET, giving rise to jet functions. 

In this matching step, the hard and jet functions appear in convolutions that are generally endpoint divergent, as is characteristic for SCET factorization theorems at subleading power. We have treated these divergences using refactorization-based subtractions, which introduce a cutoff scale $\Lambda$. While physical predictions are independent of $\Lambda$, individual components carry logarithmic $\Lambda$ dependence, and each component function has a different natural choice for which it is free of large logarithms involving this scale. For the first time, we derive the evolution equations associated with $\Lambda$ and solve them to resum these large logarithms to all orders in perturbation theory. The resulting anomalous dimensions feature a non-linear dependence on $\ln(\mu^2/\Lambda^2)$, marking a qualitative departure from the more familiar single-scale Sudakov evolution problems. The final resummed prediction is obtained by solving coupled evolution equations in both the renormalization scale $\mu$ and the subtraction scale $\Lambda$. In practice, the perturbative ingredients entering the factorization formula are evaluated at the first non-trivial order in QED (and at the relevant order in QCD required for the logarithmic accuracy). After subtracting endpoint divergences, the relevant hadronic matrix elements include the known subleading-twist light-cone distribution amplitudes of the $B$ meson, but in addition a new, genuinely non-perturbative parameter $F_-(\Lambda,\mu)$ appears, which generalizes the $B$-meson decay constant defined in HQET. Since a determination of this quantity from first principles is not yet available, we include it as a source of uncertainty in our error estimates. The main result of this first part of the paper is the resummed expression for the virtual-correction factor $\mathcal{R}_{\rm virt}$ given in \eqref{eq:Rvirt_final}, which we repeat here for the convenience of the reader:
\[
\begin{aligned}
   \mathcal{R}_{\rm virt} 
   &= \exp\bigg[ - Q_\ell\spac Q_u\,\frac{\alpha}{\pi}
    \int_{m_B}^{\mu_0}\!\frac{d\mu'}{\mu'} \int_{m_B}^{\mu'}\!\frac{d\bar\omega}{\bar\omega}\,\widetilde U_C(\mu',\bar\omega) \bigg]\,
    \frac{F_-(m_B,\mu_0)}{F_{\rm QCD}(\mu_0)} \\
   &\quad \times \Bigg\{ 1 + \frac{\alpha}{4\pi}\,\bigg[
    \frac32\,Q_\ell^2\spac\ln\frac{\mu_0^2}{m_\ell^2} 
    - \frac32\,Q_b^2\spac\ln\frac{\mu_0^2}{m_B^2} 
    - (2+z)\,Q_\ell\spac Q_b\spac\ln\frac{m_B^2}{m_\ell^2} 
    + \left( \frac{\pi^2}{12} - \frac52 \right) Q_\ell^2 \\
   &\hspace{2.67cm} 
    + \left( - \frac12 + \frac{z^2\ln z}{z-1} + z  
    - 2\spac\text{Li}_2(1-z) - \frac{\pi^2}{12} \right) Q_\ell\spac Q_b 
    - \left( 2 + 3\ln z \right) Q_b^2 \bigg] \\
   &\hspace{1.15cm} + Q_\ell\spac Q_u\,\frac{\alpha}{\pi}\,
    \int_{m_B}^{\mu_0}\!\frac{d\mu'}{\mu'}\,\frac{\widetilde U_C(\mu',m_B)}{1-\delta(\mu')} \\
   &\hspace{1.15cm} - Q_\ell\spac Q_u\,\frac{\alpha}{2\pi} 
    \left[ \frac{1}{1-\delta(\mu_0)}\,\ln\frac{\mu_0^2}{m_B\spac\omega_-(\mu_0)} 
    - h_1\big(\delta(\mu_0)\big) \right] \widetilde U_C(\mu_0,m_B) \\ 
   &\hspace{1.15cm} - Q_\ell\spac Q_u\,\frac{\alpha}{\pi}\, 
    \frac{\widetilde U_C(\mu_0,m_B)}{1-\delta(\mu_0)} 
    \int_0^\infty\!d\omega\! \int_0^\infty\!d\omega_g\,\phi_{3g}^B(\omega,\omega_g,\mu_0)
    \left[ \frac{1}{\omega_g}\,\ln\frac{\omega+\omega_g}{\omega} - \frac{1}{\omega+\omega_g} \right]
    \!\Bigg\} \,.
\end{aligned}
\]
Here $Q_\ell=-1$, $Q_u=\frac23$ and $Q_b=-\frac13$ denote the electric charges of the fermions in units of $e$, and $z=m_B/m_b$. $F_-$ and $\omega_-$ are new hadronic parameters, while $\phi_{3g}^B$ denotes the three-particle light-cone distribution amplitude of the $B$ meson. The quantities $\widetilde U_C$, $\delta$ and $h_1(\delta)$ are renormalization-group (RG) functions accomplishing the resummation of leading logarithmic corrections. The quantity $\mathcal{R}_{\rm virt}$ is renormalization-group (RG) invariant; its residual dependence on the scale $\mu_0$ results only from the truncation of the perturbative expansion.

For scales below $\lambdaqcd$, we have performed a systematic, non-perturbative matching of SCET onto a low-energy description in the framework of heavy-hadron chiral perturbation theory (HH$\chi$PT), treating the low-energy dynamics of the charged lepton in boosted heavy-lepton effective theory (bHLET). 
We have split up the decay rate into ``direct'' and ``indirect'' contributions, where the direct contribution contains the soft-photon corrections to the \bmunu\ process, with photon energies below a threshold $E_{\rm cut}$. Our RG-improved expression for this contributions given in \eqref{eq:Gammadir} reads
\[
   \Gamma_{\rm dir}(E_{\rm cut})
   = \Gamma_{\rm tree} \left(\frac{\alpha(m_Z)}{\alpha(m_B)} \right)^\frac{9}{20}\!
    \left( \frac{2 E_{\rm cut}}{m_B} \right)^{2\gamma_{\rm soft}}\! 
    \frac{e^{-2\gamma_E\spac\gamma_{\rm soft}}}{\Gamma(1+2\gamma_{\rm soft})}\,
    \mathcal{R}_{\mathrm{virt}}^2 
    \left[ 1 + \frac{Q_\ell^2\spac\alpha}{2\pi} \left( 2 - \frac{\pi^2}{3} \right) 
    + \mathcal{O}(\alpha^2) \right] .
\]
In this expression, the leading double and single QED logarithms as well as leading-logarithmic QCD effects are resummed for the first time, in the limit where the running of $\alpha$ is neglected (this can be generalized, see Appendix~\ref{app:running}). The indirect contributions corresponds to decay processes proceeding via an intermediate (off-shell) excited $B$-meson resonance. In the hadronic sector, the relevant degrees of freedom of the low-energy theory include the ($B$, $B^\ast$) heavy spin doublet of meson states as well as the light pseudoscalar mesons $\pi,K,\eta$. In this regime, structure-dependent effects arise from two sources: the $B^-\to B^{\ast-}\gamma$ transition followed by the weak decay $B^{\ast-}\to\ell^-\spac\bar\nu_\ell$ that is not chirally suppressed, and the transition $B-\to B^{\ast-}\pi^0$ followed by an electromagnetic decay (almost always $\pi^0\to\gamma\gamma$) by the on-shell pion, mediated by the chiral anomaly. This latter contribution arises as soon as $E_{\rm cut}>m_{\pi^0}$ and becomes the dominant decay mode once the veto scale exceeds 220\,MeV. 

In our phenomenological analysis we have provided numerical estimates of the QED corrections to the $B^-\to\mu^-\spac\bar\nu_\mu\spac(\gamma)$ rate, showing that they amount to a relative correction of the order of a few percent (for veto energies below the pion mass), an effect that is relevant for Belle\,II and essential for FCC-ee. The structure-dependent component to these corrections is coincidentally modest for the process at hand, becoming relevant only at a future Tera-$Z$ machine like FCC-ee. This conclusion however relies on numerical cancellations that are not obvious from the outset and should not be assumed to be generic for other decay channels. 

The framework developed here applies straightforwardly to the electron mode as well. Because of the much stronger chiral suppression in this case, the $B^-\to e^-\spac\bar\nu_e\spac(\gamma)$ decay rate is completely dominated by the $B^\ast$ contributions. The situation is qualitatively different for the tau channel: the mass of the tau lepton is not a particularly small parameter compared with $m_B$, and the appropriate effective field-theory setup for $B^-\to\tau^-\spac\bar\nu_\tau\spac(\gamma)$ decays is more closely related to heavy-to-heavy transitions, like $B\to D^{(*)}$ decays. A dedicated analysis is interesting in light of the projected future experimental accuracy, as the channel can serve as an exclusive extraction of $|V_{ub}|$ as well as a test of lepton flavor universality in conjunction with the muon channel. We leave this extension to future work.

\subsection*{Acknowledgements}

We thank Florian Bernlochner, Philipp B\"oer, Matteo Di Carlo, Daniel Jacobi, and Enrico Lunghi for useful discussions, and Gino Isidori for encouraging us to engage in this project. This research has received funding from the Cluster of Excellence PRISMA${}^+$ (EXC 2118/1, Project ID 390831469) funded by the German Research Foundation (DFG) within the Germany Excellence Strategy, and from the European Research Council (ERC) under the European Union’s Horizon 2022 Research and Innovation Program (ERC Advanced Grant agreement No.~101097780, EFT4jets). Views and opinions expressed in this work are those of the authors only and do not necessarily reflect those of the European Union or the European Research Council Executive Agency. Neither the European Union nor the granting authority can be held responsible for them.

\newpage
\begin{appendix}

\section{SCET Lagrangians}
\label{app:SCETbasics}

Here we collect the relevant terms in the SCET Lagrangian including all necessary subleading interactions for the quark and the lepton, including the mass dependence of the latter. The leading-order Lagrangian of the hard-collinear lepton and quark fields is given by
\begin{equation}
   \mathcal L_\X 
   = \sum_{f=\ell,u} \bar\X_\hc^{(f)} \left( in\cdot\mathcal D 
    + i \slashed{\mathcal D}_\perp \spac\frac{1}{i\nb\cdot\del}\,
    i\slashed{\mathcal D}_\perp \right) \frac{\nbsl}{2}\,\X_\hc^{(f)} \,,
\end{equation}
where we have defined 
\begin{equation}
   i\mathcal{D}^\mu\spac\X_\hc^{(f)} 
   = \left( i\partial^\mu + Q_f\spac\A_\hc^\mu +\G_\hc^{\mu,a}\spac t_f^a \right) \X^{(f)}_\hc \,,
\end{equation}
where $t_f^a$ are the generators of the $SU(3)_c$ representation of the fermion $f$. For the lepton, the mass gives rise to power-suppressed interactions in SCET-1, described by the Lagrangians \cite{Chay:2005ck}
\begin{equation}
   \mathcal L_{m_\ell}^{(1/2)} 
   = \bar\X_\hc^{(\ell)} \left[ 
    i\slashed{\mathcal D}_\perp, \frac{m_\ell}{i\nb\cdot\del} \right] 
    \frac{\nbsl}{2}\,\X_\hc^{(\ell)} \,, \qquad
   \mathcal L_{m_\ell}^{(1)} 
   = - m_\ell^2\,\spac\bar\X_\hc^{(\ell)}\,\frac{\nbsl}{2}\,
    \frac{1}{i\nb\cdot\del}\,\X_\hc^{(\ell)} \,.
\end{equation}
These terms generate subleading collinear interactions, responsible for introducing the lepton mass in matrix elements of SCET-1 operators of type-$C$, $D$ and $E$ that do not already contain $m_\ell$. 

Another class of important power-suppressed interactions for our calculations are those between soft and hard-collinear particles not contained in the covariant derivatives. They are contained in the subleading-power Lagrangians \cite{Beneke:2002ni,Beneke:2002ph} 
\begin{equation}
\begin{aligned}
   \mathcal{L}_{\X q}^{(1/2)} 
   &= \bar q_s\spac\Asl_\hc^\perp\spac\X_\hc + \mathrm{h.c.} \,, \\
\mathcal{L}_{\X q}^{(1)} 
   &= \bar q_s \left[ n\cdot\A_\hc 
    + \Asl_\hc^\perp\,\frac{1}{i\nb\cdot\del}\spac(i\slashed{\del}_\perp 
    + \Asl_\hc^\perp) \right] \frac{\nbsl}{2}\,\X_\hc 
    + \bar q_s \overleftarrow D^\mu\spac x_\mu^\perp\spac\Asl_\hc^\perp\spac\X_\hc 
    + \mathrm{h.c.} \,, \\ 
   \mathcal{L}_{\X q}^{(3/2)} 
   &= \bar q_s \overleftarrow D^\mu\spac x^\perp_\mu 
    \left[ n\cdot\A_\hc + \Asl_\hc^\perp\,\frac{1}{i\nb\cdot\del}\spac
     (i\slashed\del_\perp + \Asl_\hc^\perp) \right] \frac{\nbsl}{2}\,\X_\hc \\
   &\quad+ \bar q_s\spac\nb\cdot\overleftarrow D\,\frac{n\cdot x}{2}\,
    \Asl_\hc^\perp\spac\X_\hc 
    + \bar q_s \overleftarrow D^\mu \overleftarrow D^\nu\,
    \frac{x_\mu^\perp x_\nu^\perp}{2}\,\Asl_\hc^\perp\spac\X_\hc
   + \mathrm{h.c.} \,.
\end{aligned}   
\end{equation}
The soft field and its derivatives are evaluated at $x_-^\mu=\frac{n^\mu}{2}\spac \nb\cdot x$ in the usual way. The Feynman rules for interaction terms involving the coordinates $x^\mu$ explicitly implement the subleading terms in an expansion of the amplitude in the soft momentum, see Appendix~A of \cite{Beneke:2018rbh} for a detailed discussion. For the three-particle contributions we also need the interactions between soft gluons and hard-collinear quarks to leading order. They follow from \cite{Beneke:2002ph}
\begin{equation}
   \mathcal L_{\X}^{(1/2)} 
   = \bar\X_\hc\,x_\perp^\mu n^\nu F_{\mu\nu}^s\spac\frac{\nbsl}{2}\,\X_\hc \,,
\end{equation}
where the soft field-strength tensor must be evaluated at $x_-$. 

In SCET-2, the lepton Lagrangian is structurally identical, but its power-counting changes because collinear derivatives and the lepton mass are of the same order in power counting. Then the leading-power collinear Lagrangian is of the form \cite{Chay:2005ck}
\begin{equation}
    \mathcal L_\X 
    = \bar\X_\hc^{(\ell)} \left[ in\cdot\mathcal D 
     + (i \slashed{\mathcal D}_\perp - m_\ell)\,\frac{1}{i\nb\cdot\del}\,
     (i\slashed{\mathcal D}_\perp + m_\ell) \right] \frac{\nbsl}{2}\,\X_\hc^{(\ell)} \,.
\end{equation}

\section{Reduction of Dirac structures}
\label{app:all_reductions}

Loop-level matching calculations produce Dirac structures beyond those included in our operator basis. While reducible in four spacetime dimensions, care needs to be taken when dimensional regularization is employed. Here we collect the relevant technical details and scheme choices applied to our calculations.

\subsection{Scheme dependence and conversion}

When extracting the Wilson coefficients of the LEFT from the literature, we need to ensure that our reduction scheme is consistent with the one used there, which amounts to applying the appropriate scheme conversion. In matching to the LEFT, the reduction identity
\begin{equation}
    \gdirac{\gamma^\mu\gamma^\nu\gamma^\rho}{\gamma_\mu\gamma_\nu\gamma_\rho} 
    \to (16+\kappa\epsilon)\,\gdirac{\gamma^\mu}{\gamma_\mu}
\end{equation}
is used to remove redundant Dirac structures appearing beyond tree-level, where $\kappa$ is a para\-meter representing the scheme choice. To find the scheme-conversion factor between different choices for $\kappa$, we note that this parameter cancels in the combination between matching coefficients and matrix elements of the effective theory. To obtain a scheme conversion, we thus compute the one-loop matrix elements in the LEFT and express them in terms of $\kappa$. Evaluating the $b\to u\spac\ell^-\spac\bar\nu_\ell$ amplitude at one-loop order, and keeping only terms involving the reducible structures, we find
\begin{equation}
    i\cA_\mathrm{LEFT} 
    = i C_L^{[\kappa]} \left[ 1 - Q_\ell\spac(Q_b+Q_u)\,\frac{\alpha}{16\pi}\spac
     \kappa \right] \gdirac{\gamma_\mu}{\gamma^\mu} + \ldots \,,
\end{equation}
where $C_L^{[\kappa]}$ is the LEFT matching coefficient with a specific choice of $\kappa$, and the dots refer to terms independent of it. Since the product of $C_L^{[\kappa]}$ with the factor multiplying it must be independent of $\kappa$, we find that the scheme-conversion factor at one-loop order is given by
\begin{equation}
   C_L^{[\kappa']} = C_L^{[\kappa]} 
   \left[ 1 - Q_\ell\spac(Q_b+Q_u)\,\frac{\alpha}{16\pi}\spac
     (\kappa-\kappa') \right] .
\end{equation}
As discussed in the main text, our reduction identities in SCET imply the choice $\kappa'=0$. Moreover, our scheme parameter $\kappa$ relates to the one used in  \cite{Dekens:2019ept} via $\kappa=-4\spac b_\mathrm{ev}$. 

\subsection{Reduction identities in light-cone components}

We now list all the structures appearing in the matching to SCET from amplitudes involving two quarks and two leptons within our reduction scheme. They are
\begin{equation}\label{eq:reductions1}
\begin{aligned} 
 \gdirac{\frac{\slashed n\slashed\barn}{4}\gamma_\perp^\mu}{\frac{\slashed\barn \slashed n}{4} \gamma_\perp^\mu}   & = 
\gdirac{\gamma_\perp^\mu}{\frac{\slashed\barn \slashed n}{4} \gamma_\perp^\mu}\,,  \\\gdirac{\gamma_{\perp\mu}\gamma_{\perp\nu}\gamma_{\perp\rho}}{\frac{\slashed\barn \slashed n}{4} \gamma_{\perp}^\mu\gamma_{\perp}^\nu\gamma_{\perp}^\rho}  &= 4\gdirac{\gamma_\perp^\mu}{\frac{\slashed\barn \slashed n}{4} \gamma_\perp^\mu}  \\
\gdirac{\frac{\slashed \barn}{2}\gamma_\perp^\mu\gamma_\perp^\nu}{\frac{\slashed\barn \slashed n}{4} \gamma_\perp^\mu\gamma_\perp^\nu} &  = 
        4 \gdirac{\frac{\slashed\barn }{2}}{\frac{\slashed\barn \slashed n}{4}} \,,  \\ 
 \gdirac{\gamma_\perp^\mu}{\frac{\slashed\barn \slashed n}{4} \slashed k_\perp \gamma_\perp^\mu}  & = 
2\gdirac{\frac{\slashed n\slashed\barn}{4}\slashed k_\perp}{\frac{\slashed\barn \slashed n}{4}} \,, \\
 \gdirac{\frac{\slashed n\slashed\barn}{4}\gamma_\perp^\mu}{\frac{\slashed\barn \slashed n}{4} \slashed k_\perp \gamma_\perp^\mu}  & =  
 2\gdirac{\frac{\slashed\barn \slashed n}{4}\slashed k_\perp}{\frac{\slashed\barn \slashed n}{4}}\,,\\ 
 \gdirac{\frac{\slashed \barn}{2}\gamma_\perp^\mu \gamma_\perp^\nu}{\frac{\slashed\barn \slashed n}{4} \slashed k_\perp \gamma_\perp^\mu \gamma_\perp^\nu} & =
4\gdirac{\frac{\slashed\barn}{2}}{\frac{\slashed\barn \slashed n}{4} \slashed k_\perp}\,, \\
 \gdirac{\frac{\slashed n}{2}\gamma_\perp^\mu \gamma_\perp^\nu}{\frac{\slashed\barn \slashed n}{4} \slashed k_\perp \gamma_\perp^\mu \gamma_\perp^\nu}  & =
        0 \,, \\
\gdirac{\frac{\slashed n}{2}\gamma_\perp^\mu\gamma_\perp^\nu}{\frac{\slashed\barn \slashed n}{4} \gamma_\perp^\mu\gamma_\perp^\nu}  &  = 
        0 \,, 
\end{aligned}
\end{equation}
where $k$ denotes the incoming momentum of the soft spectator quark. For amplitudes with an additional soft gluon in the initial state, we also need the following reductions
\begin{equation}\label{eq:reductions2}
\begin{aligned}
 \gdirac{\gamma_\perp^\mu}{\frac{\slashed\barn \slashed n}{4} \slashed\varepsilon_\perp^g \gamma_\perp^\mu}  &=   2 \gdirac{\frac{\slashed n\slashed\barn}{4}\slashed \varepsilon_\perp^g}{\frac{\slashed\barn \slashed n}{4}} \,,\\
  \gdirac{\frac{\slashed n\slashed\barn}{4}\gamma_\perp^\mu}{\frac{\slashed\barn \slashed n}{4} \slashed \varepsilon_\perp^{g} \gamma_\perp^\mu}  &= 
        2\gdirac{\frac{\slashed n\slashed\barn}{4}\slashed \varepsilon_\perp^g}{\frac{\slashed\barn \slashed n}{4}} \,, \\
 \gdirac{\slashed\varepsilon_\perp^g \gamma_\perp^\mu \gamma_\perp^\nu}{\frac{\slashed\barn \slashed n}{4} \gamma_\perp^\mu \gamma_\perp^\nu}  &=
          4 \gdirac{\frac{\slashed n\slashed\barn}{4}\slashed \varepsilon_\perp^g}{\frac{\slashed\barn \slashed n}{4}} \,,\\
  \gdirac{\frac{\slashed n \slashed\barn}{4} \slashed\varepsilon_\perp^g \gamma_\perp^\mu\gamma_\perp^\nu}{\frac{\slashed\barn \slashed n}{4} \gamma_\perp^\mu\gamma_\perp^\nu} & =
             4 \gdirac{\frac{\slashed n\slashed\barn}{4}\slashed \varepsilon_\perp^g}{\frac{\slashed\barn \slashed n}{4}}  \,, \\
              \gdirac{\frac{\slashed n}{2}\slashed k_\perp^g \slashed\varepsilon_\perp^g \gamma_\perp^\mu \gamma_\perp^\nu}{\frac{\slashed\barn \slashed n}{4} \gamma_\perp^\mu\gamma_\perp^\nu} & =
        0\,.  \\
               \gdirac{\frac{\slashed n}{2}\slashed k^g_{\perp}\gamma_\perp^\mu}{\frac{\slashed\barn \slashed n}{4} \slashed\varepsilon_\perp^g\gamma_\perp^\mu}  &= 0  \,, \\
  \gdirac{\frac{\slashed n}{2}\slashed\varepsilon_\perp^g \gamma_\perp^\mu}{ \frac{\slashed\barn \slashed n}{4} \slashed k^g_{\perp} \gamma_\perp^\mu}  &=       0 \,, 
\end{aligned}
\end{equation}
where $k^g$ and $\varepsilon^g$  denote the incoming momentum and polarization vector of the additional soft gluon, respectively. 

\subsection{Cancellation of power-enhanced contributions}
\label{app:reds_powerenh}

As explained in the main text, power-enhanced contributions to the amplitude, as they appear in the $B_s\to\mu^+\mu^-$ decay, are absent in the charged-current decay \cite{Beneke:2017vpq}. In our calculation, they would however appear in intermediate steps if a different reduction scheme was chosen for the Dirac structures of SCET. To see the explicit cancellation, it is instructive to investigate the one-loop QED correction to the $b\to u\spac\ell^-\spac\bar\nu_\ell$ matrix element of the operator $\mathcal{O}^{V, LL}_{\ell}$ in \eqref{eq:OVLLdef}, arising from photon exchange between the up quark and the lepton, and study its hard-collinear and collinear regions. In the hard-collinear region described in SCET-1, a subset of the terms one finds is
\begin{equation}
   \mathcal A_{\ell u}^\mathrm{[hc]} 
   \supset - m_\ell\,\frac{Q_\ell\spac Q_u\spac\alpha}{8\pi\epsilon} 
   \left( \frac{1}{n\cdot k} + \frac{2\spac n\cdot p_\ell}{(n\cdot k)^2}
   - \frac{k_\perp^2}{4(n\cdot k)(\barn\cdot p_\ell)} \right)
   \gdirac{\frac{\slashed n}{2}\spac\gamma_\perp^\mu\gamma_\perp^\nu}%
          {\gamma_{\mu\perp}\gamma_{\nu\perp}} \,,
\end{equation}
where $k$ and $p_\ell$ are the momenta of the spectator and the lepton, respectively. The first term in the brackets is power-enhanced by $\order{m_b/\Lambda_\mathrm{QCD}}$ with respect to the tree-level amplitude. In our reduction scheme, the Dirac structure in this expression reduces to zero directly, but in a more general scheme, where one would write
\begin{equation}
   \gdirac{\frac{\slashed n}{2}\spac\gamma_\perp^\mu\gamma_\perp^\nu}%
          {\gamma_{\mu\perp}\gamma_{\nu\perp}}
   \to \tilde\kappa\spac\epsilon\,\gdirac{\frac{\slashed n}{2}}{} \,,
\end{equation}
this contribution would appear in the hard-collinear jet functions as a finite term proportional to $\tilde\kappa$. It would cancel, however, against the collinear matrix element in SCET-2, which is found to be given by exactly the same expression, but with the opposite sign, i.e.\ $\mathcal A_{\ell u}^\mathrm{[c]}=-\mathcal A_{\ell u}^\mathrm{[hc]}$. In the region sum, the cancellation is apparent before any reductions, whereas in the EFT calculation the cancellation occurs once the jet functions and collinear matrix elements are combined, in an analogous fashion to the cancellation of $\kappa$ in the LEFT, discussed in the previous section. Our scheme choice avoids the appearance of spurious super-leading terms with inverse powers of the spectator momentum from the start.

\section[\texorpdfstring{\boldmath Soft anomalous dimension of $F_\mp(\Lambda,\mu)$}{Soft anomalous dimension of FLambda}]{\boldmath Soft anomalous dimension of $F_\mp(\Lambda,\mu)$}
\label{app:F_anodim}

Here we outline the calculation of the anomalous dimensions for the hadronic parameters $F_\mp(\Lambda,\mu)$ in \eqref{eq:gamma_F}. To derive them, one needs to compute the UV divergences of the soft matrix elements $S_{1,2}$ defined in \eqref{eq:Sidef}. To this end, it is convenient to rewrite the $\theta_T$ functions entering these definitions in  terms of a partial derivative, as shown in \eqref{eq:miracle}. Next, we express the soft matrix elements in the form
\begin{equation}\label{eq:matrix_elem_decomp}
   S_{1,2} = \frac{\cA_1+\cA_2(\Lambda)}{\cA_3} \,,
\end{equation}
where $\cA_{1,2,3}$ denote the matrix elements 
\begin{equation}
\begin{aligned}
   \cA_1 &= \langle\spac 0\spac|\,\bar u_s\,\Gamma\,b_v\,Y_n^{(\ell)\dagger}\spac
    |B^-\rangle \,, \\
   \cA_2(\Lambda) &= - \langle\spac 0\spac|\,
    \big( \bar u_s\spac\overline{Y}_\nb^{(u)} \big)\,
    \theta_T\bigg( \frac{-i\nb\cdot\overleftarrow{\partial}_{\!\!s}}{\nb\cdot v}
    - \Lambda \bigg)\,\Gamma\spac
    \big( Y_n^{(\ell)\dagger}\,\overline{Y}_\nb^{(u)\dagger}\spac b_v \big) 
    |B^-\rangle \,, \\
   \cA_3 &= \langle\spac 0\spac|\,Y_n^{(\ell)\dagger}\,\overline{Y}_v^{(B)}\spac
    |\spac 0\spac\rangle \,,
\end{aligned}
\end{equation}
with $\Gamma=(\nbsl/\nb\cdot v)\spac P_L$ for $S_1$ and $\Gamma=(\nsl/n \cdot v)\spac P_L$ for $S_2$. 
It is convenient to present the results in terms of relative corrections to the tree-level matrix elements, i.e.\
\begin{equation}
   \cA_i = \cA_i^\mathrm{LO}\spac(1+a_i) \,.
\end{equation}
For this calculation, it is crucial to respect the appropriate $i0$ prescriptions for the Wilson lines in the $n$ and $v$ directions, as explained around \eqref{eq:WLdef}. For photons with outgoing momentum $k$, the single-emission matrix elements of the Wilson lines read 
\begin{equation}
\begin{aligned}
    \langle\gamma(k)|\,Y_n^{(\ell)\dagger}\spac|0\rangle 
    &= - Q_\ell\spac e\,\frac{n\cdot\varepsilon^\ast}{n\cdot k+v\cdot n\,\delta+i0} \,, \\
    \langle\gamma(k)|\,\overline{Y}_v^{(B)}\spac|0\rangle 
    &= Q_B\spac e\,\frac{v\cdot\varepsilon^\ast}{v\cdot k +\frac{\delta}{2}-i0} \,,
\end{aligned}
\end{equation}
where $\delta>0$ regulates IR divergences \cite{Chiu:2009yx}. With this prescription, we obtain
\begin{equation}\label{eq:softmatrixelem}
\begin{aligned}   
    a_1 & = \frac{\alpha}{4\pi}\left[ 
    \frac{Q_b^2}{\epsilon} 
    + \frac{Q_b\spac Q_u}{\epsilon}-\frac{Q_u^2}{2\epsilon}
    - Q_\ell\spac Q_b \left( \frac{1}{\epsilon^2}+\frac{1}{\epsilon}\spac\ln\frac{\mu^2}{\delta^2} \right) \right. \\ 
     & \hspace{1.5cm}\left. + \frac{Q_\ell\spac Q_u}{\epsilon}\left(\int_0^\infty d\omega\, \phi_-(\omega,\mu)\spac\ln\frac{\omega^2}{\delta^2} -2\right)
    \right]  + \frac{3 C_F\spac\alpha_s}{8\pi \epsilon} \,,  \\
   a_2 & =\left\{\begin{aligned}
    & Q_\ell\spac Q_u\,\frac{\alpha}{2\pi}\left[\frac{1}{\epsilon^2}+\frac{1}{\epsilon} \int_0^\infty\!d\omega\, \phi_-(\omega,\mu)\spac\ln \frac{\mu^2}{\omega\Lambda}
    + \frac{1}{\epsilon} \right] ; && \text{for}~~ \Gamma=\frac{\nbsl}{\nb \cdot v}\spac P_L\,,\\
      & Q_\ell\spac Q_u\,\frac{\alpha}{2\pi}\left[\frac{1}{\epsilon^2}+\frac{1}{\epsilon}\int_0^\infty d\omega\, \phi_-(\omega,\mu)\spac\ln \frac{\mu^2}{\omega\Lambda} \right] ; &&\text{for}~~ \Gamma=\frac{\nsl}{n \cdot v}\spac P_L\,,\\
    \end{aligned} \right. \\
    a_3 & = -\frac{Q_B^2\spac\alpha}{4\pi} \left( \frac{1}{\epsilon^2}+\frac{1}{\epsilon}\spac\ln\frac{\mu^2}{\delta^2}   \right) ,
\end{aligned}
\end{equation}
where only $a_2$ differs for the two Dirac structures. Using these  expressions, and introducing renormalization factors $Z_{F_\mp}(\Lambda,\mu)$ such that the renormalized matrix elements $S_1(\Lambda,\mu)=Z_{F_-}(\Lambda,\mu)\,S_1(\Lambda)$ and $S_2(\Lambda,\mu)=Z_{F_+}(\Lambda,\mu)\,S_2(\Lambda)$ are free of $1/\epsilon$ poles, we find
\begin{equation}
    Z_{F_\mp}(\Lambda) 
    = 1 + 3 C_F\,\frac{\alpha_s}{8\pi\epsilon}
     + \frac{\alpha}{4\pi} \left\{ \frac{3\spac Q_u^2}{2\epsilon} 
     + Q_\ell\spac Q_u \left[ \frac{1}{\epsilon^2} 
     + \frac{1}{\epsilon} \left( (2\pm 1) + \ln\frac{\mu^2}{\Lambda^2} \right) \right] 
     \right\} ,
\end{equation}
which leads directly to the expression \eqref{eq:gamma_F} for the anomalous dimensions of $F_\mp(\Lambda,\mu)$. Finally, the anomalous dimensions of the "unsubtracted" parameters $F_\mp(\mu)$ are obtained by setting $a_2=0$. This yields the results quoted in \eqref{eq:F_nonrefac_RG}.

\newpage
\section{RG resummation with a running QED coupling}
\label{app:running}

The derivation of the result \eqref{eq:REfinal} relied on the fact that the running of the coupling is neglected, as we have always done in this work. The formalism developed in \cite{Becher:2006nr} allows one to include the effects of the running of $\alpha$ in a systematic way. The main difference is that in this case one cannot choose $s$-dependent matching scales, because the matching conditions $\widetilde{W}_{us}(0)=\widetilde{W}_{us}(s,2\spac e^{-\gamma_E}\spac s^{-1})$ and $\widetilde{W}_{usc}(0)=\widetilde{W}_{us}(s,2\spac e^{-\gamma_E}\spac s^{-1}\spac\frac{m_\ell}{m_B})$ then involve the QED coupling evaluated at $s$-dependent values $\alpha(2\spac e^{-\gamma_E}\spac s^{-1})$ and $\alpha(2\spac e^{-\gamma_E}\spac s^{-1}\spac\frac{m_\ell}{m_B})$, respectively. This obviously complicates the inversion of the Laplace transformation. 

The way out is to use matching scales that are independent of $s$, but still eliminate large logarithms in the matching conditions for typical values $s\sim 1/(2E_{\rm cut})$. To begin, we recast the anomalous dimensions \eqref{eq:5.105} in the form
\begin{equation}
\begin{aligned}    
   \gamma_{us}(s,\mu) 
   &= \gamma_{\rm cusp}(\alpha)\,L_{us} + \gamma_{W_{us}}(\alpha) \,, \\
   \gamma_{usc}(s,\mu) 
   &= - \gamma_{\rm cusp}(\alpha)\,L_{usc} + \gamma_{W_{usc}}(\alpha) \,,
\end{aligned}
\end{equation}
where $\gamma_{\rm cusp}(\alpha)$ is the light-like cusp anomalous dimension of QED, whose explicit form can be derived from the corresponding quantity in QCD \cite{Korchemsky:1987wg,Korchemskaya:1992je} making the replacements $C_F\to Q_\ell^2$, $C_A\to 0$ and $T_F\to 1$. At one-loop order, we have 
\begin{equation}
   \gamma_{\rm cusp}(\alpha) = \gamma_{W_{us}}(\alpha) 
   = \gamma_{W_{usc}}(\alpha) = \frac{Q_\ell^2\spac\alpha}{\pi} \,,
\end{equation}
but in higher orders the three quantities can be different. Next, we introduce the QED $\beta$-function $\beta(\alpha)=d\spac\alpha(\mu)/d\ln\mu$. The perturbative expansion coefficients of the anomalous dimensions and the $\beta$-function are defined as in \eqref{eq:betagammaexp}. The one-loop coefficients are
\begin{equation}
   \gamma_{{\rm cusp},0} = \gamma_{W_{us},0} = \gamma_{W_{usc},0} = 4\spac Q_\ell^2 \,,
    \qquad
   \beta_0 = - \frac43\sum_f N_c^f\spac Q_f^2 \,,
\end{equation}
where the sum extends over all fermion species with mass below the scale $\mu$ at which the coupling is evaluated. Next, we introduce RG functions via the integrals \begin{equation}\label{eq:Sadef}
\begin{aligned}
   S(\nu,\mu) 
   &= - \int\limits_\nu^\mu\!\frac{d\mu'}{\mu'}\,
    \gamma_{\rm cusp}\big(\alpha(\mu')\big)\spac\ln\frac{\mu'}{\nu} 
    = - \int\limits_{\alpha(\nu)}^{\alpha(\mu)}\!d\alpha\,
    \frac{\gamma_{\rm cusp}(\alpha)}{\beta(\alpha)}
    \int\limits_{\alpha(\nu)}^{\alpha}\!\frac{d\alpha'}{\beta(\alpha')} \,, \\
   &= \frac{\gamma_{{\rm cusp},0}}{2\beta_0^2} \left[ \frac{4\pi}{\alpha(\nu)}
    \left( 1 - \frac{1}{r} - \ln r \right)
    + \left( \frac{\gamma_{{\rm cusp},1}}{\gamma_{{\rm cusp},0}}
    - \frac{\beta_1}{\beta_0} \right) \left( 1 - r + \ln r \right)
    + \dots \right] , \\
   a_{\rm cusp}(\nu,\mu) 
   &= - \int\limits_\nu^\mu\!\frac{d\mu'}{\mu'}\,\gamma_{\rm cusp}\big(\alpha(\mu')\big)
   = - \int\limits_{\alpha(\nu)}^{\alpha(\mu)}\!d\alpha\,
    \frac{\gamma_{\rm cusp}(\alpha)}{\beta(\alpha)} 
   = \frac{\gamma_{{\rm cusp},0}}{2\beta_0}\spac\ln r
    + \dots \,,
\end{aligned}
\end{equation}
where $r=\alpha(\mu)/\alpha(\nu)$, and we have given the leading terms in the perturbative evaluation of the two quantities (see the Appendix of \cite{Becher:2006mr}). We also define analogous functions $a_{W_{us}}(\nu,\mu)$ and $a_{W_{usc}}(\nu,\mu)$. 

The general solutions to the RG equations \eqref{eq:5.104} can now be written in the form \cite{Becher:2006nr}
\begin{equation}\label{eq:generalsolutions}
\begin{aligned}    
   \widetilde{W}_{us}(s,\mu)
   &= \exp\Big[ - 2S(\mu_0,\mu) - a_{\gamma_{us}}(\mu_0,\mu) \Big]
    \left( \frac{\mu_0^2\,s^2\spac e^{2\gamma_E}}{4}
    \right)^{-a_{\rm cusp}(\mu_0,\mu)} \widetilde{W}_{us}(s,\mu_0) \,, \\
   \widetilde{W}_{usc}(s,\mu)
   &= \exp\Big[ 2S(\mu_0',\mu) - a_{\gamma_{usc}}(\mu_0',\mu) \Big]
    \left( \frac{\mu_0^{\prime\,2}\,s^2\spac e^{2\gamma_E}\spac m_B^2}%
                {4\spac m_\ell^2}
    \right)^{a_{\rm cusp}(\mu_0',\mu)} \widetilde{W}_{usc}(s,\mu_0') \,.
\end{aligned}
\end{equation}
The fact that $s$ and $E_{\rm cut}$ are Laplace-conjugate variables suggests the choices $\mu_0=2 E_{\rm cut}$ and $\mu_0'=2 E_{\rm cut}\,\frac{m_\ell}{m_B}$ for the two matching scales, such that the initial conditions are free of large logarithms. The functions $\widetilde{W}_{us}(s,\mu_0)$ and $\widetilde{W}_{usc}(s,\mu_0')$ depend on the various scales via the logarithms
\begin{equation}
   L_{us}(\mu_0) = \ln\frac{\mu_0^2\,s^2\spac e^{2\gamma_E}}{4} \,, \qquad
   L_{usc}(\mu_0') 
   = \ln\frac{\mu_0^{\prime\,2}\,s^2\spac e^{2\gamma_E}\spac m_B^2}{4\spac m_\ell^2}  
\end{equation}
and the running couplings $\alpha(\mu_0)$ and $\alpha(\mu_0')$. In generalization of \eqref{eq:Wshortforms} we redefine 
\begin{equation}
   \widetilde{W}_{us}(s,\mu_0)
   \equiv \widetilde{W}_{us}(L_{us}(\mu_0),\mu_0) \,, \qquad
   \widetilde{W}_{usc}(s,\mu_0)
   \equiv \widetilde{W}_{usc}(L_{usc}(\mu_0'),\mu_0') \,.
\end{equation}
We can then recast the general solutions in the form
\begin{equation}
\begin{aligned}    
   \widetilde{W}_{us}(s,\mu)
   &= \exp\Big[ - 2S(\mu_0,\mu) - a_{\gamma_{us}}(\mu_0,\mu) \Big]\,
    \widetilde{W}_{us}(\partial_\eta,\mu_0) 
    \left( \frac{\mu_0^2\,s^2\spac e^{2\gamma_E}}{4}
    \right)^{\eta-a_{\rm cusp}(\mu_0,\mu)} \bigg|_{\eta=0} \,, \\
   \widetilde{W}_{usc}(s,\mu)
   &= \exp\Big[ 2S(\mu_0',\mu) - a_{\gamma_{usc}}(\mu_0',\mu) \Big]\,
    \widetilde{W}_{usc}(\partial_\sigma,\mu_0')
    \left( \frac{\mu_0^{\prime\,2}\,s^2\spac e^{2\gamma_E}\spac m_B^2}%
                {4\spac m_\ell^2}
    \right)^{\sigma+a_{\rm cusp}(\mu_0',\mu)} \bigg|_{\sigma=0} \,,
\end{aligned}
\end{equation}
which offers the advantage that the dependence on the Laplace variable $s$ now takes a simple power-law form. It is thus a simple matter to invert the Laplace transform using \eqref{eq:powerLaplace}. For the scale choices $\mu=2 E_{\rm cut}$ and $\mu_0'=2 E_{\rm cut}\,\frac{m_\ell}{m_B}$, this leads to the final expression
\begin{equation}
\begin{aligned}
   R(E_\mathrm{cut},\mu) 
   &= \exp\Big[ - 2S(\mu_0,\mu) + 2S(\mu_0',\mu) 
    - a_{\gamma_{us}}(\mu_0,\mu) - a_{\gamma_{usc}}(\mu_0',\mu) \Big] \\
   &\quad\times
    \widetilde{W}_{us}(\partial_\eta,\mu_0)\,
    \widetilde{W}_{usc}(\partial_\sigma,\mu_0')\,
    \frac{e^{-2\gamma_E\spac\big( a_{\rm cusp}(\mu_0,\mu_0') - \eta - \sigma \big)}}%
         {\Gamma\big[1+2\spac\big( a_{\rm cusp}(\mu_0,\mu_0') - \eta - \sigma \big)
          \big]} \Bigg|_{\eta=\sigma=0} \,,
\end{aligned}
\end{equation}
where we have used the relation
\begin{equation}
   a_{\rm cusp}(\mu_0,\mu) - a_{\rm cusp}(\mu_0',\mu) 
   = a_{\rm cusp}(\mu_0,\mu_0') \,.
\end{equation}

To make contact with our result \eqref{eq:REfinal}, we neglect the scale dependence of $\alpha$ and evaluate the integrals in \eqref{eq:Sadef} to obtain
\begin{equation}
   S(\nu,\mu) 
   = - \frac{Q_\ell^2\spac\alpha}{8\pi}\,\ln^2\frac{\mu^2}{\nu^2} \,, \qquad
   a_{\rm cusp}(\nu,\mu) 
   = - \frac{Q_\ell^2\spac\alpha}{2\pi}\,\ln\frac{\mu^2}{\nu^2} \,,
\end{equation}
where $a_{W_{us}}(\nu,\mu)$ and $a_{W_{usc}}(\nu,\mu)$ are equal to $a_{\rm cusp}(\nu,\mu)$ at one-loop order. Inserting the explicit values of the scales $\mu_0$ and $\mu_0'$, we then obtain
\begin{equation}
   R(E_\mathrm{cut},\mu) 
   = \left( \frac{\mu^2\spac m_B}{(2 E_{\rm cut})^2\,m_\ell} 
    \right)^{-\gamma_{\rm soft}} 
    \widetilde{W}_{us}(\partial_\eta)\,\widetilde{W}_{usc}(\partial_\sigma)\,
    \frac{ e^{-2\gamma_E\spac\big( 
    \frac{Q_\ell^2\spac\alpha}{2\pi} \ln\frac{m_B^2}{m_\ell^2} - \eta-\sigma \big)} }%
    {\Gamma\Big[1+2\spac\big( \frac{Q_\ell^2\spac\alpha}{2\pi} \ln\frac{m_B^2}{m_\ell^2} 
     - \eta-\sigma \big) \Big]} \Bigg|_{\eta=\sigma=0} \,,
\end{equation}
where the matching conditions no longer depend on the scales. At one-loop order, we have 
\begin{equation}\label{eq:WW1loop}
   \widetilde{W}_{us}(\partial_\eta)\,\widetilde{W}_{usc}(\partial_\sigma)\,
   = 1 + \frac{Q_\ell^2\spac\alpha}{2\pi} \left[ 
    \frac{\partial_\eta^2-\partial_\sigma^2}{2} + \partial_\eta + \partial_\sigma
    + 2 - \frac{\pi^2}{3} \right] + \order{\alpha^2} .
\end{equation}
But we can actually do better. Integrating the RG equations for fixed coupling yields the solutions
\begin{equation}
\begin{aligned}    
   \widetilde{W}_{us}(L_{us}) 
   = \exp\left[ \frac{Q_\ell^2\spac\alpha}{2\pi} \left( \frac{L_{us}^2}{2} + L_{us} \right) \right]
    \exp\left[ - \frac{Q_\ell^2\spac\alpha}{2\pi} \left( \frac{L_{us}^2(\mu_0)}{2} + L_{us}(\mu_0) \right) \right] \widetilde{W}_{us}\big(L_{us}(\mu_0)\big)
    \,,
\end{aligned}
\end{equation}
and similarly for $\widetilde{W}_{usc}(L_{usc})$. The fact that these expressions must be independent of the matching scales $\mu_0$ and $\mu_0'$ implies that the logarithms in the matching conditions must cancel those in the second exponential. We can thus upgrade relation \eqref{eq:WW1loop} to
\begin{equation}
   \widetilde{W}_{us}(\partial_\eta)\,\widetilde{W}_{usc}(\partial_\sigma)\,
   = \exp\left[ \frac{Q_\ell^2\spac\alpha}{2\pi} 
    \left( \frac{\partial_\eta^2-\partial_\sigma^2}{2} 
     + \partial_\eta + \partial_\sigma  \right) \right] 
    \left[ 1 + \frac{Q_\ell^2\spac\alpha}{2\pi} \left( 2 - \frac{\pi^2}{3} \right) 
     + \order{\alpha^2} \right] .
\end{equation}
We now define new variables $\eta_\pm=\eta\pm\sigma$ and derivatives $\partial_\pm=\partial/\partial\eta_\pm$, in which case the derivative operator in round parenthesis takes the form $(2\partial_+\partial_-+2\partial_+)$. Using the fact that the expression on which the differential operators act only involves $\eta_+$, we can set $\partial_-\to 0$ and obtain
\begin{equation}
\begin{aligned}    
   R(E_\mathrm{cut},\mu)  
   &= \left( \frac{\mu^2\spac m_B}{(2 E_{\rm cut})^2\,m_\ell} 
    \right)^{-\gamma_{\rm soft}} 
    \left[ 1 + \frac{Q_\ell^2\spac\alpha}{2\pi} \left( 2 - \frac{\pi^2}{3} \right) 
    + \order{\alpha^2} \right] \\
   &\quad\times \exp\bigg( \frac{Q_\ell^2\spac\alpha}{\pi}\,\partial_+ \bigg)\,
    \frac{ e^{-2\gamma_E\spac\big( 
    \frac{Q_\ell^2\spac\alpha}{2\pi} \ln\frac{m_B^2}{m_\ell^2} - \eta_+ \big)} }%
    {\Gamma\Big[1+2\spac\big( \frac{Q_\ell^2\spac\alpha}{2\pi} \ln\frac{m_B^2}{m_\ell^2} 
     - \eta_+ \big) \Big]} \Bigg|_{\eta_+=0} \,.
\end{aligned}
\end{equation}
The remaining exponential derivative operator has the effect of shifting the value of $\eta_+$ away from zero by an amount $\frac{Q_\ell^2\spac\alpha}{\pi}$, so that one exactly recovers the relation \eqref{eq:REfinal}.

\end{appendix}

\newpage
\bibliographystyle{JHEP}
\bibliography{refs}

\end{document}